\documentclass[a4paper,11pt]{article}
\usepackage[skins]{tcolorbox}
\usepackage{slashed,verbatim,subfigure}
\usepackage[numbers,sort&compress]{natbib}
\usepackage{amsmath}
\usepackage{amssymb}
\usepackage{appendix}
\usepackage{graphicx}
\usepackage{dcolumn}
\usepackage{bm}
\usepackage[colorlinks=true,linktocpage=true,
linkcolor=blue,citecolor=blue]{hyperref}
\usepackage[a4paper]{geometry}
\catcode`@11
\def\seceqaa{\@addtoreset{equation}{section}
	\def\theequation{A\arabic{equation}}}
\def\seceqbb{\@addtoreset{equation}{section}
	\def\theequation{B\arabic{equation}}}
\def\seceqcc{\@addtoreset{equation}{section}
	\def\theequation{C\arabic{equation}}}
\def\seceqdd{\@addtoreset{equation}{section}
	\def\theequation{D\arabic{equation}}}
\def\seceqee{\@addtoreset{equation}{section}
	\def\theequation{E\arabic{equation}}}
\catcode`@11
\begin{document}
\large
\title{\bf{On ${\cal M}$-Theory Dual of Large-$N$ Thermal QCD-Like Theories up to ${\cal O}(R^4)$ and $G$-Structure Classification of Underlying Non-Supersymmetric Geometries}}
\author{Vikas Yadav\footnote{email- vyadav@ph.iitr.ac.in}~~and~~Aalok Misra\footnote{email-    aalok.misra@ph.iitr.ac.in}\vspace{0.1in}\\
Department of Physics,\\
Indian Institute of Technology Roorkee, Roorkee 247667, India}
\date{}
\maketitle
\begin{abstract}
Construction of a top-down holographic dual of   thermal QCD-like theories (equivalence class of theories which are UV-conformal, IR-confining and have fundamental quarks)  {\it at intermediate 't Hooft coupling} and the $G$-structure (torsion classes) classification of the underlying geometries (in the Infra Red (IR)/non-conformal sector in particular) of the {\it non-supersymmetric} string/${\cal M}$-theory duals, have been missing in the literature. We take the first important steps in this direction by studying the ${\cal M}$ theory dual of large-$N$ thermal QCD-like theories  at intermediate gauge and 't Hooft couplings and obtaining the  ${\cal O}(l_p^6)$ corrections arising from the ${\cal O}(R^4)$ terms to the ``MQGP'' background (${\cal M}$-theory dual of large-$N$ thermal QCD-like theories at intermediate gauge/string coupling, but large 't Hooft coupling) of \cite{MQGP}. The main Physics lesson learnt is that there is a competition between non-conformal IR enhancement and Planckian and large-$N$ suppression and going to orders beyond the ${\cal O}(l_p^6)$ is necessitated if the IR enhancement wins out.  The main lesson learnt in Math is in the context of the differential geometry ($G$-structure classification) of the internal manifolds relevant to the string/${\cal M}$-theory duals of large-$N$ thermal QCD-like theories, wherein we obtain for the first time inclusive of the ${\cal O}(R^4)$ corrections in the Infra-Red (IR), the $SU(3)$-structure torsion classes of the type IIA mirror of \cite{metrics} (making contact en route with Siegel theta functions related to appropriate hyperelliptic curves, as well as the Kiepert's algorithm of solving quintics), and the $G_2/SU(4)/Spin(7)$-structure torsion classes of the seven- and eight-folds associated with its ${\cal M}$ theory uplift.

\end{abstract}

\begin{center}
{\it Dedicated by one of the authors (AM) to the memory of his  father, his ``rock".}
\end{center}

\newpage
\tableofcontents

\newpage

\section{Introduction}

The study of the dynamics of non-Abelian gauge theories at  finite temperatures is essential to studying various physical processes, like electroweak and hadronic matter and various other phenomena. No effective theory developed over the years has given a suitable explanation for the intermediate coupling regime. The results for the intermediate coupling regime have been obtained by extrapolating the results obtained via perturbation theory. In recent years, gauge/gravity duality has provided a simple, classical computational tool for understanding the strongly coupled systems and overcome the theoretical limitations in study of  non-Abelian gauge theories. In its simplest form for maximally supersymmetric $SU(N_c)$ Yang-Mills theory (${\cal N} = 4$ SYM), in the $N_c\rightarrow\infty$ limit, the gauge/gravity duality provides a tool for analysing its properties in the large `t Hooft coupling limit.  The gauge/gravity duality also allows us to study the corrections to the infinite coupling limit. These corrections appear as higher order derivative corrections  on the gravity side. The effect of these corrections to the action are incorporated in the background metric and fluxes, perturbatively by considering perturbations of the equations of motion. However, other than higher-derivative corrections quartic in the Weyl tensor, or of the Gauss-Bonnet type, in $AdS_5\times S^5$, dual to supersymmetric thermal Super Yang-Mills \cite{previous-higher-ders}, there is little known about top-down string theory duals at intermediate ’t Hooft coupling of thermal QCD-like theories. In this work, we address precisely this issue. We include  terms quartic in the eleven-dimensional Riemann curvature $R$ in the eleven-dimensional supergravity action that appear as ${\cal O}(l_p^6)( l_p$ being the 11D Planckian length)-corrections in the ${\cal M}$-theory dual of large-$N$ thermal QCD-like theories (equivalence class of theories which are UV-conformal, IR-confining and have fundamental quarks).

\noindent {\it Physics motivation behind this work}: Significance of higher order derivative corrections is not just only related to corrections to the infinite coupling limit. They also serve as the leading quantum gravity corrections to the ${\cal M}$-Theory action to study the compactifications of ${\cal M}$-Theory on compact eight-dimensional manifolds.
The study of warped compactification of ${\cal M}$-Theory on eight-dimensional compact manifolds is very interesting. Conceptually, on one hand this compactification allows for the study of three-dimensional effective theories with small amounts of supersymmetry. On the other hand it allows us to study the lifting of three-dimensional theories to four space-time dimensions for a certain class of eight-dimensional manifolds using ${\cal M}$-Theory to F-theory limit. In the past, vacua for warped compactifications of ${\cal M}$-Theory on compact eight-dimensional manifolds have been studied by including the higher derivative terms to the action. The leading quantum gravity corrections to  ${\cal M}$-Theory actions are fourth order, $R^4$, and third order, $R^3G^2$, in the eleven-dimensional Riemann curvature $R$ , where $G$ is the field strength of ${\cal M}$-Theory three form. The terms of ${\cal O}(R^4)$ have been used in past \cite{O(R^4)}, while terms involving third order have been recently analyzed in \cite{O(R^3G^2)}.   The construction of a top-down holographic dual of thermal Quantum Chromodynamics (QCD) at intermediate 't Hooft coupling has been missing in the literature. This work takes important steps to fill this gap by studying the ${\cal M}$ theory dual of large-$N$ thermal QCD-like theories at high temperatures, at intermediate gauge and 't Hooft couplings by obtaining the ${\cal O}(l_p^6)$ corrections to the ${\cal M}$-Theory uplift of \cite{metrics} as constructed in \cite{MQGP}.

\noindent{\it Mathematics motivation behind this work}: The study of the differential geometry of fluxed compatifications involving non-K\"{a}hler six-folds in Heterotic string theory via the study of $SU(3)$-structure torsion classes, was initiated in \cite{Berlin-torsion-classes}; $SU(3)$ and $G_2$-structure torsions classes of respectively six- and seven-folds in respectively type II and M-theory flux compactifications was extensively studied in  \cite{SYZ-free-delocalization,Dasgupta+Tatar-et-al,theta0-theta,Butti et al [2004]}. $SU(3)$- and $G_2$-structure torsion classes of type IIB/A holographic dual of thermal QCD-like theories  and their ${\cal M}$-theory uplift in the intermediate/large ``$r$" (the radial coordinate in the gravity dual which corrsponds to the energy on the gauge theory side), i.e., Ultra-Violet (UV)-Infra-Red (IR) interpolating region/UV region were obtained in the second reference in \cite{EPJC-2} and \cite{NPB}. In this work, for the first time, we classify the underlying six-, seven- and eight-dimensional geometries at small $r$, i.e., the IR, inclusive of the aforementioned  ${\cal O}(l_p^6)$-corrections in the $D=11$ supergravity action, as regards their $SU(3), G_2, SU(4), Spin(7)$-torsion classes (note these corrections vanish in the very large-$r$ limit, i.e., the deep UV, wherein $G$-structure approaches $G$-holonomy), both in the SYZ type IIA mirror of the type IIB holographic dual constructed in \cite{metrics} of thermal QCD-like theories, as well as its ${\cal M}$-theory uplift.

The following are the main results of this work.

\begin{itemize}
\item
The ${\cal M}$-theory dual of thermal QCD-like theories inclusive of ${\cal O}(l_p^6)$-corrections, was obtained.

\item
{\bf Proposition}:
\begin{enumerate}
\item
The non-K\"{a}hler warped six-fold $M_6$, obtained as a cone over a compact five-fold $M_5$, that appears in the type IIA background corresponding to the  the ${\cal M}$-theory uplift of thermal QCD-like theories at high temperatures, in the neighborhood of the Ouyang embedding (\ref{Ouyang}) of the type IIB flavor $D7$-branes  \cite{ouyang} (that figure in the type IIB string dual of thermal QCD \cite{metrics}) effected by working in the neighborhood of small $\theta_{1,2}$ (such as (\ref{Ouyang-theta10-theta20})), in the MQGP limit (\ref{MQGP_limit}) and {\it inclusive of the ${\cal O}(l_p^6)$ corrections} ($l_p$ being the 11D Planckian length),
\begin{itemize}
\item
is a non-complex manifold (though the deviation from $W_{1,2}^{SU(3)}=0$ being $N$-suppressed),
\item
 $W_4^{SU(3)} \sim W_5^{SU(3)}$ (upon comparison with \cite{Butti et al [2004]},  interpreted as ``almost" supersymmetric [in the large-$N$ limit]).
\end{itemize}
\item
The $G_2$-structure torsion classes of the seven-fold $M_7$ (part of the eleven-fold \\ $M_{11}(x^0,x^{1,2,3},r,\theta_1,\theta_2,\phi_1,\phi_2,\psi,x^{10})$ which is a warped product of $S^1 \times_w \mathbb{R}^3$ and a cone over  ${\cal M}$-theory-$S^1$-fibration over $M_5$: $p_1^2(M_{11}) = p_2(M_{11}) = 0$, $p_a$ being the $a$-th Pontryagin class, and  solves the $D=11$ supergravity equations of motion (\ref{eoms}))  are: $W^{G_2}_{M_7} = W^{G_2}_{14} \oplus W^{G_2}_{27}$.
\item
Inclusive of an $S^1$-valued $x^0$ at finite temperature, referred to henceforth as the thermal circle, the $SU(4)/Spin(7)$-structure torsion classes of $M_8(r,\theta_{1,2},\phi_{1,2},\psi,x^{10},x^0)$ are $W^{SU(4)/Spin(7)}_{M_8} = W_2^{SU(4)} \oplus W_3^{SU(4)} \oplus W_5^{SU(4)}/W_1^{Spin(7)} \oplus W_2^{Spin(7)}$.
\end{enumerate}
\end{itemize}
.

\subsection*{Organization of the Remainder of the Paper}

The remainder of the paper is organized as follows. Section {\bf 2} is a short review of the type IIB string theoretic dual of large-$N$ thermal QCD-like theories  as obtained in \cite{metrics}, as well as its Strominger-Yau-Zaslow type IIA mirror and the ${\cal M}$ theory uplift of the same as constructed in \cite{MQGP}.  Section {\bf 3} begins with a summary of the ${\cal O}(R^4)$ terms in $D=11$ supergravity that are considered in the remainder of the paper. The ${\cal O}(l_p^6) (l_p$ being the $D=11$ Planck length) corrections to the ${\cal M}$-Theory uplift in the ``MQGP" limit as obtained in \cite{MQGP}, near the $\psi=2n\pi, n=0, 1, 2$-branches are consequently obtained in {\bf 3.1} and for $\psi\neq2n\pi, n=0, 1, 2$ in {\bf 3.2}. There are three main lemmas in {\bf 3.1} pertaining to working in the neighborhood of $\psi=2n\pi$-branches. The first is on comparing the large-$N$ behaviors of two ${\cal O}(R^4 l_p^6)$ terms in the $D=11$ supergravity action; the second is on the ${\cal M}$-theory metric inclusive of ${\cal O}(l_p^6)$ corrections, and the third is on the consistency of setting the ${\cal O}(l_p^6)$ corrections to the ${\cal M}$-theory three-form potential, to zero. Subsection {\bf 3.2} has an analogous lemma  working in the neighbhorhood of $\psi\neq2n\pi$-coordinate patch. Section {\bf 4} discusses the  major Physics lessons learnt. Section {\bf 5}  through four sub-sections, discusses the $SU(3)/G_2/SU(4),Spin(7)$-structure torsion classes in {\bf 5.1}/{\bf 5.2}/{\bf 5.3}/{\bf 5.4}. Section {\bf 5} has five main lemmas. The first, in {\bf 5.1}, is on the type IIA metric components along the compact directions. The second, also in {\bf 5.1}, is on the underlying type IIA internal six-fold being non-complex and yet satisfying a relation of \cite{Butti et al [2004]} for supersymmetric compactification. The third, in {\bf 5.2}, is on the evaluation of the $G_2$-structure torsion classes of the relevant seven-fold which is a cone over a six-fold that is itself an ${\cal M}$-theory circle fibration over a compact five-fold. Inclusive of a ``thermal circle", the fourth, in {\bf 5.3}, is on the $SU(4)$-structure torsion classes of the underlying eight-fold. Finally, the fifth is on the evaluation of $Spin(7)$-structure torsion classes of the aforementioned eight-fold. The nine lemmas together imply the proposition stated in Section {\bf 1}. Section {\bf 6} is a summary of the results obtained in the paper and a summary of the applications of the same to Physics as obtained in \cite{Vikas+Gopal+Aalok}, \cite{Gopal+Vikas+Aalok}.   There are four supplementary appendices - a long appendix {\bf A} on the equations of motion for the metric perturbations ($f_{MN}$) and their explicit solutions obtained inclusive of the aforementioned ${\cal O}(R^4)$ terms in the IR, both near the $\psi=2n\pi, n=0, 1, 2$-branches in {\bf A.1} leading up to {\bf 3.1}, and near the $\psi\neq2n\pi, n=0, 1, 2$ coordinate patches in {\bf A.2} leading up to {\bf 3.2}.  Appendix {\bf B} has a step-by-step discussion of the Kiepert's algorithm for diagonalizing the $M_5(\theta_{1,2}, \phi_{1,2}, \psi)$ metric leading to the evaluation of $G$-structure torsion classes for
$M_6(r, \theta_{1,2}, \phi_{1,2}, \psi)$, $M_6(r, \theta_{1,2}, \phi_{1,2}, \psi)\times_w S^1(x^{10})$ and $S^1(x^0)\times_w \left(M_6(r, \theta_{1,2}, \phi_{1,2}, \psi)\times_w S^1(x^{10})\right)$. Appendix {\bf C} lists out the non-trivial ``structure constants" of the algebra of the fufnbeings/sechsbeins  in section {\bf 4}.  Appendix {\bf D} gives some calculational details relevant to showing that one can, up to ${\cal O}(l_p^6)$-corrections, consistently set the corrections at the same order in the ${\cal M}$-theory three-form potential, to zero. Finally appendix {\bf E} gives details of the $G_2$ structure torsion classes $W_{1,7}$.

\section{String/${\cal M}$-Theory Dual of Thermal QCD - A Quick Review of (and Results Related to) \cite{metrics, MQGP}}

In this section, we provide a short review of the UV complete type IIB holographic dual - {\it the only one we are aware of} - of large-$N$ thermal QCD-like theories  constructed in \cite{metrics}, its Strominger-Yau-Zaslow (SYZ) type IIA mirror at intermediate string coupling and its subsequent ${\cal M}$-Theory uplift constructed in \cite{MQGP, NPB}, as well as a summary of results in applications of the same to the study of transport coefficients and glueball-meson phenomenology.

We begin with the UV-complete type IIB holographic dual of large-$N$ thermal QCD-like theories  as constructed in  \cite{metrics} which built up on the zero-temperature Klebanov-Witten model \cite{KW}, the non-conformal Klebanov-Tseytlin model \cite{KT}, its IR completion as given in the Klebanov-Strassler model \cite{KS} and Ouyang's \cite{ouyang} inclusion  of flavor in the same,
as well as the non-zero temperature/non-extremal version of \cite{Buchel} (wherein the non-extremality function and the ten-dimensional warp factor simultaneously vanished  at the horizon radius), \cite{Gubser-et-al-finitetemp} (which was valid only at large temperatures)  and \cite{Leo-i,Leo-ii} (which addressed the IR), in the absence of flavors. The authors of \cite{metrics} considered  $N$ $D3$-branes placed at the tip of a six-dimensional conifold, $M\ D5$-branes wrapping the vanishing $S^2$ and $M\ \overline{D5}$-branes  distributed along the resolved $S^2$ and placed at the anti-podal points relative to the $M$ $D5$-branes. Denoting the average $D5/\overline{D5}$ separation  by ${\cal R}_{D5/\overline{D5}}$, roughly speaking, $r>{\cal R}_{D5/\overline{D5}}$, would correspond to the UV.    The $N_f$ flavor $D7$-branes (holomorphically embedded via Ouyang embedding \cite{ouyang} in the resolved conifold geometry) are present  in the UV, the IR-UV interpolating region and dip into the (confining) IR (without touching the $D3$-branes; the shortest $D3-D7$ string corresponding to the lightest quark). In addition, $N_f\ \overline{D7}$-branes are also  present in the UV and the UV-IR interpolating region but not the IR, for the reason given below. In the UV, there is $SU(N+M)\times SU(N+M)$ color symmetry and $SU(N_f)\times SU(N_f)$ flavor symmetry. As one goes from $r>{\cal R}_{D5/\overline{D5}}$  to $r<{\cal R}_{D5/\overline{D5}}$, there occurs a partial Higgsing of $SU(N+M)\times SU(N+M)$ to $SU(N+M)\times SU(N)$ because in the IR, i.e., at energies less than  ${\cal R}_{D5/\overline{D5}}$, the $\overline{D5}$-branes are integrated out resulting in the reduction of the rank of one of the product gauge groups (which is $SU(N + {\rm number\ of}\ D5-{\rm branes})\times SU(N + {\rm number\ of}\ \overline{D5}-{\rm branes})$). Similarly, the $\overline{D5}$-branes are ``integrated in" in the UV, resulting in the conformal Klebanov-Witten-like $SU(M+N)\times SU(M+N)$ product color gauge group \cite{KW}.  The gauge couplings, $g_{SU(N+M)}$ and $g_{SU(N)}$, were shown in \cite{KS} to flow  oppositely with the flux of the NS-NS $B$ through the vanishing $S^2$ being the obstruction to obtaining conformality which is why $M$ $\overline{D5}$-branes were included in \cite{metrics} to cancel the net $D5$-brane charge in the UV. Also, as the number $N_f$ of the flavor $D7$-branes enters the RG flow of the gauge couplings via the dilaton, their contribution therefore needs to be canceled by $N_f\ \overline{D7}$-branes. The RG flow equations for the gauge coupling $g_{SU(N+M)}$ - corresponding to the relatively higher rank gauge group - can be used to show that the same flows towards strong coupling, and the relatively lower rank $SU(N)$ gauge coupling flows towards weak coupling. One can show that the strongly coupled $SU(N+M)$ is Seiberg-like dual to weakly coupled  $SU(N-(M - N_f))$.  Under a Seiberg-like duality cascade\footnote{Even though the Seiberg duality (cascade) is applicable for supersymmetric theories, for non-supersymmetric theories such as the holographic type IIB string theory dual of \cite{metrics}, the same is effected via a radial rescaling: $r\rightarrow e^{-\frac{2\pi}{3g_sM_{\rm eff}}}r$ \cite{ouyang} under an RG flow from the UV to the IR.} all the $N\ D3$-branes are cascaded away with a finite $M$ left at the end in the IR. One will thus be left with a strongly coupled IR-confining $SU(M)$ gauge theory the finite temperature version of which is what was looked at in \cite{metrics}. So, at the end of the Seiberg-like duality cascade in the IR, the number of colors $N_c$ gets identified with $M$, which in the `MQGP' limit can be tuned to equal the value in QCD, i.e., 3. Now, $N_c$ can be written as the sum of the effective number $N_{\rm eff}(r)$ of $D3$-branes and the effective number $M_{\rm eff}$ of the fractional $D3$-branes:  $N_c = N_{\rm eff}(r) + M_{\rm eff}(r)$; $N_{\rm eff}(r)$ is defined via $\tilde{F}_5\equiv dC_4 + B_2\wedge F_3 = {\cal F}_5 + *{\cal F}_5$ where ${\cal F}_5\equiv N_{\rm eff}{\rm Vol}({\rm Base\ of\ Resolved\ Warped\ Deformed\ Conifold})$, and $M_{\rm eff}$ is defined via $M_{\rm eff} = \int_{S^3}\tilde{F}_3 (= F_3 - \tau H_3)$ (the $S^3$ being dual to $\ e_\psi\wedge\left(\sin\theta_1 d\theta_1\wedge d\phi_1 - B_1\sin\theta_2\wedge d\phi_2\right)$, wherein $B_1$ is an asymmetry factor defined in \cite{metrics}, and $e_\psi\equiv d\psi + \cos ~\theta_1~d\phi_1 + \cos ~\theta_2~d\phi_2$).  
(See \cite{metrics,M(r)N_f(r)-Dasgupta_et_al} for details.).  The finite temperature on the gauge/brane side is effected in \cite{metrics} in the gravitational dual via a black hole in the latter. Turning on of the temperature (in addition to requiring a finite separation between the $M\ D5$-branes and $M\ \overline{D5}$-branes so as to provide a natural energy scale to demarcate the UV) corresponds in the gravitational dual to having a non-trivial resolution parameter of the conifold. IR confinement on the brane/gauge theory side, like the KS model \cite{KS}, corresponds to having a non-trivial deformation (in addition to the aforementioned resolution) of the conifold geometry in the gravitational dual.  The gravity dual is hence given by a  resolved warped deformed conifold wherein the $D3$-branes and the $D5$-branes are replaced by fluxes in the IR, and the back-reactions are included in the 10D warp factor as well as fluxes.  Hence, the type IIB model of \cite{metrics} is an ideal holographic dual of thermal QCD-like theories  because: (i) it is UV conformal (with the Landau poles being absent), (ii) it is IR confining, (iii) the quarks transform in the fundamental representation of flavor and color groups, and (iv) it  is defined for the entire range of temperature - both low (i.e., $T<T_c$ corresponding to a vanishing horizon radius in the gravitational dual ) and high (i.e., $T>T_c$ corresponding to non-vanishing horizon radius in the gravitational dual).


Now, we give a brief review the type IIA Stominger-Yau-Zaslow (SYZ) mirror \cite{syz} of \cite{metrics} and its ${\cal M}$-Theory uplift at intermediate gauge coupling, as constructed in \cite{MQGP}. Now, to construct a holographic dual of thermal QCD-like theories, one would have to consider intermediate gauge coupling (as well as finite number of colors) $-$ dubbed as the `MQGP limit' defined in \cite{MQGP} as follows:
\begin{equation}
\label{MQGP_limit}
g_s\stackrel{<}{\sim}1, M, N_f \equiv {\cal O}(1),\ N \gg1,\ \frac{g_s M^2}{N}\ll1.
\end{equation}
 From the perspective of gauge-gravity duality, this therefore requires looking at the strong-coupling/non-perturbative limit of string theory - ${\cal M}$ theory. 

The ${\cal M}$-Theory uplift of the type IIB holographic dual of \cite{metrics} was constructed in \cite{MQGP} by working out the SYZ type IIA mirror of \cite{metrics} implemented via a triple T duality along a delocalized special Lagrangian (sLag) $T^3$ $-$ which could be identified with the $T^2$-invariant sLag of \cite{M.Ionel and M.Min-OO (2008)} with a large base ${\cal B}(r,\theta_1,\theta_2)$ \cite{NPB,EPJC-2} \footnote{Consider $D5$-branes wrapping the resolved $S^2$ of a resolved conifold geometry as in \cite{Zayas-Tseytlin}, which, globally, breaks  SUSY \cite{Franche-thesis}.  As in \cite{SYZ-free-delocalization}, to begin with, a delocalized SYZ mirror is constructed wherein the pair of $S^2$s are replaced by a pair of $T^2$s, and the correct T-duality coordinates are identified. Then,  when uplifting the mirror to ${\cal M}$ theory, it was found  that a $G_2$-structure  can be chosen that is in fact, free, of the aforementioned delocalization. For the delocalized SYZ mirror of the resolved warped deformed conifold uplifted to ${\cal M}$-Theory with $G_2$  in \cite{MQGP}, the idea is precisely the same. Also, as shown in the second reference of \cite{EPJC-2} and \cite{NPB}, the type IIB/IIA $SU(3)$ structure torsion classes in the MQGP limit and in the UV/UV-IR interpolating region (and as will be shown in {\bf Sec. 4} of this paper, also in the IR and inclusive of ${\cal O}(l_p^6)$ corrections), satisfy the same relationships as satisfied by corresponding supersymmetric conifold geometries \cite{Butti et al [2004]}.} Let us explain the basic idea. Consider the aforementioned $N$ D3-branes  oriented along $x^{0, 1, 2, 3}$ at the tip of conifold and the $M\ D5$-branes parallel to these $D3$-branes as well as wrapping the vanishing $S^2(\theta_1,\phi_1)$. A single T-dual along $\psi$  yields $N\ D4$-branes wrapping the $\psi$ circle and $M\ D4$-branes straddling a pair of orthogonal $NS5$-branes. This pair of $NS5$-branes
correspond to the vanishing $S^2(\theta_1,\phi_1)$ and the blown-up $S^2(\theta_2,\phi_2)$ with a non-zero resolution parameter $a$ - the radius of the blown-up $S^2(\theta_2,\phi_2)$. Now, two more T-dualities along $\phi_1$ and $\phi_2$, convert the aforementioned pair of orthogonal $NS5$-branes into a pair of orthogonal Taub-NUT spaces, the $N\ D4$-branes into $N$ color $D6$-branes and the $M$ straddling $D4$-branes also to color $D6$-branes. Similarly, in the presence of the aforementioned $N_f$ flavor $D7$-branes, oriented parallel to the $D3$-branes and ``wrapping" a non-compact four-cycle $\Sigma^{(4)}(r, \psi, \theta_1, \phi_1$), upon T-dualization yield $N_f$ flavor $D6$-branes ``wrapping" a non-compact three-cycle  $\Sigma^{(3)}(r, \theta_1, \phi_2$). An uplift to ${\cal M}$-Theory will convert  the $D6$-branes to KK monopoles, which are variants of Taub-NUT spaces.  All the branes are hence converted into geometry and fluxes, and one  ends up with ${\cal M}$-Theory on a $G_2$-structure manifold. Similarly, one may perform identical three T-dualities on the gravity dual on the type IIB side, which is a resolved warped-deformed conifold with fluxes,  to obtain another $G_2$ structure manifold, yielding the ${\cal M}$-Theory uplift of  \cite{MQGP,NPB}.


To the best of our knowledge, \cite{MQGP} is the only  holographic ${\cal M}$-Theory dual of thermal QCD that is able to:
\begin{itemize}
\item
 yield a deconfinement temperature $T_c$ from a Hawking-Page phase transition at vanishing baryon chemical potential consistent with the very recent lattice QCD results in the heavy quark \cite{Misra+Gale_Conformal_Anomaly}
limit

\item
yield a conformal anomaly variation with temperature compatible with the very recent lattice results at  high ($T>T_c$) {\it and} low ($T<T_c$) temperatures \cite{Misra+Gale_Conformal_Anomaly}

\item
Condensed Matter Physics: inclusive of  the non-conformal corrections, obtain:
\begin{enumerate}
\item
 a lattice-compatible shear-viscosity-to-entropy-density ratio (first reference in \cite{EPJC-2})

\item
temperature variation of a variety of transport coefficients including the bulk-viscosity-to-shear-viscosity ratio,  diffusion coefficient, speed of sound (the last reference in \cite{EPJC-2}), electrical and thermal conductivity and the Wiedemann-Franz law (first reference in \cite{EPJC-2});
\end{enumerate}

\item
Particle Phenomenology:  obtain:
\begin{enumerate}
\item
 lattice compatible glueball spectroscopy \cite{Sil+Yadav+Misra-glueball}

\item
meson spectroscopy (first reference of \cite{mesons_0E++-to-mesons-decays})

\item
glueball-to-meson decay widths (second reference of \cite{mesons_0E++-to-mesons-decays})
\end{enumerate}

\item
Mathematics:  provide, for the first time,  an $SU(3)$-structure (for type IIB (second reference of \cite{EPJC-2})/IIA \cite{NPB}   holographic dual) and $G_2$-structure  \cite{NPB}  torsion classes of the six- and seven-folds in the UV-IR interpolating region/UV, relevant to type string/${\cal M}$-Theory holographic duals of thermal QCD-like theories at high temperatures.

\end{itemize}

\section{${\cal O}(l_p^6)$ Corrections to the Background of \cite{MQGP} in the MQGP Limit}

In this section, we discuss how the equations of motion (EOMs) starting from $D=11$ supergravity action inclusive of the ${\cal O}(R^4)$ terms in the same (which provide the ${\cal O}(l_p^6)$ corrections to the leading order terms in the action), are obtained and how the same are solved. The actual EOMs are given in Appendix {\bf A} - EOMs in {\bf A.1} obtained near the $\psi=2n\pi, n=0, 1, 2$ coordinate patches (wherein $G^{\cal M}_{rM}, M\neq r$  and $G^{\cal M}_{x^{10}N}, N\neq x^{10}$ vanish) and EOMs in {\bf A.2} obtained away from the same. The solutions of the EOMs  are similarly split across subsections {\bf 3.1} and {\bf 3.2}.

Let us begin with a discussion on the ${\cal O}(R^4)$-corrections to the ${\cal N}=1, D=11$ supergravity action. There are two ways of understanding the origin of these corrections. One is in the context of the effects of $D$-instantons  in IIB supergravity/string theory via the four-graviton scattering amplitude \cite{Green and Gutperle}. The other is $D=10$ supersymmetry \cite{Green and Vanhove}. Let us discuss both in some detail.

\begin{itemize}
\item
Let us first look at interactions that are induced at leading order in an instanton background   in both, the supergravity and the string descriptions, including a one-instanton correction to the tree-level  and one loop $R^4$ terms \cite{Green and Gutperle}. The bosonic zero modes are parameterised by the coordinates corresponding to the position of the $D$-instanton. The fermionic zero modes are generated by the broken supersymmetries.  The physical closed-string states can be expressed in terms of a light-cone scalar superfield
$\Phi(x,\theta)$, $\theta^{a(=1,...,8)}$ being an ${\bf 8}_s$ $SO(8)$ spinor. The 16-component (indexed by $A$) broken supersymmetry chiral spinor can be decomposed under $SO(8)$ into $\eta^a, \dot{\eta}^{\dot{a}}$. The Grassmann parameters
are fermionic  supermoduli  corresponding to zero modes of $\lambda$ - the dilatino - and
must be integrated over together with the bosonic zero modes,
$y^\mu$. The simplest open-string world-sheet that arises in a D-brane
process is the disk diagram.  An instanton carrying some zero  modes corresponds,  at lowest
order, to a disk world-sheet with open-string states attached to the boundary.  An instanton carrying some zero  modes corresponds,  at lowest order, to a disk
world-sheet with open-string states attached to the boundary. The one-instanton terms in the supergravity effective action  can be deduced by considering on-shell amplitudes in the instanton background.      The
integration over the fermionic moduli absorbs  the sixteen  independent
fermionic zero modes. The authors of \cite{Green and Gutperle}  considered a contact term proportional to
$\lambda^{16}$ arising  in IIB supergravity from the nonlocal Green function which at  long distances looks like  a  momentum-independent term in the S-matrix with sixteen
external on-shell dilatinos:
 \begin{equation*}
e^{2\pi i  W_1} \epsilon^{A_1\cdots A_{16}} \lambda^{A_1}\dots
\lambda^{A_{16}}.
\end{equation*}
The same result is also
obtained in string theory from diagrams with sixteen disconnected
disks with  a single  dilatino vertex operator  and a single open-string fermion state
attached to each one.   The overall factor of $e^{2\pi i  W_1}$,
which  is characteristic of the stringy D-instanton , is evaluated at $\chi =\Re W_1 =0$ in the string calculation. Consider now amplitudes with four external gravitons. The leading
term  in supergravity is again one in which each graviton is associated with
four fermionic zero modes. Integration over $y^\mu$  generates a nonlocal four-graviton interaction. In the corresponding string calculation  the world-sheet  consists of four disconnected disks to each of which is  attached a single closed-string graviton vertex and four fermionic open-string vertices. Writing the polarization tensor as $\zeta^{\mu_r\nu_r} =
\zeta^{(\mu_r} \tilde \zeta^{\nu_r)}$, and evaluating the fermionic integrals  in a
special frame as described in \cite{Green and Gutperle}, in terms of its $SO(8)$ components, one obtains the following result for $\langle h\rangle_4$  \footnote{In both, string theory and supergravity, the four-graviton scattering  result is given as an integral of the product of  four factors of ``$\langle h\rangle_4$"
defined as the tadpole associated with the disk with four fermion zero modes coupled to the graviton and the self dual fourth-rank antisymmetric tensor:
\begin{eqnarray}
\langle h \rangle_4 &=&    \bar\epsilon_0\gamma^{\rho\mu\tau }\epsilon_0 \;
 \bar\epsilon_0\gamma^{\lambda\nu\tau }\epsilon_0 \; \zeta_{\mu\nu}k_\rho
 k_\lambda ,
 \label{gravitonamp}
\end{eqnarray}
$\epsilon_0$ corresponding to the broken supersymmetry - the only covariant combination of four $\epsilon_0$'s, two physical momenta and the physical polarization tensor.} :
\begin{eqnarray}
\langle h\rangle_4 &=&
-{1\over 2}\eta_{a}\gamma^{ij}_{ab}\eta_{b} \;\dot
\eta_{\dot{a}}\gamma^{mn}_{\dot{a}\dot{b}}\dot \eta_{\dot{b}}  R_{ijmn}
\label{fourferms}
\end{eqnarray}
where: $\gamma^{ij}_{ab} = \frac{1}{2}\gamma^{[i}_{a\dot{a}}\gamma^{j]}_{\dot{a}b}, \gamma^i$ being the generators of $Cl_8$ Clifford algebra and $R_{ijmn} \equiv k_ik_m\zeta_{(j}\tilde \zeta_{n)}$ is the linearized curvature. The result  contains two parity-conserving
 terms \footnote{The integral over the dotted and undotted spinors in the four-graviton scattering amplitude
factorizes and can be evaluated by using,
\begin{equation}
  \int d^8\eta^a \eta^{a_1}\cdots\eta^{a_8}=\epsilon^{a_1
  \cdots a_8}\;,\quad\qquad \;\int d^8\dot \eta^{\dot{a}}
\dot \eta^{\dot{a}_1}\cdots\dot \eta^{\dot{a}_8}=\epsilon^{\dot{a}_1
  \cdots \dot{a}_8}.
\end{equation}
Substituting into the four-graviton scattering amplitude the following tensors appear
\begin{eqnarray}
  \epsilon_{a_1a_2\cdots a_8}\gamma^{i_1j_1}_{a_1a_2}\cdots
  \gamma^{i_4j_4}_{a_7a_8}&=&t^{i_1j_1\cdots
  i_4j_4}={t}_8^{i_1j_1\cdots i_4j_4}+{1\over2}\epsilon^{i_1j_1\cdots
i_4j_4}\nonumber\\
  \epsilon_{\dot{a}_1\dot{a}_2\cdots
\dot{a}_8}\gamma^{i_1j_1}_{\dot{a}_1\dot{a}_2}\cdots
  \gamma^{i_4j_4}_{\dot{a}_7\dot{a}_8}&=&t_8^{i_1j_1\cdots
i_4j_4}={t}^{i_1j_1\cdots i_4j_4}-{1\over2}\epsilon^{i_1j_1\cdots
i_4j_4},
\end{eqnarray}
$t_8$ symbol defined in (\ref{t_8}).}
\begin{eqnarray}
A_4(\{\zeta^{(r)}_h\}) &=  & C  e^{2i\pi  W_1}   \int d^{10}y
e^{i \sum_r
k_r\cdot y}  \nonumber\\
&&\times\left({t}^{i_1j_1\cdots i_4j_4}{t}_{m_1n_1\cdots
  m_4n_4}-{1\over 4}\epsilon^{i_1j_1\cdots i_4j_4}\epsilon_{m_1n_1\cdots
  m_4n_4}\right)  R_{i_1j_1}^{m_1n_1}
R_{i_2j_2}^{m_2n_2}R_{i_3j_3}^{m_3n_3}
R_{i_4j_4}^{m_4n_4} .\nonumber\\
& &
 \end{eqnarray}

\item
In \cite{Green and Vanhove}, it is shown that the eleven-dimensional ${\cal O}(R^4)$ corrections have an independent motivation based on supersymmetry in ten dimensions.
This was shown to follow from its relation to the term  $C^{(3)}\wedge X_8$ in the ${\cal M}$-theory
effective action which is known to arise from a variety of
arguments, e.g. anomaly cancellation \cite{Horava and Witten}. The expression
$X_8$ is the eight-form in the curvatures that is inherited from the term in type IIA
superstring theory \cite{Vafa and Witten} which is given by
\begin{equation}-   \int d^{10}x  B\wedge  X_8 =  -
{1\over 2}
\int  d^{10}x \sqrt{-g^{A(10)}}\epsilon_{10} B X_8,
\end{equation}
where
\begin{equation}
\label{X_8-def}
X_8 = {1 \over 192} \left( {\rm tr}\ R^4 -
{1\over 4} ({\rm tr}\ R^2)^2\right).
\end{equation}

There are two independent ten-dimensional $N=1$ super-invariants which contain an odd-parity term
(\cite{Tseytlin} and previous authors):
\begin{equation}
I_3= t_8 {\rm tr}\ R^4 - {1\over 4} \epsilon_{10}B {\rm tr}\ R^4
\end{equation}
 and:
\begin{equation}
I_4= t_8 ({\rm tr}\ R^2)^2 - {1\over 4} \epsilon_{10}B ({\rm tr}\ R^2)^2.
\end{equation}
 Using the fact that
\begin{equation}
t_8t_8R^4=24t_8{\rm tr}(R^4)-6t_8({\rm tr}\ R^2)^2,
\end{equation}
 it follows that  the particular linear combination,
\begin{equation}
\label{t_8^2R^4}
I_3 - {1\over 4}I_4 = {1\over 24} t_8 t_8 R^4 - 48 \epsilon_{10}B\ X_8
\end{equation}
contains both the ten-form $B\wedge X_8$ and $t_8 t_8 R^4$. The $R$ refers to the curvature two-form, $\epsilon_{10}$ is the ten-dimensional Levi-Civita symbol and the $t_8$ symbol is defined as follows:
{\footnotesize
\begin{eqnarray}
\label{t_8}
t_8^{N_1\dots N_8}   &=& \frac{1}{16} \big( -  2 \left(   G^{ N_1 N_3  }G^{  N_2  N_4  }G^{ N_5   N_7  }G^{ N_6 N_8  }
 + G^{ N_1 N_5  }G^{ N_2 N_6  }G^{ N_3   N_7  }G^{  N_4   N_8   }
 +  G^{ N_1 N_7  }G^{ N_2 N_8  }G^{ N_3   N_5  }G^{  N_4 N_6   }  \right) \nonumber \\
 & &  +
 8 \left(  G^{  N_2     N_3   }G^{ N_4    N_5  }G^{ N_6    N_7  }G^{ N_8   N_1   }
  +G^{  N_2     N_5   }G^{ N_6    N_3  }G^{ N_4    N_7  }G^{ N_8   N_1   }
  +   G^{  N_2     N_5   }G^{ N_6    N_7  }G^{ N_8    N_3  }G^{ N_4  N_1   }
\right) \nonumber \\
& &  - (N_1 \leftrightarrow  N_2) -( N_3 \leftrightarrow  N_4) - (N_5 \leftrightarrow  N_6) - (N_7 \leftrightarrow  N_8) \big),
\end{eqnarray}
}
wherein $G^{M_1M_2}$ is the metric inverse.

\end{itemize}

The ${\cal N}=1, D=11$ supergravity action inclusive of ${\cal O}(l_p^6)$ terms, is hence given by:
\begin{eqnarray}
\label{D=11_O(l_p^6)}
& & \hskip -0.5in {\cal S}_{D=11} = \frac{1}{2\kappa_{11}^2}\Biggl[\int_{M_{11}}\sqrt{G}R + \int_{\partial M_{11}}\sqrt{h}K -\frac{1}{2}\int_{M_{11}}\sqrt{G}G_4^2
-\frac{1}{6}\int_{M_{11}}C_3\wedge G_4\wedge G_4\nonumber\\
& & \hskip -0.5in  + \frac{\left(4\pi\kappa_{11}^2\right)^{\frac{2}{3}}}{{(2\pi)}^4 3^2.2^{13}}\Biggl(\int_{\cal{M}} d^{11}\!x \sqrt{G^{\cal M}}\left(J_0-\frac{1}{2}E_8\right) + 3^2.2^{13}\int C_3 \wedge X_8 + \int t_8 t_8 G^2 R^3 + \cdot\Biggr)\Biggr] - {\cal S}^{\rm ct},
\end{eqnarray}
where:
\begin{eqnarray}
\label{J0+E8-definitions}
& & \hskip -0.8in J_0  =3\cdot 2^8 (R^{HMNK}R_{PMNQ}{R_H}^{RSP}{R^Q}_{RSK}+
{1\over 2} R^{HKMN}R_{PQMN}{R_H}^{RSP}{R^Q}_{RSK})\nonumber\\
& & \hskip -0.8inE_8  ={ 1\over 3!} \epsilon^{ABCM_1 N_1 \dots M_4 N_4}
\epsilon_{ABCM_1' N_1' \dots M_4' N_4' }{R^{M_1'N_1'}}_{M_1 N_1} \dots
{R^{M_4' N_4'}}_{M_4 N_4},\nonumber\\
& & \hskip -0.8in t_8t_8G^2R^3 = t_8^{M_1...M_8}t^8_{N_1....N_8}G_{M_1}\ ^{N_1 PQ}G_{M_2}\ ^{N_2}_{\ \ PQ}R_{M_3M_4}^{\ \ \ \ N_3N_4}R_{M_5M_6}^{\ \ \ \ N_5N_6}R_{M_7M_8}^{\ \ \ \ N_7N_8};\nonumber\\
& & \hskip -0.8in\kappa_{11}^2 = \frac{(2\pi)^8 l_p^{9}}{2};
\end{eqnarray}
$\kappa_{11}^2$ being related to the eleven-dimensional Newtonian coupling constant, and $G=dC$ with $C$ being the ${\cal M}$-theory three-form potential with the four-form $G$ being the associated four-form field strength.

In the spirit of completion of the 1-loop ${\cal O}(R^4)$ in the presence of NS-NS $B$ in type IIA compatible with T duality, and hence defining the torsionful spin connection, $\Omega_\pm\equiv\Omega\pm\frac{1}{2}{\cal H}, {\cal H}^{ab} = {\cal H}^{ab}_\mu dx^\mu$, and $\overline{X_8} \equiv \frac{X_8\left(R(\Omega_+)\right) + X_8\left(R(\Omega_-)\right)}{2}$, where $R(\Omega_+) = R(\Omega) + \frac{1}{2}d{\cal H} + \frac{1}{4}{\cal H}\wedge{\cal H}$, the ten dimensional $\overline{X}_8$ shifts by an exact form \cite{O(R^3G^2)} \footnote{To be consistent with the notation of the rest of the paper, we have dropped the $\hat{}$ over eleven-dimensional objects in (\ref{BX8-CX8}); when wedged with $C$ it will be understood that the objects like the metric, curvature, etc. are eleven-dimensional and when wedged with $B$, ten dimensional.}:
\begin{eqnarray}
\label{eq:x8shift}
\overline{X}_{8} &=& \frac{1}{ 192 (2\pi)^4}\bigg[  \left( {\rm tr}R^4 -\frac{1}{4}({\rm tr}R^2)^2 \right) + \nonumber \\
&&\quad d \,\, \bigg( \frac{1}{2} {\rm tr} \left( {\cal H} \nabla {\cal H} R^2 + {\cal H} R \nabla {\cal H} R
+ {\cal H} R^2 \nabla {\cal H} \right) -\frac{1}{8} \left( {\rm tr}R^2 \, {\rm tr} {\cal H} \nabla {\cal H}
+ 2 {\rm tr} {\cal H} R \, {\rm tr} R \nabla {\cal H} \right)    \nonumber \\
&&\qquad \frac{1}{16 }  {\rm tr} \left( 2 {\cal H}^3 (\nabla {\cal H} R + R \nabla {\cal H}) + {\cal H} R {\cal H}^2 \nabla {\cal H}
+ {\cal H} \nabla {\cal H} {\cal H}^2 R \right)\nonumber\\
&& \qquad - \frac{1}{2}    \left( {\rm tr} {\cal H} \nabla {\cal H} \,  {\rm tr} R {\cal H}^2 +  {\rm tr} R \nabla {\cal H} \, {\rm tr} {\cal H}^3
+ {\rm tr} \nabla {\cal H} {\cal H}^2 \, {\rm tr} {\cal H} R \right)    + \nonumber \\
&&\qquad \frac{1}{32 } {\rm tr} \nabla {\cal H} {\cal H}^5   -\frac{1}{192}  {\rm tr} \nabla {\cal H} {\cal H}^2 \, {\rm tr} {\cal H}^3  + 
 \frac{1}{16 }  {\rm tr} {\cal H} (\nabla {\cal H})^3  -\frac{1}{64} {\rm tr} {\cal H} \nabla {\cal H} \, {\rm tr} (\nabla {\cal H})^2  \bigg) \bigg].
\end{eqnarray}
Defining the $O(1,10)$-valued one-form ${\cal G}^{abc} \equiv 4 G_{\mu\nu\rho\lambda}dx^\mu e^{a\nu} e^{b\rho}e^{c\lambda}$, the ${\cal M}$-theory uplift of the first two lines of  (\ref{eq:x8shift}) of type IIA, yields \cite{O(R^3G^2)} \footnote{Strictly speaking, (\ref{BX8-CX8}) is valid when 
$M_{11}$ is a trivial $S^1$ fibration over an $M_{10}$ and $G_{\mu\nu\rho x^{10}}\neq0, G_{\mu\nu\rho\lambda}=0$. We, near the $\psi=2n\pi$-coordinate patches, have $M_{11}$ as a warped product of the ${\cal M}$-theory circle and $M_{10}$, which for a delocalized (IR-valued in this paper) value of $r$ can be thought of as a trivial circle fibration. The $G_{\mu\nu\rho\lambda}$ arising from $A^{\rm IIA}\wedge H^{\rm IIA}$, via $\int G\wedge *G$, results in a UV-divergent contribution which is canceled off by an appropriate boundary flux term \cite{Glueball-Roorkee}.}:
\begin{eqnarray}
\label{BX8-CX8}
B_2\wedge\overline X_8&\longrightarrow&\frac{1}{192(2\pi)^4}\Bigl[
 C\wedge\left({\rm tr} R^4-\frac{1}{4}({\rm tr} R^2)^2\right)\nonumber\\
&&+ G\wedge\Bigl(\frac{1}{4}\left(
 R^{ab} R^{bc}{\cal G}^{cde}\nabla{\cal G}^{dae}
+2 R^{ab}{\cal G}^{bce} R^{cd}\nabla{\cal G}^{dae}
+ R^{ab} R^{bc}\nabla{\cal G}^{cde}{\cal G}^{dae}\right)\nonumber\\
&&-\frac{1}{24}\left({\rm tr} R^2\wedge{\cal G}^{abe}\nabla{\cal G}^{bae}
+6 R^{ab}{\cal G}^{bae} R^{cd}{\cal G}^{dce}\right)+\cdots\Bigr)\Bigr].
\label{eq:CPolift}
\end{eqnarray}
In this paper, we restrict ourselves only to the first line in (\ref{BX8-CX8}). Given that the same was shown to vanish \cite{MQGP}, perhaps to be T-duality invariant, the sum of the terms in the second and third lines of (\ref{BX8-CX8}) too yield zero. We have not proven the same.

 The action in (\ref{D=11_O(l_p^6)}) is holographically renormalizable by construction of appropriate counter terms ${\cal S}^{\rm ct}$. This is seen as follows. It can be shown \cite{Gopal+Vikas+Aalok} that  the bulk on-shell $D=11$ supergravity action inclusive of ${\cal O}(R^4)$-corrections is given by:
\begin{equation}
\label{on-shell-D=11-action-up-to-beta}
\hskip -0.3in S_{D=11}^{\rm on-shell} = -\frac{1}{2}\Biggl[-2S_{\rm EH}^{(0)} + 2 S_{\rm GHY}^{(0)}+ \beta\left(\frac{20}{11}S_{\rm EH} - 2\int_{M_{11}}\sqrt{-g^{(1)}}R^{(0)}
+ 2 S_{\rm GHY} - \frac{2}{11}\int_{M_{11}}\sqrt{-g^{(0)}}g_{(0)}^{MN}\frac{\delta J_0}{\delta g_{(0)}^{MN}}\right)\Biggr].
\end{equation}
The UV divergences of the various terms in (\ref{on-shell-D=11-action-up-to-beta}) are summarized below:
\begin{eqnarray}
\label{UV_divergences}
& &\left. \int_{M_{11}}\sqrt{-g}R\right|_{\rm UV-divergent},\ \left.\int_{\partial M_{11}}\sqrt{-h}K\right|_{\rm UV-divergent} \sim r_{\rm UV}^4 \log r_{\rm UV},\nonumber\\
& & \left.\int_{M_{11}} \sqrt{-g}g^{MN}\frac{\delta J_0}{\delta g^{MN}}\right|_{\rm UV-divergent} \sim
\frac{r_{\rm UV}^4}{\log r_{\rm UV}}.
\end{eqnarray}
It can be shown \cite{Gopal+Vikas+Aalok} that an appropriate linear combination of the boundary  terms: $\left.\int_{\partial M_{11}}\sqrt{-h}K\right|_{r=r_{\rm UV}}$ and $\left.\int_{\partial M_{11}}\sqrt{-h}h^{mn}\frac{\partial J_0}{\partial h^{mn}}\right|_{r=r_{\rm UV}}$ serves as the appropriate counter terms to cancel the UV divergences (\ref{UV_divergences}) \footnote{For consistency, one needs to impose the following relationship between the UV-valued effective number of flavor $D7$-branes of the parent type IIB dual, $N_f^{\rm UV}$ and $\log r_{\rm UV}$: $N_f^{\rm UV} = \frac{\left(\log r_{\rm UV}\right)^{\frac{15}{2}}}{\log N}$.}.

The EOMS are:
\begin{eqnarray}
\label{eoms}
& & R_{MN} - \frac{1}{2}g_{MN}{\cal R} - \frac{1}{12}\left(G_{MPQR}G_N^{\ PQR} - \frac{g_{MN}}{8}G_{PQRS}G^{PQRS} \right)\nonumber\\
 & &  = - \beta\left[\frac{g_{MN}}{2}\left( J_0 - \frac{1}{2}E_8\right) + \frac{\delta}{\delta g^{MN}}\left( J_0 - \frac{1}{2}E_8\right)\right],\nonumber\\
& & d*G = \frac{1}{2} G\wedge G +3^22^{13} \left(2\pi\right)^{4}\beta X_8,\nonumber\\
& &
\end{eqnarray}
where \cite{Becker-sisters-O(R^4)}:
\begin{equation}
\label{beta-def}
\beta \equiv \frac{\left(2\pi^2\right)^{\frac{1}{3}}\left(\kappa_{11}^2\right)^{\frac{2}{3}}}{\left(2\pi\right)^43^22^{12}} \sim l_p^6,
\end{equation}
$R_{MNPQ}, R_{MN}, {\cal R}$  in  (\ref{D=11_O(l_p^6)})/(\ref{eoms}) being respectively the elven-dimensional Riemann curvature tensor, Ricci tensor and the Ricci scalar.

Now, one sees that if one makes an ansatz:
\begin{eqnarray}
\label{ansaetze}
& & \hskip -0.8ing_{MN} = g_{MN}^{(0)} +\beta g_{MN}^{(1)},\nonumber\\
& & \hskip -0.8inC_{MNP} = C^{(0)}_{MNP} + \beta C_{MNP}^{(1)},
\end{eqnarray}
then symbolically, one obtains:
\begin{eqnarray}
\label{deltaC=0consistent}
& & \beta \partial\left(\sqrt{-g}\partial C^{(1)}\right) + \beta \partial\left[\left(\sqrt{-g}\right)^{(1)}\partial C^{(0)}\right] + \beta\epsilon_{11}\partial C^{(0)} \partial C^{(1)} = {\cal O}(\beta^2) \sim 0 [{\rm up\ to}\ {\cal O}(\beta)].
\nonumber\\
& & \end{eqnarray}
One can see that one can find a consistent set of solutions to (\ref{deltaC=0consistent}) wherein $C^{(1)}_{MNP}=0$ up to ${\cal O}(\beta)$. This will be shown after (\ref{ M-theory-metric-psi=2npi-patch}).  Assuming that one can do so, henceforth we will  define:
\begin{eqnarray}
\label{fMN-definitions}
\delta g_{MN} =\beta g^{(1)}_{MN} = G_{MN}^{\rm MQGP} f_{MN}(r),
\end{eqnarray}
no summation implied. The first equation in (\ref{eoms}) will be denoted by ${\rm EOM}_{MN}$ in Appendix A. Appendix A has all the EOMs listed. The discussion in the same is divided into two sub-sections: the EOMs and their solutions for $f_{MN}$s are worked out  for the $\psi=0, 2\pi, 4\pi$-branches in {\bf A.1} and for
$\psi=\psi_0\neq2n\pi, n=0, 1, 2$ in {\bf A.2}.

One can show:
{\footnotesize
\begin{eqnarray}
\label{delta J_0}
 \delta J_0 & \stackrel{\rm MQGP,\ IR}{\xrightarrow{\hspace*{1.5cm}}} & 3\times 2^8 \delta R^{HMNK} R_H^{\ RSP}\Biggl(R_{PQNK}R^Q_{\ RSM} + R_{PSQK}R^Q_{\ MNR}  \nonumber\\
 & & + 2\left[R_{PMNQ}R^Q_{\ RSK} + R_{PNMQ}R^Q_{\ SRK}\right]\Biggr) \nonumber\\
& &  \equiv 3\times 2^8 \delta R^{HMNK}\chi_{HMNK}\nonumber\\
& & = -\delta g_{\tilde{M}\tilde{N}}\Biggl[ g^{M\tilde{N}} R^{H\tilde{N}NK}\chi_{HMNK}
+ g^{N\tilde{N}} R^{HM\tilde{M}K}\chi_{HMNK} + g^{K\tilde{M}}R^{HMN\tilde{N}}\chi_{HMNK}
\nonumber\\
& & + \frac{1}{2}\Biggl(g^{H\tilde{N}}[D_{K_1},D_{N_1}]\chi_H^{\tilde{M}N_1K_1} +
g^{H\tilde{N}}D_{M_1}D_{N_1} \chi_H^{M_1[\tilde{N}_1\tilde{M}]}  - g^{H\tilde{H}} D_{\tilde{H}}D_{N_1}\chi_H^{\tilde{N}[N_1\tilde{M}]}\Biggr)\Biggr],\nonumber\\
& &
\end{eqnarray}
}
where:
{\footnotesize
\begin{eqnarray}
\label{chi-def}
& & \chi_{HMNK} \equiv R_H^{\ \ RSP}\left[R_{PQNK} R^Q_{\ \ RSM}
+ R_{PSQK} R^Q_{\ \ MNR} + 2\left(R_{PMNQ} R^Q_{\ \ RSK} + R_{PNMQ}R^Q_{\ \ SRK}\right)\right].
\end{eqnarray}
}
Further:
{\footnotesize
\begin{eqnarray}
\label{deltaE8}
& & \delta E_8 \sim -\frac{2}{3}\delta g_{\tilde{M}\tilde{N}} g^{N_1^\prime\tilde{N}} \epsilon^{ABCM_1N_1,,,M_4N_4}\epsilon_{ABCM_1^\prime N_1^\prime...M_4^\prime N_4^\prime} R^{M_1^\prime\tilde{M}}_{\ \ \ \ \ M_1N_1}R^{M_2^\prime N_2^\prime}_{\ \ \ \ \ M_2N_2}R^{M_3^\prime N_3^\prime}_{\ \ \ \ \ M_2N_2}R^{M_4^\prime N_4^\prime}_{\ \ \ \ \ M_4N_4}\nonumber\\
& & + \frac{\delta g_{\tilde{M}\tilde{N}}}{3}\Biggl[ 2 \epsilon^{ABCM_1\tilde{N},,,M_4N_4}\epsilon_{ABCM_1^\prime N_1^\prime...M_4^\prime N_4^\prime}
g^{N_1^\prime\tilde{N}_1}g^{M_1^\prime\tilde{M}}D_{\tilde{N}_1}D_{M_1}\left(R^{M_2^\prime N_2^\prime}_{\ \ \ \ \ M_2N_2}R^{M_3^\prime N_3^\prime}_{\ \ \ \ \ M_2N_2}R^{M_4^\prime N_4^\prime}_{\ \ \ \ \ M_4N_4}\right)\nonumber\\
& & + \epsilon^{ABCM_1N_1,,,M_4N_4}\epsilon_{ABCM_1^\prime N_1^\prime...M_4^\prime N_4^\prime}
g^{N_1^\prime\tilde{N}_1}g^{M_1^\prime\tilde{M}}[D_{\tilde{N}_1},D_{M_1}]\left(R^{M_2^\prime N_2^\prime}_{\ \ \ \ \ M_2N_2}R^{M_3^\prime N_3^\prime}_{M_2N_2}R^{M_4^\prime N_4^\prime}_{\ \ \ \ \ M_4N_4}\right)\nonumber\\
& & - 2 \epsilon^{ABCM_1\tilde{M},,,M_4N_4}\epsilon_{ABCM_1^\prime N_1^\prime...M_4^\prime N_4^\prime}
g^{N_1^\prime\tilde{N}}g^{M_1^\prime\tilde{L}}D_{\tilde{L}_1}D_{M_1}\left(R^{M_2^\prime N_2^\prime}_{\ \ \ \ \ M_2N_2}R^{M_3^\prime N_3^\prime}_{\ \ \ \ \ M_2N_2}R^{M_4^\prime N_4^\prime}_{\ \ \ \ \ M_4N_4}\right)\Biggr],
\end{eqnarray}
}
where, e.g.,  \cite{Tseytlin-epsilonD^2R^4-kroneckerdeltaR^4}
{\footnotesize
\begin{eqnarray}
\label{epsilonD^2R^4}
& &  \epsilon^{ABCM_1M_2,,,M_8}\epsilon_{ABCM_1^\prime M_2^\prime...M_8^\prime}R^{M_1^\prime M_2^\prime}_{\ \ \ \ \ M_1M_2}R^{M_3^\prime M_4^\prime}_{\ \ \ \ \ M_3M_4}R^{M_5^\prime M_6^\prime}_{M_5M_6}R^{M_7^\prime M_8^\prime}_{\ \ \ \ \ M_7M_8}\nonumber\\
& &  = -3!8!
\delta^{M_1}_{N_1]}...\delta^{M_8}_{M_8^\prime]}R^{M_1^\prime M_2^\prime}_{\ \ \ \ \ M_1M_2}R^{M_3^\prime M_4^\prime}_{\ \ \ \ \ M_3M_4}R^{M_5^\prime M_6^\prime}_{M_5M_6}R^{M_7^\prime M_8^\prime}_{\ \ \ \ \ M_7M_8}.
\end{eqnarray}
}

Writing: $T_{MN} \equiv G_M^{\ \ PQR}G_{NPQR} - \frac{g_{MN}}{8}G^2$, the ${\cal O}(l_p^6)$ ``perturbations"  $T_{MN}^{(1)}$ therein will be given by:
\begin{eqnarray}
\label{TMN-first-order-in-beta-1}
& & T_{MN}^{(1)} = {\cal T}^{(1)}_{MN} + {\cal T}^{(2)}_{MN} -\frac{g_{MN}}{2}\delta g^{PP^\prime} {\cal T}^{(3)}_{PP^\prime},
\end{eqnarray}
where:
\begin{eqnarray}
\label{TMN-first-order-in-beta-2}
& & {\cal T}^{(1)}_{MN} \equiv 3 \delta g^{PP^\prime}g^{QQ^\prime}G_{MPQR}G_{NP^\prime Q^\prime R^\prime} \equiv \delta g^{PP^\prime} {\cal C}_{MNPP^\prime},\nonumber\\
& & {\cal T}^{(2)}_{MN} \equiv -\frac{\delta g_{MN}}{8}G_{PQRS}G^{PQRS} \equiv - \frac{\delta g_{MN}}{8} G^2,\nonumber\\
& & {\cal T}^{(3)}_{PQ} \equiv g^{QQ^\prime}g^{RR^\prime}g^{SS^\prime}G_{PQRS}G_{P^\prime Q^\prime R^\prime S^\prime}.
\end{eqnarray}

In the IR (i.e. small-$r$ limit), the various EOMs, denoted by ${\rm EOM}_{MN}$ henceforth, corresponding to perturbation of the first equation of (\ref{eoms}) up to ${\cal O}(\beta)$, and their solutions, have been obtained in appendix {\bf A}:  near the $\psi=2n\pi, n=0, 1, 2$-patches in {\bf A.1}, and near the $\psi\neq2n\pi, n=0, 1, 2$-patch (wherein, unlike  $\psi=2n\pi, n=0, 1, 2$-patches, some $G^{\cal M}_{rM}, M\neq r$ and $G^{\cal M}_{x^{10}N}, N\neq x^{10}$ components are non-zero) in {\bf A.2}. The EOMs are obtained by expanding the coefficients of $f^{(n)}_{MN}, n=0,1,2$ near $r=r_h$ and retaining the LO terms in the powers of $(r-r_h)$ in the same, and then performing a large-$N$-large-$|\log r_h|$-$\log N$ expansion.  It is shown in {\bf A.1} that near the $\psi=2n\pi, n=0, 1, 2$-patches the EOMs reduce to fifteen independent EOMs and four consistency checks, and in {\bf A.2} in $\psi=\psi_0$ (arbitrary but different from $2n\pi, n=0, 1, 2$)-branches to seven independent EOMs and one consistency check equation. In the following pair of subsections - {\bf 3.1} and {\bf 3.2} - we present the final results for the ${\cal M}$-theory metric components up to ${\cal O}(\beta)$.

\subsection{Near $\psi=2n\pi, n=0, 1, 2$-Coordinate Patches and Near $r=r_h$}

In this sub-section, we will obtain the EOMs and their solutions, in the IR, near the $\psi=0, 2\pi, 4\pi$-coordinate patches for the ${\cal M}$ theory black hole solution dual to thermal QCD-like theories at high temperature:
\begin{eqnarray}
\label{TypeIIA-from-M-theory-Witten-prescription-T>Tc}
\hskip -0.1in ds_{11}^2 & = & e^{-\frac{2\phi^{\rm IIA}}{3}}\Biggl[\frac{1}{\sqrt{h(r,\theta_{1,2})}}\left(-g(r) dt^2 + \left(dx^1\right)^2 +  \left(dx^2\right)^2 +\left(dx^3\right)^2 \right)
\nonumber\\
& & \hskip -0.1in+ \sqrt{h(r,\theta_{1,2})}\left(\frac{dr^2}{g(r)} + ds^2_{\rm IIA}(r,\theta_{1,2},\phi_{1,2},\psi)\right)
\Biggr] + e^{\frac{4\phi^{\rm IIA}}{3}}\left(dx^{11} + A_{\rm IIA}^{F_1^{\rm IIB} + F_3^{\rm IIB} + F_5^{\rm IIB}}\right)^2,
\end{eqnarray}
where $A_{\rm IIA}^{F^{\rm IIB}_{i=1,3,5}}$ are the type IIA RR 1-forms obtained from the triple T/SYZ-dual of the type IIB $F_{1,3,5}^{\rm IIB}$ fluxes in the type IIB holographic dual of \cite{metrics}, and $g(r) = 1 - \frac{r_h^4}{r^4}$.
For simplicity, we will be restricting to the Ouyang embedding:
\begin{equation}
\label{Ouyang}
\left(r^6 + 9 a^2 r^4\right)^{\frac{1}{4}}e^{\frac{i}{2}\left(\psi - \phi_1 - \phi_2\right)}\sin \frac{\theta_1}{2}\sin\frac{\theta_2}{2} = \mu,
\end{equation}
 $\mu$ being the Ouyang embedding parameter assuming $|\mu|\ll r^{\frac{3}{2}}$, effected, e.g., by working in the neighborhood of:
\begin{equation}
\label{Ouyang-theta10-theta20}
\theta_1=\frac{\alpha_{\theta_1}}{N^{\frac{1}{5}}}, \theta_2 = \frac{\alpha_{\theta_2}}{N^{\frac{3}{10}}}; \alpha_{\theta_{1,2}}\equiv{\cal O}(1)
\end{equation}
 (wherein an explicit $SU(3)$-structure for the type IIB dual of \cite{metrics} and its delocalized SYZ  type IIA mirror \cite{MQGP}, and an explicit $G_2$-structure for its ${\cal M}$-Theory uplift \cite{MQGP} was worked out in \cite{NPB}). Note, using (\ref{e12}) - (\ref{e^6}) and arguments similar to the ones given in \cite{SYZ-free-delocalization}, one can show that our results are independent of any delocalization in $\theta_{1,2}$.   The EOMs, corresponding to the ${\cal O}(l_p^6)$ variation of (\ref{eoms}) (via the substitution of (\ref{ansaetze}) into (\ref{eoms})) will be labelled as ${\rm EOM}_{MN}$ in this section and appendix {\bf A}.

\noindent This brings us to the first main lemma of this paper:

\noindent{\it Lemma 1}: In the neighborhood of the Ouyang embedding  of flavor $D7$-branes  \cite{ouyang} (that figure in the type IIB string dual of thermal QCD-like theories  at high temperatures\cite{metrics}) effected by working in the neighborhood of small $\theta_{1,2}$ (assuming a vanishingly small Ouyang embedding parameter), in the MQGP limit (\ref{MQGP_limit}), $\lim_{N\rightarrow\infty}\frac{E_8}{J_0}=0, \lim_{N\rightarrow\infty} \frac{t_8t_8G^2R^3}{E_8}=0$.

\noindent{\it Proof}: 
\begin{itemize}
\item
One can show that the leading-order-in-$N$ contribution to $J_0$ is given by
:
\begin{eqnarray}
\label{J0-1}
J_0 = \frac{1}{2} R^{\phi_2r \theta_1r}  R_{r \psi\theta_1r}  R_{\phi_2}^{\ \ r \phi_1r}  R^{\psi}_{r \phi_1r} -R^{\phi_2r \theta_1r}
   R_{\phi_1r\theta_1r}  R_{\phi_2}^{\ \ r \phi_1r}  R^{\theta_1}_{\ \ r\theta_1r} ,
\end{eqnarray}
where, e.g., near (\ref{Ouyang-theta10-theta20}),
\begin{eqnarray}
\label{RMNPQs}
R^{\phi_2r \theta_1r}  & \sim & \frac{ \left(\frac{1}{N}\right)^{5/4} \left(9 a^2+r^2\right) \left(r^4-{r_h}^4\right) \alpha _{\theta _1}^3 \sum_{n_1,n_2,n_3:n_1+n_2+n_3=6}a^{2n_1}r^{2n_2}r_h^{2n_3}}{{g_s}^{5/4}
   {N_f}^2 r^4 \left(r^2-3 a^2\right)^3 \left(6 a^2+r^2\right)^3 \log ^3(N) \alpha _{\theta _2}^2}\nonumber\\
   R_{\phi_1r \theta_1r} & \sim & -\frac{{g_s}^{7/4} M N^{11/20} {N_f}^{5/3} \log ^{\frac{5}{3}}(N) \log (r)\sum_{n_1,n_2,n_3:n_1+n_2+n_3=6}a^{2n_1}r^{2n_2}r_h^{2n_3}}{ r^4 \left(r^2-3 a^2\right) \left(6
   a^2+r^2\right) \left(9 a^2+r^2\right) \left(r^4-{r_h}^4\right) \alpha _{\theta _1} \alpha _{\theta _2}^2}\nonumber\\
 R_{r \psi\theta_1r} & \sim & -\frac{{g_s}^{7/4} M N^{3/20} {N_f}^{5/3} \log ^{\frac{5}{3}}(N) \log (r) \alpha _{\theta _1}\sum_{n_1,n_2,n_3:n_1+n_2+n_3=6}a^{2n_1}r^{2n_2}r_h^{2n_3} }{ r^4 \left(r^2-3
   a^2\right) \left(6 a^2+r^2\right) \left(9 a^2+r^2\right) \left(r^4-{r_h}^4\right) \alpha _{\theta _2}^2}\nonumber\\
R_{\phi_2}^{\ \ r \phi_1r} & \sim & -\frac{a^4 \left(\frac{1}{N}\right)^{21/20} \left(9 a^2+r^2\right)^2 \left(\frac{1}{\log (N)}\right)^{4/3}
   \left(r^4-{r_h}^4\right)^2 \alpha _{\theta _2}\Sigma_1}{{g_s} {N_f}^{4/3} r^6 \left(r^2-3 a^2\right)^2 \left(6 a^2+r^2\right)^2 \alpha _{\theta _1}^2}\nonumber\\
   R^{\theta_1}_{\ \ r \theta_1r} & \sim & \frac{ \sum_{n_1,n_2,n_3:n_1+n_2+n_3=6}a^{2n_1}r^{2n_2}r_h^{2n_3}}{r^2 \left(r^2-3 a^2\right)^2 \left(6 a^2+r^2\right) \left(9 a^2+r^2\right) \left(r^4-{r_h}^4\right)},\nonumber\\
R^{\psi}_{\ \ r \phi_1r} &  \sim & -\frac{N^{2/5}\sum_{n_1,n_2,n_3:n_1+n_2+n_3=6}a^{2n_1}r^{2n_2}r_h^{2n_3}}{r^2 \left(r^2-3 a^2\right)^2 \left(6 a^2+r^2\right) \left(9 a^2+r^2\right) \left(r^4-{r_h}^4\right) \alpha _{\theta _1}^2},\nonumber\\
& &
\end{eqnarray}
where $\Sigma_1$ is defined in (\ref{Sigma_1-3-def}), and $ \sum_{n_1,n_2,n_3:n_1+n_2+n_3=6}a^{2n_1}r^{2n_2}r_h^{2n_3} =  -81 a^8
   r^4+243 a^8 {r_h}^4+27 a^6 r^6-36 a^6 r^2 {r_h}^4+15 a^4 r^8-27 a^4 r^4 {r_h}^4+a^2 r^{10}-2 a^2 r^6 {r_h}^4$. In (\ref{RMNPQs}) and henceforth, $r/a/r_h$ in fact would imply $\frac{r/a/r_h}{{\cal R}_{D5/\overline{D5}}}$ (see the last reference in \cite{EPJC-2}).  Substituting (\ref{RMNPQs})  into (\ref{J0-1}), one therefore obtains:
\begin{eqnarray}
\label{J0-2}
& & \hskip -0.5in J_0 \sim \frac{ a^{10} \left(\frac{1}{{\log N}}\right)^{8/3} M \left(\frac{1}{N}\right)^{7/4} \left(9 a^2+r^2\right) \left(r^4-{r_h}^4\right)
   \log (r) \Sigma_1\sum_{n_1,n_2,n_3:n_1+n_2+n_3=6}a^{2n_1}r^{2n_2}r_h^{2n_3}
  }{ \sqrt{{g_s}} {N_f}^{5/3} r^{16} \left(3 a^2-r^2\right)^8 \left(6 a^2+r^2\right)^7 \alpha _{\theta _2}^3}\nonumber\\
& &  \sim \left(\frac{1}{N}\right)^{7/4}.
\end{eqnarray}
For an arbitrary but small $\theta_{1,2}$, one can show that:
\begin{eqnarray}
\label{J0-arb-small-theta12}
& & \hskip -0.6in  J_0 \nonumber\\
& & \hskip -0.6in \sim  \frac{ a^{10} \left(\frac{1}{{\log N}}\right)^{8/3} M \left(\frac{1}{N}\right)^{29/20} \left(9 a^2+r^2\right) \left(r^4-{r_h}^4\right)
   \log (r)\left(19683
   \sqrt{6} \sin^6\theta_1+6642 \sin^2{\theta _2} \sin^3{\theta _1}-40 \sqrt{6} \sin^4{\theta _2}\right)
  }{ \sqrt{{g_s}} {N_f}^{5/3} r^{16} \left(3 a^2-r^2\right)^8 \left(6 a^2+r^2\right)^7 \sin^3{\theta _2}}\nonumber\\
  & & \hskip -0.6in \times \sum_{n_1,n_2,n_3:n_1+n_2+n_3=6}a^{2n_1}r^{2n_2}r_h^{2n_3}.
\end{eqnarray}

\item
For evaluating the contribution of $E_8$, (\ref{epsilonD^2R^4}), one notes that one needs to pick out eight of the eleven space-time indices (and anti-symmetrize appropriately). Let us consider\\ $R^{M_1N_1}_{\ \ \ \ \ M_1N_1} R^{M_2N_2}_{\ \ \ \ \ M_2N_2}R^{M_3N_3}_{\ \ \ \ \ M_3N_3}R^{M_4N_4}_{\ \ \ \ \ M_4N_4}$ which will be one of the kinds of terms one will obtain using (\ref{epsilonD^2R^4}).
After a very long and careful computation, one can then show that for arbitrary small $\theta_{1,2}$ and not just restricted to (\ref{Ouyang-theta10-theta20}), the above contributes a $\frac{1}{N^2}$ via the following most dominant term in the MQGP limit:
\begin{eqnarray}
\label{E8-dominant-large-N}
E_8 \ni R^{tx^1}_{\ \ \ tx^1}R^{x^2x^3}_{\ \ \ x^2x^3}R^{r\theta_1}_{\ \ \ r\theta_1}\left(R^{\psi x^{10}}_{\ \ \ \psi x^{10}}
+ R^{\phi_1\psi}_{\ \ \ \phi_1\psi} +  R^{yz}_{\ \ \ \phi_2\psi}\right) \left(\sim {\cal O}\left(\frac{1}{N^2}\right)\right).
\end{eqnarray}
A similar computation for the other types of summands in (\ref{epsilonD^2R^4}) yield a similar $N$ dependence. Consequently, $\frac{E_8}{J_0}\sim\frac{1}{N^\alpha}, \alpha>0$. 

\item
Summing first w.r.t. $M_{3,4}, N_{3,4}$ in (\ref{J0+E8-definitions}), one obtains 
\begin{equation}
\label{G2R3-ii}
\frac{\chi_1(r,\langle\theta_{1,2}\rangle)}{N^{31/20}} G_{M_1}^{\ N_1 MN} G_{M_2}^{\ N_2}\ _{MN} R_{M_5 M_6}^{\ \ N_5N_6} R_{M_7M_8}^{\ \ N_7N_8} t_{N_1 N_2 x^0\theta_2N_5N_6N_7N_8} t^{M_1M_2x^0\phi_2M_5M_6M_7M_8}
\end{equation}
 as the LO term in $N$. Summing w.r.t. $M_{5,6}, N_{5,6}$, one obtains
 \begin{equation}
 \label{G2R3-ii}
\frac{\chi_2(r; \langle \theta_{1,2}\rangle)}{N^{31/10}}
  G_{M_1}^{N_1MN} G_{M_2}^{\ N_2}\ _{MN} R_{M_7M_8}^{\ \ N_7 N_8} t_{N_1N_2x^0\theta_2N_7N_8} t^{M_1M_2x^0\phi_2M_7M_8}
\end{equation}  
up to LO in $N$. Finally, summing w.r.t. $M_{7,8}, N_{7,8}$ one obtains: 
 \begin{equation}
 \label{G2R3-iii} 
  \frac{\chi_3(r,\langle \theta_{1,2}\rangle)}{N^{93/20}}
  G_{M_1}^{\ N_1MN} G_{M_2}^{\ N_2}\ _{MN} t_{N_1N_2x^0\theta_2x^0\theta_2x^0\theta_2} t^{M_1M_2x^0\phi_2x^0\phi_2x^0\phi_2},
\end{equation}
and restricted to (\ref{Ouyang})$\cap$(\ref{Ouyang-theta10-theta20}) one obtains:
\begin{equation}
\label{G2R3-iv}
\left.t_8t_8G^2R^3\right|_{(\ref{Ouyang})\cap(\ref{Ouyang-theta10-theta20})}
\sim \frac{\chi_4(r;\langle \theta_{1,2}\rangle}{N^{111/20}},
\end{equation}
which is large-$N$ suppressed as compared to the $J_0$ and $E_8$.

Hence, (\ref{J0-2})-(\ref{J0-arb-small-theta12}) along with (\ref{G2R3-iv}), one proves Lemma 1, and obtains the following hierarchy: 
\begin{equation}
\label{Hierarchy-G2R3-E8-J0}
t_8^2G^2R^3 < E_8 < J_0.
\end{equation}
\end{itemize}
Henceforth, $E_8, t_8^2G^2R^3$ (and their variations) will be disregarded as compared to $J_0$ (and its variation) in the MQGP limit. 

\noindent We now come to the second lemma of this paper:

\noindent{\it Lemma 2}: The ${\cal O}(\beta)$-corrected ${\cal M}$-theory metric of \cite{MQGP} in the MQGP limit near the $\psi=2n\pi, n=0, 1, 2$-branches up to ${\cal O}((r-r_h)^2)$ [and up to ${\cal O}((r-r_h)^3)$ for some of the off-diagonal components along the delocalized $T^3(x,y,z)$] - the components which do not receive an ${\cal O}(\beta)$ corrections, are not listed in (\ref{ M-theory-metric-psi=2npi-patch}) - is given below:
{\footnotesize
\begin{eqnarray}
\label{ M-theory-metric-psi=2npi-patch}
 \hskip -0.5in G_{tt} & = & G^{\rm MQGP}_{tt}\Biggl[1 + \frac{1}{4}  \frac{4 b^8 \left(9 b^2+1\right)^3 \left(4374 b^6+1035 b^4+9 b^2-4\right) \beta  M \left(\frac{1}{N}\right)^{9/4} \Sigma_1
   \left(6 a^2+  {r_h}^2\right) \log (  {r_h})}{27 \pi  \left(18 b^4-3 b^2-1\right)^5  \log N ^2   {N_f}   {r_h}^2
   \alpha _{\theta _2}^3 \left(9 a^2+  {r_h}^2\right)} (r-  {r_h})^2\Biggr]
\nonumber\\
G_{x^{1,2,3}x^{1,2,3}} & = &  G^{\rm MQGP}_{x^{1,2,3}x^{1,2,3}}
\Biggl[1 - \frac{1}{4} \frac{4 b^8 \left(9 b^2+1\right)^4 \left(39 b^2-4\right) M \left(\frac{1}{N}\right)^{9/4} \beta  \left(6 a^2+{r_h}^2\right) \log
   ({r_h})\Sigma_1}{9 \pi  \left(3 b^2-1\right)^5 \left(6 b^2+1\right)^4 \log N ^2 {N_f} {r_h}^2 \left(9 a^2+{r_h}^2\right) \alpha
   _{\theta _2}^3} (r - {r_h})^2\Biggr]\nonumber\\
G_{rr} & = & G^{\rm MQGP}_{rr}\Biggl[1 + \Biggl(- \frac{2 \left(9 b^2+1\right)^4 b^{10} M   \left(6 a^2+{r_h}^2\right) \left((r-{r_h})^2+{r_h}^2\right)\Sigma_1}{3 \pi
   \left(-18 b^4+3 b^2+1\right)^4 \log N  N^{8/15} {N_f} \left(-27 a^4+6 a^2 {r_h}^2+{r_h}^4\right) \alpha _{\theta
   _2}^3}\nonumber\\
& & +{C_{zz}}^{(1)}-2 {C_{\theta_1z}}^{(1)}+2 {C_{\theta_1x}}^{(1)}\Biggr)\beta\Biggr]\nonumber\\
 G_{\theta_1x} & = & G^{\rm MQGP}_{\theta_1x}\Biggl[1 + \Biggl(
- \frac{\left(9 b^2+1\right)^4 b^{10} M  \left(6 a^2+{r_h}^2\right) \left((r-{r_h})^2+{r_h}^2\right)
   \Sigma_1}{3 \pi  \left(-18 b^4+3 b^2+1\right)^4 \log N  N^{8/15} {N_f} \left(-27 a^4+6 a^2
   {r_h}^2+{r_h}^4\right) \alpha _{\theta _2}^3}+{C_{\theta_1x}}^{(1)}
\Biggr)\beta\Biggr]\nonumber\\
G_{\theta_1z} & = & G^{\rm MQGP}_{\theta_1z}\Biggl[1 + \Biggl(\frac{16 \left(9 b^2+1\right)^4 b^{12}    \left(\frac{(r-{r_h})^3}{{r_h}^3}+1\right)\Sigma_1}{243\sqrt{2}
   \pi ^3 \left(1-3 b^2\right)^{10} \left(6 b^2+1\right)^8 {g_s}^{9/4} \log N ^4 N^{7/6} {N_f}^3 \left(-27 a^4 {r_h}+6 a^2
   {r_h}^3+{r_h}^5\right) \alpha _{\theta _1}^7 \alpha _{\theta _2}^6}+C_{\theta_1z}^{(1)}\Biggr)\beta\Biggr]\nonumber\\
   G_{\theta_2x} & = & G^{\rm MQGP}_{\theta_2x}\Biggl[1 + \Biggl(
   \frac{16 \left(9 b^2+1\right)^4 b^{12} \left(\frac{(r-{r_h})^3}{{r_h}^3}+1\right) \Sigma_1}{243\sqrt{2} \pi ^3 \left(1-3
   b^2\right)^{10} \left(6 b^2+1\right)^8 {g_s}^{9/4} \log N ^4 N^{7/6} {N_f}^3 \left(-27 a^4 {r_h}+6 a^2
   {r_h}^3+{r_h}^5\right) \alpha _{\theta _1}^7 \alpha _{\theta _2}^6}+C_{\theta_2x}^{((1)}\Biggr)\beta\Biggr]\nonumber\\
G_{\theta_2y} & = & G^{\rm MQGP}_{\theta_2y}\Biggl[1 +  \frac{3 b^{10} \left(9 b^2+1\right)^4 M \beta \left(6 a^2+{r_h}^2\right) \left(1-\frac{(r-{r_h})^2}{{r_h}^2}\right) \log
   ({r_h}) \Sigma_1}{\pi  \left(3 b^2-1\right)^5 \left(6 b^2+1\right)^4 \log N ^2 N^{7/5} {N_f} \left(9 a^2+{r_h}^2\right) \alpha
   _{\theta _2}^3}\Biggr]\nonumber\\
G_{\theta_2z} & = & G^{\rm MQGP}_{\theta_2z}\Biggl[1 + \Biggl(\frac{3 \left(9 b^2+1\right)^4 b^{10} M  \left(6 a^2+{r_h}^2\right) \left(1-\frac{(r-{r_h})^2}{{r_h}^2}\right) \log
   ({r_h})\Sigma_1}{\pi  \left(3 b^2-1\right)^5 \left(6 b^2+1\right)^4 {\log N}^2 N^{7/6} {N_f} \left(9 a^2+{r_h}^2\right) \alpha
   _{\theta _2}^3} +{C_{\theta_2z}}^{(1)}\Biggr)\beta\Biggr]\nonumber\\
G_{xy} & = & G^{\rm MQGP}_{xy}\Biggl[1 + \Biggl(\frac{3 \left(9 b^2+1\right)^4 b^{10} M  \left(6 a^2+{r_h}^2\right) \left(\frac{(r-{r_h})^2}{{r_h}^2}+1\right) \log
   ({r_h}) \alpha _{\theta _2}^3\Sigma_1}{\pi  \left(3 b^2-1\right)^5 \left(6 b^2+1\right)^4 \log N ^2 N^{21/20} {N_f} \left(9
   a^2+{r_h}^2\right) \alpha _{\theta _{2 l}}^6}+C_{xy}^{(1)}\Biggr)\beta\Biggr]\nonumber\\
G_{xz}  & = & G^{\rm MQGP}_{xz}\Biggl[1 + \frac{18 b^{10} \left(9 b^2+1\right)^4 M   \left(6 a^2+{r_h}^2\right)
   \left(\frac{(r-{r_h})^2}{{r_h}^2}+1\right) \log ^3({r_h}) \Sigma_1}{\pi  \left(3b^2-1\right)^5 \left(6 b^2+1\right)^4 \log N ^4 N^{5/4} {N_f} \left(9 a^2+{r_h}^2\right) \alpha
   _{\theta _2}^3}\Biggr]\nonumber\\
G_{yy} & = & G^{\rm MQGP}_{yy}\Biggl[1  - \frac{3 b^{10} \left(9 b^2+1\right)^4 M \left(\frac{1}{N}\right)^{7/4}   \left(6 a^2+{r_h}^2\right) \log ({r_h})\Sigma_1
   \left(\frac{(r-{r_h})^2}{r_h^2}+1\right)}{\pi  \left(3 b^2-1\right)^5 \left(6 b^2+1\right)^4 \log N ^2 {N_f} {r_h}^2 \left(9
   a^2+{r_h}^2\right) \alpha _{\theta _2}^3}\Biggr]\nonumber\\
 G_{yz} & = & G^{\rm MQGP}_{yz}\Biggl[1 + \Biggl(\frac{64 \left(9 b^2+1\right)^8 b^{22} M \left(\frac{1}{N}\right)^{29/12}  \left(6 a^2+{r_h}^2\right)
   \left(\frac{(r-{r_h})^3}{{r_h}^3}+1\right) \log ({r_h})\Sigma_3 }{27 \pi ^4 \left(3 b^2-1\right)^{15} \left(6 b^2+1\right)^{12}
   {g_s}^{9/4} \log N ^6  {N_f}^4 {r_h}^3 \left({r_h}^2-3 a^2\right) \left(9 a^2+{r_h}^2\right)^2 \alpha
   _{\theta _1}^7 \alpha _{\theta _2}^9} +C_{yz}^{(1)}\Biggr)\beta\Biggr]\nonumber\\
G_{zz} & = & G^{\rm MQGP}_{zz}\Biggl[1 + \Biggl(C_{zz}^{(1)}-\frac{b^{10} \left(9 b^2+1\right)^4 M \left({r_h}^2-\frac{(r-{r_h})^3}{{r_h}}\right) \log ({r_h})
   \Sigma_1}{27 \pi ^{3/2} \left(3 b^2-1\right)^5 \left(6 b^2+1\right)^4 \sqrt{{g_s}} \log N ^2 N^{23/20} {N_f} \alpha
   _{\theta _2}^5}\Biggr)\beta\Biggr]\nonumber\\
G_{x^{10}x^{10}} & = & G^{\rm MQGP}_{x^{10}x^{10}}\Biggl[1 -\frac{27 b^{10} \left(9 b^2+1\right)^4 M \left(\frac{1}{N}\right)^{5/4} \beta  \left(6 a^2+{r_h}^2\right)
   \left(1-\frac{(r-{r_h})^2}{{r_h}^2}\right) \log ^3({r_h}) \Sigma_1}{\pi  \left(3 b^2-1\right)^5 \left(6 b^2+1\right)^4 \log N ^4
   {N_f} {r_h}^2 \left(9 a^2+{r_h}^2\right) \alpha _{\theta _2}^3}\Biggr],
\end{eqnarray}
   }
\noindent where $\Sigma_{1,3}$ are defined in (\ref{Sigma_1-3-def}), and $G^{\rm MQGP}_{MN}$ are the ${\cal M}$ theory metric components in the MQGP limit at ${\cal O}(\beta^0)$
\cite{mesons_0E++-to-mesons-decays}. The explicit dependence on $\theta_{1,2}$ of the ${\cal M}$-theory metric components up to ${\cal O}(\beta)$, using (\ref{Ouyang-theta10-theta20}), is effected by the replacemements:
$\alpha_{\theta_1}\rightarrow N^{\frac{1}{5}}\sin\theta_{1},\ \alpha_{\theta_2}\rightarrow N^{\frac{3}{10}}\sin\theta_{2}$ in (\ref{ M-theory-metric-psi=2npi-patch}). Also, see footnote 5.

\noindent We now present the third lemma of this paper:

\noindent{\it Lemma 3}: $C^{(1)}_{MNP}=0$ up to ${\cal O}(\beta)$ is a consistent solution of (\ref{eoms}).

\noindent{\it Proof}: The eleven-fold $M_{11}$ in the ${\cal M}$ theory uplift as obtained in \cite{MQGP} is a warped product of $S^1(x^0)\times \mathbb{R}_{\rm conformal}$
and $M_7(r,\theta_{1,2},\phi_{1,2},\psi,x^{10})$, the latter being a cone over $M_6(\theta_{1,2},\phi_{1,2},\psi,x^{10})$ where $M_6(\theta_{1,2},\phi_{1,2},\psi,x^{10})$ has the following nested fibration structure:
{\footnotesize
\begin{equation*}
\hskip -0.4in
\label{M_6}
\begin{array}{cc}
&{\cal M}_6(\theta_{1,2},\phi_{1,2},\psi,x_{10})   \longleftarrow   S^1(x^{10}) \\
&\downarrow  \\
& \hskip 0.7in {\cal M}_5(\theta_{1,2},\phi_{1,2},\psi) \longleftarrow  {\cal M}_3(\phi_1,\phi_2,\psi)  \\
&\downarrow  \\
 &{\cal B}_2(\theta_1,\theta_2)
\end{array}.
\end{equation*}
}
As shown in \cite{MQGP}, $p_1^2(M_{11}) = p_2(M_{11}) = 0$ up to ${\cal O}(\beta^0)$ where $p_a$ is the $a$-th Pontryagin class of $M_{11}$. This hence implies that $X_8=0$ up to ${\cal O}(\beta^0)$.

Now, (\ref{deltaC=0consistent}) implies:
\begin{eqnarray}
\label{C1=0-II}
& & \sum_{N,P\in\left\{t,x^{1,2,3},r,\theta_{1,2},\phi_{1,2},\psi,x^{10}\right\}}\beta\partial_M\left(\sqrt{-g^{(0)}}g^{(0)NP}g_{NP}^{(0)}f_{NP}\partial^{[M}C^{M_1M_2M_3]}_{(0)}\right) \sim 0,
\end{eqnarray}
where the ``$(0)$" implies the ${\cal O}(\beta^0)$-terms of \cite{MQGP,NPB,mesons_0E++-to-mesons-decays} and the $\sim$ implies equality up to ${\cal O}(\beta^2)$ corrections. For simplicity we  work near the $\psi=2n\pi, n=0,1,2$-branches (resulting in the decoupling of $M_5(t,x^{1,2,3},r)$ and $M_6(\theta_{1,2},\phi_{1,2},\psi,x^{10})$ and $g_{MN}^{(0)}$ being diagonal for $M=r,x^{10}$ \cite{NPB}) restricted to the Ouyang embedding (effected by the delocalized limit wherein one works in the neighborhood of $\theta_{10}=\frac{\alpha_{\theta_1}}{N^{\frac{1}{5}}}, \theta_{20} = \frac{\alpha_{\theta_2}}{N^{\frac{3}{10}}}$ (see footnote 5) wherein, as also mentioned in {\bf 3.1}, an explicit $SU(3)$-structure for the type IIB dual as well as its delocalized Strominger-Yau-Zaslow (SYZ)  type IIA mirror as string theory duals of large-$N$ thermal QCD-like theories, and an explicit $G_2$-structure for its ${\cal M}$-theory uplift \cite{MQGP}, was worked out in \cite{NPB}; using (\ref{e12}) - (\ref{e^6}) and arguments similar to the ones given in \cite{SYZ-free-delocalization}, one can show that our results are independent of any delocalization in $\theta_{1,2}$). Using the non-zero components of $C_{MNP}: C_{\theta_{1,2}\ \phi_{1,2}/\psi\ x^{10}}$ \cite{MQGP}, one can show that (\ref{C1=0-II}) implies:
\begin{eqnarray}
\label{C1=0-III}
& & \sum_{N,P\in\left\{t,x^{1,2,3},r,\theta_{1,2},\phi_{1,2},\psi,x^{10}\right\}}\beta\partial_r\left(\sqrt{-g^{(0)}}g^{(0)NP}g_{NP}^{(0)}f_{NP}g_{(0)}^{rr}\partial_rC^{M_1M_2 x^{10}}_{(0)}\right) \delta^{M_3}_{x^{10}}\sim 0,\nonumber\\
& &
\end{eqnarray}
where $M_1,M_2 = \theta_{1,2},\phi_{1,2},\psi$ or precisely $\theta_{1,2}, x, y, z$ where the delocalized $T^3(x, y, z)$ coordinates are defined near $r=r_0\in$IR as \cite{MQGP} \footnote{As explained in \cite{Knauf-thesis}, the $T^3$-valued $(x, y, z)$ are defined via:
\begin{eqnarray}
\label{xyz-definitions}
& & \phi_1 = \phi_{10} + \frac{x}{\sqrt{h_2}\left[h(r_0,\theta_{10,20})\right]^{\frac{1}{4}} \sin\theta_{10}\ r_0},\nonumber\\
& & \phi_2 = \phi_{20} + \frac{y}{ \sqrt{h_4}
\left[h( r_0,\theta_{10,20})\right]^{\frac{1}{4}}\sin\theta_{20}\ r_0}\nonumber\\
& & \psi = \psi_0 + \frac{z}{\sqrt{h_1} \left[h( r_0,\theta_{10,20})
\right]^{\frac{1}{4}}\ r_0},
\end{eqnarray}
and one works up to linear order in $(x, y, z)$. Up to linear order in $r$, i.e., in the IR, it can be shown \cite{theta0-theta} that $\theta_{10,20}$ can be promoted to global coordinates $\theta_{1,2}$ in all the results in the paper.}:
\begin{eqnarray}
\label{xyz-defs}
& & \hskip -0.3in  x = \sqrt{h_2}\left[h(r_0,\theta_{10,20})\right]^{\frac{1}{4}} \sin\theta_{10}\ r_0 \phi_1,\ y = \sqrt{h_4}
\left[h( r_0,\theta_{10,20})\right]^{\frac{1}{4}}\sin\theta_{20}\ r_0 \phi_2,\ z=\sqrt{h_1} \left[h( r_0,\theta_{10,20})
\right]^{\frac{1}{4}}\ r_0 \psi,\nonumber\\
& &
\end{eqnarray}
$h$ being the delocalized warp factor \cite{metrics}:
\begin{eqnarray}
\label{eq:h}
&& h(r_0,\theta_{10,20}) =\frac{L^4}{r_0^4}\Bigg[1+\frac{3g_sM_{\rm eff}^2}{2\pi N}{\rm log}r_0\left\{1+\frac{3g_sN^{\rm eff}_f}{2\pi}\left({\log} r_0+\frac{1}{2}\right)+\frac{g_sN^{\rm eff}_f}{4\pi}{\rm log}\left({\sin}\frac{\theta_{10}}{2}
{\rm sin}\frac{\theta_{20}}{2}\right)\right\}\Bigg],\nonumber\\
\end{eqnarray}
wherein $M_{\rm eff}$ was defined in Section {\bf 2} and $N_f^{\rm eff}$ is defined via the type IIB axion $C_0 = \frac{N_f^{\rm eff}}{4\pi}(\psi - \phi_1 - \phi_2)$ (by standard monodromy arguments); the squashing factors are defined below \cite{metrics}:
\begin{equation}
\label{h_{1,2,4}-defs}
h_1 = \frac{1}{9} + {\cal O}\left(\frac{g_sM^2}{N}\right),\ h_2 = \frac{1}{6} + {\cal O}\left(\frac{g_sM^2}{N}\right),\
h_4 = h_2 + \frac{4a^2}{r_0^2},
\end{equation}
($a$ being the radius of the blown-up $S^2$).

One immediately notes from (\ref{C1=0-III}) that (\ref{C1=0-II}) is identically satisfied for $M_{1,2,3}\in x^{0,1,2,3}, r$. The set of $\ ^5C_2$ equations (\ref{C1=0-III}) for $M_{1,2}\in \theta_{1,2}, x, y, z,$ and  $M_3=x^{10}$ are considered in Appendix {\bf D} where one sees that in the IR: $r = \chi r_h, \chi = {\cal O}(1)$ [and $a$(the resolution parameter)$=\left(b  + {\cal O}\left(\frac{g_sM^2}{N}\right)\right)r_h$ \cite{EPJC-2}], all ten of these equations substituting in the solutions for $f_{MN}$ from {\bf 3.1}, reduce to:
{\footnotesize
\begin{eqnarray}
\label{10-C^1-EOMs}
& & \beta N_f^{\alpha_f}\left(\log N\right)^{\alpha_{\log N}}\left(N^{\alpha_1} {\cal F}^{M_1M_2}_{\theta_1x}(b,\chi, \alpha_{\theta_{1,2}}, r_h)C^{(1)}_{\theta_1x} + N^{\alpha_2}\sum_{(M, N)=(z, z), (\theta_1, z), (\theta_2, z)}{\cal F}^{M_1M_2}_{MN}(b,\chi, \alpha_{\theta_{1,2}}, r_h)C^{(1)}_{MN}\right)  = 0,\nonumber\\
& &
\end{eqnarray}
}
where $\alpha_f=2, 3; \alpha_{\log N}=1, 2; \alpha_1>\alpha_2$ and $C^{(1)}_{MN}$ are the constants of integration appearing in the solutions (\ref{ M-theory-metric-psi=2npi-patch}) to the ${\cal O}(\beta)$-corrections to the ${\cal M}$ theory metric components of \cite{MQGP, NPB, mesons_0E++-to-mesons-decays} (\ref{ M-theory-metric-psi=2npi-patch}). Hence, up ${\cal O}(\beta)$ and LO in $N$, (\ref{C1=0-II}) is identically satisfied if:
\begin{equation}
\label{C1_68=0}
C^{(1)}_{\theta_1x} = 0\ {\rm in}\ (\ref{ M-theory-metric-psi=2npi-patch}),
\end{equation}
and up to ${\cal O}(\beta)$ and NLO in $N$ (assuming as in (\ref{ansatz-b}), $b \sim \frac{1}{\sqrt{3}}$), additionally:
\begin{eqnarray}
\label{constraints-C1_MN-NLO-N}
& & \left\{{\cal F}^{M_1M_2}_{\theta_1z, \theta_2z, zz}=0\right\},\nonumber\\
& & {\rm implying}:\nonumber\\
& &  C^{(1)}_{zz} = 2 C^{(1)}_{\theta_1z},\ C^{(1)}_{\theta_2z} = 0\ {\rm in}\ (\ref{ M-theory-metric-psi=2npi-patch}).
\end{eqnarray}
One therefore sees that one can consistently set $C^{(1)}_{MNP}=0$ up to ${\cal O}(\beta)$.

\subsection{Near $\psi\neq2n\pi, n=0, 1, 2$, and Near $r=r_h$}

In this sub-section we will be looking at the EOMs and their solutions near the $\psi=\psi_0\neq2n\pi, n=0, 1, 2$-branches (wherein some $G^{\cal M}_{rM}, M\neq r$ and $G^{\cal M}_{x^{10}N}, N\neq x^{10}$ components are non-zero) near $r=r_h$.

One can show that leading-order-in-$N$ contribution to $J_0$ for small $\theta_{1,2}$ (i.e. corresponding to the Ouyang embedding of type IIB $D7$-branes' embedding \cite{ouyang} for vanishing small $\mu_{\rm Ouyang}$) is given by:
\begin{eqnarray}
\label{J0-psineq0}
& & J_0 \sim \frac{1}{2} R^{x^{10}t\theta_2t}R_{t\theta_1\theta_2t}R_{x^{10}}^{\ \ tx^{10}t}R^{\theta_1}_{\ \ tx^{10}t},
\end{eqnarray}
where:
{
\begin{eqnarray}
& & R^{x^{10}t\theta_2t}  \sim  \frac{ a^2 N^{\frac{17}{20}} \sin{\theta _1} \sin^2{\theta _2} \left(9 a^2 r^4+9 a^2 {r_h}^4+r^6+r^2 {r_h}^4\right)}{
   ({g_s}-1) {g_s}^{5/4} {\log N}^2 M  {N_f}^4 \sin\left(\frac{\psi_0}{2}\right)  r \left(6 a^2+r^2\right)
   \left(r^4-{r_h}^4\right)^2 \log (r)}\nonumber\\
& & R_{t\theta_1\theta_2t} \sim \frac{ {g_s}^{5/2} \left(\frac{1}{{\log N}}\right)^{4/3} M^2 {N_f}^{8/3} \sin^2\left(\frac{\psi_0}{2}\right)  \left(9 a^2+r^2\right)
   \left(r^4-{r_h}^4\right) \left(r^4+{r_h}^4\right) \log (r)}{N^{3/2} r^6 \sin^3{\theta _1} \sin^3{\theta
   _2} \left(6 a^2+r^2\right)}\nonumber\\
& & R_{x^{10}}^{\ \ tx^{10}t} \sim -\frac{ \left(\frac{1}{{\log N}}\right)^{4/3} \sin^4\left(\frac{\psi_0}{2}\right)  \left(9 a^2+r^2\right) \left(r^4+{r_h}^4\right)}{
   {N_f}^{4/3} \sin^4\phi_{20} r^2 \sin^4{\theta _2} \left(6 a^2+r^2\right) \left(r^4-{r_h}^4\right)
   \log (r)}\nonumber\\
& & R^{\theta_1}_{\ \ tx^{10}t} \sim \frac{ \left(\frac{1}{N}\right)^{3/4} \sin\left(\frac{\psi_0}{2}\right)  \sin^2{\theta _1} \left(9 a^2+r^2\right) \left(r^4-{r_h}^4\right)
   \left(r^4+{r_h}^4\right)}{ {g_s}^{11/4} M {N_f}^2 \sin^2\phi_{20} r^6 \sin{\theta _2} \left(6 a^2+r^2\right) \log
   ^2(r)}.
\end{eqnarray}
}
yields:
\begin{eqnarray}
\label{J0}
& & \hskip -0.8in J_0 \sim -\frac{ a^2 \left(\frac{1}{{\log N}}\right)^{14/3}  \sin^6\left(\frac{\psi_0}{2}\right) \left(9 a^2+r^2\right)^4
   \left(r^4+{r_h}^4\right)^4}{ N^{7/5}({g_s}-1) {g_s}^{3/2} {N_f}^{14/3} \sin^6\phi_{20} r^{15}
\sin^6{\theta _2}
   \left(6 a^2+r^2\right)^4 \left(r^4-{r_h}^4\right) \log ^3(r)}.
\end{eqnarray}

\noindent We now arrive at the fourth lemma of this paper:

{\it Lemma 4}: The following is the final result as regards the ${\cal O}(\beta)$-corrected ${\cal M}$-theory metric of \cite{MQGP} in the MQGP limit in the
$\psi\neq 2n\pi, n=0, 1, 2$-branches, e.g., near  (\ref{Ouyang-theta10-theta20}), up to ${\cal O}((r-r_h)^2)$ - the components which do not receive an ${\cal O}(\beta)$ corrections, are not listed in (\ref{ M-theory-metric-psi-neq-2npi-patch}):
{\scriptsize
\begin{eqnarray}
\label{ M-theory-metric-psi-neq-2npi-patch}
G_{\theta_1\theta_2} & = & G_{\theta_1\theta_2}^{\rm MQGP}\Biggl[1 + \Biggl(\frac{\kappa_{\theta_1\theta_2}  \kappa_b ^2 \sin^2\left(\frac{\psi_0}{2}\right) {r_h}^2   \left(\frac{1}{N}\right)^{2 \alpha +\frac{3}{2}}}{
   \sqrt{{g_s}} {N_f}^6 \sin^2\phi_{20} (r-{r_h})^2 \alpha _{\theta _2}^2}+{C_{\theta_1\theta_2}}^{(1)}\Biggr)\beta\Biggr]\nonumber\\
G_{yy} & = & G_{yy}^{\rm MQGP}\Biggl[1 + \Biggl(-\frac{256 N^{3/5} \sin^2\phi_{20} {r_h}^4 \alpha _{\theta _2}^2}{9 ({g_s}-1)
   \left({r_h}^2-3 a^2\right)^2 \log ^2(N) \alpha _{\theta _1}^6 \log ^2\left(9 a^2 {r_h}^4+{r_h}^6\right)}\nonumber\\
& &\times  \Biggl\{\frac{\kappa_{yy}   N^{3/10} \sin^2\left(\frac{\psi_0}{2}\right)   \left(9
   a^2+{r_h}^2\right) (\log ({r_h})-1) \left(\left(9 a^2+{r_h}^2\right) \log \left(9 a^2 {r_h}^4+{r_h}^6\right)-8 \left(6
   a^2+{r_h}^2\right) \log ({r_h})\right)^2}{ \sqrt{{g_s}} {N_f}^6 \sin^2\phi_{20} {r_h}^2 \left(6 a^2+{r_h}^2\right)^3
   (r-{r_h})^2 \log ^4({r_h}) \alpha _{\theta _2}^2 \log ^8\left(9 a^2 {r_h}^4+{r_h}^6\right)}\nonumber\\
& & + {C_{\theta_1\theta_2}}^{(1)}\Biggr\}\Biggr)\beta\Biggr]\nonumber\\
G_{\theta_1y} & = & G_{\theta_1y}^{\rm MQGP}\Biggl[1 + \frac{\kappa_{\theta_1y} \beta  {g_s} \kappa_b  M 16\sin^4\left(\frac{\psi_0}{2}\right) {r_h} \beta  N^{-\alpha }}{ {N_f}^5 \sin^2\phi_{20}
   (r-{r_h})^3 \log ^{\frac{9}{2}}({r_h}) \alpha _{\theta _1}^4 \alpha _{\theta _2}^3}\Biggr]\nonumber\\
G_{\theta_1z} & = & G_{\theta_1z}^{\rm MQGP}\Biggl[1 + \frac{\kappa_{\theta_1z} \beta  {g_s}^{15/2} \kappa_b  M \sin^4\left(\frac{\psi_0}{2}\right) {r_h} \beta  N^{-\alpha }}{ {g_s}^{13/2} {N_f}^5
   \sin^2\phi_{20} (r-{r_h})^3 \log ^{\frac{9}{2}}({r_h}) \alpha _{\theta _1}^4 \alpha _{\theta _2}^3}\Biggr]\nonumber\\
G_{xy} & = & G_{xy}^{\rm MQGP}[1 + \frac{\kappa_{xy} \beta  \sqrt{{g_s}} (3 {g_s}-4) \kappa_b  \left(\frac{1}{N}\right)^{2/5} \sin^4\left(\frac{\psi_0}{2}\right)  \beta  N^{-\alpha
   }}{({g_s}-1)^2 {N_f}^6 \sin^2\phi_{20} \log (N) (r-{r_h}) \log ^{\frac{11}{2}}({r_h}) \alpha _{\theta _1} \alpha
   _{\theta _2}^2}]\nonumber\\
 G_{yz}  & = & G_{yz}^{\rm MQGP}\Biggl[1 + \frac{\kappa_{yz} \beta  {g_s} \kappa_b ^2 \log N  M  \sin^2\left(\frac{\psi_0}{2}\right) {r_h}^3 \beta  N^{-2 \alpha -\frac{11}{10}} \log
   ^2({r_h})}{ ({g_s}-1) {N_f}^5 \sin^2\phi_{20} (r-{r_h})^3 \alpha _{\theta _1}^3 \alpha _{\theta _2}^3}\Biggr]\nonumber\\
G_{xz} & = & G_{xz}^{\rm MQGP}\Biggl[1 + \Biggl(C_{xz}^{(1)} -\frac{\kappa_{xz} \tilde{\Sigma}_2\sqrt{2} \pi ^{11/2}  \sqrt{{g_s}} (3 {g_s}-4) \kappa_b  \sin^4\left(\frac{\psi_0}{2}\right) \beta  N^{-\alpha -\frac{2}{5}}
  }{ ({g_s}-1)^2 {N_f}^6 \sin^2\phi_{20}
   (r-{r_h}) \log ^{\frac{13}{2}}({r_h}) \alpha _{\theta _1} \alpha _{\theta _2}^2}\Biggr)\beta\Biggr],\nonumber\\
& &
\end{eqnarray}
}
where $\kappa_{\theta_1\theta_2,\ \theta_1y,\ \theta_1z,\ xy,\ xz,\ yz}\ll1,\ \kappa_{yy}\sim{\cal O}(1)$ and $\tilde{\Sigma_2}$ is defined in (\ref{Sigmatilde_2-def}) and $\alpha\in\mathbb{Z}^+$ appearing via (\ref{ansatz-b}). Analogous to working near the $\psi=2n\pi$-coordinate patches, the explicit dependence on $\theta_{1,2}$ of the ${\cal M}$-theory metric components up to ${\cal O}(\beta)$, using (\ref{Ouyang-theta10-theta20}), is effected by the replacemements:
$\alpha_{\theta_1}\rightarrow N^{\frac{1}{5}}\sin\theta_{1},\ \alpha_{\theta_2}\rightarrow N^{\frac{3}{10}}\sin\theta_{2}$ in (\ref{ M-theory-metric-psi-neq-2npi-patch}). Also, see footnote 5. The Physics implication of (\ref{ M-theory-metric-psi-neq-2npi-patch}) is similar to (\ref{IR-beta-N-suppressed-logrh-rh-neg-exp-enhanced}) arising from (\ref{ M-theory-metric-psi=2npi-patch}).
\newpage

\section{Physics Lessons Learnt - IR-enhancement large-$N$/Planckian-suppresion competition and When ${\cal O}(l_p^6)$ Is (Not) Enough}

Based on the results of this paper and its applications as discussed in detail in \cite{Vikas+Gopal+Aalok}, \cite{Gopal+Vikas+Aalok}, we now discuss the Physics lessons learnt as a consequence of working out the ${\cal O}(R^4)
/{\cal  O}(l_p^6)$ corrections to the ${\cal M}$-theory dual of large-$N$ thermal QCD-like theories.

The main Physics-related take-away of Section {\bf 3}, e.g. from (\ref{ M-theory-metric-psi=2npi-patch}), can be abstracted from the following table:
\begin{table}[h]
\begin{center}
\begin{tabular}{|c|c|c|c|} \hline
S. No. & $G^{\cal M}_{MN}$ & IR Enhancement Factor & $N$ Suppression \\
& &  $\frac{\left(\log {\cal R}_h\right)^m}{{\cal R}_h^n}, m,n\in\mathbb{Z}^+$ & Factor \\
& & in the ${\cal O}(R^4)$ Correction & in the ${\cal O}(R^4)$ Correction   \\ \hline
1 & $G^{\cal M}_{\mathbb{R}^{1,3}}$ & $\log {\cal R}_h$ & ${N^{-\frac{9}{4}}}$ \\  \hline
2 & $G^{\cal M}_{rr, \theta_1x}$ & 1 & ${N^{-\frac{8}{15}}}$ \\  \hline
3 & $G^{\cal M}_{\theta_1z,\theta_2x}$ & ${{\cal R}_h^{-5}}$ & ${N^{-\frac{7}{6}}}$ \\  \hline
4 &  $G^{\cal M}_{\theta_2y}$ & $\log {\cal R}_h$ & ${N^{-\frac{7}{5}}}$ \\ \hline
5 &  $G^{\cal M}_{\theta_2z}$ & $\log {\cal R}_h$ & ${N^{-\frac{7}{6}}}$
\\ \hline
6 &  $G^{\cal M}_{xy}$ & $\log {\cal R}_h$ & ${N^{-\frac{21}{20}}}$ \\  \hline
7 &  $G^{\cal M}_{xz}$ &  $\left(\log {\cal R}_h\right)^3$ & ${N^{-\frac{5}{4}}}$ \\ \hline
8 &  $G^{\cal M}_{yy}$ & $\log {\cal R}_h$ & ${N^{-\frac{7}{4}}}$ \\  \hline
9 &  $G^{\cal M}_{yz}$ & $\frac{\log {\cal R}_h}{{\cal R}_h^7}$ & ${N^{-\frac{29}{12}}}$ \\ \hline
10 &  $G^{\cal M}_{zz}$ & $\log {\cal R}_h$ & ${N^{-\frac{23}{20}}}$ \\ \hline
11  &  $G^{\cal M}_{x^{10}x^{10}}$ & $\frac{\log {\cal R}_h^3}{{\cal R}_h^2}$ & ${N^{-\frac{5}{4}}}$ \\ \hline
\end{tabular}
\end{center}
\caption{IR Enhancement vs. large-$N$ Suppression in ${\cal O}(R^4)$-Corrections in the M-theory Metric in the $\psi=2n\pi, n=0,1,2$ Patches; ${\cal R}_h \equiv \frac{r_h}{{\cal R}_{D5/\overline{D5}}}\ll1$, ${\cal R}_{D5/\overline{D5}}$ being the $D5-\overline{D5}$ separation}
\end{table}

 One notes that in the IR: $r = \chi r_h, \chi\equiv {\cal O}(1)$, and up to ${\cal O}(\beta)$:
\begin{equation}
\label{IR-beta-N-suppressed-logrh-rh-neg-exp-enhanced}
f_{MN} \sim \beta\frac{\left(\log {\cal R}_h\right)^{m}}{{\cal R}_h^n N^{\beta_N}},\ m\in\left\{0,1,3\right\},\ n\in\left\{0,2,5,7\right\},\
\beta_N>0.
\end{equation}
Now, $|{\cal R}_h|\ll1$. As estimated in \cite{Bulk-Viscosity}, $|\log {\cal R}_h|\sim N^{\frac{1}{3}}$, implying there is a competition between Planckian and large-$N$ suppression and infra-red enhancement arising from $m,n\neq0$ in (\ref{IR-beta-N-suppressed-logrh-rh-neg-exp-enhanced}). One could choose a heirarchy: $\beta\sim e^{-\gamma_\beta N^{\gamma_N}}, \gamma_\beta,\gamma_N>0: \gamma_\beta N^{\gamma_N}>7N^{\frac{1}{3}} + \left(\frac{m}{3} - \beta_N\right)\log N$ (ensuring that the IR-enhancement does not overpower Planckian suppression - we took the ${\cal O}(\beta)$ correction to $G^{\cal M}_{yz}$, which had the largest IR enhancement, to set a lower bound on $\gamma_{\beta,N}$/Planckian suppression). If $\gamma_\beta N^{\gamma_N}\sim7N^{\frac{1}{3}}$, then one will be required to go to a higher order in $\beta$. This hence answers the question, when one can truncate at ${\cal O}(\beta)$.

\section{Differential Geometry or (IR) $G$-Structure Torsion Classes of Non-Supersymmetric String/${\cal M}$-Theory duals Including the ${\cal O}(R^4)$ Corrections }

The use of $G$-structure torsion classes is a very useful tool for classifying, specially non-K\"{a}hler geometries. A complete classification of the $SU(n)$ structures  relevant to non-supersymmetric string vacua, does not exist \cite{Magdalena-2013}. In the literature, in the context of $SU(3)$-structure manifolds, classes of maximally symmetric non-supersymmetric vacua that break supersymmetry in a controllable way, have been constructed, e.g. \cite{GKP} wherein the first vacuum of this type was obtained by compactifying type IIB/F-theory with $O3$ planes
on conformally CY manifolds  ($SU(3)$-structure manifolds with
$W_1 = W_2 = W_3=0$ and $3W_4 = 2 W_5$);  type II vacua of this type were studied in \cite{Camara+Grana} and classified using calibrations in
\cite{Luest-et-al_2008}, and similar solutions in heterotic string theory were obtained in \cite{Luest-et-al_2010} - see \cite{thesis-Held} for $G_2$ structures relevant to non-supersymmetric vacua in heterotic(${\cal M}$-)SUGRA.

{\it A classification of $SU(3)/G_2/Spin(7)/Spin(4)$ structures relevant to {\it non-supersymmetric} (UV-complete) string theoretic dual of large-$N$ thermal QCD-like theories, and its ${\cal M}$-theory uplift, has been missing in the literature. This is what we aim at achieving in this section.}

Using the results for Ricci scalar of $M_6(r,\theta_1,\theta_2,\phi_1,\phi_2,\psi), M_7(r,\theta_1,\theta_2,\phi_1,\phi_2,\psi,x^{10})$,\\ $M_8(x^0,r,\theta_1,\theta_2,\phi_1,\phi_2,\psi,x^{10})$ that figure in the string/${\cal M}$-theory dual of large-$N$ thermal QCD-like theories in this work, in terms of the:
\begin{enumerate}
\item
 {\bf $SU(3)$-structure torsion classes} \cite{Bedulli+Vezzoni}, it is observed:
\begin{eqnarray}
\label{Ricci-scalar-SU3}
& & \hskip -0.3in  R(M_6(r,\theta_1,\theta_2,\phi_1,\phi_2,\psi)) = 15 |W_1|^2 - |W_2|^2 - |W_3|^2 + 8\langle W_5,W_4 \rangle - 2|W_4|^2 + 4 d*\left(W_4 + W_5\right) \neq 0\nonumber\\
& &
\end{eqnarray}
($\langle , \rangle$ denoting Mukai pairing),
\item
{\bf $G_2$-structure torsion classes} \cite{Bryant-G2}, it is observed::
\begin{eqnarray}
\label{Ricci-scalar-G2}
& & \hskip -0.3in R(M_7(r,\theta_1,\theta_2,\phi_1,\phi_2,\psi,x^{10})) = 12 \delta W_7 + \frac{21}{8}W_1^2 + 30 |W_7|^2 - \frac{1}{2}|W_{14}|^2 - \frac{1}{2}|W_{27}|^2,\nonumber\\
& &
\end{eqnarray}

\item
{\bf $Spin(7)$-structure torsion classes} \cite{Spin7-Ivanov}, it is observed:
\begin{eqnarray}
\label{Ricci-scalar-Spin(7)}
R(M_8(x^0,r,\theta_1,\theta_2,\phi_1,\phi_2,\psi,x^{10})) = \frac{49}{18}||\theta||^2 - \frac{1}{12}||T||^2 + \frac{7}{2}\delta\theta,
\end{eqnarray}
where: $\theta = \frac{1}{7}*\left(\delta\Phi\wedge\Phi\right), T = -\delta\Phi - \frac{7}{6}*\left(\theta\wedge\Phi\right)$, $\Phi$ being the $Spin(7)$ fundamental four-form. Note, that the eight-fold $M_8(x^0,r,\theta_1,\theta_2,\phi_1,\phi_2,\psi,x^{10})$ admits a $Spin(7)$ structure if
$p_1^2(M_8) - 4 p_2(M_8) + 8 \chi(M_8) = 0$ \cite{Spin7-Ivanov}, $p_a(M_8)$ being the $a$-th Pontryagin class of $M_8$. Given that $M_8$ could be thought of as elliptic/$T^2(x^0,x^{10})$ fibration over $M_6(r,\theta_1,\theta_2,\phi_1,\phi_2,\psi)$, using the Kunneth formula one sees that
$\chi(M_8)=\chi(T^2)\chi(M_6)=0$. In the delocalized limit, also modifying the arguments of \cite{MQGP} (which showed $X_8=0$ as $p_1^2(M_{11}=\mathbb{R}^3\times M_8) = p_2(M_{11}) = 0$), one can show that the $p_1^2(M_8) = p_2(M_8)=0$.
\end{enumerate}

In this section, we will derive in the IR near the $\psi=2n\pi, n=0, 1, 2$-branches the non-zero $SU(3)$-structure torsion classes  of the six-fold relevant to the type IIA mirror, the $G_2$-structure torsion classes of the seven-fold and the $SU(4)$- and $Spin(7)$-structure torsion classes of the eight-fold relevant to the ${\cal M}$-Theory uplift of the type IIA mirror. As in Section {\bf 3}, for simplicity, we work near the Ouyang embedding (assuming a very small Ouyang embedding parameter). But as mentioned later, using (\ref{e12}) - (\ref{e^6}), based on arguments of \cite{SYZ-free-delocalization}, one can see that the results of Table 1 in Section {\bf  5}, will still remain valid for arbitrary $\theta_{1,2}$. For arbitrary $\psi$, using the results of subsection {\bf 3.2}, it is expected that the results of Table 1 in Section {\bf 5} will go through, though the co-frames will be considerably modified. We postpone this discussion to a later work.

\subsection{$SU(3)$-Structure Torsion Classes of the Type IIA Mirror}

Generically for $SU(n>2)$-structures, the intrinsic torsion decomposes into five torsion classes $W_{i=1,...,5}$ \cite{Chiossi-Salamon}, i.e.,
\begin{equation}
\label{T_SU(n)}
T \in \Lambda^1\otimes su(n)^\perp = \bigoplus_{i=1}^5W_i.
\end{equation}
 The adjoint representation $15$ of $SO(6)$  decomposes under  $SU(3)$ as
$15 = 1 + 8 + 3 + \bar{3}$. Thus, $su(3)^\perp\sim  1 \oplus 3 \oplus \bar{3}$, and:
\begin{eqnarray*}
& & T \in \Lambda^1 \otimes su(3)^\perp = (3 \oplus \bar{3}) \otimes (1 \oplus 3 \oplus \bar{3)}
\nonumber\\
& & = (1 \oplus 1) \oplus (8 \oplus 8) \oplus (6 \oplus \bar{6}) \oplus (3 \oplus \bar{3}) \oplus (3 \oplus \bar{3})^\prime\nonumber\\
& & \equiv W_1\oplus W_2\oplus W_3\oplus W_4\oplus W_5.
\end{eqnarray*}
The $SU(3)$ structure torsion classes  can be defined in terms of J, $ \Omega $, dJ, $ d{\Omega}$ and
the contraction operator  $\lrcorner : {\Lambda}^k T^{\star} \otimes {\Lambda}^n
T^{\star} \rightarrow {\Lambda}^{n-k} T^{\star}$,  $J$ being given by:
$$ J  =  e^1 \wedge e^2 + e^3 \wedge e^4 + e^5 \wedge e^6$$ (the metric being understood to be given in terms of the coframes as: $ds_6^2 = \sum_{i=1}^6 \left(de^a\right)^2$),
and the (3,0)-form $ \Omega $ being given by
$$ \Omega  =  ( e^1 + ie^2) \wedge (e^3 +
ie^4) \wedge (e^5 + ie^6). $$
\begin{eqnarray}
\label{W_12345}
& & W_1=W_1^+ + W_1^-\ with:\nonumber\\
& &  d {\Omega}_+ \wedge J = {\Omega}_+ \wedge dJ = W_1^+ J\wedge J\wedge J,\nonumber\\
& & d {\Omega}_- \wedge J = {\Omega}_- \wedge dJ = W_1^- J \wedge J \wedge J;\nonumber\\
& & (d{\Omega}_+)^{(2,2)}=W_1^+ J \wedge J + W_2^+ \wedge J,\nonumber\\
& & (d{\Omega}_-)^{(2,2)}=W_1^- J \wedge J + W_2^- \wedge J;\nonumber\\
& & W_3=dJ^{(2,1)} -[J \wedge W_4]^{(2,1)},\nonumber\\
& & W_4 =\frac{1}{2} J\lrcorner dJ,\nonumber\\
& & W_5  = \frac{1}{2} {\Omega}_+\lrcorner d{\Omega}_+.
\end{eqnarray}

We now proceed to work out the coframes $\left\{e^a\right\}$. This brings us to the next lemma:

\noindent {\it Lemma 5}: The non-zero components of the type IIA metric near $\psi=0,2\pi,4\pi$ coordinate patch, e.g. near  (\ref{Ouyang-theta10-theta20}), obtained from the ${\cal M}$ theory metric of  Sec. {\bf 3} inclusive of the ${\cal O}(R^4)$ corrections are given by:
{\scriptsize
\begin{eqnarray}
\label{GIIA+beta}
& & G^{\rm IIA}_{\theta_1x} = -\frac{{g_s}^{7/4} \log N  M N^{11/20} {N_f}
   \left(r^2-3 a^2\right) \log (r)}{3 \sqrt{2} \pi ^{5/4} r^2 \alpha
   _{\theta _1} \alpha _{\theta _2}^2} -\frac{\beta  {{\cal C}_{\theta_1x}}^{(1)} {g_s}^{7/4} \log N  M N^{11/20}
   {N_f} \left({r_h}^2-3 a^2\right) \log ({r_h})}{3 \sqrt{2}
   \pi ^{5/4} {r_h}^2 \alpha _{\theta _1} \alpha _{\theta
   _2}^2}\nonumber\\
   & & G^{\rm IIA}_{\theta_1z} = \frac{{g_s}^{7/4}
   \log N  M N^{3/20} {N_f} \alpha _{\theta _1} \left(r^2-3
   a^2\right) \log (r)}{2 \sqrt{2} \pi ^{5/4} r^2 \alpha _{\theta _2}^2} + \frac{\beta  {C_{\theta_1z}}^{(1)} {g_s}^{7/4} \log N  M N^{3/20}
   {N_f} \alpha _{\theta _1} \left(r^2-3 a^2\right) \log (r)}{2
   \sqrt{2} \pi ^{5/4} r^2 \alpha _{\theta _2}^2}\nonumber\\
   & & G^{\rm IIA}_{\theta_2x} = \frac{{g_s}^{7/4} M N^{13/20} {N_f} \log (r) \left(36
   a^2 \log (r)+r\right)}{3 \sqrt{2} \pi ^{5/4} r \alpha _{\theta _2}^3} + \frac{\beta  {{\cal C}_{\theta_2x}}^{(1)} {g_s}^{7/4} M N^{13/20} {N_f} \log (r)
   \left(36 a^2 \log (r)+r\right)}{3 \sqrt{2} \pi ^{5/4} r \alpha _{\theta
   _2}^3}\nonumber\\
   & & G^{\rm IIA}_{\theta_2y} = \frac{\sqrt{2} \sqrt[4]{\pi } \sqrt[4]{{g_s}} N^{7/20} \alpha _{\theta
   _2}}{9 \alpha _{\theta _1}^2}-\frac{\beta  {{\cal C}_{\theta_2y}}(1)
   {g_s}^{7/4} M \sqrt[4]{N} \sqrt{\frac{1}{{N_f}^{4/3}}}
   {N_f}^{5/3} \alpha _{\theta _1}^2 \log (r) \left(36 a^2 \log
   (r)+r\right)}{2 \sqrt{2} \pi ^{5/4} r \alpha _{\theta _2}^3}\nonumber\\
   & & G^{\rm IIA}_{\theta_2z} = -\frac{{g_s}^{7/4} M \sqrt[4]{N} {N_f} \alpha
   _{\theta _1}^2 \log (r) \left(36 a^2 \log (r)+r\right)}{2 \sqrt{2} \pi
   ^{5/4} r \alpha _{\theta _2}^3}  - \frac{\beta  {{\cal C}_{\theta\psi}}^{(1)} {g_s}^{7/4} M \sqrt[4]{N}
   {N_f} \alpha _{\theta _1}^2
   \log (r) \left(36 a^2 \log (r)+r\right)}{2 \sqrt{2} \pi ^{5/4} r \alpha
   _{\theta _2}^3}\nonumber\\
   & & G^{\rm IIA}_{xx} = 1-\frac{27 b^{10} \left(6 b^2+1\right) \left(9 b^2+1\right)^3 \beta  M
   \left(\frac{1}{N}\right)^{5/4} \left(19683 \sqrt{6} \alpha _{\theta
   _1}^6+6642 \alpha _{\theta _2}^2 \alpha _{\theta _1}^3-40 \sqrt{6}
   \alpha _{\theta _2}^4\right) \log ^3({r_h})}{2 \pi  \left(3
   b^2-1\right)^5 {N_f} {r_h}^2 \alpha _{\theta _2}^3 \left(6 b^2
   \log N +\log N \right)^4}\nonumber\\
   & & G^{\rm IIA}_{yy} =  1 + \frac{27 \left(9 b^2+1\right)^3 \beta  b^{10} M
   \left(\frac{1}{N}\right)^{5/4} \left(-19683 \sqrt{6} \alpha _{\theta
   _1}^6-6642 \alpha _{\theta _2}^2 \alpha _{\theta _1}^3+40 \sqrt{6}
   \alpha _{\theta _2}^4\right) \log ^3({r_h})}{2 \pi  \left(3
   b^2-1\right)^5 \left(6 b^2+1\right)^3 \log N ^4 {N_f}
   {r_h}^2 \alpha _{\theta _2}^3}\nonumber\\
   & & G^{\rm IIA}_{zz} = \frac{2
   N^{3/5}}{27 \alpha _{\theta _2}^2} + \frac{2 \beta  C_{zz}^{(1)} N^{3/5}}{27 \alpha _{\theta _2}^2}\nonumber\\
   & & G^{\rm IIA}_{xy}  =  \frac{2 \sqrt{\frac{2}{3}}
   N^{7/10}}{9 \alpha _{\theta _1}^2 \alpha _{\theta _2}} + \frac{2 \sqrt{\frac{2}{3}} \beta  {{\cal C}_{\phi_1\phi_2}}^{(1)} N^{7/10}}{9 \alpha
   _{\theta _1}^2 \alpha _{\theta _2}}\nonumber\\
   & & G^{\rm IIA}_{xz} = -\frac{4 N}{81 \alpha _{\theta _1}^2 \alpha _{\theta
   _2}^2} + \frac{2 b^{10} \left(9 b^2+1\right)^3 \beta  M \sqrt[4]{\frac{1}{N}}
   \left(19683 \sqrt{6} \alpha _{\theta _1}^6+6642 \alpha _{\theta _2}^2
   \alpha _{\theta _1}^3-40 \sqrt{6} \alpha _{\theta _2}^4\right) \log
   ^3({r_h})}{3 \pi  \left(3 b^2-1\right)^5 \left(6 b^2+1\right)^3
   \log N ^4 {N_f} {r_h}^2 \alpha _{\theta _1}^2 \alpha
   _{\theta _2}^5}\nonumber\\
   & & G^{\rm IIA}_{yz} =  -\frac{\sqrt{\frac{2}{3}} N^{3/10}}{3 \alpha _{\theta _2}} - \frac{\sqrt{\frac{2}{3}} \beta  {{\cal C}_{\phi_2\psi}}^{(1)} N^{3/10}}{3 \alpha
   _{\theta _2}}.
\end{eqnarray}
}

To work out the co-frames corresponding to (\ref{GIIA+beta}), one diagonalizes
$G_{mn}(r,\theta_{1,2},\phi_{1,2},\psi)$ or equivalently $G_{\tilde{m}\tilde{n}}(\theta_{1,2},\phi_{1,2},\psi)$ for which one needs to solve the following secular equation - a quintic:
\begin{equation}
\label{quintic-secular-D5}
P(x) \equiv x^5 + A x^4 + B x^3 + C x^2 + F x + G= 0,
\end{equation}
where:
\begin{eqnarray}
\label{ABCFG}
& & A = -\frac{2 N^{3/5} (\beta  {\cal C}_{zz}^{(1)}+1)}{27 \alpha _{\theta _2}^2},\nonumber\\
& & B = -\frac{16 N^2}{6561 \alpha _{\theta _1}^4 \alpha _{\theta _2}^4,}\nonumber\\
& & C = \frac{16 N^2 (\beta  {\cal C}_{zz}^{(1)}-2 \beta  {{\cal C}_{yz}}^{(1)})}{6561 \alpha _{\theta _1}^4 \alpha _{\theta _2}^4}\nonumber\\
& & F = \frac{32 \sqrt{\pi } \sqrt{{g_s}} N^{27/10}}{531441 \alpha _{\theta _1}^8 \alpha _{\theta _2}^2},\nonumber\\
& & G = -\frac{2 {g_s}^4 {\log N}^2 M^2 N^{12/5} {N_f}^2  }{19683 \pi ^2
   r^4 {r_h}^4 \alpha _{\theta _1}^6 \alpha _{\theta _2}^4}\nonumber\\
   & & \times \left({r_h}^2 \left(3 a^2-r^2\right) \log (r) (\beta  ({{\cal C}_{zz}}^{(1)}-2
   {{\cal C}_{\theta_1z}}^{(1)})-1)+\beta  {\cal C}_{\theta_1x}^{(1)} r^2 \left(3 a^2-{r_h}^2\right) (\beta  {{\cal C}_{zz}}^{(1)}+1) \log ({r_h})\right)\nonumber\\
   & & \times \left(\beta
{{\cal C}_{\theta_1x}}^{(1)} r^2 \left(3 a^2-{r_h}^2\right) \log
   ({r_h})+{r_h}^2 \left(3 a^2-r^2\right) \log (r)\right)
\end{eqnarray}

Using Umemura's result \cite{Umemura} on expressing the roots of an algebraic polynomial of  degree $n$ in terms of Siegel theta functions
of genus $g(>1)=[(n+2)/2]$ :  $\theta\left[\begin{array}{c} \mu\\
\nu
\end{array}\right](z,\Omega)$ for $\mu,\nu\in{\bf R}^g, z\in {\bf C}^g$ and $\Omega$ being a complex symmetric
$g\times g$ period matrix of the hyperelliptic curve $Y^2 = P({\cal Z})$ with $Im(\Omega)>0$, and defined as follows:
$$
\theta\left[\begin{array}{c} \mu\\
\nu
\end{array}\right](z,\Omega)=\sum_{n\in{\bf Z}^g}e^{i\pi(n+\mu)^T\Omega(n+\mu)+2i\pi(n+\mu)^T(z+\nu)}.$$
Hence for a quintic, one needs to use Siegel theta functions of genus three. The period matrix $\Omega$ will be defined as follows:
$$\Omega_{ij}=\sigma^{ik}\rho_{kj}$$
where $$\sigma_{ij}\equiv\oint_{A_j}d{\cal Z} \frac{{\cal Z}^{i-1}}{\sqrt{{\cal Z}({\cal Z}-1)P({\cal Z})}}$$ and
$$\rho_{ij}\equiv\oint_{B_j}\frac{{\cal Z}^{i-1}}{\sqrt{{\cal Z}({\cal Z}-1)({\cal Z}-2)P({\cal Z})}},$$
$\{A_i\}$ and $\{B_i\}$ being a canonical basis of cycles satisfying: $A_i\cdot A_j=B_i\cdot B_j=0$ and
$A_i\cdot B_j=\delta_{ij}$; $\sigma^{ij}$ are normalization constants determined by:
$\sigma^{ik}\sigma_{kj} = \delta^i_j$. Umemura's result then is that a root:
$$\hskip-0.3in\frac{1}{2\left(\theta\left[\begin{array}{ccc}
\frac{1}{2} & 0 & 0  \\
0 & 0 & 0   \end{array}\right](0,\Omega)\right)^4
\left(\theta\left[\begin{array}{ccc}
\frac{1}{2} & \frac{1}{2} & 0  \\
0 & 0 & 0  \end{array}\right](0,\Omega)\right)^4}$$
$$\hskip-0.3in\times\Biggl[\left(\theta\left[\begin{array}{ccc}
\frac{1}{2} & 0 & 0  \\
0 & 0 & 0   \end{array}\right](0,\Omega)\right)^4\left(\theta\left[\begin{array}{ccc}
\frac{1}{2} & \frac{1}{2} & 0 \\
0 & 0 & 0  \end{array}\right](0,\Omega)\right)^4$$
$$\hskip-0.3in+ \left(\theta\left[\begin{array}{ccc}
0 & 0 & 0  \\
0 & 0 & 0   \end{array}\right](0,\Omega)\right)^4
\left(\theta\left[\begin{array}{ccc}
0 & \frac{1}{2} &  0 \\
0 & 0 & 0   \end{array}\right](0,\Omega)\right)^4$$
$$\hskip-0.3in- \left(\theta\left[\begin{array}{ccc}
0 & 0 & 0  \\
\frac{1}{2} & 0 & 0   \end{array}\right](0,\Omega)\right)^4
\left(\theta\left[\begin{array}{ccc}
0 & \frac{1}{2} & 0  \\
\frac{1}{2} & 0 & 0  \end{array} \right](0,\Omega)\right)^4\Biggr].$$
However, using the results of
\cite{Zhivkov}, one can express the roots of a quintic in terms of derivatives of genus-two Siegel theta functions as follows:
{\scriptsize
$$\hskip-0.6in x_0 = \left[\frac{\sigma_{22}\frac{d}{dz_1}\theta\left[\begin{array}{cc}
\frac{1}{2}&\frac{1}{2} \\
0&\frac{1}{2}
\end{array}\right]\left((z_1,z_2),\Omega\right)
- \sigma_{21}\frac{d}{dz_2}\theta\left[\begin{array}{cc}
\frac{1}{2}&\frac{1}{2} \\
0&\frac{1}{2}
\end{array}\right]\left((z_1,z_2),\Omega\right) }
{\xi \theta\left[\begin{array}{cc}
\frac{1}{2}&0 \\
0&\frac{1}{2}
\end{array}\right]\left((z_1,z_2),\Omega\right)
\theta\left[\begin{array}{cc}
\frac{1}{2}&0 \\
0&0
\end{array}\right]\left((z_1,z_2),\Omega\right)
\theta\left[\begin{array}{cc}
0&\frac{1}{2} \\
\frac{1}{2}&0
\end{array}\right]\left((z_1,z_2),\Omega\right)
\theta\left[\begin{array}{cc}
0 & \frac{1}{2} \\
0&0
\end{array}\right]\left((z_1,z_2),\Omega\right)
}\right]_{z_1=z_2=0},$$

$$\hskip-0.6in x_1 = \left[\frac{\sigma_{22}\frac{d}{dz_1}\theta\left[\begin{array}{cc}
0&\frac{1}{2} \\
0&\frac{1}{2}
\end{array}\right]\left((z_1,z_2),\Omega\right)
- \sigma_{21}\frac{d}{dz_2}\theta\left[\begin{array}{cc}
0&\frac{1}{2} \\
0&\frac{1}{2}
\end{array}\right]\left((z_1,z_2),\Omega\right) }
{\xi \theta\left[\begin{array}{cc}
0&0 \\
0&\frac{1}{2}
\end{array}\right]\left((z_1,z_2),\Omega\right)
\theta\left[\begin{array}{cc}
0&0 \\
0&0
\end{array}\right]\left((z_1,z_2),\Omega\right)
\theta\left[\begin{array}{cc}
\frac{1}{2}&\frac{1}{2} \\
0&0
\end{array}\right]\left((z_1,z_2),\Omega\right)
\theta\left[\begin{array}{cc}
0 & \frac{1}{2} \\
\frac{1}{2}&0
\end{array}\right]\left((z_1,z_2),\Omega\right)
}\right]_{z_1=z_2=0},$$

$$\hskip-0.6in x_2 = \left[\frac{\sigma_{22}\frac{d}{dz_1}\theta\left[\begin{array}{cc}
0&\frac{1}{2} \\
\frac{1}{2}&\frac{1}{2}
\end{array}\right]\left((z_1,z_2),\Omega\right)
- \sigma_{21}\frac{d}{dz_2}\theta\left[\begin{array}{cc}
0&\frac{1}{2} \\
\frac{1}{2}&\frac{1}{2}
\end{array}\right]\left((z_1,z_2),\Omega\right) }
{\xi \theta\left[\begin{array}{cc}
0&0 \\
\frac{1}{2}&\frac{1}{2}
\end{array}\right]\left((z_1,z_2),\Omega\right)
\theta\left[\begin{array}{cc}
0&0 \\
0&0
\end{array}\right]\left((z_1,z_2),\Omega\right)
\theta\left[\begin{array}{cc}
\frac{1}{2}&\frac{1}{2} \\
0&0
\end{array}\right]\left((z_1,z_2),\Omega\right)
\theta\left[\begin{array}{cc}
0 & \frac{1}{2} \\
0&0
\end{array}\right]\left((z_1,z_2),\Omega\right)
}\right]_{z_1=z_2=0},$$

$$\hskip-0.6in x_3 = \left[\frac{\sigma_{22}\frac{d}{dz_1}\theta\left[\begin{array}{cc}
\frac{1}{2} &0 \\
\frac{1}{2}&\frac{1}{2}
\end{array}\right]\left((z_1,z_2),\Omega\right)
- \sigma_{21}\frac{d}{dz_2}\theta\left[\begin{array}{cc}
\frac{1}{2} &0 \\
\frac{1}{2}&\frac{1}{2}
\end{array}\right]\left((z_1,z_2),\Omega\right) }
{\xi \theta\left[\begin{array}{cc}
\frac{1}{2} &\frac{1}{2}  \\
\frac{1}{2}&\frac{1}{2}
\end{array}\right]\left((z_1,z_2),\Omega\right)
\theta\left[\begin{array}{cc}
0&0 \\
0&0
\end{array}\right]\left((z_1,z_2),\Omega\right)
\theta\left[\begin{array}{cc}
0&0 \\
\frac{1}{2} &0
\end{array}\right]\left((z_1,z_2),\Omega\right)
\theta\left[\begin{array}{cc}
\frac{1}{2}  & 0 \\
0&0
\end{array}\right]\left((z_1,z_2),\Omega\right)
}\right]_{z_1=z_2=0},$$

$$\hskip-0.6in x_4 = \left[\frac{\sigma_{22}\frac{d}{dz_1}\theta\left[\begin{array}{cc}
\frac{1}{2} &0 \\
\frac{1}{2}&0
\end{array}\right]\left((z_1,z_2),\Omega\right)
- \sigma_{21}\frac{d}{dz_2}\theta\left[\begin{array}{cc}
\frac{1}{2} &0 \\
\frac{1}{2}&0
\end{array}\right]\left((z_1,z_2),\Omega\right) }
{\xi \theta\left[\begin{array}{cc}
\frac{1}{2}&\frac{1}{2} \\
\frac{1}{2}&\frac{1}{2}
\end{array}\right]\left((z_1,z_2),\Omega\right)
\theta\left[\begin{array}{cc}
0&0 \\
0&\frac{1}{2}
\end{array}\right]\left((z_1,z_2),\Omega\right)
\theta\left[\begin{array}{cc}
0&0 \\
\frac{1}{2}&\frac{1}{2}
\end{array}\right]\left((z_1,z_2),\Omega\right)
\theta\left[\begin{array}{cc}
\frac{1}{2} & 0 \\
0&\frac{1}{2}
\end{array}\right]\left((z_1,z_2),\Omega\right)
}\right]_{z_1=z_2=0},$$
}
where:
{\footnotesize
$$\hskip -0.4in\xi \equiv -A\pi^2\sum_{m=1}^5 \left[\frac{\sigma_{22}\frac{d}{dz_1}\theta\left[\eta_m\right]\left((z_1,z_2),\Omega\right)
- \sigma_{21}\frac{d}{dz_2}\theta\left[\eta_m\right]\left((z_1,z_2),\Omega\right) }
{\frac{d}{dz_1}\theta\left[\eta_m\right]\left((z_1,z_2),\Omega\right)\frac{d}{dz_2}\theta\left[\eta_6\right]\left((z_1,z_2),\Omega\right) - \frac{d}{dz_2}\theta\left[\eta_m\right]\left((z_1,z_2),\Omega\right)\frac{d}{dz_1}\theta\left[\eta_6\right]\left((z_1,z_2),\Omega\right)  }\right]_{z_1=z_2=0},$$
\begin{eqnarray*}
\label{eta_m}
& & \hskip -0.7in \left[\eta_1\right] \equiv \left[\begin{array}{cc} \frac{1}{2} & \frac{1}{2} \\ 0 & \frac{1}{2} \end{array}\right],\ \ \ \left[\eta_2\right] \equiv \left[\begin{array}{cc} 0 & \frac{1}{2} \\ 0 & \frac{1}{2} \end{array}\right],\ \ \  \left[\eta_3\right] \equiv \left[\begin{array}{cc}0 & \frac{1}{2} \\  \frac{1}{2} & \frac{1}{2} \end{array}\right],\ \ \ \left[\eta_4\right] \equiv \left[\begin{array}{cc} \frac{1}{2} & 0 \\  \frac{1}{2} & \frac{1}{2} \end{array}\right],\ \ \  \left[\eta_5\right] \equiv \left[\begin{array}{cc} \frac{1}{2} & 0 \\  \frac{1}{2} & 0 \end{array}\right],\ \ \  \left[\eta_6\right] \equiv \left[\begin{array}{cc} \frac{1}{2} & \frac{1}{2} \\ \frac{1}{2} & 0 \end{array}\right].\nonumber\\
\end{eqnarray*}
}
The symmetric period matrix corresponding to the hyperelliptic
curve $w^2=P(z)$ is given by:
$$\hskip-0.3in\left(\begin{array}{cc}
\Omega_{11} & \Omega_{12} \\
\Omega_{12} & \Omega_{22}
\end{array}\right)=\frac{1}{\sigma_{11}\sigma_{22}-\sigma_{12}\sigma_{21}}\left(\begin{array}{cc}
\sigma_{22} & -\sigma_{12} \\
-\sigma_{21} & \sigma_{11}
\end{array}\right)\left(\begin{array}{cc}
\rho_{11} & \rho_{12} \\
\rho_{21} & \rho_{22}
\end{array}\right),$$
where $\sigma_{ij}=\int_{{\cal Z}_*{A_j}}\frac{{\cal Z}^{i-1}d{\cal Z}}{\sqrt{P({\cal Z})}}$ and
$\rho_{ij}=\int_{z_*{B_j}}\frac{{\cal Z}^{i-1}d{\cal Z}}{\sqrt{P({\cal Z})}}$ where ${\cal Z}$ maps the $A_i$ and $B_j$ cycles to the
${\cal Z}-$plane.  However, both results are not amenable to actual calculations due to the non-trivial period matrix computations.

The quintic (\ref{quintic-secular-D5}) is solved using the Kiepert's algorithm described very nicely in the wonderful book \cite{Bruce-King-Beyond-Quartic}. The details of the same are given in Appendix {\bf B}. Utilising the results of Appendix {\bf B}, we will now work out the $SU(3)$ torsion classes of $M_6(r,\theta_{1,2},\phi_{1,2},\psi)$ and prove the following lemma:

{\it Lemma 6}: In the neighborhood of $(\theta_{10}=\frac{\alpha_{\theta_1}}{N^{\frac{1}{5}}}, \theta_{20} = \frac{\alpha_{\theta_{20}}}{N^{\frac{3}{10}}}, \psi=2n\pi), n=0, 1, 2$, the $SU(3)$-structure torsion classes $W_{i=1,2,3,4,5}^{M_6}\neq0$ (implying $M_6$ is a non-complex manifolds) with $W_4 \sim W_5$.

\noindent{\it Proof}:  In the neighborhood of $(\theta_{10}=\frac{\alpha_{\theta_1}}{N^{\frac{1}{5}}}, \theta_{20} = \frac{\alpha_{\theta_{20}}}{N^{\frac{3}{10}}}, \psi=2n\pi), n=0, 1, 2$, in the MQGP limit (\ref{MQGP_limit}), inverting the co-frames of Appendix {\bf C} :
\begin{eqnarray}
\label{Theta_ia+X_a+Y_a+Z_a}
& & d\theta_{i=1/2} = \sum_{a=2}^6\Theta_{ia}e^a,\nonumber\\
& & dx = \sum_{a=2}^6{\cal X}_ae^a,\nonumber\\
& & dy = \sum_{a=2}^6{\cal Y}_ae^a,\nonumber\\
& & dz = \sum_{a=2}^6{\cal Z}_ae^a,
\end{eqnarray}
and (\ref{e12}) - (\ref{e^6}):
\begin{eqnarray}
\label{e^am}
& & e^a = e^{a\theta_1}(r)d\theta_1 + e^{a\theta_2}(r)d\theta_2 + e^{ax}(r)dx + e^{ay}(r)dy + e^{az}(r)dz,
\end{eqnarray}
and defining:
\begin{equation}
\label{e1}
e^1 = \sqrt{G^{\cal M}_{rr}}dr,
\end{equation}
one notes that:
\begin{equation}
\label{dea}
de^a = \Omega_{ab}e^1\wedge e^b,
\end{equation}
where the ``structure constants" $\Omega_{ab}$ are defined as under:
\begin{eqnarray}
\label{Omega_ab}
& & \Omega_{ab} \equiv \frac{\left(e^{a\theta_1}\ ^\prime(r)\Theta_{1b} + e^{a\theta_2}\ ^\prime(r)\Theta_{2b} + e^{ax}\ ^\prime(r){\cal X}_b + e^{ay}\ ^\prime(r){\cal Y}_b + e^{az}\ ^\prime(r){\cal Z}_b\right)}{\sqrt{G^{\cal M}_{rr}}}.
\end{eqnarray}
The components of $\Omega_{ab}$s after a small-$\beta$ large-$N$ small-$a$ expansion are given in (\ref{Omega_2a}). The two-form associated with the almost complex structure is given by:
\begin{equation}
\label{J-ea}
J = e^{12} + e^{34} + e^{56},
\end{equation}
and the nowhere vanishing $(3,0)-$form $\Omega$ is given by:
\begin{equation}
\label{Omega}
\Omega = \left(e^1 + i e^2\right)\wedge\left(e^3 + i e^4\right)\wedge\left(e^5 + i e^6\right) \equiv \Omega_+ + i \Omega_-,
\end{equation}
where $e^{a_1....a_p} \equiv e^{a_1}\wedge ....e^{a_p}$. The five $SU(3)$-structure torsion classes are denoted by $W_{1,2,3,4,5}$.

One sees:
{\footnotesize
\begin{eqnarray}
\label{dJ+dOmega}
& & dJ = \Omega_{32}e^{124} - \Omega_{42}e^{312} + \Omega_{52}e^{126} - \Omega_{62}e^{512} + \left(\Omega_{35} + \Omega_{62}\right)e^{154} + \left(\Omega_{36} - \Omega_{54}\right)e^{146}\nonumber\\
& & + \left(\Omega_{33} + \Omega_{44}\right)e^{134} + \left(\Omega_{45} + \Omega_{63}\right)e^{315} + \left(-\Omega_{46} - \Omega_{53}\right)e^{316}
+ \left(\Omega_{55} + \Omega_{66}\right)e^{156},\nonumber\\
& & d\Omega_+ = -\Biggl(\Omega_{23}e^{1345} + \Omega_{26}e^{1645}
+ \Omega_{55}e^{2415} + \Omega_{22}e^{1236}+ \Omega_{24}e^{1436} + \Omega_{25}e_{1536} + \left(\Omega_{22} \Omega_{44} + \Omega_{55}\right)e^{1245}
\nonumber\\
& & + \left(\Omega_{35} - \Omega_{46}\right)e^{2165} + \left(\Omega_{34} + \Omega_{56}\right)e_{2416}  - \left(\Omega_{33} + \Omega_{66}\right)e^{2136} + \left(\Omega_{53} - \Omega_{64}\right)e^{2413} - \left(\Omega_{43} + \Omega_{65}\right)e^{2135}\Biggr),\nonumber\\
& & d\Omega_- = \Omega_{24}e^{1435} + \Omega_{26}e^{1635} - \Omega_{23}e^{1346} - \Omega_{25} e^{1546} - \Omega_{66}e^{2416} + \left(\Omega_{22} + \Omega_{33} + \Omega_{55}\right)e^{1235}\nonumber\\
& & + \left(-\Omega_{34} + \Omega_{65}\right)e_{2145} + \left(-\Omega_{36} - \Omega_{45}\right)e^{2165} + \left(\Omega_{54} + \Omega_{63}\right)e^{2314} +
\left(\Omega_{56} - \Omega_{43}\right)e^{2316} - \left(\Omega_{22} + \Omega_{44}\right)e^{1246},\nonumber\\
& &
\end{eqnarray}
}
implying:
\begin{eqnarray}
\label{W_4-1}
2 W_4 & = & J \lrcorner dJ = \Omega_{32}e^4 - \Omega_{42}e^3 + \Omega_{52}e^6 - \Omega_{62}e^5 + \left(\Omega_{33} + \Omega_{44}\right)e^1  + \left(\Omega_{55} + \Omega_{66}\right)e^1.\nonumber\\
& &
\end{eqnarray}
Now, substituting (\ref{Omega_2a}) - (\ref{Omega_6a}), one sees that the ${\cal O}(l_p^0)$ terms in $\Omega_{62}$ goes like $\left(12.5 - 43.6 \frac{a^2}{r^2}\right)\sqrt{1-\frac{r_h^4}{r^4}}$ which assuming $a = r_h\left(0.6 + \frac{g_sM^2}{N}\right)$ \cite{Sil+Yadav+Misra-glueball}, vanishes for  $r\sim 1.25 r_h$. Similarly, the ${\cal O}(l_p^0)$ term in $\Omega_{24}$ can be proportional to the ${\cal O}(l_p^0)$ term in $-\Omega_{42}$, i.e., ${\cal O}(1)\frac{a^2}{r^2}$ for $r\sim0.5\sqrt{4+{\cal O}(1)}a$. Thus:
\begin{equation}
\label{W_4-2}
W_4 \approx \frac{\Omega_{32}e^4 + \Omega_{52}e^6  + \Omega_{66}e^1}{2}.
\end{equation}
\begin{eqnarray}
\label{W_5-1}
2 W_5 & = & \Omega_+ \lrcorner d\Omega_+ = \Omega_{23}e^4 - \Omega_{25}e^6
+ \left(\Omega_{[43]} + \Omega_{[65]}\right)e^2 + \Omega_{24}e^3 \nonumber\\
& &   + \left(2\Omega_{22} + \Omega_{33} + \Omega_{44} + 2 \Omega_{55} + \Omega_{66}\right)e^1.
\end{eqnarray}
Now, one sees that the ${\cal O}(l_p^0)$ terms in
$\left(\Omega_{[43]} + \Omega_{[65]}\right)e^2$ for the aforementioned IR-valued $r$ would vanish for $\frac{\alpha_{\theta_1}}{\alpha_{\theta_2}} \sim \frac{1.3}{\alpha_rg_s^{\frac{7}{8}}\sqrt{M N_f}}$, where $|\log r| = \alpha_r N^{\frac{1}{3}}$ \cite{Bulk-Viscosity}. Also, from (\ref{Omega_2a}) - (\ref{Omega_6a}), one sees that:
$2\Omega_{22} + \Omega_{33} + \Omega_{44} + 2 \Omega_{55} + \Omega_{66} \approx \Omega_{66}$. Thus:
\begin{equation}
\label{W_5-2}
W_5 \sim \frac{\Omega_{23}e^4 - \Omega_{25}e^6
 + \Omega_{66}e^1}{2}\stackrel{|\Omega_{23,25}| < |\Omega_{66}|}{\longrightarrow}\frac{\Omega_{66}e^1}{2}\sim W_4.
\end{equation}
This mimics supersymmetric \cite{Butti et al [2004], Keshav_et_al_SUSY_type-II_torsion} IIA mirror though for a non-complex manifold - see below.
As:
\begin{eqnarray}
\label{W_1^+-1}
& & \Omega_+\wedge dJ = -\left(\Omega_{(46)} + \Omega_{(53)}\right)\frac{J^3}{6}
\equiv W_1^+ J^3,
\end{eqnarray}
implying:
\begin{equation}
\label{W_1^+-2}
W_1^+ = -\frac{\left(\Omega_{(46)} + \Omega_{(53)}\right)}{6}.
\end{equation}
Similarly, as:
\begin{eqnarray}
\label{W_1^-1}
& & \Omega_-\wedge dJ = \left(\Omega_{(36)} - \Omega_{(45)}\right)\frac{J^3}{6} \equiv W_1^- J^3,
\end{eqnarray}
implying:
\begin{equation}
\label{W_1^-2}
W_1^- = \frac{\left(\Omega_{(36)} - \Omega_{(45)}\right)}{6}.
\end{equation}
Also, using the notation: $E^{i_1}\wedge....E^{i_p}\wedge {\bar E}^{j_1}\wedge....{\bar E}^{j_q} \equiv E^{i_1....i_p{\bar j}_1....{\bar j}_q}$, one notes that:
\begin{eqnarray}
\label{dJ-(2,1)}
& & \left(dJ\right)^{(2,1)} = \frac{1}{4}\left(\Omega_{32} - i \Omega_{42}\right)E^{1{\bar 1}2} + \frac{1}{4}\left(\Omega_{52} - \Omega_{62}\right) +   \frac{1}{8}\left(-2\Omega_{45} + 2\Omega_{63} + \Omega_{46} + \Omega_{53}\right)E^{21{\bar 3}}\nonumber\\
& & + \frac{1}{8}\left(-\Omega_{45} + \Omega_{63} + i \Omega_{46} + i \Omega_{53}\right)E^{{\bar 2}13} + \frac{i}{4}\left(\Omega_{55} + \Omega_{66}\right)E^{3{\bar 3}1}.
\end{eqnarray}
Therefore:
{\footnotesize
\begin{eqnarray}
\label{W_3}
& & W_3 = \left(dJ\right)^{(2,1)} - \left(J\wedge W_4\right)^{(2,1)} =
\left(\frac{\Omega_{32}}{4} - \frac{\Omega_{32}}{4} + i \frac{\Omega_{42}}{2}\right)E^{1{\bar 1}2} + \left[\frac{1}{4}\left(\Omega_{52} - \Omega_{62}\right)-\frac{1}{2}\left(\Omega_{52} - i \Omega_{62}\right)\right]E^{1{\bar 1}3}\nonumber\\
& & \frac{1}{8}\left(-2\Omega_{45} + 2\Omega_{63} + \Omega_{46} + \Omega_{53}\right)E^{21{\bar 3}} + \frac{1}{8}\left(-\Omega_{45} + \Omega_{63} + i\Omega_{46} + i\Omega_{53}\right)E^{{\bar 2}13} -\frac{1}{2}\left(\Omega_{52} - i \Omega_{62}\right) E^{2{\bar 2}3} \nonumber\\
& &- i\left(\Omega_{33} + \Omega_{44} + \Omega_{55} + \Omega_{66}\right)E^{2{\bar 2}1} + \left(i\Omega_{42} - \Omega_{32}\right)E^{3{\bar 3}2} \nonumber\\
& & + \left[\frac{i}{4}\left(\Omega_{55} + \Omega_{66}\right) - \frac{i}{2}\left(\Omega_{33} + \Omega_{44} + \Omega_{55} + \Omega_{66}\right)\right]E^{3{\bar 3}1}.
\end{eqnarray}
}

To determine $W_2^{\pm}$, one notes that:
{\footnotesize
\begin{eqnarray}
\label{W2+-1}
& & - \left(d\Omega_+\right)^{2,2} = \left(\frac{i\Omega_{23} - \Omega_{24}}{8}\right)
E^{2{\bar 2}{\bar 1}3} + \left(\frac{i\Omega_{23} + \Omega_{24}}{8}\right)E^{2{\bar 2}1{\bar 3}} - \left(\frac{\Omega_{26} + i\Omega_{25}}{8}\right)E^{3{\bar 3}1{\bar 2}}\nonumber\\
& & \left(\frac{\Omega_{26} - i\Omega_{25}}{8}\right)E^{3{\bar 3}{\bar 1}2} +
\left(\frac{- 2 \Omega_{22} + \Omega_{33} - \Omega_{44} + \Omega_{66}}{8}
+ \frac{i}{8}\left(\Omega_{(34)} + \Omega_{(56)}\right)\right)E^{1{\bar 1}2{\bar 3}}\nonumber\\
& & + \left(\frac{2 \Omega_{22} - \Omega_{33} + \Omega_{44} - \Omega_{66}}{8}
- \frac{i}{8}\left(\Omega_{[34]} + \Omega_{[56]}\right)\right)E^{1{\bar 1}{\bar 2}3}
- \frac{\left(\Omega_{35} - \Omega_{46}\right)}{4}E^{1{\bar 1}3{\bar 3}}
+ \frac{1}{4}\left(\Omega_{53} - \Omega_{64}\right)E^{1{\bar 1}2{\bar 2}}\nonumber\\
& & = - \left(W_1^+ J^2 + W_2^+\wedge J\right),
\end{eqnarray}
}
and this implies:
\begin{eqnarray}
\label{W_2+-2}
& & W_2^+ = \alpha_1 E^{1{\bar 1}} + \beta_1 E^{2{\bar 2}} + \gamma_1 E^{3{\bar 3}}
+ \alpha_2 E^{1{\bar 2}} + \beta_2 E^{2{\bar 3}} + \gamma_2 E^{1{\bar 3}} + \alpha_3 E^{{\bar 1}2} + \beta_3 E^{{\bar 2}3} + \gamma_3 E^{{\bar 1}3},\nonumber\\
& &
\end{eqnarray}
where:
\begin{eqnarray}
\label{W_2+-3}
& & \alpha_1 = -i\left[\frac{W_1^+ + \Omega_{53} - \Omega_{64}}{2}\right] + \frac{i}{4}
\left(\Omega_{[53]} - \Omega_{[64]} + W_1^+\right)\nonumber\\
& & \beta_1 = -\frac{i}{4}\left(\Omega_{[53]} - \Omega_{[64]} + W_1^+\right)\nonumber\\
& & \gamma_1 = -\frac{i}{4}\left(W_1^+ - \Omega_{[53]} + \Omega_{[64]}\right),\nonumber\\
& & \alpha_2 = - \frac{\Omega_{26} -i\Omega_{25}}{8},\nonumber\\
& & \gamma_2 = \frac{\Omega_{24} + i \Omega_{23}}{8}\nonumber\\
& & \beta_2 = \frac{-2\Omega_{22} + \Omega_{33} - \Omega_{44} + \Omega_{66}}{8}
+ \frac{i}{8}\left(\Omega_{[56]} + \Omega_{[34]}\right),\nonumber\\
& & \alpha_3 = \frac{\Omega_{26} - i \Omega_{25}}{8},\nonumber\\
& & \beta_3 = \frac{2\Omega_{22} - 3 \Omega_{33} + \Omega_{44} - \Omega_{66}}{8}
+ \frac{i}{8}\left(\Omega_{[43]} + \Omega_{[65]}\right).
\end{eqnarray}

\subsection{$G_2$-Structure Torsion Classes of the Seven-Fold in the ${\cal M}$-Theory Uplift}

Given that the adjoint of $SO(7)$ decomposes under $G_2$ as ${\bf 21}\rightarrow{\bf 7}\oplus{\bf 14}$ where ${\bf 14}$ is the adjoint representation of $G_2$, one obtains:
\begin{equation}
\label{T-G2}
T \in \Lambda^1 \otimes g_2^\perp = W_1 \oplus W_{14} \oplus W_{27} \oplus W_7.
\end{equation}

\noindent We now present the seventh lemma:

 {\it Lemma 7}: In the neighborhood of $(\theta_{10}=\frac{\alpha_{\theta_1}}{N^{\frac{1}{5}}}, \theta_{20} = \frac{\alpha_{\theta_{20}}}{N^{\frac{3}{10}}}, \psi=2n\pi), n=0, 1, 2$, the $G_2$-structure torsion classes of $M_7$ - a cone over a six-fold which is an ${\cal M}$-theory $S^1$-fibration over a compact five-fold $M_5(\theta_1,\theta_2,\phi_1,\phi_2,\psi)$ - are given by: $W^{G_2}_{M_7} =W_{14} \oplus W_{27}$.

\noindent{\it Proof}: Now,  near the $\psi=0,2\pi,4\pi$-branches, the ${\cal M}$-Theory coframe $e^7 = \sqrt{G^{\cal M}_{x^{10}x^{10}}} dx^{10}$. Further, the three-form $\Phi$ corresponding to a $G_2$ structure is given by \cite{Phi-G_2}:
\begin{eqnarray}
\label{Phi-G_2}
& & \Phi = e^{-\Phi^{\rm IIA}}f_{abc}e^{abc} + e^{-\frac{2}{3}\Phi^{\rm IIA}}J\wedge dx^{10}\nonumber\\
& & = e^{-\Phi^{\rm IIA}}\left(e^{135} - e^{146} - e^{236} - e^{245}\right)
+ \frac{e^{-\frac{2}{3}\Phi^{\rm IIA}}}{\sqrt{G^{\cal M}_{x^{10}x^{10}}}}\left(e^{3417} + e^{5617}\right)\nonumber\\
\end{eqnarray}
{\footnotesize
\begin{eqnarray}
\label{dPhi}
& & d\Phi = - \frac{e^{-\Phi^{\rm IIA}}\Phi_{\rm IIA}^\prime}{\sqrt{G^{\cal M}_{rr}}}\left(-e^{1236} - e^{1245}\right) - \frac{2}{3}\frac{e^{-\frac{2}{3}\Phi^{\rm IIA}}\Phi_{\rm IIA}^\prime}{\sqrt{G^{\cal M}_{rr}G^{\cal M}_{x^{10}x^{10}}}} J \wedge e^{17}\nonumber\\
& & e^{-\Phi^{IIA}}\Biggl(-\Omega_{24}e^{1436} - \Omega_{25}e^{1536}  - \Omega_{66}e^{2316} - \Omega_{23}e^{1345} - \Omega_{26}e^{1645}\nonumber\\
& &-\left(\Omega_{33} + \Omega_{22}\right)e^{1236} - \left(\Omega_{22} + \Omega_{44} + \Omega_{55}\right)e^{1245} + \left(\Omega_{35} - \Omega_{46}\right)e^{2156} + \left(\Omega_{53} - \Omega_{64}\right)e^{2314} \nonumber\\
& & -\left(\Omega_{43} + \Omega_{65}\right)e^{2315} + \left(\Omega_{34} + \Omega_{56}\right)e^{2146}\Biggr)\nonumber\\
& & + \frac{e^{-\frac{2}{3}\Phi^{\rm IIA}}}{\sqrt{G^{\cal M}_{x^{10}x^{10}}}}\Biggl(
\Omega_{32}e^{1247} - \Omega_{42}e^{3127} + \Omega_{52}e^{1267} - \Omega_{62}e^{5127} + \left(\Omega_{35} + \Omega_{64}\right)e^{1547} + \left(\Omega_{36} - \Omega_{54}\right)e^{1467}\nonumber\\
& & \left(\Omega_{33} + \Omega_{44}\right)e^{1347} + \left(\Omega_{45} + \Omega_{63}\right)e^{3157} - \left(\Omega_{46} + \Omega_{53}\right)e^{3167}
+ \left(\Omega_{55} + \Omega_{66}\right)e^{1567}\Biggr)\nonumber\\
& & = 4 W_1 *_7\Phi - 3 W_7\wedge\Phi - *_7W_{27}.
\end{eqnarray}
}
Similarly:
\begin{eqnarray}
\label{d*Phi}
& & d*_7\Phi = - \frac{e^{-\Phi^{\rm IIA}}\Phi_{\rm IIA}^\prime}{\sqrt{G^{\cal M}_{rr}}}
\left(e^{12467} - e^{12357}\right) - \frac{2}{3}\frac{e^{-\frac{2}{3}\Phi^{\rm IIA}}\Phi_{\rm IIA}^\prime}{\sqrt{G^{\cal M}_{rr}}}e^{13456}\nonumber\\
& & e^{-\Phi^{\rm IIA}}\Biggl(\Omega_{23}e^{13467} + \Omega_{25}e^{15467} - \Omega_{43}e^{21367} - \Omega_{44}e^{21467} - \Omega_{45}e^{21567}\nonumber\\
& & + \Omega_{22}e^{12467} + \Omega_{63}e^{24137} + \Omega_{65}e^{24157} + \Omega_{66}e^{24167} - \Omega_{22}e^{12357} - \Omega_{24}e^{14357}\nonumber\\
& & -\Omega_{26}e^{16357} + \Omega_{33}e^{21357} + \Omega_{34}e^{21457} + \Omega_{36}e^{21657} - \Omega_{54}e^{23147}\nonumber\\
& & - \Omega_{55}e^{23157} - \Omega_{56}e^{23167} + \frac{e^{24617} - e^{23517}}{2\sqrt{G^{\cal M}_{rr}G^{\cal M}_{x^{10}x^{10}}}}\Biggr)\nonumber\\
& & + e^{-\frac{2}{3}\Phi^{\rm IIA}}\Biggl(\Omega_{32}e^{12456} + \Omega_{33}e^{13456} - \Omega_{42}e^{31256} - \Omega_{44}e^{31456}\nonumber\\
& & + \Omega_{52}e^{34126} + \Omega_{55}e^{34156} - \Omega_{62}e^{34512} - \Omega_{66}e^{34516}\Biggr)\nonumber\\
& & = - 4 W_7 \wedge *_7 \Phi - 2 *_7 W_{14}.
\end{eqnarray}
One hence obtains (see App. {\bf E} for details):
\begin{eqnarray}
\label{G2-Ws}
& & W_1 = W_7 = 0,\nonumber\\
& & W_{27} = - *_7 d\Phi,\nonumber\\
& & W_{14} = - \frac{1}{2}*_7d*_7\Phi.
\end{eqnarray}

\subsection{$SU(4)$-Structure Torsion Classes of the Eight-Fold in the ${\cal M}$-Theory Uplift}

The $SU(4)$-structure torsion classes are given by \cite{Gauntlett_et_al}:
{\footnotesize
\begin{eqnarray}
\label{SU(4)-torsion-classes-i}
& & \Lambda^1\otimes su(4)^\perp = \left({\bf 4} \oplus {\bf\bar  4}\right)\otimes \left({\bf 1} \oplus {\bf 6} \oplus {\bf\bar  6}\right) = \left({\bf 4}\oplus {\bf\bar 4}\right)
\oplus \left({\bf 20} \oplus {\bf\overline{20}}\right) \oplus \left({\bf 20} \oplus {\bf\overline{20}}\right) \oplus \left({\bf 4} \oplus {\bf\bar  4}\right) \oplus \left({\bf 4} \oplus {\bf\bar  4}\right)
\nonumber\\
& & = W_1 \oplus W_2 \oplus W_3 \oplus W_4 \oplus W_5,
\end{eqnarray}
}
where:
\begin{eqnarray}
\label{SU(4)-torsion-classes-ii}
& & dJ_4 = W_1 \lrcorner {\bar\Omega}_4 + W_3 + W_4\wedge J_4 + {\rm c.c.}\nonumber\\
& & d\Omega_4 = \frac{8i}{3} W_1 \wedge J_4^2 + W_2 \wedge J_4 + {\bar W}_5 \wedge \Omega_4.
\end{eqnarray}

\noindent We are now set to present the eighth lemma:

 {\it Lemma 8}: In the neighborhood of $(\theta_{10}=\frac{\alpha_{\theta_1}}{N^{\frac{1}{5}}}, \theta_{20} = \frac{\alpha_{\theta_{20}}}{N^{\frac{3}{10}}}, \psi=2n\pi), n=0, 1, 2$, the $SU(4)$-structure torsion classes of $M_8(r,\theta_{1,2},\phi_{1,2},\psi,x^{10},x^0)$ are $W^{SU(4)}_{M_8} = W_2^{SU(4)} \oplus W_3^{SU(4)} \oplus W_5^{SU(4)}$.

{\it Proof}: Near  the $\psi=0,2\pi,4\pi$-branches, defining:
\begin{eqnarray}
\label{e6&e0}
& & e^7 = \sqrt{G^{\cal M}_{x^{10}x^{10}}} dx^{10},\nonumber\\
& & e^0 = \sqrt{G^{\cal M}_{00}} dx^0,
\end{eqnarray}
using which construct $E^4 = e^7 + i e^0$. Defining:
{\footnotesize
\begin{eqnarray}
\label{Zeta-abc}
& & \hskip -0.5in \zeta_{12a} \equiv \frac{\Biggl[\left(e^{1\theta_1}\ ^\prime + i e^{2\theta_1}\ ^\prime\right)\Theta_{1a}  +  \left(e^{1\theta_2}\ ^\prime + i e^{2\theta_2}\ ^\prime\right)\Theta_{2a} + \left(e^{1x}\ ^\prime + i e^{2x}\ ^\prime\right){\cal X}_{a} + \left(e^{1y}\ ^\prime + i e^{2y}\ ^\prime\right){\cal Y}_{a} + \left(e^{1z}\ ^\prime + i e^{2z}\ ^\prime\right){\cal Z}_{a}\Biggr]}{\sqrt{G^{\cal M}_{rr}}},\nonumber\\
& & \hskip -0.5in \zeta_{34a} \equiv \frac{\Biggl[\left(e^{3\theta_1}\ ^\prime + i e^{4\theta_1}\ ^\prime\right)\Theta_{13}  +  \left(e^{3\theta_2}\ ^\prime + i e_{4\theta_2}\ ^\prime\right)\Theta_{2a} + \left(e^{3x}\ ^\prime + i e^{4x}\ ^\prime\right){\cal X}_{a} + \left(e^{3y}\ ^\prime + i e^{4y}\ ^\prime\right){\cal Y}_{a} + \left(e^{3z}\ ^\prime + i e^{4z}\ ^\prime\right){\cal Z}_{a}\Biggr]}{\sqrt{G^{\cal M}_{rr}}},\nonumber\\
& & \hskip -0.5in \zeta_{56a} \equiv \frac{\Biggl[\left(e^{5\theta_1}\ ^\prime + i e^{2\theta_1}\ ^\prime\right)\Theta_{1a}  +  \left(e^{5\theta_2}\ ^\prime + i e^{6\theta_2}\ ^\prime\right)\Theta_{2a} + \left(e^{5x}\ ^\prime + i e^{6x}\ ^\prime\right){\cal X}_{a} + \left(e^{5y}\ ^\prime + i e^{6y}\ ^\prime\right){\cal Y}_{a} + \left(e^{5z}\ ^\prime + i e^{6z}\ ^\prime\right){\cal Z}_{a}\Biggr]}{\sqrt{G^{\cal M}_{rr}}},\nonumber\\
\end{eqnarray}
}
where $a=2,...,6$, one obtains:
{\footnotesize
\begin{eqnarray}
\label{dE1}
& & dE^1 = \frac{i}{2}\frac{\zeta_{122}}{\sqrt{G^{\cal M}_{rr}}}E^1\wedge{\bar E}^1
+ \frac{\zeta_{12(3-i4)}}{\sqrt{G^{\cal M}_{rr}}}E^1\wedge E^2 +  \frac{\zeta_{12(3+i4)}}{\sqrt{G^{\cal M}_{rr}}}{\bar E}^1\wedge {\bar E}^2  +  \frac{\zeta_{12(3-i4)}}{\sqrt{G^{\cal M}_{rr}}}E^1\wedge E^3\nonumber\\
& &  + \frac{\zeta_{12(3+i4)}}{\sqrt{G^{\cal M}_{rr}}}{\bar E}^1\wedge {\bar E}^3 +
  + \frac{\zeta_{12(3+i4)}}{\sqrt{G^{\cal M}_{rr}}}E^1\wedge{\bar E}^2
  + \frac{\zeta_{12(3-i4)}}{\sqrt{G^{\cal M}_{rr}}}{\bar E}^1\wedge E^2
  + \frac{\zeta_{12(5+i6)}}{\sqrt{G^{\cal M}_{rr}}}{\bar E}^1\wedge E^3
  + \frac{\zeta_{12(5-i6)}}{\sqrt{G^{\cal M}_{rr}}}E^1\wedge{\bar E}^3,\nonumber\\
  & &
\end{eqnarray}
}
etc., where, e.g., $\zeta_{12(3\pm i4)} \equiv \zeta_{123} \pm i \zeta_{124}$.
One obtains:
\begin{eqnarray}
\label{W134}
& & W_1 = W_4 = 0,\nonumber\\
& & W_3 + {\bar W}_3 = dJ.
\end{eqnarray}
Writing:
\begin{equation}
\label{Omega4-Omega3}
\Omega_4 = \Omega_3\wedge E^4,
\end{equation}
one obtains:
\begin{eqnarray}
\label{dOmega4}
& & d\Omega_4 = \frac{i}{2}\frac{\zeta_{122}}{\sqrt{G^{\cal M}_{rr}}}E^{1{\bar 1}234} +
\frac{i}{4}\frac{\zeta_{12(3+i4)}}{\sqrt{G^{\cal M}_{rr}}}E^{{\bar 1}{\bar 2}234} +
\frac{\zeta_{12(3+i4)}}{4\sqrt{G^{\cal M}_{rr}}}E^{{\bar 1}{\bar 3}234} + \frac{\zeta_{12(3+i4)}}{4\sqrt{G^{\cal M}_{rr}}}E^{1{\bar 2}234}\nonumber\\
& & + \frac{\zeta_{12(5-i6)}}{4\sqrt{G^{\cal M}_{rr}}}E^{1{\bar 3}234}
- \Biggl(\frac{i}{4}\frac{\zeta_{34(3+i4)}}{\sqrt{G^{\cal M}_{rr}}}E^{1{\bar 1}{\bar 2}34}
+ \frac{\zeta_{23(3+i4)}}{4\sqrt{G^{\cal M}_{rr}}}E^{1{\bar 1}{\bar 3}34}
+ \frac{\zeta_{34(3-i4)}}{4\sqrt{G^{\cal M}_{rr}}}E^{1{\bar 1}234}    \Biggr)
\nonumber\\
 & & + \frac{\zeta_{564(3+i4)}}{\sqrt{G^{\cal M}_{rr}}}E^{12{\bar 1}{\bar 2}34}
 + \frac{\zeta_{56(3+i4)}}{4\sqrt{G^{\cal M}_{rr}}}E^{12{\bar 1}{\bar 3}4} +
 \frac{\zeta_{56(5+i6)}}{4\sqrt{G^{\cal M}_{rr}}}E^{12{\bar 1}34} \nonumber\\
& &   +
 \biggl(\frac{G^{\cal M}_{x^{10}x^{10}}\ ^\prime}{8\sqrt{G^{\cal M}_{x^{10}x^{10}} G^{\cal M}_{rr}}} - i \frac{G^{\cal M}_{00}\ ^\prime}{8\sqrt{G^{\cal M}_{00} G^{\cal M}_{rr}}}\biggr)\left(E^1 + {\bar E}^{\bar 1}\right)\wedge\Omega_4\nonumber\\
 & & - \biggl( \frac{G^{\cal M}_{x^{10}x^{10}}\ ^\prime}{8\sqrt{G^{\cal M}_{x^{10}x^{10}} G^{\cal M}_{rr}}}+ i  \frac{G^{\cal M}_{00}\ ^\prime}{8\sqrt{G^{\cal M}_{00} G^{\cal M}_{rr}}}\biggr)\Omega_3\wedge E^{{\bar 1}{\bar 4}} \nonumber\\
& &  - \biggl(\frac{G^{\cal M}_{x^{10}x^{10}}\ ^\prime}{8\sqrt{G^{\cal M}_{x^{10}x^{10}} G^{\cal M}_{rr}}}  + i \frac{G^{\cal M}_{00}\ ^\prime}{8\sqrt{G^{\cal M}_{00} G^{\cal M}_{rr}}}\biggr)\Omega_3\wedge E^{1{\bar 4}}\nonumber\\
& & = W_2\wedge J_4 + W_5 \wedge \Omega_4,
\end{eqnarray}
implying:
{\footnotesize
\begin{eqnarray}
\label{W5+W2}
W_5 & = & \biggl(\frac{G^{\cal M}_{x^{10}x^{10}}\ ^\prime}{8\sqrt{G^{\cal M}_{x^{10}x^{10}} G^{\cal M}_{rr}}} - i \frac{G^{\cal M}_{00}\ ^\prime}{8\sqrt{G^{\cal M}_{00} G^{\cal M}_{rr}}}\biggr)E^1 + \biggl(\frac{G^{\cal M}_{x^{10}x^{10}}\ ^\prime}{8\sqrt{G^{\cal M}_{x^{10}x^{10}} G^{\cal M}_{rr}}} - i \frac{G^{\cal M}_{00}\ ^\prime}{8\sqrt{G^{\cal M}_{00} G^{\cal M}_{rr}}} + \frac{i}{2}\frac{\zeta_{122}}{\sqrt{G^{\cal M}_{rr}}}\biggr)E^{\bar 1}\nonumber\\
& & + \frac{\zeta_{12(3+i4)}}{4\sqrt{G^{\cal M}_{rr}}}E^{\bar 2} + \frac{\zeta_{12(5-i6)}}{4\sqrt{G^{\cal M}_{rr}}}E^{\bar 3};\nonumber\\
W_2  & = &  - i \frac{\zeta_{12(3+i4)}}{4\sqrt{G^{\cal M}_{rr}}}E^{134} +
\frac{\zeta_{12(3+i4)}}{4\sqrt{G^{\cal M}_{rr}}}E^{{\bar 1}24} - i\frac{\zeta_{34(3+i4)}}{4\sqrt{G^{\cal M}_{rr}}}E^{{\bar 2}34} - \frac{\zeta_{56(3+i4)}}{4\sqrt{G^{\cal M}_{rr}}}
E^{2{\bar 3}4}\nonumber\\
& & + \frac{\alpha_1}{4\sqrt{G^{\cal M}_{rr}}}E^{3{\bar 3}4} + \frac{\alpha_1}{4\sqrt{G^{\cal M}_{rr}}}E^{2{\bar 2}4}  + \frac{\alpha_1}{4\sqrt{G^{\cal M}_{rr}}}E^{1{\bar 1}4},
\end{eqnarray}
}
where:
\begin{eqnarray}
\label{alpha123}
& & \alpha_1 = i \left(\zeta_{23(3+i4)} + \zeta_{56(3+i4)}\right) = - \alpha_2,\nonumber\\
& & \alpha_3 = -i\left(\zeta_{56(3+i4)} - \zeta_{23(3+i4)}\right).
\end{eqnarray}

\subsection{$Spin(7)$-Structure Torsion Classes of the Eight-Fold  in the ${\cal M}$-Theory Uplift}

The $Spin(7)$-torsion classes are given by:
\begin{equation}
\label{Spin(7)}
\Lambda^1\otimes Spin(7)^\perp = \left({\bf 8} \oplus {\bf 7}\right) \otimes {\bf 7} = {\bf 8}
\oplus {\bf 48} = W_1 \oplus W_2,
\end{equation}
where:
\begin{equation}
\label{W_1 Spin(7)}
W_1 = \Psi \lrcorner d\Psi,
\end{equation}
where $\Psi$ is a $Spin(7)$-invariant self-dual four-form:
\begin{eqnarray}
\label{Psi}
& & \Psi = e^{1234} + e^{1256} + e^{1278} + e^{3456} + e^{3478} + e^{5678} + e^{1357}\nonumber\\
& & - e^{1268} - e^{1458} - e^{1467} - e^{2358} - e^{2367} - e^{2457} + e^{2468}.
\end{eqnarray}

\noindent We now present the final lemma:

{\it Lemma 9}: In the neighborhood of $(\theta_{10}=\frac{\alpha_{\theta_1}}{N^{\frac{1}{5}}}, \theta_{20} = \frac{\alpha_{\theta_{20}}}{N^{\frac{3}{10}}}, \psi=2n\pi), n=0, 1, 2$, the $Spin(7)$-torsion classes are given by: $W^{Spin(7)}_{M_8} = W_1^{Spin(7)} \oplus W_2^{Spin(7)}$.

\noindent{\it Proof}: Hence:
{\footnotesize
\begin{eqnarray}
\label{dPsi}
& & d\Psi =\sum_{a=2}^6\Biggl[ \Omega_{3a}\left(e^{1a456} + e^{1a478} + e^{21a58} + e^{21a67}\right) - \Omega_{4a}\left(e^{21a68} + e^{31a56} + e^{31a78} - e^{21a57}\right)\nonumber\\
& & + \Omega_{5a}\left(e^{341a6} + e^{1a678} - e^{231a8} - e^{241a7}\right) - \Omega_{6a}\left(e^{3451a} + e^{51a78} + e^{231a7} - e^{241a8}\right) \nonumber\\
& & + \Omega_{7a}\left(e^{341a8} + e^{561a8} + e^{2361a}\right)   + \Omega_{2a}\left( - e^{1a358} - e^{1a367} - e^{1a457} + e^{1a468}\right)\Biggr] \nonumber\\
& & + \frac{G^{\cal M}_{x^{10}x^{10}}\ ^\prime}{2\sqrt{G^{\cal M}_{x^{10}x^{10}}G^{\cal M}_{rr}}}e^{24517} - \frac{G^{\cal M}_{x^{0}x^{0}}\ ^\prime}{2\sqrt{G^{\cal M}_{x^{0}x^{0}}G^{\cal M}_{rr}}}\left(e^{34718} + e^{56718} - e^{23518} + e^{24618}\right),
\end{eqnarray}
}
implying $W_1 = \sum_{a=1}^8 \Lambda_a\left(\left\{\Omega_{bc}\right\}\right) e^a$ and non-trivial $W_2$.

Now, $W_2$ is identified with the space of three-forms: $\Lambda^3_{48} = \left\{\gamma\in\Lambda^:\gamma\wedge\Psi=0\right\}$ \cite{Puhle-Spin(7)}.  From (\ref{Psi}), one sees that such a three-form will be given by:
\begin{eqnarray}
\label{three-form-W2_48}
\gamma & = & \sum_{a,n\neq1: 21\ {\rm components}}\alpha_{[ab]}e^{1ab} + \sum_{c,d\neq1,2: 15\ {\rm components}}\beta_{[cd]}e^{2cd} + \sum_{e,f\neq1,2,3: 10\ {\rm components}}\kappa_{[ef]}e^{3ef} \nonumber\\
& &  + \sum_{g,h\neq1,2,3,5: 3\ {\rm componentsi}}\omega_{[gh]}e^{5gh},
\end{eqnarray}
along with one constraint on the 49 $\alpha_{[ab]}, \beta_{[cd]}, \kappa_{[ef]}, \omega_{[gh]}$ coefficients in (\ref{three-form-W2_48}).

\section{Summary and Future Directions}

Finite (gauge/'t Hooft) coupling top-down non-conformal holography is a largely unexplored territory in the field of gauge-gravity duality. The only
Ultra Violet-complete top-down holographic dual of thermal QCD-like theories  that we are aware of, was proposed in \cite{metrics}. Later, the type IIA mirror of
the same at intermediate gauge/string coupling was constructed and the ${\cal M}$ theory uplift of the same were constructed in \cite{MQGP,NPB}.  Other than higher-derivative corrections quartic in the Weyl tensor, or of the Gauss-Bonnet type, in $AdS_5\times S^5$, dual to supersymmetric thermal Super Yang-Mills \cite{previous-higher-ders}, there is little known about top-down string theory duals at intermediate ’t Hooft coupling of thermal QCD-like theories. This paper fills this gap by working out the ${\cal M}$ theory dual of thermal QCD-like theories  at intermediate 't Hooft coupling in the IR.

The following is  a summary of the important results obtained in this work.

\begin{enumerate}
\item
We work out the ${\cal O}(l_p^6)$ corrections to the ${\cal M}$-Theory metric worked out in \cite{MQGP,NPB} arising from the ${\cal O}(R^4)$ terms in $D=11$ supergravity. We realize that in the MQGP limit of \cite{MQGP}, the contribution from the $J_0$ (and its variation) dominate over the contribution from $E_8$ and its variation as a consequence of which $E_8$ has been disregarded. The computations have been partitioned into two portions - one near the $\psi=2n\pi, n=0, 1, 2$ patches and the other away from the same (wherein there is no decoupling of the radial direction, the six angles and the ${\cal M}$-Theory circle).

\item
We also note that there is a close connection between finite ('t Hooft) coupling effects in the IR and non-conformality (which being effected via the effective number of the fractional $D3$-branes, vanishes in the UV) as almost all corrections to the ${\cal M}$-Theory metric components of \cite{MQGP,NPB} in the IR arising from the aforementioned ${\cal O}(R^4)$ terms in $D=11$ supergravity action, vanish when the number of fractional $D3$-branes is set to zero.

\item
The importance of the higher derivative corrections arises from the competition between the non-conformal Infra-Red enhancement $\frac{\left(\log {\cal R}_h\right)^m}{{\cal R}_h^n}, {\cal R}_h\equiv\frac{r_h}{{\cal R}_{D5/\overline{D5}}}, m=0,1,2,3, n=0,$ and the Planckian and large-$N$ suppression $\frac{l_p^6}{N^{\beta_N}}, \beta_N>0$ in the ${\cal O}(l_p^6)$ corrections to the ${\cal M}$-theory dual \cite{MQGP,NPB} of thermal QCD-like theories. As $|\log {\cal R}_h|\sim N^{\frac{1}{3}}$ \cite{Bulk-Viscosity}, for appropriate values of $N$, it may turn out that this correction may become of ${\cal O}(1)$, and thereby very significant. This would also then imply that one will need to consider higher order corrections beyond ${\cal O}(l_p^6)$.

\item
\begin{itemize}
\item
On the mathematical side, using Lemmas 1 - 9 of sections {\bf 4} and {\bf 5}, the main result of this work, in addition to providing for the first time the ${\cal O}(l_p^6)$-corrections to the ${\cal M}$-theory dual of thermal QCD-like theories  of \cite{MQGP}, is  Proposition 1 stated in Section {\bf 1}. We work out the fundamental two-form and the nowhere vanishing holomorphic three-form of the six-fold obtained by an ${\cal M}$-Theory circle reduction of the ${\cal M}$-Theory dual obtained. This enabled us to work out the $SU(3)$-structure torsion classes of the aforementioned six-fold $M_6(r,\theta_{1,2},\phi_{1,2},\psi)$ relevant to the type IIA SYZ mirror, the $G_2$-structure torsion classes  of the seven-fold $M_7(r,\theta_{1,2},\phi_{1,2},\psi,x^{10})$ as well as the $SU(4)$-structure and $Spin(7)$-structure torsion classes of the eight-fold $M_8(x^0,r,\theta_{1,2},\phi_{1,2},\psi,x^{10})$ relevant to the ${\cal M}$-Theory uplift.  Table 1 summarizes the $G$-structure torsion classes' results.  Table 2 summarizes the $G$-structure torsion classes' results.
\begin{table}[h]
\begin{center}
\begin{tabular}{|c|c|c|c|}\hline
S. No. & Manifold & $G$-Structure & Non-Trivial Torsion Classes \\ \hline
1. & $M_6(r,\theta_1,\theta_2,\phi_1,\phi_2,\psi)$ & $SU(3)$ & $T^{\rm IIA}_{SU(3)} = W_1 \oplus W_2 \oplus W_3 \oplus W_4 \oplus W_5: W_4 \sim W_5$ \\ \hline
2. & $M_7(r,\theta_1,\theta_2,\phi_1,\phi_2,\psi,x^{10})$ & $G_2$ & $T^{\cal M}_{G_2} = W_{14} \oplus W_{27}$ \\ \hline
3. & $M_8(x^0,r,\theta_1,\theta_2,\phi_1,\phi_2,\psi,x^{10})$ & $SU(4)$ & $T^{\cal M}_{SU(4)} = W_2 \oplus W_3 \oplus W_5$ \\ \hline
4. & $M_8(x^0,r,\theta_1,\theta_2,\phi_1,\phi_2,\psi,x^{10})$ & $Spin(7)$ & $T^{\cal M}_{Spin(7)} = W_1 \oplus W_2$ \\ \hline
\end{tabular}
\end{center}
\caption{IR $G$-Structure Classification of Six-/Seven-/Eight-Folds in the type IIA/${\cal M}$-Theory Duals of Thermal QCD-Like Theories (at High Temperatures)}
\end{table}

\item
Along the Ouyang embedding (for very small [modulus of the] Ouyang embedding parameter) effected, e.g., near the $\psi=2n\pi, n=0, 1, 2$-patches in the MQGP limit, the large-base of the delocalized $T^2$-invariant sLag-fibration relevant to constructing the delocalized SYZ type IIA mirror in \cite{MQGP, NPB} of the type IIB dual of thermal QCD-like theories  in \cite{metrics} manifests itself in the ${\cal O}(R^4)$ corrections to the co-frames that diagonalize the mirror six-fold metric in the following sense. It is only the constant of integration ${\cal C}^{(1)}_{zz}$ appearing in the solution to the $f_{zz}$ EOM corresponding to the (delocalized version of the) $U(1)$-fiber $S^1(\psi)$ - part of the (delocalized version of) $T^3(\psi,\phi_1,\phi_2)$ orthogonal to the aforementioned large base $B_3(r,\theta_1,\theta_2)$ - that determines in the MQGP limit, the aforementioned $l_p^6$ corrections in the IR to the MQGP results of the co-frames and hence $G$-structure torsion classes.

\end{itemize}

\item {\bf Brief summary of published) applications of the results of this paper}: {\it We had decided to first work out applications of the results obtained in this paper to a variety of issues in Physics also including comparison (for some of the issues) with experiments/phenomenological data available, and after successfully doing so in \cite{Vikas+Gopal+Aalok}, \cite{Gopal+Vikas+Aalok}, submit an abridged version of the original version of this work (that was posted on the arXiv last year, arXiv:2004.07259[hep-th], cross-listed with math.dg), to ATMP.}

\begin{itemize}
\item
As an application of the results of this paper modified to a thermal ${\cal M}$-theory dual of thermal QCD-like theories at low temperatures, we now summarize in the context of  {\it ${\cal M}\chi$PT}, the main result of \cite{Vikas+Gopal+Aalok}  (involving both the authors):
\begin{itemize}

\item
{\it ${\cal O}(R^4)$-large-$N$ connection}: In the context of low energy coupling constants (LECs) of the $SU(3)\ \chi$PT Lagrangian in the chiral limit at ${\cal O}(p^4)$, as shown in detail in \cite{Vikas+Gopal+Aalok} (and briefly explained in Section {\bf 4}) as an application of the ${\cal O}(R^4)$ corrections to the ${\cal M}$-theory uplift of large-$N$ thermal QCD-like theories, matching the values of the one-loop renormalized coupling constants up to ${\cal O}(p^4)$ with experimental/lattice results shows that there is an underlying connection between large-$N$ suppression and higher derivative corrections.

\item
{\it ${\cal M}\chi$PT and Flavor Memory}: As shown in \cite{Vikas+Gopal+Aalok}  (involving both the authors), matching the phenomenological value of the 1-loop renormalized coupling constant corresponding to the ${\cal O}(p^4)$ $SU(3)\ \chi$PT Lagrangian term ``$\left(\nabla_\mu U^\dagger\nabla^\mu U\right)^2$", with the value obtained from the type IIA dual of thermal QCD-like theories inclusive of the aforementioned ${\cal O}(R^4)$ corrections, required the ${\cal O}(R^4)$ corrections arising from the contributions arising from the corrections to the metric along the compact $S^3$ part of the non-compact four-cycle ``wrapped" by the flavor $D7$-branes of the parent type IIB theory,  to have a definite sign (negative). The thermal supergravity background dual to type IIB (solitonic) $D3$-branes at low temperatures, includes $\mathbb{R}^2\times S^1(\frac{1}{M_{\rm KK}})$. By taking the $M_{\rm KK}\rightarrow0$ limit (to recover a boundary four-dimensional QCD-like theory after compactifying on the base of a $G_2$-structure cone),  remarkably, via a delicate cancelation between some of the aforementioned contributions arising from the ${\cal O}(R^4)$ metric corrections with a resultant contribution solely along the vanishing $S^2$ (with the abovementioned $S^3$, an $S^1$ fibration over the vanishing $S^2$) of the parent type IIB suviving,  we {\it derive} and hence verify the M$\chi$PT requirement of the sign. We also referred to this as ``Flavor Memory" in \cite{Gopal+Vikas+Aalok}  (involving both the authors).
\end{itemize}

\item
 As an application of the results of this paper as well as the same modified to ${\cal M}$-theory duals of thermal QCD-like theories at respectively high and low temperatures, we now summarize in the context of obtaining {\it Deconfinement temperature}, the main result of \cite{Gopal+Vikas+Aalok}  (involving both the authors):
:
\begin{itemize}
\item
{\it UV-IR Mixing and Flavor Memory}: Performing a semiclassical computation \cite{Witten-Hawking-Page-Tc} in \cite{Gopal+Vikas+Aalok} (involving both the authors),  by matching the actions at the deconfinement temperature of the ${\cal M}$-theory uplifts of the thermal and black-hole backgrounds at the UV cut-off, it was shown that one obtains a relationship in the IR between the ${\cal O}(R^4)$ corrections to the ${\cal M}$-theory metric  along the ${\cal M}$-theory circle in the thermal background and the ${\cal O}(R^4)$ correction to a specific combination of the  ${\cal M}$-theory metric components along the compact part of the four-cycle ``wrapped" by the flavor $D7$-branes of the parent type IIB (warped resolved deformed) conifold geometry - the latter referred to as ``Flavor Memory" in the context of ${\cal M}\chi$PT above.

\item
{\it Non-Renormalization of $T_c$}: We further showed in \cite{Gopal+Vikas+Aalok} that the LO result for $T_c$ also holds even after inclusion of the ${\cal O}(R^4)$ corrections. The dominant contribution from the ${\cal O}(R^4)$ terms in the large-$N$ limit arises from the $t_8t_8R^4$ terms, which from a type IIB perspective in the zero-instanton sector, correspond to the tree-level contribution at ${\cal O}\left((\alpha^\prime)^3\right)$ as well as one-loop contribution to four-graviton scattering amplitude and  obtained from integration of the fermionic zero modes. As from the type IIB perspective, the $SL(2,\mathbb{Z})$ completion of these $R^4$ terms \cite{Green and Gutperle} suggests that they are not renormalized  perturbatively beyond one loop in the zero-instanton sector, this therefore suggests the non-renormalization of $T_c$ at all loops in ${\cal M}$-theory at ${\cal O}(R^4)$.

\item
{\it $T_c$ from Entanglement Entropy}: With an obvious generalization of \cite{Tc-EE} to ${\cal M}$-theory, the entanglement entropy between two regions by dividing one of the spatial coordinates of the thermal ${\cal M}$-theory background into a segment of finite length ${\it l}$ and its complement, was also calculated in \cite{Gopal+Vikas+Aalok}. Like \cite{Tc-EE}, there are two RT surfaces - connected and disconnected. There is a critical value of ${\it l}$ - denoted by ${\it l_{crit}}$  - such that if $l<{\it l_{crit}}$, corresponding to the confined phase then it is the connected surface that dominates the entanglement entropy, and if $l>{\it l_{crit}}$ corresponding to the deconfined phase then it is the disconnected surface that dominates the entanglement entropy. This is interpreted as confinement-deconfinement phase transition.

\item
{\it Non-Renormalization of $T_c$ from Entanglement Entropy}: Remarkably, when evaluating the deconfinement temperature from an entanglement entropy computation in the thermal gravity dual, due to an exact and delicate cancelation between the ${\cal O}(R^4)$ corrections from a subset of the abovementioned metric components, one sees that there are consequently no corrections to $T_c$ at quartic order in the curvature supporting the conjecture made in  on the basis of a semiclassical computation.
\end{itemize}
\end{itemize}

\item
{\bf Future directions}:

\begin{itemize}
\item {\it Math}:
\begin{itemize}
\item
{\bf Almost Contact Metric Structure, Contact Structure and $SU(3)/SU(2)$ structure  from $G_2$ structure}: Using the $G_2$ structure seven-fold $M_7$ of the ${\cal M}$-theory uplift of large-$N$ thermal QCD-like theories  inclusive of ${\cal O}(R^4)$ corrections as obtained in this work, equipped with a positive form $\varphi$ and the $G_2$ metric $g_{G_2}$, it can be shown that the same is equipped with an Almost Contact Metric Structure (ACMS) $(J,R,\sigma,g_{G_2})$ \cite{ACMS-1,ACMS-SU2}, $R$ being a unit vector field with $J$ being a vector-valued one-form on $M_7$: $J^i_j = -\varphi^i_{jk}R^k$ and $\sigma$ being a one-form: $\sigma_i = g^{G_2}_{ij}R^j$. It will be very interesting to explicitly construct the $R$ and hence $J$ and $\sigma$, and verify if the ACMS so obtained is also a contact structure \cite{AC3MS+SU3}. Using results of \cite{Cvetic_et_al_G2_SU3_SU2}, it will also be extremely interesting to explicitly obtain an embedding of $SU(3)$ and $SU(2)$ structures in $G_2$ structure and using the results of \cite{ACMS-2},  an Almost Contact 3-Structure (AC3S) \cite{AC3MS+SU3}.

\item
For simplicity, we worked out the aforementioned $G$-Structures near the $\psi=2n\pi, n=0, 1, 2$-branches restricted to small-parameter Ouyang embedding, but as mentioned towards the beginning of Section {\bf 4}, using (\ref{e12}) - (\ref{e^6}) and ideas of  \cite{SYZ-free-delocalization}, the results of Table 1 are independent of angular delocalization in $\theta_{1,2}$. As regards independence of the results of Table 1 of $\psi$-delocalization, using the results of Section {\bf 3}, one sees that the secular equation needed to be solved to diagonalize the $M_7(r,\theta_{1,2},\phi_{1,2},\psi,x^{10})$ will be a septic ${\cal P}_7=0$.
Hence, one needs to use Siegel theta functions of genus four. The period matrix $\Omega$ will be defined as follows:
$$\Omega_{ij}=\sigma^{ik}\rho_{kj}$$
where $$\sigma_{ij}\equiv\oint_{A_j}d{\cal Z} \frac{{\cal Z}^{i-1}}{\sqrt{{\cal Z}({\cal Z}-1){\cal P}_7({\cal Z})}}$$ and
$$\rho_{ij}\equiv\oint_{B_j}\frac{{\cal Z}^{i-1}}{\sqrt{{\cal Z}({\cal Z}-1)({\cal Z}-2){\cal P}_7({\cal Z})}},$$
$\{A_i\}$ and $\{B_i\}$ being a canonical basis of cycles satisfying: $A_i\cdot A_j=B_i\cdot B_j=0$ and
$A_i\cdot B_j=\delta_{ij}$; $\sigma^{ij}$ are normalization constants determined by:
$\sigma^{ik}\sigma_{kj} = \delta^i_j$. Umemura's result then is that a root is given by:
$$\hskip-0.3in\frac{1}{2\left(\theta\left[\begin{array}{cccc}
\frac{1}{2} & 0 & 0 & 0 \\
0 & 0 & 0 & 0  \end{array}\right](0,\Omega)\right)^4
\left(\theta\left[\begin{array}{cccc}
\frac{1}{2} & \frac{1}{2} & 0 & 0 \\
0 & 0 & 0 & 0 \end{array}\right](0,\Omega)\right)^4}$$
$$\hskip-0.3in\times\Biggl[\left(\theta\left[\begin{array}{cccc}
\frac{1}{2} & 0 & 0 & 0 \\
0 & 0 & 0 & 0  \end{array}\right](0,\Omega)\right)^4\left(\theta\left[\begin{array}{cccc}
\frac{1}{2} & \frac{1}{2} & 0 & 0 \\
0 & 0 & 0  \end{array}\right](0,\Omega)\right)^4$$
$$\hskip-0.3in+ \left(\theta\left[\begin{array}{cccc}
0 & 0 & 0 & 0 \\
0 & 0 & 0 & 0  \end{array}\right](0,\Omega)\right)^4
\left(\theta\left[\begin{array}{cccc}
0 & \frac{1}{2} &  0 & 0\\
0 & 0 & 0 & 0  \end{array}\right](0,\Omega)\right)^4$$
$$\hskip-0.3in- \left(\theta\left[\begin{array}{cccc}
0 & 0 & 0 & 0 \\
\frac{1}{2} & 0 & 0 & 0  \end{array}\right](0,\Omega)\right)^4
\left(\theta\left[\begin{array}{cccc}
0 & \frac{1}{2} & 0 & 0 \\
\frac{1}{2} & 0 & 0 & 0  \end{array} \right](0,\Omega)\right)^4\Biggr].$$
However, it will again be a non-trivial task to evaluate the period integrals. Alternatives will be deferred to a later work.
\end{itemize}

\item {\it Physics}:
In the context of intermediate 't Hooft coupling top-down holography, there is no known literature on applying gauge-gravity duality techniques to studying the perturbative regime of thermal QCD-like theories  so as to be able to explain, e.g., low-frequency peaks expected to occur in spectral functions associated with transport coefficients, from ${\cal M}$ theory. In higher dimensional (Gauss-Bonnet or quartic in the Weyl tensor) holography, in the past couple of
years using previously known results, it has been shown by the (Leiden-)MIT-Oxford collaboration  \cite{previous-higher-ders}
that one obtains low frequency peaks in correlation/spectral functions of energy momentum tensor, per unit frequency, obtained
from the dissipative (i.e. purely imaginary) quasi-normal modes. As an extremely crucial application of the results of our paper,
 for the first time,  spectral/correlation functions involving the energy momentum
tensor with the inclusion of the ${\cal O}(l_p^6)$ corrections in the ${\cal M}$ theory (uplift) metric of \cite{MQGP} can be evaluated and hence one would be able to make direct connection between previous results in perturbative thermal QCD-like theories  (e.g., \cite{Moore+Saremi}) as well as QCD plasma in RHIC
experiments, and ${\cal M}$ theory. Further, the temperature dependence of the speed of sound, the attenuation constant and bulk viscosity can also be obtained from its
solution, as well as the ${\cal O}(l_p^6)$ and the non-conformal corrections to the conformal results thereof. One could see if one could reproduce the known weak-coupling result from ${\cal M}$ theory that the ratio of the bulk and
shear viscosities goes like the square of the deviation of the square of the speed of sound from its conformal
value (the last reference in \cite{EPJC-2}). Generically, the dissipative quasi-normal modes in the spectral functions at low frequencies can be investigated to study the existence of peaks at low frequencies in transport coefficients, thus making direct contact with perturbative QCD results as well as (QCD plasma in) RHIC experiments.
\end{itemize}
\end{enumerate}

\section*{Acknowledgements}

VY is supported by a Senior Research Fellowship (SRF) from the University Grants Commission,
Govt. of India. AM was supported by IIT Roorkee and the Council of Scientific and Industrial Research via a grant CSR-1477-PHY. He
would also like to thank the theory group at McGill University, and Keshav Dasgupta in particular, for the wonderful hospitality and support during various stages of the work.

\appendix

\section{Equations of Motion}
\setcounter{equation}{0} \seceqaa

This appendix discusses the form of the equations of motion (EOMs) satisfied by $f_{MN}$s inclusive of the ${\cal O}(l_p^6 R^4)$ corrections to the ${\cal M}$-Theory uplift of \cite{MQGP} as well as their solutions. ${\bf A.1}$ works  out the same near the $\psi=2n\pi, n=0, 1, 2$-patches, and {\bf A.2}  near the $\psi\neq2n\pi, n=0, 1, 2$-patch. The  EOMs have been obtained expanding the coefficients of $f^{(p)}_{MN}, p=0,1,2$ near $r=r_h$ and retaining the LO terms in the powers of $(r-r_h)$ in the same, and then performing a large-$N$-large-$|\log r_h|$-$\log N$ expansion the resulting LO terms are written out. We should keep in  mind that near the $\psi=\psi_0\neq2n\pi, n=0, 1, 2$-patch, some $G^{\cal M}_{rM}, M\neq r$ and $G^{\cal M}_{x^{10}N}, N\neq x^{10}$ components are non-zero, making this exercise much more non-trivial.

As the EOMs are too long, they have not been explicitly typed but their forms have been written out. The solutions of the EOMs are discussed in detail.

\subsection{EOMs for $f_{MN}$ and Their Solutions Near $\psi=2n\pi, n=0, 1, 2$-Branches and Near $r=r_h$}

Working in the IR,  the EOMs near the $\psi=0\ 2\pi\ 4\pi$-branches near $r=r_h$ as described in Section {\bf 5.2}, can be written as follows:
{\footnotesize
\begin{eqnarray}
\label{IR-psi=2nPi-EOMs}
& & {\rm EOM}_{MN}:\nonumber\\
& &  \sum_{p=0}^2\sum_{i=0}^2a_{MN}^{(p,i)}\left(r_h, a, N, M, N_f, g_s, \alpha_{\theta_{1,2}}\right)(r-r_h)^if_{MN}^{(p)}(r) +
\beta {\cal F}_{MN}\left(r_h, a, N, M, N_f, g_s, \alpha_{\theta_{1,2}}\right)(r-r_h)^{\alpha_{MN}^{\rm LO}} = 0,\nonumber\\
& &
\end{eqnarray}
}
where $M, N$ run over the $D=11$ coordinates,   $f^{(p)}_{MN}\equiv \frac{d^p f_{MN}}{dr^p}, p=0, 1, 2$, $\alpha_{MN}^{\rm LO}=0, 1, 2, 3$ denotes the leading order (LO) terms in powers of $r-r_h$ in the IR when the ${\cal O}(\beta)$-terms are expanded in a Taylor series about $r=r_h$, and ${\cal F}_{\theta_1\theta_2} = {\cal F}_{\theta_2\theta_2} = 0$ (i.e. EOM$_{\theta_1\theta_2}$ and EOM$_{\theta_2\theta_2}$ are homogeneous up to ${\cal O}(\beta)$).

In the EOMs in this appendix and their solutions in Section {\bf 3}:
{\scriptsize
\begin{eqnarray}
\label{Sigma_1-3-def}
& & \hskip -0.8in\Sigma_1 \equiv 19683
   \sqrt{6} \alpha _{\theta _1}^6+6642 \alpha _{\theta _2}^2 \alpha _{\theta _1}^3-40 \sqrt{6} \alpha _{\theta _2}^4\stackrel{\rm Global}{\longrightarrow} N^{\frac{6}{5}}\left(19683
   \sqrt{6} \sin^6\theta_1+6642 \sin^2{\theta _2} \sin^3{\theta _1}-40 \sqrt{6} \sin^4{\theta _2}\right),
\nonumber\\
& & \hskip -0.8in\Sigma_2\equiv  \left(387420489 \sqrt{2} \alpha _{\theta _1}^{12}+87156324 \sqrt{3}
   \alpha _{\theta _2}^2 \alpha _{\theta _1}^9+5778054 \sqrt{2} \alpha _{\theta _2}^4 \alpha _{\theta _1}^6-177120 \sqrt{3} \alpha _{\theta
   _2}^6 \alpha _{\theta _1}^3+1600 \sqrt{2} \alpha _{\theta _2}^8\right)\nonumber\\
& &  \hskip -0.8in\stackrel{\rm Global}{\longrightarrow} N^{\frac{12}{5}}\left(387420489 \sqrt{2} \sin^{12}_{\theta _1}+87156324 \sqrt{3}
   \sin^2_{\theta _2} \sin^9{\theta _1}+5778054 \sqrt{2} \sin^4{\theta _2} \sin^6{\theta _1}-177120 \sqrt{3} \sin^6{\theta} \sin^3_{\theta _1}+1600 \sqrt{2} \sin^8{\theta _2}\right).
\end{eqnarray}
}

The following EOMs' solutions will be obtained assuming $f_{\theta_1y}''(r)=0$. One can show that one hence ends up 15 independent EOMs and four that serve as consistency checks. We now discuss all below.

(i) ${\rm EOM}_{tt}$:

{\footnotesize
\begin{eqnarray}
& & \frac{4 \left(9 b^2+1\right)^3 \left(4374 b^6+1035 b^4+9 b^2-4\right) \beta  b^8 M \left(\frac{1}{N}\right)^{9/4} \Sigma_1
   \left(6 a^2+  {r_h}^2\right) \log (  {r_h})}{27 \pi  \left(18 b^4-3 b^2-1\right)^5  \log N ^2   {N_f}   {r_h}^2
   \alpha _{\theta _2}^3 \left(9 a^2+  {r_h}^2\right)}-\frac{6 \left(  {r_h}^2-2 a^2\right)  {f_t}(r)}{  {r_h}
   \left(  {r_h}^2-3 a^2\right) (r-  {r_h})}\nonumber\\
    & & -\frac{32 \sqrt{2} \left(9 b^2+1\right)^4 \beta  b^{12}
   \left(\frac{1}{N}\right)^{3/20}\Sigma_1 (r-  {r_h})}{81 \pi ^3 \left(1-3 b^2\right)^{10} \left(6 b^2+1\right)^8
     {g_s}^{9/4}  \log N ^4 N^{61/60}   {N_f}^3   {r_h}^4 \alpha _{\theta _1}^7 \alpha _{\theta _2}^6 \left(-27 a^4+6
   a^2   {r_h}^2+  {r_h}^4\right)}+2  {f_t}''(r)=0,
\end{eqnarray}
}
where $\Sigma_1$ is defined in (\ref{Sigma_1-3-def}).

As, the solution to the differential equation:
\begin{equation}
2  {f_t}''(r)+\frac{ {\Gamma_{f_{t1}}}  {f_t}(r)}{r-  {r_h}}+ \Gamma_{f_{t2}} (r-  {r_h})+ \Gamma_{f_{t3}}=0,
\end{equation}
is given by:
{\scriptsize
\begin{eqnarray}
& & \frac{1}{2} \Biggl[-\frac{1}{ {\Gamma_{f_{t1}}}}\Biggl\{(r-  {r_h})^2 \Biggl( {\Gamma_{f_{t1}}}  {\Gamma_{f_{t2}}} (r-  {r_h}) I_1\left(\sqrt{2}
   \sqrt{ {\Gamma_{f_{t1}}} (  {r_h}-r)}\right) G_{1,3}^{2,1}\left(\left.\frac{\sqrt{ {\Gamma_{f_{t1}}}
   (  {r_h}-r)}}{\sqrt{2}},\frac{1}{2}\right|
\begin{array}{c}
 -\frac{3}{2} \\
 -\frac{1}{2},\frac{1}{2},-\frac{5}{2} \\
\end{array}
\right)\nonumber\\
& & + {\Gamma_{f_{t1}}}  {\Gamma_{f_{t3}}} I_1\left(\sqrt{2} \sqrt{ {\Gamma_{f_{t1}}} (  {r_h}-r)}\right)
   G_{1,3}^{2,1}\left(\left.\frac{\sqrt{ {\Gamma_{f_{t1}}} (  {r_h}-r)}}{\sqrt{2}},\frac{1}{2}\right|
\begin{array}{c}
 -\frac{1}{2} \\
 -\frac{1}{2},\frac{1}{2},-\frac{3}{2} \\
\end{array}
\right)\nonumber\\
& & +K_1\left(\sqrt{2} \sqrt{ {\Gamma_{f_{t1}}} (  {r_h}-r)}\right) \Biggl[\sqrt{2} \sqrt{ {\Gamma_{f_{t1}}} (  {r_h}-r)}
   \left(2  {\Gamma_{f_{t2}}} I_4\left(\sqrt{2} \sqrt{ {\Gamma_{f_{t1}}} (  {r_h}-r)}\right)- {\Gamma_{f_{t1}}}  {\Gamma_{f_{t3}}} \,
   _0\tilde{F}_1\left(;3;\frac{1}{2}  {\Gamma_{f_{t1}}} (  {r_h}-r)\right)\right)\nonumber\\
   & & +8  {\Gamma_{f_{t2}}} I_3\left(\sqrt{2}
   \sqrt{ {\Gamma_{f_{t1}}} (  {r_h}-r)}\right)\Biggr]\Biggr)\Biggr\} \nonumber\\
& &    +\sqrt{2} c_1 \sqrt{ {\Gamma_{f_{t1}}} (  {r_h}-r)}
   I_1\left(\sqrt{2} \sqrt{ {\Gamma_{f_{t1}}} (  {r_h}-r)}\right)-\sqrt{2} c_2 \sqrt{ {\Gamma_{f_{t1}}} (  {r_h}-r)}
   K_1\left(\sqrt{2} \sqrt{ {\Gamma_{f_{t1}}} (  {r_h}-r)}\right)\Biggr].
\end{eqnarray}
}
To prevent the occurency of a (logarithmic) singularity at $r=r_h$, one sets: $c_2=0$ which yields:

\begin{tcolorbox}[enhanced,width=7in,center upper,size=fbox,
    drop shadow southwest,sharp corners]
\begin{equation}
f_t(r) = \frac{1}{4}  {\Gamma_{f_{t3}}} (r-  {r_h})^2 + {\cal O}\left((r-r_h)^2\right),
\end{equation}
where:
{\footnotesize
\begin{eqnarray}
\label{ft3}
\Gamma_{f_{t3}} \equiv \frac{4 b^8 \left(9 b^2+1\right)^3 \left(4374 b^6+1035 b^4+9 b^2-4\right) \beta  M \left(\frac{1}{N}\right)^{9/4} \Sigma_1
   \left(6 a^2+  {r_h}^2\right) \log (  {r_h})}{27 \pi  \left(18 b^4-3 b^2-1\right)^5  \log N ^2   {N_f}   {r_h}^2
   \alpha _{\theta _2}^3 \left(9 a^2+  {r_h}^2\right)},\nonumber\\
& &
\end{eqnarray}
}
\end{tcolorbox}
where $\Sigma_1$ is defined in (\ref{Sigma_1-3-def}).

(ii) ${\rm EOM}_{x^1x^1}$:

{\footnotesize
\begin{eqnarray}
& &-\frac{6   {r_h} \left(57 a^4+14 a^2   {r_h}^2+  {r_h}^4\right)
   f(r)}{\left(  {r_h}^2-3 a^2\right) \left(6 a^2+  {r_h}^2\right) \left(9 a^2+  {r_h}^2\right) (r-  {r_h})}+2 f''(r)
\nonumber\\
& & -\frac{4 \left(9 b^2+1\right)^4 \left(39 b^2-4\right) \beta  b^8 M \left(\frac{1}{N}\right)^{9/4} \Sigma_1\left(6
   a^2+  {r_h}^2\right) \log (  {r_h})}{9 \pi  \left(3 b^2-1\right)^5 \left(6 b^2+1\right)^4  \log N ^2   {N_f}
     {r_h}^2 \alpha _{\theta _2}^3 \left(9 a^2+  {r_h}^2\right)}\nonumber\\
     & & -\frac{32 \sqrt{2} \left(9 b^2+1\right)^4 \beta  b^{12}
   \left(\frac{1}{N}\right)^{3/20} \Sigma_1 (r-  {r_h})}{81 \pi ^3 \left(1-3 b^2\right)^{10} \left(6 b^2+1\right)^8
     {g_s}^{9/4}  \log N ^4 N^{61/60}   {N_f}^3   {r_h}^4 \alpha _{\theta _1}^7 \alpha _{\theta _2}^6 \left(-27 a^4+6
   a^2   {r_h}^2+  {r_h}^4\right)}=0.
\nonumber\\
& &
\end{eqnarray}
}
This yields:

\begin{tcolorbox}[enhanced,width=7in,center upper,size=fbox,
    drop shadow southwest,sharp corners]
\begin{equation}
f(r) = \frac{1}{4}  \gamma_{f_2} (r - {r_h})^2 + {\cal O}\left((r-r_h)^3\right),
\end{equation}
where:
\begin{equation}
\gamma_{f_2} \equiv -\frac{4 b^8 \left(9 b^2+1\right)^4 \left(39 b^2-4\right) M \left(\frac{1}{N}\right)^{9/4} \beta  \left(6 a^2+{r_h}^2\right) \log
   ({r_h})\Sigma_1}{9 \pi  \left(3 b^2-1\right)^5 \left(6 b^2+1\right)^4 \log N ^2 {N_f} {r_h}^2 \left(9 a^2+{r_h}^2\right) \alpha
   _{\theta _2}^3}.
\end{equation}
\end{tcolorbox}

(iii) ${\rm EOM}_{\theta_1x}$:

{\footnotesize
\begin{eqnarray}
\label{f-theta1phi1-i}
& &  -3
    {f_{\theta_1z}}''(r)+2  {f_{\theta_1x}}''(r)-3  {f_{\theta_2y}}''(r) -\frac{4 \left(9 b^2+1\right)^4 \beta  b^{10} M \sqrt[5]{\frac{1}{N}} \Sigma_1 \left(6
   a^2+  {r_h}^2\right)}{3 \pi  \left(-18 b^4+3 b^2+1\right)^4  \log N  \sqrt[3]{N}   {N_f} \alpha
   _{\theta _2}^3 \left(  {r_h}^2-3 a^2\right) \left(9 a^2+  {r_h}^2\right)}\nonumber\\
   & & -\frac{32 \sqrt{2} \left(9
   b^2+1\right)^4 \beta  b^{12} \left(\frac{1}{N}\right)^{3/20} \Sigma_1 (r-  {r_h})}{81 \pi ^3
   \left(1-3 b^2\right)^{10} \left(6 b^2+1\right)^8   {g_s}^{9/4}  \log N ^4 N^{61/60}   {N_f}^3
     {r_h}^4 \alpha _{\theta _1}^7 \alpha _{\theta _2}^6 \left(-27 a^4+6 a^2   {r_h}^2+  {r_h}^4\right)}
     =0.
\end{eqnarray}
}
Choosing the two constants of integration obtained by solving (\ref{f-theta1phi1-i}) in such a way that the Neumann b.c. at $r=r_h: f_{\theta_1x}^\prime(r=r_h)=0$, one obtains:

\begin{tcolorbox}[enhanced,width=7in,center upper,size=fbox,
    drop shadow southwest,sharp corners]
{\footnotesize
\begin{eqnarray}
\label{f-theta1phi1-ii}
f_{\theta_1x}(r) = \left(- \frac{\left(9 b^2+1\right)^4 b^{10} M  \left(6 a^2+{r_h}^2\right) \left((r-{r_h})^2+{r_h}^2\right)
   \Sigma_1}{3 \pi  \left(-18 b^4+3 b^2+1\right)^4 \log N  N^{8/15} {N_f} \left(-27 a^4+6 a^2
   {r_h}^2+{r_h}^4\right) \alpha _{\theta _2}^3}+{C_{\theta_1x}}^{(1)}\right)\beta + {\cal O}(r-r_h)^3
\end{eqnarray}
}
\end{tcolorbox}

(iv) ${\rm EOM}_{\theta_1y}$:

{\footnotesize
\begin{eqnarray}
\label{EOM-theta1phi2}
& & \hskip -0.8in -6 f''(r)+2  {f_{zz}}''(r)-2
    {f_{x^{10} x^{10}}}''(r)-3  {f_{\theta_1z}}''(r)-3  {f_{\theta_2y}}''(r)-3
    {f_{xz}}''(r)-2  {f_t}''(r) \nonumber\\
& & \hskip -0.8in\frac{32 \sqrt{2} \left(9 b^2+1\right)^4 \beta  b^{12}
   \left(\frac{1}{N}\right)^{3/20}\Sigma_1
   (r-  {r_h})}{81 \pi ^3 \left(3 b^2-1\right)^{10} \left(6
   b^2+1\right)^8   {g_s}^{9/4}  \log N ^4 N^{61/60}
     {N_f}^3   {r_h}^4 \alpha _{\theta _1}^7 \alpha _{\theta
   _2}^6 \left(  {r_h}^2-3 a^2\right) \left(9
   a^2+  {r_h}^2\right)}=0.
\end{eqnarray}
}
The equation (\ref{EOM-theta1phi2}) can be shown to be equivalent to a decoupled second order EOM for $f_{xz}$. Then, expanding the solution around the horizon and requiring the constant of integration $C_{xz}^{(1)}$ appearing in the ${\cal O}(r-r_h)^0$ term to satisfy:
\begin{eqnarray}
\label{fphi1psi-i}
& & -\frac{32 \left(9 b^2+1\right)^4 b^{12} \beta  \left(19683 \sqrt{3} \alpha _{\theta _1}^6+3321 \sqrt{2} \alpha
   _{\theta _2}^2 \alpha _{\theta _1}^3-40 \sqrt{3} \alpha _{\theta _2}^4\right)}{729 \pi ^3 \left(1-3
   b^2\right)^{10} \left(6 b^2+1\right)^8 {g_s}^{9/4} \log N ^4 N^{7/6} {N_f}^3 \left(-27 a^4 {r_h}+6
   a^2 {r_h}^3+{r_h}^5\right) \alpha _{\theta _1}^7 \alpha _{\theta _2}^6}\nonumber\\
   & & -\frac{4 \left(9 b^2+1\right)^4
   b^{10} M {r_h}^2 \beta  \log ({r_h}) \Sigma_1}{81 \pi ^{3/2} \left(3 b^2-1\right)^5 \left(6
   b^2+1\right)^4 \sqrt{{g_s}} \log N ^2 N^{23/20} {N_f} \alpha _{\theta _2}^5}+C_{xz}^{(1)} = 0,
\end{eqnarray}
one obtains:

\begin{tcolorbox}[enhanced,width=7in,center upper,size=fbox,
    drop shadow southwest,sharp corners]
\begin{eqnarray}
& & f_{xz}(r) = \frac{18 b^{10} \left(9 b^2+1\right)^4 M \beta  \left(6 a^2+{r_h}^2\right)
   \left(\frac{(r-{r_h})^2}{{r_h}^2}+1\right) \log ^3({r_h}) \Sigma_1}{\pi  \left(3
   b^2-1\right)^5 \left(6 b^2+1\right)^4 \log N ^4 N^{5/4} {N_f} \left(9 a^2+{r_h}^2\right) \alpha
   _{\theta _2}^3}.
\end{eqnarray}
\end{tcolorbox}

(v) ${\rm EOM}_{\theta_1z}$:

{\footnotesize
\begin{eqnarray}
\label{f-theta1psi-i}
2  {f_{\theta_1z}}''(r)-\frac{32 \sqrt{2} b^{12} \left(9 b^2+1\right)^4 \beta  \left(\frac{1}{N}\right)^{3/20} \Sigma_1
   (r-  {r_h})}{81 \pi ^3 \left(1-3 b^2\right)^{10} \left(6 b^2+1\right)^8   {g_s}^{9/4}  \log N ^4 N^{61/60}
     {N_f}^3   {r_h}^4 \alpha _{\theta _1}^7 \alpha _{\theta _2}^6 \left(-27 a^4+6 a^2   {r_h}^2+  {r_h}^4\right)}=0.
\end{eqnarray}
}
Choosing the two constants of integration obtained by solving (\ref{f-theta1psi-i}) in such a way that the Neumann b.c. at $r=r_h: f_{\theta_1z}^\prime(r=r_h)=0$, one obtains:

\begin{tcolorbox}[enhanced,width=7in,center upper,size=fbox,
    drop shadow southwest,sharp corners]
{\footnotesize
\begin{eqnarray}
\label{f-theta1psi-ii}
& & \hskip -0.3in f_{\theta_1z}(r) = \left(\frac{16 \left(9 b^2+1\right)^4 b^{12} \left(\frac{1}{N}\right)^{3/20}  \left(\frac{(r-{r_h})^3}{{r_h}^3}+1\right) \left(19683
   \sqrt{3} \alpha _{\theta _1}^6+3321 \sqrt{2} \alpha _{\theta _2}^2 \alpha _{\theta _1}^3-40 \sqrt{3} \alpha _{\theta _2}^4\right)}{243
   \pi ^3 \left(1-3 b^2\right)^{10} \left(6 b^2+1\right)^8 {g_s}^{9/4} \log N ^4 N^{61/60} {N_f}^3 \left(-27 a^4 {r_h}+6 a^2
   {r_h}^3+{r_h}^5\right) \alpha _{\theta _1}^7 \alpha _{\theta _2}^6}+C_{\theta_1z}^{(1)}\right)\beta+ {\cal O}(r-r_h)^3.\nonumber\\
& &
\end{eqnarray}
}
\end{tcolorbox}

(vi) ${\rm EOM}_{\theta_2x}$:

{\footnotesize
\begin{eqnarray}
2  {f_{\theta_2x}}''(r)-\frac{32 \sqrt{2} b^{12} \left(9 b^2+1\right)^4 \beta  \left(\frac{1}{N}\right)^{3/20} \Sigma_1 (r-  {r_h})}{81 \pi ^3
   \left(1-3 b^2\right)^{10} \left(6 b^2+1\right)^8   {g_s}^{9/4}  \log N ^4 N^{61/60}   {N_f}^3   {r_h}^4 \alpha _{\theta
   _1}^7 \alpha _{\theta _2}^6 \left(-27 a^4+6 a^2   {r_h}^2+  {r_h}^4\right)} = 0.
\end{eqnarray}
}

\begin{tcolorbox}[enhanced,width=7in,center upper,size=fbox,
    drop shadow southwest,sharp corners]
{\footnotesize
\begin{eqnarray}
\label{f-theta2phi1-i}
& & \hskip -0.3in f_{\theta_2x}(r) = \left(\frac{16 \left(9 b^2+1\right)^4 b^{12} \left(\frac{1}{N}\right)^{3/20}   \left(\frac{(r-{r_h})^3}{{r_h}^3}+1\right) \left(19683 \sqrt{3}
   \alpha _{\theta _1}^6+3321 \sqrt{2} \alpha _{\theta _2}^2 \alpha _{\theta _1}^3-40 \sqrt{3} \alpha _{\theta _2}^4\right)}{243 \pi ^3 \left(1-3
   b^2\right)^{10} \left(6 b^2+1\right)^8 {g_s}^{9/4} \log N ^4 N^{61/60} {N_f}^3 \left(-27 a^4 {r_h}+6 a^2
   {r_h}^3+{r_h}^5\right) \alpha _{\theta _1}^7 \alpha _{\theta _2}^6}+C_{\theta_2x}^{((1)}\right)\beta\nonumber\\
& &
\end{eqnarray}
}
\end{tcolorbox}

(vii) ${\rm EOM}_{\theta_2y}$:

{\footnotesize
\begin{eqnarray}
\label{f-theta2phi2-i}
& & 2  {f_{\theta_2y}}''(r) + \frac{12 \left(9 b^2+1\right)^4 \beta  b^{10} M \left(\frac{1}{N}\right)^{7/5} \Sigma_1 \left(6
   a^2+  {r_h}^2\right) \log (  {r_h})}{\pi  \left(3 b^2-1\right)^5 \left(6 b^2+1\right)^4  \log N ^2   {N_f}
     {r_h}^2 \alpha _{\theta _2}^3 \left(9 a^2+  {r_h}^2\right)}\nonumber\\
     & & -\frac{32 \sqrt{2} \left(9 b^2+1\right)^4 \beta  b^{12}
   \left(\frac{1}{N}\right)^{3/20} \Sigma_1 (r-  {r_h})}{81 \pi ^3 \left(1-3 b^2\right)^{10} \left(6 b^2+1\right)^8
     {g_s}^{9/4}  \log N ^4 N^{61/60}   {N_f}^3   {r_h}^4 \alpha _{\theta _1}^7 \alpha _{\theta _2}^6 \left(-27
   a^4+6 a^2   {r_h}^2+  {r_h}^4\right)}=0.
\end{eqnarray}
}

Choosing the two constants of integration obtained by solving (\ref{f-theta2phi2-i}) in such a way that the Neumann b.c. at $r=r_h: f_{\theta_2y}^\prime(r=r_h)=0$, and requiring the constant of integration $C_{\theta_2y}^{(1)}$ that figures in the ${\cal O}(r-r_h)^0$-term to satisfy:
{\footnotesize
\begin{eqnarray}
\label{Ctheta2phi2}
\frac{16 \left(9 b^2+1\right)^4 b^{12}  \left(19683 \sqrt{3} \alpha _{\theta _1}^6+3321 \sqrt{2} \alpha _{\theta _2}^2 \alpha
   _{\theta _1}^3-40 \sqrt{3} \alpha _{\theta _2}^4\right)}{243 \pi ^3 \left(1-3 b^2\right)^{10} \left(6 b^2+1\right)^8 {g_s}^{9/4}
   \log N ^4 N^{7/6} {N_f}^3 \left(-27 a^4 {r_h}+6 a^2 {r_h}^3+{r_h}^5\right) \alpha _{\theta _1}^7 \alpha _{\theta
   _2}^6}+C_{\theta_2y}^{(1)} = 0,
\end{eqnarray}
}
 one obtains:

\begin{tcolorbox}[enhanced,width=7in,center upper,size=fbox,
    drop shadow southwest,sharp corners]
{\footnotesize
\begin{eqnarray}
\label{f-theta2phi2-ii}
f_{\theta_2y} = \frac{3 b^{10} \left(9 b^2+1\right)^4 M \beta  \left(6 a^2+{r_h}^2\right) \left(1-\frac{(r-{r_h})^2}{{r_h}^2}\right) \log
   ({r_h}) \Sigma_1}{\pi  \left(3 b^2-1\right)^5 \left(6 b^2+1\right)^4 \log N ^2 N^{7/5} {N_f} \left(9 a^2+{r_h}^2\right) \alpha
   _{\theta _2}^3}+ {\cal O}\left((r-r_h)^3\right).
\end{eqnarray}
}
\end{tcolorbox}

(viii) ${\rm EOM}_{\theta_2z}$

{\footnotesize
\begin{eqnarray}
\label{f-theta2z-i}
& & \frac{12 \left(9 b^2+1\right)^4 \beta  b^{10} M \sqrt{\frac{1}{N}} \Sigma_1 \left(6 a^2+{r_h}^2\right) \log
   ({r_h})}{\pi  \left(3 b^2-1\right)^5 \left(6 b^2+1\right)^4 {\log N}^2 N^{2/3} {N_f} {r_h}^2 \alpha
   _{\theta _2}^3 \left(9 a^2+{r_h}^2\right)}\nonumber\\
& & -\frac{32 \sqrt{2} \left(9 b^2+1\right)^4 \beta  b^{12}
   \left(\frac{1}{N}\right)^{3/20} \Sigma_1 (r-{r_h})}{81 \pi ^3 \left(1-3 b^2\right)^{10} \left(6 b^2+1\right)^8
   {g_s}^{9/4} {\log N}^4 N^{61/60} {N_f}^3 {r_h}^4 \alpha _{\theta _1}^7 \alpha _{\theta _2}^6 \left(-27
   a^4+6 a^2 {r_h}^2+{r_h}^4\right)}+2 {f_{\theta_2z}}''(r)=0.\nonumber\\
& &
\end{eqnarray}
}

Choosing the two constants of integration obtained by solving (\ref{f-theta2z-i}) in such a way that the Neumann b.c. at $r=r_h: f_{\theta_2z}^\prime(r=r_h)=0$, one obtains:

\begin{tcolorbox}[enhanced,width=7in,center upper,size=fbox,
    drop shadow southwest,sharp corners]
{\footnotesize
\begin{eqnarray}
\label{f-theta2z-ii}
& & \hskip -0.3in f_{\theta_2z} = \left(\frac{3 \left(9 b^2+1\right)^4 b^{10} M   \left(6 a^2+{r_h}^2\right) \left(1-\frac{(r-{r_h})^2}{{r_h}^2}\right) \log
   ({r_h}) \left(19683 \sqrt{6} \alpha _{\theta _1}^6+6642 \alpha _{\theta _2}^2 \alpha _{\theta _1}^3-40 \sqrt{6} \alpha _{\theta
   _2}^4\right)}{\pi  \left(3 b^2-1\right)^5 \left(6 b^2+1\right)^4 {\log N}^2 N^{7/6} {N_f} \left(9 a^2+{r_h}^2\right) \alpha
   _{\theta _2}^3}+{C_{\theta_2z}}^{(1)}\right)\beta.\nonumber\\
& &
\end{eqnarray}
}
\end{tcolorbox}

(ix) ${\rm EOM}_{xx}$:

{\footnotesize
\begin{eqnarray}
\label{f-phi1phi1-i}
& & {f_{zz}}(r)-2  {f_{\theta_1z}}(r)+2  {f_{\theta_1\phi_{1}}}(r)- {f_r}(r)\nonumber\\
& & +\frac{81 \left(9 b^2+1\right)^4 \beta  b^{10} M \left(\frac{1}{N}\right)^{53/20} \alpha _{\theta _1}^4 \left(19683
   \sqrt{6} \alpha _{\theta _1}^6+6642 \alpha _{\theta _2}^2 \alpha _{\theta _1}^3-40 \sqrt{6} \alpha _{\theta
   _2}^4\right) \left(  {r_h}^2-3 a^2\right)^2 \left(6 a^2+  {r_h}^2\right) \log (  {r_h})}{16 \pi  \left(3
   b^2-1\right)^5  \log N ^2   {N_f} \left(6 a b^2+a\right)^4 \alpha _{\theta _2} \left(9
   a^2+  {r_h}^2\right)}=0.\nonumber\\
   & &
\end{eqnarray}
}
Substituting (\ref{f-theta1phi1-ii}), (\ref{f-theta1psi-ii}) and (\ref{f-psipsi-ii}) into (\ref{f-phi1phi1-i}), one obtains:

\begin{tcolorbox}[enhanced,width=7in,center upper,size=fbox,
    drop shadow southwest,sharp corners]
{\footnotesize
\begin{eqnarray}
\label{f-phi1phi1-ii}
f_r(r) & = & \Biggl(- \frac{2 \left(9 b^2+1\right)^4 b^{10} M   \left(6 a^2+{r_h}^2\right) \left((r-{r_h})^2+{r_h}^2\right)\Sigma_1}{3 \pi
   \left(-18 b^4+3 b^2+1\right)^4 \log N  N^{8/15} {N_f} \left(-27 a^4+6 a^2 {r_h}^2+{r_h}^4\right) \alpha _{\theta
   _2}^3}\nonumber\\
& & +{C_{zz}}^{(1)}-2 {C_{\theta_1z}}^{(1)}+2 {C_{\theta_1x}}^{(1)}\Biggr)\beta + {\cal O}(r-r_h)^3.
\end{eqnarray}
}
\end{tcolorbox}

(x) ${\rm EOM}_{xy}$:

{\footnotesize
\begin{eqnarray}
& & 2  {f_{xy}}''(r) + \frac{12 \left(9 b^2+1\right)^4 \beta  b^{10} M \left(\frac{1}{N}\right)^{21/20} \Sigma_1 \left(6 a^2+  {r_h}^2\right) \log
   (  {r_h})}{\pi  \left(3 b^2-1\right)^5 \left(6 b^2+1\right)^4  \log N ^2   {N_f}   {r_h}^2 \alpha _{\theta _2}^3
   \left(9 a^2+  {r_h}^2\right)}\nonumber\\
   & & -\frac{32 \sqrt{2} \left(9 b^2+1\right)^4 \beta  b^{12} \left(\frac{1}{N}\right)^{3/20}
  \Sigma_1 (r-  {r_h})}{81 \pi ^3 \left(1-3 b^2\right)^{10} \left(6 b^2+1\right)^8   {g_s}^{9/4}  \log N ^4
   N^{61/60}   {N_f}^3   {r_h}^4 \alpha _{\theta _1}^7 \alpha _{\theta _2}^6 \left(-27 a^4+6 a^2
     {r_h}^2+  {r_h}^4\right)}= 0.
\end{eqnarray}
}
Choosing the two constants of integration obtained by solving (\ref{f-phi1phi2-i}) in such a way that the Neumann b.c. at $r=r_h: f_{xy}^\prime(r=r_h)=0$, one obtains:

\begin{tcolorbox}[enhanced,width=7in,center upper,size=fbox,
    drop shadow southwest,sharp corners]
{\footnotesize
\begin{eqnarray}
\label{f-phi1phi2-i}
f_{xy}(r) =\left(\frac{3 \left(9 b^2+1\right)^4 b^{10} M  \left(6 a^2+{r_h}^2\right) \left(\frac{(r-{r_h})^2}{{r_h}^2}+1\right) \log
   ({r_h}) \alpha _{\theta _2}^3\Sigma_1}{\pi  \left(3 b^2-1\right)^5 \left(6 b^2+1\right)^4 \log N ^2 N^{21/20} {N_f} \left(9
   a^2+{r_h}^2\right) \alpha _{\theta _{2 l}}^6}+C_{xy}^{(1)}\right)\beta+ {\cal O}\left(r-r_h\right)^3.
\end{eqnarray}
}
\end{tcolorbox}

(xi) EOM$\  _{xz}$:

{\footnotesize
\begin{eqnarray}
& & -8  {f_t}''(r)+\frac{24 \left(9 b^2+1\right)^4 \beta  b^{10} M \left(\frac{1}{N}\right)^{3/4} \Sigma_1 \left(9
   a^2+  {r_h}^2\right) \log (  {r_h})}{\pi ^{3/2} \left(3 b^2-1\right)^5 \left(6 b^2+1\right)^4 \sqrt{  {g_s}}
    \log N ^2   {N_f} \alpha _{\theta _1}^2 \alpha _{\theta _2}^5}\nonumber\\
    & & -\frac{64 \sqrt{2} \left(9 b^2+1\right)^4 \beta
   b^{12} \left(\frac{1}{N}\right)^{3/20} \Sigma_1 (r-  {r_h})}{81 \pi ^3 \left(1-3 b^2\right)^{10} \left(6
   b^2+1\right)^8   {g_s}^{9/4}  \log N ^4 N^{61/60}   {N_f}^3   {r_h}^4 \alpha _{\theta _1}^7 \alpha _{\theta
   _2}^6 \left(-27 a^4+6 a^2   {r_h}^2+  {r_h}^4\right)}=0.
\end{eqnarray}
}
The solution is given as under:
\begin{tcolorbox}[enhanced,width=7in,center upper,size=fbox,
    drop shadow southwest,sharp corners]
\begin{equation}
f_t(r) = \frac{1}{16}  \gamma_{f_{t2}}  (r-  {r_h})^2 + {\cal O}\left((r-  {r_h})^3\right),
\end{equation}
where:
\begin{eqnarray}
\gamma_{f_{t2}} \equiv \frac{24 b^{10} \left(9 b^2+1\right)^4 \beta  M \left(\frac{1}{N}\right)^{3/4} \Sigma_1 \left(9
   a^2+  {r_h}^2\right) \log (  {r_h})}{\pi ^{3/2} \left(3 b^2-1\right)^5 \left(6 b^2+1\right)^4 \sqrt{  {g_s}}
    \log N ^2   {N_f} \alpha _{\theta _1}^2 \alpha _{\theta _2}^5}.
\end{eqnarray}
\end{tcolorbox}
Consistency with (\ref{ft3}) requires
(as in \cite{mesons_0E++-to-mesons-decays} wherein $r_h\sim N^{-\alpha}, \alpha>0$):
\begin{equation}
r_h = \frac{\sqrt[8]{\pi } \sqrt[4]{4374 b^6+1035 b^4+9 b^2-4} \sqrt[8]{{g_s}} \left(\frac{1}{N}\right)^{3/8} \sqrt{\alpha
   _{\theta _1} \alpha _{\theta _2}}}{3 \sqrt[4]{2} \sqrt{b} \left(9 b^2+1\right)^{3/4}}.
\end{equation}
Note $4374 b^6+1035 b^4+9 b^2-4>0$ for $b$ given as in (\ref{ansatz-b}).

(xii) ${\rm EOM}_{yy}$:

{\footnotesize
\begin{eqnarray}
\label{f-phi2phi2-i}
& & 2
    {f_{\phi_{2}\phi_{2}}}''(r) + \frac{12 \left(9 b^2+1\right)^4 \beta  b^{10} M \left(\frac{1}{N}\right)^{7/4} \Sigma_1 \left(6 a^2+  {r_h}^2\right) \log
   (  {r_h})}{\pi  \left(3 b^2-1\right)^5 \left(6 b^2+1\right)^4  \log N ^2   {N_f}   {r_h}^2 \alpha _{\theta _2}^3
   \left(9 a^2+  {r_h}^2\right)}\nonumber\\
   & & -\frac{32 \sqrt{2} \left(9 b^2+1\right)^4 \beta  b^{12} \left(\frac{1}{N}\right)^{3/20}
  \Sigma_1 (r-  {r_h})}{81 \pi ^3 \left(1-3 b^2\right)^{10} \left(6 b^2+1\right)^8   {g_s}^{9/4}  \log N ^4 N^{61/60}
     {N_f}^3   {r_h}^4 \alpha _{\theta _1}^7 \alpha _{\theta _2}^6 \left(-27 a^4+6 a^2   {r_h}^2+  {r_h}^4\right)}=0
\end{eqnarray}
}
Choosing the two constants of integration obtained by solving (\ref{f-phi2phi2-i}) in such a way that the Neumann b.c. at $r=r_h: f_{yy}^\prime(r=r_h)=0$, and choosing the constant of integration $C_{yy}^{(1)}$ appearing in the ${\cal O}(r-r_h)^0$-term  to satisfy:
{\footnotesize
\begin{eqnarray}
\label{Cphi2phi2[1]}
\frac{16 \left(9 b^2+1\right)^4 b^{12}  \left(19683 \sqrt{3} \alpha _{\theta _1}^6+3321 \sqrt{2} \alpha _{\theta _2}^2 \alpha _{\theta
   _1}^3-40 \sqrt{3} \alpha _{\theta _2}^4\right)}{243 \pi ^3 \left(1-3 b^2\right)^{10} \left(6 b^2+1\right)^8 {g_s}^{9/4} \log N ^4
   N^{7/6} {N_f}^3 \left(-27 a^4 {r_h}+6 a^2 {r_h}^3+{r_h}^5\right) \alpha _{\theta _1}^7 \alpha _{\theta _2}^6}+C_{yy}^{(1)} = 0,
\end{eqnarray}
}
one obtains:

\begin{tcolorbox}[enhanced,width=7in,center upper,size=fbox,
    drop shadow southwest,sharp corners]
{\footnotesize
\begin{eqnarray}
\label{f-phi2phi2-ii}
& & \hskip -0.8in f_{yy}(r) = - \frac{3 b^{10} \left(9 b^2+1\right)^4 M \left(\frac{1}{N}\right)^{7/4} \beta  \left(6 a^2+{r_h}^2\right) \log ({r_h})\Sigma_1
   \left(\frac{(r-{r_h})^2}{h^2 r^2}+1\right)}{\pi  \left(3 b^2-1\right)^5 \left(6 b^2+1\right)^4 \log N ^2 {N_f} {r_h}^2 \left(9
   a^2+{r_h}^2\right) \alpha _{\theta _2}^3}+ {\cal O}\left(\left(r-r_h\right)^3\right).\nonumber\\
   & &
\end{eqnarray}
}
\end{tcolorbox}

(xiii) ${\rm EOM}_{yz}$:
{\footnotesize
\begin{eqnarray}
\label{f-phi2psi-i}
2  {f_{\phi_{2}\psi}}''(r)-\frac{128 \sqrt{2} b^{22} \left(9 b^2+1\right)^8 \beta ^2 M \left(\frac{1}{N}\right)^{3/5} \Sigma_1{}^2
   \left(6 a^2+  {r_h}^2\right) (r-  {r_h}) \log (  {r_h})}{27 \pi ^4 \left(3 b^2-1\right)^{15} \left(6 b^2+1\right)^{12}
     {g_s}^{9/4}  \log N ^6 N^{109/60}   {N_f}^4   {r_h}^6 \alpha _{\theta _1}^7 \alpha _{\theta _2}^9
   \left(  {r_h}^2-3 a^2\right) \left(9 a^2+  {r_h}^2\right)^2}=0.
\end{eqnarray}
}
Choosing the two constants of integration obtained by solving (\ref{f-x10x10-i}) in such a way that the Neumann b.c. at $r=r_h: f_{x^{10}x^{10}}^\prime(r=r_h)=0$, one obtains:

\hskip -0.5in
\begin{tcolorbox}[enhanced,width=7in,center upper,size=fbox,
    drop shadow southwest,sharp corners]
{\footnotesize
\begin{eqnarray}
\label{f-phi2psi-ii}
& & \hskip -0.3in f_{yz}(r) =\Biggl( \frac{64 \left(9 b^2+1\right)^8 b^{22} M \left(\frac{1}{N}\right)^{3/5}  \left(6 a^2+{r_h}^2\right)
   \left(\frac{(r-{r_h})^3}{{r_h}^3}+1\right) \log ({r_h}) }{27 \pi ^4 \left(3 b^2-1\right)^{15} \left(6 b^2+1\right)^{12}
   {g_s}^{9/4} \log N ^6 N^{109/60} {N_f}^4 {r_h}^3 \left({r_h}^2-3 a^2\right) \left(9 a^2+{r_h}^2\right)^2 \alpha
   _{\theta _1}^7 \alpha _{\theta _2}^9}\nonumber\\
& & \hskip -0.3in \times \left(387420489 \sqrt{2} \alpha _{\theta _1}^{12}+87156324 \sqrt{3}
   \alpha _{\theta _2}^2 \alpha _{\theta _1}^9+5778054 \sqrt{2} \alpha _{\theta _2}^4 \alpha _{\theta _1}^6-177120 \sqrt{3} \alpha _{\theta
   _2}^6 \alpha _{\theta _1}^3+1600 \sqrt{2} \alpha _{\theta _2}^8\right)+C_{yz}^{(1)}\Biggr)\beta+ {\cal O}(r-r_h)^3.\nonumber\\
& &
   \end{eqnarray}
}
\end{tcolorbox}

(xiv) ${\rm EOM}_{zz}$:

{\footnotesize
\begin{eqnarray}
\label{f-psipsi-i}
& & 2  {f_{zz}}''(r)-\frac{32 \sqrt{2} \left(9 b^2+1\right)^4 \beta  b^{12} \left(\frac{1}{N}\right)^{3/20} \Sigma_1
   (r-  {r_h})}{81 \pi ^3 \left(1-3 b^2\right)^{10} \left(6 b^2+1\right)^8   {g_s}^{9/4}  \log N ^4 N^{61/60}
     {N_f}^3   {r_h}^4 \alpha _{\theta _1}^7 \alpha _{\theta _2}^6 \left(-27 a^4+6 a^2
     {r_h}^2+  {r_h}^4\right)}\nonumber\\
 & &   +\frac{4 \left(9 b^2+1\right)^4 \beta  b^{10} M \left(\frac{1}{N}\right)^{23/20}
  \Sigma_1 (r-  {r_h}) \log (  {r_h})}{9 \pi ^{3/2} \left(3 b^2-1\right)^5 \left(6 b^2+1\right)^4
   \sqrt{  {g_s}}  \log N ^2   {N_f}   {r_h} \alpha _{\theta _2}^5}=0.
\end{eqnarray}
}
Choosing the two constants of integration obtained by solving (\ref{f-psipsi-i}) in such a way that the Neumann b.c. at $r=r_h: f_{zz}^\prime(r=r_h)=0$, one obtains:

\begin{tcolorbox}[enhanced,width=7in,center upper,size=fbox,
    drop shadow southwest,sharp corners]
{\footnotesize
\begin{eqnarray}
\label{f-psipsi-ii}
f_{zz}(r) = \left(C_{zz}^{(1)}-\frac{b^{10} \left(9 b^2+1\right)^4 M   \left({r_h}^2-\frac{(r-{r_h})^3}{{r_h}}\right) \log ({r_h})
   \Sigma_1}{27 \pi ^{3/2} \left(3 b^2-1\right)^5 \left(6 b^2+1\right)^4 \sqrt{{g_s}} \log N ^2 N^{23/20} {N_f} \alpha
   _{\theta _2}^5}\right)\beta+ {\cal O}(r-r_h)^3.
\end{eqnarray}
}
\end{tcolorbox}

(xv) ${\rm EOM}_{x^{10}x^{10}}$:
{\footnotesize
\begin{eqnarray*}
\label{f-x10x10-i}
& & \frac{4 \left(9 b^2+1\right)^3 \beta  b^8 M \left(\frac{1}{N}\right)^{5/4} \Sigma_1 \left(6 a^2+  {r_h}^2\right) \left(9
   b^4 (27  \log N +16)+3 b^2 (9  \log N -8)-8\right) \log ^3(  {r_h})}{\pi  \left(3 b^2-1\right)^5 \left(6
   b^2+1\right)^4  \log N ^5   {N_f}   {r_h}^2 \alpha _{\theta _2}^3 \left(9 a^2+  {r_h}^2\right)}\nonumber\\
   & & - \frac{4 \left(9
   b^2+1\right)^4 \beta  b^{10} M \left(\frac{1}{N}\right)^{23/20} \Sigma_1 (r-  {r_h}) \log (  {r_h})}{9 \pi ^{3/2}
   \left(3 b^2-1\right)^5 \left(6 b^2+1\right)^4 \sqrt{  {g_s}}  (\log N) ^2   {N_f}   {r_h} \alpha _{\theta _2}^5}+2
    {f_{x^{10} x^{10}}}''(r)=0.
\end{eqnarray*}
}
Choosing the two constants of integration obtained by solving (\ref{f-x10x10-i}) in such a way that the Neumann b.c. at $r=r_h: f_{x^{10}x^{10}}^\prime(r=r_h)=0$, and requiring the constant of integration $C_{x^{10}x^{10}}^{(1)}$ appearing in the ${\cal O}(r-r_h)^0$ to satisfy:
\begin{eqnarray}
\frac{\left(9 b^2+1\right)^4 b^{10} M {r_h}^2 \beta  \log ({r_h}) \Sigma_1}{27 \pi ^{3/2} \left(3 b^2-1\right)^5 \left(6 b^2+1\right)^4
   \sqrt{{g_s}} \log N ^2 N^{23/20} {N_f} \alpha _{\theta _2}^5}+C_{x^{10}x^{10}}^{(1)} = 0,
\end{eqnarray}
 one obtains:

\begin{tcolorbox}[enhanced,width=7in,center upper,size=fbox,
    drop shadow southwest,sharp corners]
{\footnotesize
\begin{eqnarray}
\label{f-x10x10-ii}
f_{x^{10}x^{10}} = -\frac{27 b^{10} \left(9 b^2+1\right)^4 M \left(\frac{1}{N}\right)^{5/4} \beta  \left(6 a^2+{r_h}^2\right)
   \left(1-\frac{(r-{r_h})^2}{{r_h}^2}\right) \log ^3({r_h}) \Sigma_1}{\pi  \left(3 b^2-1\right)^5 \left(6 b^2+1\right)^4 \log N ^4
   {N_f} {r_h}^2 \left(9 a^2+{r_h}^2\right) \alpha _{\theta _2}^3} + {\cal O}(r-r_h)^3.
\end{eqnarray}
}
\end{tcolorbox}

The remaining EOMS provide consistency checks and are listed below:

\begin{itemize}
\item ${\rm EOM}_{rr}$:

\begin{eqnarray}
\label{consistency-EOMrr-i}
& &  \frac{3  {\alpha_h} \left(9 b^2+1\right)^3 \beta  b^{10} M \Sigma_1}{\pi
   \left(3 b^2-1\right)^5 \left(6 b^2+1\right)^3  \log N ^2 N^{11/12}   {N_f}   {r_h}^2 \alpha
   _{\theta _2}^3}-\frac{ {f_{\theta_1z}}''(r)}{4}-\frac{ {f_{\theta_2y}}''(r)}{4}-\frac{ {f_{xz}}''(r)}{4} = 0.\nonumber\\
& &
\end{eqnarray}

\item ${\rm EOM}_{\theta_1\theta_1}$
{\footnotesize
\begin{eqnarray}
\label{consistency-EOMtheta1theta1-i}
& & \hskip -0.8in \frac{ {f_{zz}}(r)}{2}- {f_{yz}}(r)+\frac{ {f_{yy}}(r)}{2}\nonumber\\
& & \hskip -0.8in -\frac{32 \sqrt{2} \sqrt[4]{\pi } \left(9 b^2+1\right)^3 \beta  b^{12} \left(\frac{1}{N}\right)^{7/10}
   \left(-19683 \alpha _{\theta _1}^6+216 \sqrt{6} \alpha _{\theta _2}^2 \alpha _{\theta _1}^3+530 \alpha
   _{\theta _2}^4\right)\Sigma_1 \left(6 a^2   {r_h}+  {r_h}^3\right)
   (r-  {r_h})^2}{14348907 \left(1-3 b^2\right)^4   {g_s}^{7/4}  \log N  N^{4/5} \alpha _{\theta _1}^8
   \left(9 a^2+  {r_h}^2\right) \left(6 b^2   {r_h}+  {r_h}\right)^3 \left(108 b^2   {N_f}
     {r_h}^2+  {N_f}\right)^2}=0.\nonumber\\
     & &
\end{eqnarray}
}

\item ${\rm EOM}_{\theta_1\theta_2}$
{\scriptsize
\begin{eqnarray}
\label{consistency-EOMtheta1theta2-i}
-\frac{441 N^{3/10} \left(2   {r_h}^2 \alpha _{\theta _1}^3  {f_{x^{10} x^{10}}}(r)+  {r_h}^2 \alpha
   _{\theta _1}^3  {f_{\theta_2y}}(r)\right)}{512 \alpha _{\theta _2}^3 \left(  {r_h}^2-3 a^2\right) \log
   (  {r_h})}-\frac{3 \sqrt{\frac{3}{2}}   {g_s}^{3/2} M \sqrt[10]{N}   {N_f}   {r_h} \left(9
   a^2+  {r_h}^2\right) \left(108 b^2   {r_h}^2+1\right)^2  {f_r}(r) (r-  {r_h})}{\pi ^{3/2}
   \alpha _{\theta _1}^3 \left(-18 a^4+3 a^2   {r_h}^2+  {r_h}^4\right)} = 0.
\end{eqnarray}
}

\item ${\rm EOM}_{\theta_2\theta_2}$:

\begin{equation}
\label{consistency-EOMtheta2theta2-i}
 {f_{zz}}(r)- {f_{x^{10} x^{10}}}(r)-2  {f_{\theta_1z}}(r)- {f_r}(r) = 0.
\end{equation}

One can show that by requiring:
\begin{eqnarray}
\label{consistency-EOMtheta1theta1-ii}
& & C_{zz}^{(1)} - 2 C_{\theta_1z}^{(1)} + 2 C_{\theta_1x}^{(1)} =0,\nonumber\\
& & C_{zz}^{(1)} - 2 C_{yz} = 0,\nonumber\\
& & \left|\Sigma_1\right|\ll1,\nonumber\\
& & \frac{2 b^{10} \left(9 b^2+1\right)^4 M {r_h}^2 \beta  \left(6 a^2+{r_h}^2\right) \Sigma_1}{3 \pi
   \left(-18 b^4+3 b^2+1\right)^4 \log N  N^{8/15} {N_f} \left(-27 a^4+6 a^2 {r_h}^2+{r_h}^4\right) \alpha
   _{\theta _2}^3}-2 {C_{\theta_1x}}^{(1)} = 0,\nonumber\\
& &
\end{eqnarray}
 (\ref{consistency-EOMrr-i}) -(\ref{consistency-EOMtheta2theta2-i}) will automatically be satisfied.

\end{itemize}

\subsection{$\psi\neq2n\pi, n=0, 1, 2$ near $r=r_h$}

Working in the IR,  the EOMs  near $r=r_h$ and up to LO in $N$, can be written as follows:
{\footnotesize
\begin{eqnarray}
\label{IR-psi=2nPi-EOMs}
& & {\rm EOM}_{MN}:\nonumber\\
& &  \sum_{p=0}^2\sum_{i=0}^2b_{MN}^{(p,i)}\left(r_h, a, N, M, N_f, g_s, \alpha_{\theta_{1,2}}\right)(r-r_h)^if_{MN}^{(p)}(r) +
\beta \frac{{\cal H}_{MN}\left(r_h, a, N, M, N_f, g_s, \alpha_{\theta_{1,2}}\right)}{(r-r_h)^{\gamma_{MN}^{\rm LO}}} = 0,\nonumber\\
& &
\end{eqnarray}
}
\noindent where as in {\bf A.1}, $M, N$ run over the $D=11$ coordinates,   $f^{(p)}_{MN}\equiv \frac{d^p f_{MN}}{dr^p}, p=0, 1, 2$, $\gamma_{MN}^{\rm LO}=1, 2$ denotes the leading order (LO) terms in powers of $r-r_h$ in the IR when the ${\cal O}(\beta)$-terms are Laurent-expanded about $r=r_h$.

One can show that a set of ten linearly independent EOMs for the ${\cal O}(l_p^6 R^4)$ corrections to the MQGP metric, with the simplifying assumtion ${f_{\theta_1\theta_1}} = {f_{\theta_1x^{10}}} = {f_{x^{10}x^{10}}} =0$,  reduce to the following set of seven equations and one that serves as a consistency check.

(a) ${\rm EOM}_{tt}$:

\begin{eqnarray}
\label{EOMttII-i}
\alpha_{tt}^{f_{\theta_1\theta_2}^\prime} (r-{r_h})^2 f_{\theta_1\theta_2}'(r)+\frac{{\alpha_{tt}^\beta} \beta }{r-{r_h}}=0,
\end{eqnarray}
where:
{\footnotesize
\begin{eqnarray}
\label{defs-67}
& & \alpha_{tt}^{f_{\theta_1\theta_2}^\prime} \equiv \frac{12\times4 a^2 \left(\frac{1}{N}\right)^{2/5} \sin^2\left(\frac{\psi_0}{2}\right) \left(9 a^2+{r_h}^2\right) \log ({r_h})}{\pi  ({g_s}-1) {g_s}
   \sin^2\phi_{20} \left(6 a^2+{r_h}^2\right) \alpha _{\theta _2}^2}\nonumber\\
   & & \alpha_{tt}^\beta \equiv \frac{8192\times16 \pi ^{9/2} a^2 \sqrt[10]{\frac{1}{N}} \sin^4\left(\frac{\psi_0}{2}\right) \beta  \left(9 a^2+{r_h}^2\right)^2 (\log ({r_h})-1) \left(\left(9
   a^2+{r_h}^2\right) \log \left(9 a^2 {r_h}^4+{r_h}^6\right)-8 \left(6 a^2+{r_h}^2\right) \log ({r_h})\right)^2}{729\times16
   ({g_s}-1) {g_s}^{3/2} {N_f}^6 \sin^4\phi_{20} {r_h}^2 \left(6 a^2+{r_h}^2\right)^4 \log ^3({r_h}) \alpha _{\theta
   _2}^4 \log ^8\left(9 a^2 {r_h}^4+{r_h}^6\right)}.\nonumber\\
& &
\end{eqnarray}
}
whose solution is given by:

{\footnotesize
\begin{eqnarray}
\label{EOMttII-ii}
& & \hskip -0.8in f_{\theta_1\theta_2}(r) = \biggl(\frac{\upsilon_{\theta_1\theta_2}  N^{3/10} \sin^2\left(\frac{\psi_0}{2}\right)  \left(9 a^2+{r_h}^2\right) (\log ({r_h})-1) \left(\left(9 a^2+{r_h}^2\right) \log
   \left(9 a^2 {r_h}^4+{r_h}^6\right)-8 \left(6 a^2+{r_h}^2\right) \log ({r_h})\right)^2}{ \sqrt{{g_s}} {N_f}^6
   \sin^2\phi_{20} {r_h}^2 \left(6 a^2+{r_h}^2\right)^3 (r-{r_h})^2 \log ^4({r_h}) \alpha _{\theta _2}^2 \log ^8\left(9 a^2
   {r_h}^4+{r_h}^6\right)}\nonumber\\
& & \hskip -0.8in +{C_{\theta_1\theta_2}}^{(1)}\biggr)\beta  = \left(\frac{\tilde{\upsilon}_{\theta_1\theta_2} \left(1-3 b^2\right)^2 \left(9 b^2+1\right)   N^{3/10}\sin^2\left(\frac{\psi_0}{2}\right) \beta }{ \left(6 b^2+1\right)^3 \sqrt{{g_s}}
   {N_f}^6 \sin^2\phi_{20} {r_h}^2 (r-{r_h})^2 \log ^9({r_h}) \alpha _{\theta
   _2}^2}+{C_{\theta_1\theta_2}}^{(1)}\right)\beta + {\cal O}\left(\frac{1}{N^{7/10}}\right),
\end{eqnarray}
where $\upsilon_{\theta_1\theta_2}\sim{\cal O}(1),\ \tilde{\upsilon}_{\theta_1\theta_2}\ll1$.
}

Assuming:
\begin{equation}
\label{ansatz-b}
b = \frac{1}{\sqrt{3}} - \kappa_b r_h^2\left(\log r_h\right)^{\frac{9}{2}} N^{-\frac{9}{10} - \alpha },
\end{equation}
one obtains:

\begin{tcolorbox}[enhanced,width=7in,center upper,size=fbox,
    drop shadow southwest,sharp corners]
\begin{eqnarray}
\label{f67II}
f_{\theta_1\theta_2}(r) = \left(\frac{\tilde{\tilde{\upsilon}}_{\theta_1\theta_2}   \kappa_b ^2 \sin^2\left(\frac{\psi_0}{2}\right) {r_h}^2 \left(\frac{1}{N}\right)^{2 \alpha +\frac{3}{2}}}{
   \sqrt{{g_s}} {N_f}^6 \sin^2\phi_{20} (r-{r_h})^2 \alpha _{\theta _2}^2}+{C_{\theta_1\theta_2}}^{(1)}\right)\beta,
\end{eqnarray}
where $\tilde{\tilde{\upsilon}}_{\theta_1\theta_2} \ll1$.
\end{tcolorbox}

(b) ${\rm EOM}_{ry}$

\begin{equation}
\label{EOM59II-i}
\alpha_{ry}^{\theta_1\theta_2} f_{\theta_1\theta_2}(r)+\alpha_{ry}^{yy} f_{yy}(r)=0,
\end{equation}
where:
{\footnotesize
\begin{eqnarray}
\label{defs-59}
& & \alpha_{ry}^{\theta_1\theta_2} \equiv \frac{7 \pi ^{17/4} \left(108 a^2+{r_h}\right) \alpha _{\theta _1}^4 \alpha _{\theta _2} \log ^4\left(9 a^2
   {r_h}^4+{r_h}^6\right)}{768 \sqrt{3} ({g_s}-1)^2 {g_s}^{19/4} M^3 \left(\frac{1}{N}\right)^{3/20} {N_f}^3
   \sin^4\phi_{20} {r_h}^2 \log ^3({r_h})},\nonumber\\
& & \alpha_{ry}^{yy} \equiv \frac{7 \sqrt{3} \pi ^{17/4} \log N ^2 \left(\frac{1}{N}\right)^{9/20} \left(108 a^2+{r_h}\right) \left({r_h}^2-3 a^2\right)^2
   (2 \log ({r_h})+1)^2 \alpha _{\theta _1}^{10} \log ^6\left(9 a^2 {r_h}^4+{r_h}^6\right)}{65536 ({g_s}-1)
   {g_s}^{19/4} M^3 {N_f}^3 \sin^6\phi_{20} {r_h}^6 \log ^5({r_h}) \alpha _{\theta _2}}, \nonumber\\
& &
\end{eqnarray}
}
and one obtains:
\begin{tcolorbox}[enhanced,width=7.3in,center upper,size=fbox,
    drop shadow southwest,sharp corners]
{\footnotesize
\begin{eqnarray}
\label{f99}
& & \hskip -0.3in f_{yy}(r) = -\frac{256 N^{3/5} \sin^2\phi_{20} {r_h}^4 \alpha _{\theta _2}^2}{9 ({g_s}-1)
   \left({r_h}^2-3 a^2\right)^2 \log ^2(N) \alpha _{\theta _1}^6 \log ^2\left(9 a^2 {r_h}^4+{r_h}^6\right)}\nonumber\\
& & \hskip -0.3in \times  \left(\frac{\upsilon_{yy}   N^{3/10} \sin^2\left(\frac{\psi_0}{2}\right)   \left(9
   a^2+{r_h}^2\right) (\log ({r_h})-1) \left(\left(9 a^2+{r_h}^2\right) \log \left(9 a^2 {r_h}^4+{r_h}^6\right)-8 \left(6
   a^2+{r_h}^2\right) \log ({r_h})\right)^2}{ \sqrt{{g_s}} {N_f}^6 \sin^2\phi_{20} {r_h}^2 \left(6 a^2+{r_h}^2\right)^3
   (r-{r_h})^2 \log ^4({r_h}) \alpha _{\theta _2}^2 \log ^8\left(9 a^2 {r_h}^4+{r_h}^6\right)}+{C_{\theta_1\theta_2}}^{(1)}\right)\beta,\nonumber\\
& &
\end{eqnarray}
}
\hskip -5.5in where $\upsilon_{yy}\sim{\cal O}(1)$.
\end{tcolorbox}

Even though $f_{yy}(r)$ is numerically suppressed as the same is ${\cal O}\left(10^{-7}\right)$ apart from an ${\cal O}\left(l_p^6\right)$-suppression - the latter of course common to most $f_{MN}$s -  $f_{yy}(r)$, near $r=r_h$ for ${\cal O}(1)$
${C_{\theta_1\theta_2}}^{(1)}$, goes like $\frac{\frac{N^{\frac{9}{10}}}{r_h^2\log^{11} r_h}}{(r-r_h)^2}$.  To ensure $f_{yy}$ remains finite one has to forego the assumption that ${C_{\theta_1\theta_2}}^{(1)}$ is ${\cal O}(1)$. Around a chosen $(\psi_0,\phi_{20})$, writing $r = r_h + \epsilon_r, \epsilon<<r_h$ close to the horizon, by assuming ${C_{\theta_1\theta_2}}^{(1)} = {C_{\theta_1\theta_2}}^{(1)}(\psi_0,\phi_{20})$:
{\footnotesize
\begin{eqnarray}
\label{finite-f99}
& & \frac{\delta_{\theta_1\theta_2}\left(9 a^2+{r_h}^2\right) \left({r_h}^2-3 a^2\right)^2\sin^2\left(\frac{\psi_0}{2}\right)}{ {\epsilon_r}^3 \sqrt{{g_s}}
   \left(\frac{1}{N}\right)^{3/10} {N_f}^6 {r_h}^2 \left(6 a^2+{r_h}^2\right)^3 \log ^9({r_h}) \alpha _{\theta _2}^2\sin^2\phi_{20}}+{C_{\theta_1\theta_2}}^{(1)}(\psi_0,\phi_{20})=0,
\end{eqnarray}
}
(wherein $\delta_{\theta_1\theta_2}\ll1$) which would imply one can consistently set $f_{yy}(r)=0$ up to ${\cal O}(\beta)$. The idea is that for every chosen value of
$(\psi_0,\phi_{20})$, once upgraded to a local uplift, using the ideas similar to
\cite{SYZ-free-delocalization}, one can show that the same will correspond to a $G_2$ structure.

(c) ${\rm EOM}_{x^1x^1}$

\begin{eqnarray}
\label{EOM22II-i}
\frac{\alpha_{tt}^\beta \beta }{4 \left(1-\frac{r}{{r_h}}\right) (r-{r_h})}+\alpha_{x^1x^1}^{\theta_1\phi_2}  f_{\theta_1y}(r) (r-{r_h})=0,
\end{eqnarray}
where:
\begin{eqnarray}
\label{defs-EOM22II}
\alpha_{x^1x^1}^{\theta_1\phi_2} \equiv -\frac{\sqrt{\frac{3 \pi }{2}} \left({r_h}^2-3 a^2\right) \left(9 a^2+{r_h}^2\right) \alpha _{\theta _1}^4 \log ^2\left(9 a^2
   {r_h}^4+{r_h}^6\right)}{32 ({g_s}-1) {g_s}^{5/2} M N {N_f} \sin^2\phi_{20} \left(6 a^2+{r_h}^2\right) \log
   ({r_h}) \alpha _{\theta _2}},
\end{eqnarray}
and obtain:
{\footnotesize
\begin{eqnarray}
\label{EOM22II-ii}
& & \hskip -0.8in f_{\theta_1y}(r) \nonumber\\
& & \hskip -0.8in = -\frac{\tilde{\upsilon}_{\theta_1y} a^2 \beta  {g_s} M N^{9/10} 16 \sin^4\left(\frac{\psi_0}{2}\right)\beta  \left(9 a^2+{r_h}^2\right) (\log ({r_h})-1)
   \left(\left(9 a^2+{r_h}^2\right) \log \left(9 a^2 {r_h}^4+{r_h}^6\right)-8 \left(6 a^2+{r_h}^2\right) \log
   ({r_h})\right)^2}{{N_f}^5 \sin^2\phi_{20} \left(6 a^2+{r_h}^2\right)^3 \left({r_h}^3-3 a^2 {r_h}\right) (r-{r_h})^3
   \log ^2({r_h}) \alpha _{\theta _1}^4 \alpha _{\theta _2}^3 \log ^{10}\left(9 a^2 {r_h}^4+{r_h}^6\right)}
   \nonumber\\
   & &  \hskip -0.8in = \frac{\upsilon_{\theta_1y} b^2 \left(3 b^2-1\right) \left(9 b^2+1\right) \beta  {g_s} M N^{9/10} \sin^4\left(\frac{\psi_0}{2}\right) \beta }{ \left(6
   b^2+1\right)^3 {N_f}^5 \sin^2\phi_{20} {r_h} (r-{r_h})^3 \log ^9({r_h}) \alpha _{\theta _1}^4 \alpha _{\theta _2}^3},
\end{eqnarray}
}
where $\tilde{\upsilon}_{\theta_1y}\sim{\cal O}(100), \upsilon_{\theta_1y}\ll1$, yielding:

\begin{tcolorbox}[enhanced,width=7in,center upper,size=fbox,
    drop shadow southwest,sharp corners]
\begin{eqnarray}
\label{EOM22II-iii}
& & f_{\theta_1y}(r) = \frac{\tilde{\tilde{\upsilon}}_{\theta_1y}\sqrt{2} \pi ^4 \beta  {g_s} \kappa_b  M 16\sin^4\left(\frac{\psi_0}{2}\right) {r_h} \beta  N^{-\alpha }}{{N_f}^5 \sin^2\phi_{20}
   (r-{r_h})^3 \log ^{\frac{9}{2}}({r_h}) \alpha _{\theta _1}^4 \alpha _{\theta _2}^3}.
\end{eqnarray}
\hskip -2in where $\tilde{\tilde{\upsilon}}_{\theta_1y}\ll1$.
\end{tcolorbox}

(d) ${\rm EOM}_{\theta_1z}$

\begin{equation}
\label{EOM610II-i}
\alpha_{y x^{10}}^{\theta_1\phi_2}  f_{\theta_1y}(r) + \alpha_{y x^{10}}^{\theta_1z}  f_{\theta_1z}(r) + \alpha_{y x^{10}}^{yy}  f_{yy}(r) = 0,
\end{equation}
where:
\begin{eqnarray}
\label{defs-EOM610II}
& & \alpha_{y x^{10}}^{\theta_1\phi_2}  \equiv -\frac{\upsilon_{\theta_1z} \log N  \left(\frac{1}{N}\right)^{3/10} \left({r_h}^2-3 a^2\right) ( \log ({r_h})+1) \alpha _{\theta
   _1}^9 \log ^4\left(9 a^2 {r_h}^4+{r_h}^6\right)}{{g_s}^{13/2} ({g_s}-1) M^4 {N_f}^5 \sin^4\phi_{20}
   2\sin\left(\frac{\psi_0}{2}\right)  {r_h}^2 \log ^5({r_h})}\nonumber\\
   & & \alpha_{y x^{10}}^{\theta_1z}  = - \alpha_{y x^{10}}^{\theta_1\phi_2}  = \alpha_{y x^{10}}^{yy} ,
\end{eqnarray}
and obtain:
\begin{eqnarray}
\label{EOM610II-ii}
& & f_{\theta_1z}(r) = \frac{\upsilon_{\theta_1z}\log N  \left(\frac{1}{N}\right)^{3/10} \left({r_h}^2-3 a^2\right) (2 \log ({r_h})+1) \alpha _{\theta _1}^9 \log
   ^4\left(9 a^2 {r_h}^4+{r_h}^6\right)}{ ({g_s}-1) {g_s}^{13/2} M^4 {N_f}^5 \sin^4\phi_{20} 2\sin\left(\frac{\psi_0}{2}\right)  {r_h}^2 \log
   ^5({r_h})},\nonumber\\
\end{eqnarray}
$\upsilon_{\theta_1z}\sim{\cal O}(1)$, yielding:

\begin{tcolorbox}[enhanced,width=7in,center upper,size=fbox,
    drop shadow southwest,sharp corners]
\begin{eqnarray}
   & & f_{\theta_1z}(r) = \frac{\tilde{\upsilon}_{\theta_1z}\times16 \sqrt{2} \pi ^4 \beta  {g_s}^{15/2} \kappa_b  M \sin^4\left(\frac{\psi_0}{2}\right) {r_h} \beta  N^{-\alpha }}{ {g_s}^{13/2} {N_f}^5
   \sin^2\phi_{20} (r-{r_h})^3 \log ^{\frac{9}{2}}({r_h}) \alpha _{\theta _1}^4 \alpha _{\theta _2}^3}
\end{eqnarray}
where $\tilde{\upsilon}_{\theta_1z}\ll1$.
\end{tcolorbox}

(e1) ${\rm EOM}_{xy}$

\begin{eqnarray}
\label{EOM89II-i}
\frac{\alpha_{tt}^\beta \beta }{{R_{\frac{tt}{r\phi_1}}} (r-{r_h})^2}+\alpha_{r\phi_1}^{f_{xy}^{\prime\prime}} (r-{r_h}) f_{xy}''(r)+\alpha_{r\phi_1}^{f_{xy}^\prime}
   (r-{r_h}) f_{xy}'(r)=0,
\end{eqnarray}
where:
{\scriptsize
\begin{eqnarray}
\label{defs-EOM89II}
& & \hskip -0.8in {R_{\frac{tt}{r\phi_1}}} \equiv \frac{3 \left(\frac{1}{N}\right)^{3/5} {r_h} \left(9 a^2+{r_h}^2\right) \alpha _{\theta _2} \log ^3\left(9 a^2
   {r_h}^4+{r_h}^6\right)}{160 \sqrt{\pi } \sqrt{{g_s}} \sin\phi_{10} \sin\left(\frac{\psi_0}{2}\right)  \alpha _{\theta _1} \left(24 ({g_s}-1)^2 \left(6
   a^2+{r_h}^2\right) \log ({r_h})+(2 {g_s}-3) \left(9 a^2+{r_h}^2\right) \log \left(9 a^2 {r_h}^4+{r_h}^6\right)\right)},\nonumber\\
   & & \hskip -0.8in \alpha_{r\phi_1}^{f_{xy}^{\prime\prime}}\equiv \frac{5 ({g_s}-1) \sin\phi_1 2\sin\left(\frac{\psi_0}{2}\right)  {r_h} \left(9 a^2+{r_h}^2\right) }{729 \sqrt{6 \pi } {g_s}^{3/2} {N_f} \sin^2\phi_{20} \left(6 a^2+{r_h}^2\right) \log
   ({r_h}) \alpha _{\theta _1} \alpha _{\theta _2}^3 \log ^2\left(9 a^2 {r_h}^4+{r_h}^6\right)} \Biggl(112 ({g_s}-1) {g_s} {N_f} \psi ^2 \log
   ({r_h}) \left(81 \alpha _{\theta _1}^3+5 \sqrt{6} \alpha _{\theta _2}^2\right)\nonumber\\
& & \hskip -0.8in -\frac{1}{{r_h}^2}\Biggl\{243 \alpha _{\theta _1}^3 \log ^2\left(9 a^2
   {r_h}^4+{r_h}^6\right)\nonumber\\
& & \hskip -0.8in \times \Biggl[\log ({r_h}) \left(6 a^2 ({g_s} \log N  {N_f}-4 \pi )+4 {g_s} {N_f}
   \left({r_h}^2-3 a^2\right) \log \left(\frac{1}{4} \alpha _{\theta _1} \alpha _{\theta _2}\right)+{r_h}^2 (8 \pi -{g_s} (2
   \log N +3) {N_f})\right)+{g_s} {N_f} \left(3 a^2-{r_h}^2\right) \left(\log N -2 \log \left(\frac{1}{4} \alpha
   _{\theta _1} \alpha _{\theta _2}\right)\right)\nonumber\\
& & \hskip -0.8in +18 {g_s} {N_f} \left({r_h}^2-3 a^2 (6 {r_h}+1)\right) \log
   ^2({r_h})\Biggr]\Biggr\}\Biggr),
\end{eqnarray}
}
which yields:
{\footnotesize
\begin{eqnarray}
\label{EOM89II-ii}
& & \hskip -0.8in f_{xy}(r) = \frac{e^{-\frac{\alpha_{r\phi_1}^{f_{xy}^\prime} r}{\alpha_{r\phi_1}^{f_{xy}^{\prime\prime}}}} \left(\alpha_{tt}^\beta \alpha_{r\phi_1}^{f_{xy}^\prime}\ ^2 \beta  (r-{r_h})
   e^{\frac{\alpha_{r\phi_1}^{f_{xy}^\prime} {r_h}}{\alpha_{r\phi_1}^{f_{xy}^{\prime\prime}}}} {Ei}\left(\frac{\alpha_{r\phi_1}^{f\ _{xy}^\prime}
   (r-{r_h})}{\alpha_{r\phi_1}^{f_{xy}^{\prime\prime}}}\right)-\alpha_{tt}^\beta \alpha_{r\phi_1}^{f_{xy}^{\prime\prime}} \alpha_{r\phi_1}^{f_{xy}^\prime} \beta  e^{\frac{\alpha_{r\phi_1}^{f_{xy}^\prime}
   r}{\alpha_{r\phi_1}^{f_{xy}^{\prime\prime}}}}+2 \alpha_{r\phi_1}^{f_{xy}^{\prime\prime}}\ ^3 c_1^{(89)} {R_{\frac{tt}{r\phi_1}}} ({r_h}-r)\right)}{2 \alpha_{r\phi_1}^{f_{xy}^{\prime\prime}}\ ^2
   \alpha_{r\phi_1}^{f\ _{xy}^\prime} {R_{\frac{tt}{r\phi_1}}} (r-{r_h})}+c_2^{(89)}\nonumber\\
& & \hskip -0.8in = -\frac{\alpha_{tt}^\beta \beta }{2 (\alpha_{r\phi_1}^{f_{xy}^{\prime\prime}} {R_{\frac{tt}{r\phi_1}}}) (r-{r_h})}+\Biggl(\frac{\alpha_{tt}^\beta \alpha_{r\phi_1}^{f_{xy}^\prime} \beta  \log
   (r-{r_h})}{2 \alpha_{r\phi_1}^{f_{xy}^{\prime\prime}}\ ^2 {R_{\frac{tt}{r\phi_1}}}}+\frac{\gamma
\alpha_{tt}^\beta \alpha_{r\phi_1}^{f_{xy}^\prime} \beta }{2 \alpha_{r\phi_1}^{f_{xy}^{\prime\prime}}\ ^2
   {R_{\frac{tt}{r\phi_1}}}}-\frac{\alpha_{tt}^\beta \alpha_{r\phi_1}^{f_{xy}^\prime} \beta  \log \left(\frac{\alpha_{r\phi_1}^{f_{xy}^{\prime\prime}}}{\alpha_{r\phi_1}^{f_{xy}^\prime}}\right)}{4
   \alpha_{r\phi_1}^{f_{xy}^{\prime\prime}}\ ^2 {R_{\frac{tt}{r\phi_1}}}}+\frac{\alpha_{tt}^\beta \alpha_{r\phi_1}^{f_{xy}^\prime} \beta  \log
   \left(\frac{\alpha_{r\phi_1}^{f_{xy}^\prime}}{\alpha_{r\phi_1}^{f_{xy}^{\prime\prime}}}\right)}{4 \alpha_{r\phi_1}^{f_{xy}^{\prime\prime}}\ ^2 {R_{\frac{tt}{r\phi_1}}}}\nonumber\\
& & \hskip -0.8in-\frac{\alpha_{r\phi_1}^{f_{xy}^{\prime\prime}} c_1
   e^{-\frac{\alpha_{r\phi_1}^{f_{xy}^\prime} {r_h}}{\alpha_{r\phi_1}^{f_{xy}^{\prime\prime}}}}}{\alpha_{r\phi_1}^{f_{xy}^\prime}}+c_2\Biggr)+O\left(r-{r_h}\right)  \sim \frac{ b^2 \left(3 b^2-1\right) \beta  \sqrt{N} \sin^4\left(\frac{\psi_0}{2}\right) \beta  \left(3 b^2 \left(8 {g_s}^2-10
   {g_s}-1\right)+4 {g_s}^2-6 {g_s}+1\right)}{\left(6 b^2+1\right)^3 ({g_s}-1)^2 \sqrt{{g_s}} \log N  {N_f}^6
   \sin^2\phi_{20} {r_h}^2 (r-{r_h}) \log ^{10}({r_h}) \alpha _{\theta _1} \alpha _{\theta _2}^2},\nonumber\\
\end{eqnarray}
}
implying:

\begin{tcolorbox}[enhanced,width=7in,center upper,size=fbox,
    drop shadow southwest,sharp corners]
\begin{eqnarray}
\label{EOM89II-iii}
 & & \hskip -0.8in f_{xy}(r) = \frac{\upsilon_{xy} \beta  \sqrt{{g_s}} (3 {g_s}-4) \kappa_b  \left(\frac{1}{N}\right)^{2/5} \sin^4\left(\frac{\psi_0}{2}\right)  \beta  N^{-\alpha
   }}{ ({g_s}-1)^2 {N_f}^6 \sin^2\phi_{20} \log (N) (r-{r_h}) \log ^{\frac{11}{2}}({r_h}) \alpha _{\theta _1} \alpha
   _{\theta _2}^2},
\end{eqnarray}
where $\upsilon_{xy}\ll1$.
\end{tcolorbox}

(e2) ${\rm EOM}_{\theta_1x}$ (consistency)

\begin{equation}
\label{EOM68II-i}
\alpha_{\theta_1x}^{f_{xy}^\prime}\  f_{xy}'(r)+\frac{\alpha_{\theta_1x}^\beta  \beta }{(r-{r_h})^2}=0,
\end{equation}
where:
{\scriptsize
\begin{eqnarray}
\label{defs-EOM68II}
& &  \hskip -0.8in \alpha_{\theta_1x}^{f_{xy}^\prime}\  \equiv -\frac{4{g_s}^{5/4} \log N  M {N_f} \sin^2\left(\frac{\psi_0}{2}\right) \left({r_h}^2-3 a^2\right) \left(9 a^2+{r_h}^2\right) (2 \log ({r_h})+1)}{36
   \sqrt{2} \pi ^{7/4} \left(\frac{1}{N}\right)^{13/20} \sin^2\phi_{20} {r_h}^2 \left(6 a^2+{r_h}^2\right) \log ({r_h}) \alpha _{\theta
   _1} \alpha _{\theta _2}^4}\nonumber\\
& &  \hskip -0.8in \alpha_{\theta_1x}^\beta  \equiv \frac{ a^2 {g_s}^{3/4} M N^{23/20} \sin^6\left(\frac{\psi_0}{2}\right)  \beta  \left(9 a^2+{r_h}^2\right)^2 (\log ({r_h})-1) \left(\left(9
   a^2+{r_h}^2\right) \log \left(9 a^2 {r_h}^4+{r_h}^6\right)-8 \left(6 a^2+{r_h}^2\right) \log ({r_h})\right)^2}{2187 \sqrt{3}
   ({g_s}-1) {N_f}^5 \sin^4\phi_{20} {r_h}^3 \left(6 a^2+{r_h}^2\right)^4 \log ^2({r_h}) \alpha _{\theta _1}^2 \alpha _{\theta
   _2}^6 \log ^{10}\left(9 a^2 {r_h}^4+{r_h}^6\right)},\nonumber\\
& &
\end{eqnarray}
}
which obtains as its LHS:
\begin{eqnarray}
\label{EOM68II-ii}
& & \frac{\upsilon_{\theta_1x} b^2 \left(1-3 b^2\right)^2 \left(9 b^2+1\right)^2 \beta  {g_s}^{3/4} M N^{23/20} \sin\left(\frac{\psi_0}{2}\right) \beta }{
   ({g_s}-1) {N_f}^5 {r_h} \left(6 b^2 +1\right)^4\sin^4\phi_{20} (r-{r_h})^2 \log ^9({r_h}) \alpha _{\theta _1}^2 \alpha
   _{\theta _2}^6} \nonumber\\
   & & = \frac{\tilde{\upsilon}_{\theta_1x} \beta  {g_s}^{3/4} \kappa_b ^2 M \sin^6\left(\frac{\psi_0}{2}\right)  {r_h}^3 \beta  \left(\frac{1}{N}\right)^{2 \alpha
   +\frac{13}{20}}}{ ({g_s}-1) {N_f}^5 \sin^4\phi_{20} (r-{r_h})^2 \alpha _{\theta _1}^2 \alpha _{\theta _2}^6},
\end{eqnarray}
where $\upsilon_{\theta_1x}, \tilde{\upsilon}_{\theta_1x}\ll1$ that in the MQGP limit, is vanishingly small.

(f) ${\rm EOM}_{r\theta_1}$

\begin{eqnarray}
\label{EOM56II-i}
& & a_{r\theta_1}^{\theta_1y} f_{\theta_1y}(r)+a_{r\theta_1}^{xy} f_{xy}(r)+a_{r\theta_1}^{yz}  f_{yz}(r)+\frac{a_{r\theta_1}^\beta  \beta }{(r-{r_h})^2}=0,
\end{eqnarray}
where:
{\scriptsize
\begin{eqnarray}
\label{EOM56II-i}
& &  \hskip -0.8in a_{r\theta_1}^{\theta_1y} \sim -\frac{\log N  \left(108 a^2+{r_h}\right) \left({r_h}^2-3 a^2\right) (2 \log ({r_h})+1) \alpha _{\theta _1}^7 \log
   ^4\left(9 a^2 {r_h}^4+{r_h}^6\right)}{ ({g_s}-1) {g_s}^3 M^2 \sqrt[5]{\frac{1}{N}} {N_f}^2 \sin^4\phi_{20} {r_h}^4 \log
   ^3({r_h}) \alpha _{\theta _2}^2},\nonumber\\
   & & \hskip -0.8in a_{r\theta_1}^{xy} \sim \frac{ \sin^2\left(\frac{\psi_0}{2}\right) \left(108 a^2+{r_h}\right) \alpha _{\theta _1}^6 \log ^2\left(9 a^2 {r_h}^4+{r_h}^6\right)}{
   {g_s}^3 M^2 \left(\frac{1}{N}\right)^{2/5} {N_f}^2 \sin^4\phi_{20} {r_h}^2 \log ^2({r_h}) \alpha _{\theta _2}^2},\nonumber\\
   & & \hskip -0.8in a_{r\theta_1}^{yz}  \sim -\frac{  \sin^2\left(\frac{\psi_0}{2}\right) \left(108 a^2+{r_h}\right) \alpha _{\theta _1}^6 \log ^2\left(9 a^2 {r_h}^4+{r_h}^6\right)}{
   {g_s}^3 M^2 \left(\frac{1}{N}\right)^{2/5} {N_f}^2 \sin^4\phi_{20} {r_h}^2 \log ^2({r_h}) \alpha _{\theta _2}^2},\nonumber\\
   & & \hskip -0.8in a_{r\theta_1}^\beta  \sim -\frac{ a^2 {g_s}^{3/4} M N^{13/20} \sin\phi_{10}  \sin^5\left(\frac{\psi_0}{2}\right) \beta  \left(9 a^2+{r_h}^2\right)^2 (\log ({r_h})-1)
   \left(\left(9 a^2+{r_h}^2\right) \log \left(9 a^2 {r_h}^4+{r_h}^6\right)-8 \left(6 a^2+{r_h}^2\right) \log
   ({r_h})\right)^2}{ ({g_s}-1) {N_f}^5 \sin^4\phi_{20} {r_h}^3 \left(6 a^2+{r_h}^2\right)^4 (r-{r_h})^2
   \log ^2({r_h}) \alpha _{\theta _1} \alpha _{\theta _2}^5 \log ^{10}\left(9 a^2 {r_h}^4+{r_h}^6\right)},\nonumber\\
& &
\end{eqnarray}
}
that yields:
\begin{eqnarray}
\label{EOM56II-ii}
& & f_{yz}(r) = \frac{\upsilon_{yz} b^2 \left(1-3 b^2\right)^2 \left(9 b^2+1\right) \beta  {g_s} \log N  M N^{7/10}  \sin^2\left(\frac{\psi_0}{2}\right) \beta }{\left(6
   b^2+1\right)^3 ({g_s}-1) {N_f}^5 \sin^2\phi_{20} {r_h} (r-{r_h})^3 \log ^7({r_h}) \alpha _{\theta _1}^3 \alpha _{\theta _2}^3}\nonumber\\
\end{eqnarray}
where $\upsilon_{yz}\ll1$, implying:

\begin{tcolorbox}[enhanced,width=7in,center upper,size=fbox,
    drop shadow southwest,sharp corners]
\begin{eqnarray}
\label{EOM56II-iii}
   & & f_{yz}(r) = \frac{\tilde{\upsilon}_{yz}\beta  {g_s} \kappa_b ^2 \log N  M  \sin^2\left(\frac{\psi_0}{2}\right) {r_h}^3 \beta  N^{-2 \alpha -\frac{11}{10}} \log
   ^2({r_h})}{ ({g_s}-1) {N_f}^5 \sin^2\phi_{20} (r-{r_h})^3 \alpha _{\theta _1}^3 \alpha _{\theta _2}^3}.
\end{eqnarray}
where $\tilde{\upsilon}_{yz}\ll1$.
\end{tcolorbox}

(g) ${\rm EOM}_{xx}$

\begin{equation}
\label{EOM88II-i}
a_{xx}^{f_{xy}^\prime}  f_{xy}'(r) + a_{xx}^{f_{xz}^\prime}    f_{xz}'(r) = 0,
\end{equation}
where:
{\scriptsize
\begin{eqnarray}
& &  \hskip -0.86in a_{xx}^{f_{xy}^\prime}  \sim \nonumber\\
& &  \hskip -0.86in -\frac{ \left(2 \pi ^3 \alpha _{\theta _1}^2 \alpha _{\theta _2}^4 \log ^2\left(9 a^2 {r_h}^4+{r_h}^6\right)+360 ({g_s}-1)
   {g_s}^3 M^2 {N_f}^2 \sin^2\left(\frac{\psi_0}{2}\right) \log ^2({r_h}) \alpha _{\theta _2}^2+243\times4 \sqrt{6} ({g_s}-1) {g_s}^3 M^2 {N_f}^2
   \sin^2\left(\frac{\psi_0}{2}\right) \log ^2({r_h}) \alpha _{\theta _1}^3\right)}{ ({g_s}-1) {g_s}^{7/2} M^2
   \left(\frac{1}{N}\right)^{7/10} {N_f}^2 \sin^2\left(\frac{\psi_0}{2}\right) \log ^2({r_h}) \alpha _{\theta _1}^6 \alpha _{\theta _2}^2 \log \left(9 a^2
   {r_h}^4+{r_h}^6\right)},\nonumber\\
   & &  \hskip -0.86in a_{xx}^{f_{xz}^\prime}    \equiv \nonumber\\
& &  \hskip -0.86in -\frac{ {r_h} \left(9 a^2+{r_h}^2\right) \alpha _{\theta _2}^2 }{ ({g_s}-1) {g_s}^{7/2} M^2
   \left(\frac{1}{N}\right)^{7/10} {N_f}^2 \sin^2\phi_{20} \psi ^2 \left(6 a^2+{r_h}^2\right) \log ^2({r_h}) \alpha _{\theta
   _1}^6 \log ^3\left(9 a^2 {r_h}^4+{r_h}^6\right)}\nonumber\\
& &  \hskip -0.86in \times \Biggl(\frac{\kappa_{xx}^{(1)} ({g_s}-1) {g_s}^3 M^2 {N_f}^2
   \sin^2\phi_{20} \sin^2\left(\frac{\psi_0}{2}\right) \left(6 a^2+{r_h}^2\right) \log ^2({r_h}) \left(4 \alpha _{\theta _2}^2-27 \sqrt{6} \alpha _{\theta
   _1}^3\right) \log ^2\left(9 a^2 {r_h}^4+{r_h}^6\right)}{\left(9 a^2 {r_h}+{r_h}^3\right) \alpha _{\theta
   _2}^4}\nonumber\\
& & \hskip -0.86in +\frac{\kappa_{xx}^{(2)} ({g_s}-1)^2 {g_s}^3 M^2 {N_f}^2 \sin^6\left(\frac{\psi_0}{2}\right) \log ({r_h}) \alpha _{\theta _1}^6 \log \left(9 a^2
   {r_h}^4+{r_h}^6\right)}{{r_h} \alpha _{\theta _2}^6}+\frac{\kappa_{xx}^{(3)}\times64 ({g_s}-1)^3 {g_s}^3 M^2 {N_f}^2 \sin^6\left(\frac{\psi_0}{2}\right)
   \left(6 a^2+{r_h}^2\right) \log ^2({r_h}) \alpha _{\theta _1}^6}{\left(9 a^2 {r_h}+{r_h}^3\right) \alpha _{\theta
   _2}^6}\nonumber\\
& & \hskip -0.86in -\frac{14 \pi ^3 \sin^2\phi_{20} \left(6 a^2+{r_h}^2\right) \alpha _{\theta _1}^2 \log ^4\left(9 a^2
   {r_h}^4+{r_h}^6\right)}{9 a^2 {r_h}+{r_h}^3}\Biggr),
\end{eqnarray}
}
which yields:

\begin{tcolorbox}[enhanced,width=7in,center upper,size=fbox,
    drop shadow southwest,sharp corners]
\hskip -1in
{\footnotesize
\begin{eqnarray}
\label{EOM88II-ii}
& &  f_{xz}(r) =\left( C_{xz}^{(1)} -\frac{\upsilon_{xz}\tilde{\Sigma}_2\sqrt{2} \pi ^{11/2}  \sqrt{{g_s}} (3 {g_s}-4) \kappa_b  \sin^4\left(\frac{\psi_0}{2}\right) \beta  N^{-\alpha -\frac{2}{5}}
  }{ ({g_s}-1)^2 {N_f}^6 \sin^2\phi_{20}
   (r-{r_h}) \log ^{\frac{13}{2}}({r_h}) \alpha _{\theta _1} \alpha _{\theta _2}^2}\right)\beta.\nonumber\\
& &
\end{eqnarray}
}
where $\upsilon_{xz}\ll1$:
{\footnotesize
\begin{eqnarray}
\label{Sigmatilde_2-def}
& & \hskip -0.9in\tilde{\Sigma}_2 \equiv \frac{ \left(40 ({g_s}-1) {g_s}^3 M^2 {N_f}^2 \sin^2\left(\frac{\psi_0}{2}\right) \alpha _{\theta _2}^2+108 \sqrt{6} ({g_s}-1) {g_s}^3 M^2 {N_f}^2 \sin^2\left(\frac{\psi_0}{2}\right)
   \alpha _{\theta _1}^3+8 \pi ^3 \alpha _{\theta _1}^2 \alpha _{\theta _2}^4\right)}{ \left(-4 ({g_s}-1) {g_s}^3 M^2 {N_f}^2 \psi
   ^2 \alpha _{\theta _2}^2+108 \sqrt{6} ({g_s}-1) {g_s}^3 M^2 {N_f}^2 \sin^2\left(\frac{\psi_0}{2}\right) \alpha _{\theta _1}^3+8 \pi ^3 \alpha _{\theta _1}^2
   \alpha _{\theta _2}^4\right)}
\end{eqnarray}
}
\end{tcolorbox}

\section{The Kiepert's Algorithm for Solving the Quintic (\ref{quintic-secular-D5}) and Diagonalization of the SYZ type IIA Mirror Inclusive of ${\cal O}(R^4)$ Corrections}
\setcounter{equation}{0} \seceqcc

In this appendix, we give the details pertaining to solving the quintic (\ref{quintic-secular-D5}) to help in obtaining the $G$-structure torsion classes of six-, seven- and eight-folds in Section {\bf 4}. This appendix is based on techniques and results summarized in \cite{Bruce-King-Beyond-Quartic}, and laid out as a five-step algorithm in this appendix.

\begin{itemize}
\item {\bf Step 1}
Consider the Tschirnhausen transformation to convert general quintic (\ref{quintic-secular-D5}) to the principal quintic:
\begin{equation}
\label{principal-quintic}
z^5 + 5 a z^2 + 5 b z + c =0,
\end{equation}
where:
\begin{equation}
\label{z}
z = x^2 - u x +v.
\end{equation}
In (\ref{z}) $u$ is determined by:
\begin{equation}
\label{u}
2 A^4+u \left(4 A^3-13 A B+15 P\right)+u^2 \left(2 A^2-5 B\right)-8 A^2 B+10 A P+3 B^2-10 F = 0,
\end{equation}
whose root, e.g., near (\ref{Ouyang-theta10-theta20}) that we work with is:
\begin{equation}
\label{u-root}
u = \frac{4 i N}{27 \sqrt{15} \alpha _{\theta _1}^2 \alpha _{\theta _2}^2} +  \frac{13 \beta  C_{zz}^{(1)} N^{3/5}}{135 \alpha _{\theta _2}^2}.
\end{equation}
The global small-$\theta_{1,2}$-uplift of (\ref{u-root}) will be: $u = \frac{i\kappa_u^{\beta^0}}{\sin^2{\theta _1} \sin^2{\theta _2}} +  \frac{\kappa_u^{ \beta}
C_{zz}^{(1)} }{\sin^2{\theta _2}}$.
Using (\ref{u-root}), $v$ is given by:
\begin{eqnarray}
\label{v}
& & v = \frac{- A u - A^2 + 2 B}{5}\nonumber\\
& & = -\frac{32 N^2}{32805 \alpha _{\theta _1}^4 \alpha _{\theta _2}^4}+\frac{8 i \beta  C_{zz}^{(1)} N^{8/5}}{3645 \sqrt{15} \alpha _{\theta
   _1}^2 \alpha _{\theta _2}^4}.
\end{eqnarray}
The global small-$\theta_{1,2}$-uplift of (\ref{v}) will be: $ -\frac{\kappa_v^{\beta^0}}{
\sin^4 {\theta _1} \sin ^4{\theta _2}}+\frac{\kappa_v^{\beta} i \beta  C_{zz}^{(1)} }{\sin\theta_1^2 \sin^4{\theta _2}}$.
The constants $a, b$ and $c$ in (\ref{principal-quintic}) are given by:
{\footnotesize
\begin{eqnarray}
\label{a+b+c}
& & a = \frac{1}{5} \left(F \left(3 A u+2 B+4 u^2\right)-P \left(A u^2+B u+P+u^3\right)-G (2 A+5 u)-10 v^3\right)\nonumber\\
   & & b = \frac{1}{5} \left(-10 a v-G \left(4 A u^2+3 B u+P+5 u^3\right)+F \left(B u^2+F+P u+u^4+4 u^3\right)-5 v^4\right)\nonumber\\
   &  & c = -F\left(u^5 + A u^4 + B u^3 + C u^2 + F u + G\right) - v^5 - 5 a v^2 - 5 b v.
\end{eqnarray}
}
It should be noted that the vanishingly small numerical pre-factors appearing in (\ref{a+b+c}) are compensated by very large powers of $N$.

\item {\bf Step 2}
To transform the principal quintic to the Brioschi quintic:
\begin{equation}
\label{Brioschi}
y^5 - 10 Z y^3 + 45 Z^2 y - Z^2 = 0,
\end{equation}
via the Tschirnhausen transformation:
\begin{equation}
\label{principal-to-Brioschi-quintic}
z_k = \frac{\lambda + \mu y_k}{\frac{y_k^2}{Z} - 3},
\end{equation}
$\lambda$ in (\ref{principal-to-Brioschi-quintic}) is determined by the quadratic:
\begin{equation}
\label{lambda-quadratic-i}
\lambda ^2 \left(a^4+a b c-b^3\right)-\lambda  \left(11 a^3 b-a c^2+2 b^2 c\right)-27 a^3 c+64 a^2 b^2-b c^2 = 0.
\end{equation}
Defining:
\begin{eqnarray}
\label{f-T-Z}
& & f \equiv u v \left(u^{10}+11 u^5 v^5-v^{10}\right) \nonumber\\
   & & T \equiv u^{30}+522 u^{25} v^5-10005 u^{20} v^{10}-10005 u^{10} v^{20}-522 u^5 v^{25}+v^{30}\nonumber\\
  & & Z \equiv \frac{f^5}{T^2},
\end{eqnarray}
one determines:
{\footnotesize
\begin{eqnarray}
\label{mu}
& & \mu \equiv \sqrt{\frac{\lambda b + c}{Z a}}.
\end{eqnarray}
}

\item
{\bf Step 3}:
We now discuss the transformation of the Brioschi quintic to the Jacobi sextic:
\begin{equation}
\label{Jacobi-sextic}
s^6 - 10 f s^3 + H s + 5 f^2 = 0,
\end{equation}
where:
\begin{eqnarray}
\label{H}
& & H \equiv -u^{20}+228 u^{15} v^5-494 u^{10} v^{10}-228 u^5 v^{15}-v^{20}..
\end{eqnarray}
Defining:
\begin{eqnarray}
\label{Delta+g2+g3}
& & \Delta \equiv \frac{1}{Z}\nonumber\\
& & g_2 \equiv \frac{\left(\frac{1 - 1728 Z}{Z^2}\right)^{\frac{1}{3}}}{12}
\nonumber\\
& & g_3 \equiv \sqrt{\frac{g_2^3 - \Delta}{27}},
\end{eqnarray}
one solves the cubic:
\begin{equation}
\label{cubic}
x^3 - \frac{g_2}{4} x - \frac{g_3}{4} = 0.
\end{equation}
The roots of (\ref{cubic}), e.g., near (\ref{Ouyang-theta10-theta20}) are given by:
{\footnotesize
\begin{eqnarray}
\label{roots-e1+e2+e3}
& & {\cal E}^i = \kappa_{{\cal E}^i,\ \mathbb{C}}\ ^{\beta^0} N^{5/3} \sqrt[3]{\frac{1}{\alpha _{\theta _1}^{10} \alpha _{\theta _2}^{10}}}+\frac{\kappa_{{\cal E}^i,\ \mathbb{C}}\ ^{\beta} \sqrt{\beta }
   \sqrt{C_{zz}^{(1)}} N^{22/15}}{\sqrt[3]{\alpha _{\theta _1}^7 \alpha _{\theta _2}^{10}}},\nonumber\\
      & &
\end{eqnarray}
}
where $i=1, 2, 3$ and $|\kappa_{{\cal E}^i, \mathbb{C}}\ ^{\beta^0/\beta}|\ll1$. The  global small-$\theta_{1,2}$-uplift of (\ref{roots-e1+e2+e3}) is ${\cal E}^i = \tilde{\kappa}_{{\cal E}^i,\ \mathbb{C}}\ ^{\beta^0}  \sqrt[3]{\frac{1}{\sin^{10}{\theta _1} \sin^{10}{\theta _2}}}+\frac{\tilde{\kappa}_{{\cal E}^i,\ \mathbb{C}}\ ^{\beta} \sqrt{\beta }
   \sqrt{C_{zz}^{(1)}} }{\sqrt[3]{\sin^7{\theta _1} \sin^{10}{\theta _2}}}$
Defining, $L \equiv \frac{\sqrt[4]{{\cal E}^1-{\cal E}^3}-\sqrt[4]{{{\cal E}^1}-{\cal E}^2}}{\sqrt[4]{{\cal E}^1-{\cal E}^2}+\sqrt[4]{{\cal E}^1-{\cal E}^3}}$, e.g., near (\ref{Ouyang-theta10-theta20}), $L = -1+(2.2\, +0.4 i) \sqrt[8]{\beta } \sqrt[4]{C_{zz}^{(1)}} \sqrt[20]{\frac{1}{N}} \sqrt[4]{\alpha _{\theta _1}}$, whose global small-$\theta_{1,2}$-uplift will be:
$L = -1+\kappa_{L,\ \mathbb{C}} \sqrt[8]{\beta } \sqrt[4]{C_{zz}^{(1)}}\sqrt[4]{\sin{\theta _1}}$.

The Jacobi nome $q$ is defined as:
\begin{equation}
\label{q-i}
q = \sum_{j=0}^\infty q_j \left(\frac{L}{2}\right)^{4j+1},
\end{equation}
wherein e.g., near (\ref{Ouyang-theta10-theta20}):
\begin{eqnarray}
\label{qj}
& & q_0 = 1\nonumber\\
& & q_1 = 2\nonumber\\
& & q_2 = 15\nonumber\\
& & q_3 = 150\nonumber\\
& & q_4 = 1707\nonumber\\
& & q_5 = 20,910\nonumber\\
& & q_6 = 268,616\nonumber\\
& & q_7 = 3,567,400\nonumber\\
& & q_8 = 48,555,069\nonumber\\
& & q_9 = 673,458,874\nonumber\\
& & q_{10} = 9,481,557,398\nonumber\\
& & q_{11} = 135,119,529,972\nonumber\\
& & q_{12} = 1,944,997,539,623\nonumber\\
& & q_{13} = 28,235,172,753,886\nonumber\\
& & .......
\end{eqnarray}
The value of $q$, e.g., near (\ref{Ouyang-theta10-theta20}),  appears to converge to a form:
\begin{eqnarray}
\label{q-ii}
q = -0.7 + \sqrt[4]{\beta } \sqrt{C_{zz}^{(1)}} {\cal C}_q \sqrt[10]{\frac{1}{N}} \sqrt{\alpha _{\theta _1}},
\end{eqnarray}
whose  global small-$\theta_{1,2}$-uplift will be: $q = -0.7 + \sqrt[4]{\beta } \sqrt{C_{zz}^{(1)}} {\cal C}_q \sqrt{\sin_{\theta _1}}$
The roots,  e.g., near (\ref{Ouyang-theta10-theta20}),  are given by:
\begin{enumerate}
\item

\begin{eqnarray}
\label{roots-sextic-i}
& & \sqrt{s_\infty} = \frac{\sqrt{5}}{\Delta^{\frac{1}{6}}}\frac{\sum_{j:-\infty}^\infty (-)^jq^{\frac{5(6j+1)^2}{12}}}{\sum_{j:-\infty}^\infty (-)^j q^{\frac{(6 j + 1)^2}{12}}};
\end{eqnarray}
appears to converge to:
{\footnotesize
\begin{eqnarray}
\label{sqrt[sinfty]}
& & \sqrt{S_\infty} = \frac{(11.4\, +18031 i) \left(\alpha _{\theta _1} \alpha _{\theta _2}\right){}^{10/3}}{N^{5/3}}-\frac{(160.2\, +266468 i) \sqrt[4]{\beta }
   \sqrt{C_{zz}^{(1)}} {\cal C}_q \alpha _{\theta _1}^{23/6} \alpha _{\theta _2}^{10/3}}{N^{53/30}},
\end{eqnarray}
}
whose  global small-$\theta_{1,2}$-uplift will be: $\kappa_{S_\infty}^{\beta^0} \left(\sin{\theta _1} \sin{\theta _2}\right){}^{10/3}-\kappa_{S_\infty}^{\beta} \sqrt[4]{\beta }
   \sqrt{C_{zz}^{(1)}} {\cal C}_q \sin^{23/6}{\theta _1} \sin^{10/3}{\theta _2}$.

\item
\begin{eqnarray}
\label{sqrt[Sk]}
& & \sqrt{S_k} = - \frac{1}{\Delta^{\frac{1}{6}}}\frac{\sum_{j:-\infty}^\infty(-)^j\varepsilon^{k(6j+1)^2}q^{\frac{(6j+1)^2}{60}}}{\sum_{j:-\infty}^\infty(-)^jq^{\frac{(6j+1)^2}{12}}},
\end{eqnarray}
e.g., near (\ref{Ouyang-theta10-theta20}), yielding:
 \begin{eqnarray}
  \label{sqrt[S_0]}
   & & S_i = \frac{\kappa_{S_i,\ \mathbb{C}}^{\beta^0} \left(\alpha _{\theta _1} \alpha _{\theta
   _2}\right){}^{10/3}}{N^{5/3}} + \frac{\kappa_{S_i,\ \mathbb{C}}^{\beta}\sqrt{C_{zz}^{(1)}} {\cal C}_q \sqrt[4]{\beta } \alpha _{\theta _2}^{10/3} \alpha
   _{\theta _1}^{23/6}}{N^{53/30}},
\end{eqnarray}
$i=0,1,...,8$, whose  global small-$\theta_{1,2}$-uplift will be: $S_i =  -\kappa_{S_i}^{\beta^0} \left(\sin{\theta _1} \sin{\theta _2}\right){}^{10/3} + \kappa_{S_i}^{\beta} \sqrt[4]{\beta } \sqrt{C_{zz}^{(1)}} {\cal C}_q \sin^{23/6}{\theta _1} \sin^{10/3}{\theta
   _2}$. It turns out that seven of the nine $S_i$s, have $|\kappa_{S_i,\ \mathbb{C}}^{\beta^0,\beta}|\gg1$ and the remaining two have moduli much less than unity.

\end{enumerate}
The roots of the Jacobi sextic are related to those of the Brioschi quintic via:
\begin{equation}
\label{yk-Sk}
y_k = \frac{1}{\sqrt{5}}\left(S_\infty - S_k\right)\left(S_{k+2} - S_{k+3}\right)\left(S_{k+4} - S_{k+1}\right),
\end{equation}
yielding, e.g., near (\ref{Ouyang-theta10-theta20}),
\begin{eqnarray}
\label{yk}
& & y_i = \frac{\kappa_{y_i,\ \mathbb{C}}^{\beta^0} \alpha _{\theta _1}^5 \alpha _{\theta _2}^5}{N^{5/2}}-\frac{\kappa_{y_i,\ \mathbb{C}}^{\beta}
   \sqrt[4]{\beta } \sqrt{C_{zz}^{(1)}} {\cal C}_q \alpha _{\theta _1}^{11/2} \alpha _{\theta _2}^5}{N^{13/5}},\nonumber\\
& &
\end{eqnarray}
$i=0, 1,...,5$ and $|\kappa_{y_i,\ \mathbb{C}}^{\beta^0,\beta}|\gg1$. It should be noted that the large numerical factors in the numerators of (\ref{sqrt[S_0]}) and (\ref{yk}) are balanced off by the large powers of $N$ - numerically taken to be $10^2-10^3$ -  in the denominators of the same. The global small-$\theta_{1,2}$-uplift of (\ref{yk}) will be: $y_i = \tilde{\kappa}_{y_i,\ \mathbb{C}}^{\beta^0} \sin^5{\theta _1} \sin^5{\theta _2}-\kappa_{y_i,\ \mathbb{C}}^{\beta}
   \sqrt[4]{\beta } \sqrt{C_{zz}^{(1)}} {\cal C}_q \sin^{11/2}{\theta _1} \sin^5{\theta _2}$.

\item {\bf Step 4}:
Hence, the roots of the principal quintic using $z_k = \frac{\lambda + \mu y_k}{\frac{y_k^2}{Z} - 3}$ can be obtained. One notes that the same involve vanishing small numerical prefactors accompanied by very large powers of $N$.

\item
{\bf Step 5}:
Now, finally the roots $x_k$ of the original quintic are given by:
\begin{eqnarray}
\label{x_k-z_k}
x_k = \frac{-({zk}-v) \left(A u^2+B u+C+u^3\right)-(A+2 u) ({z_k}-v)^2-G}{({z_k}-v) \left(2 A v+B+3 u^2\right)+A u^3+B u^2+F+C u+u^4+({z_k}-v)^2}.
\end{eqnarray}
The above, e.g., near (\ref{Ouyang-theta10-theta20}), yields the six roots, five of which are:
{\footnotesize
\begin{eqnarray}
\label{x0234}
& & x_{i\neq1} = \frac{0.07
   N^{3/5}}{\alpha _{\theta _2}^2} +\frac{\kappa_{x_i}^{\beta^0} r^6 \alpha _{\theta _2}^{61}
   \alpha _{\theta _1}^{64}}{{g_s}^6 \log N ^3 M^3 N^{311/10} {N_f}^3 \log ^3(r)} + \frac{\kappa_{x_i}^{\beta} \sqrt{C_{zz}^{(1)}} {\cal C}_q N^{9/10} \sqrt[4]{\beta
   }}{\alpha _{\theta _2}^2 \alpha _{\theta _1}^{3/2}},\nonumber\\
   \end{eqnarray}
   and,
   \begin{eqnarray}
   \label{x1}
   & & x_1 = \frac{0.07
   N^{3/5}}{\alpha _{\theta _2}^2}  + \frac{\kappa_{x_1}^{\beta^0} r^6 \alpha _{\theta _2}^{61} \alpha
   _{\theta _1}^{64}}{{g_s}^6 \log N ^3 M^3 N^{311/10} {N_f}^3 \log ^3(r)} - \frac{\kappa_{x_1}^{\beta}\sqrt{C_{zz}^{(1)}} {\cal C}_q \sqrt{N} \sqrt[4]{\beta } \sqrt{\alpha _{\theta
   _1}}}{ \alpha _{\theta _2}^2} ,\nonumber\\
   \end{eqnarray}
}
wherein $|\kappa_{x_k}^{\beta^0}|\gg1, |\kappa_{x_k}^{\beta}|\ll1$ where $\kappa_{x_i}^{\beta^0}\approx \frac{10^{1-3}}{\left(\kappa_{x_k}^{\beta}\right)^2}, k=0, 1, ...,4$, whose global small-$\theta_{1,2}$-uplift are given as under:
\begin{eqnarray}
\label{x01234-global}
& & x_i =  \frac{\tilde{\kappa}_{x_i}^{\beta^0}}{\sin^2{\theta _2}} +\frac{\tilde{\kappa}_{x_i}^{\beta^0} r^6 \sin^{61}_{\theta _2}
   \sin^{64}{\theta _1}}{{g_s}^6 \log N ^3 M^3  {N_f}^3 \log ^3(r)} + \frac{\tilde{\kappa}_{x_i}^{\beta} \sqrt{C_{zz}^{(1)}} {\cal C}_q  \sqrt[4]{\beta
   }}{\sin^2{\theta _2} \sin^{3/2}{\theta _1}}.
\end{eqnarray}
Strictly speaking, one ought to also consider the $\sqrt{\beta}, \beta^{\frac{3}{4}}, \beta$ terms in $x_{0,1,2,3,4}$. Their forms however is extremely cumbersome. To capture the essence of the results that one gets if one were to actually do so, in the following what is being assumed is that one is working near the type IIB Ouyang's $D7$-brane embedding coordinate patches effected via delocalization parameters $\alpha_{\theta_{1,2}}\sim{\cal O}(1)$ and small values of $\theta_{1,2}$, and setting $N\sim10^2, g_s\sim0.1, M=N_f=3$, and $r$ in the IR estimated as $r\left(\in {\rm IR}\right)\sim N^{-\frac{f_r}{3}}$ with $f_r\sim1$ \cite{Vikas+Gopal+Aalok}, \cite{Gopal+Vikas+Aalok}. One  can then show, e.g., in $x_0$ that by working in the neighborhood of $(\theta_{10}=\frac{\alpha_{\theta_1}}{N^{\frac{1}{5}}}, \theta_{20} = \frac{\alpha_{\theta_{20}}}{N^{\frac{3}{10}}}, \psi=2n\pi), n=0, 1, 2$, one obtains the following $\beta$-dependent terms: ${\cal O}(10^{-14})\sqrt[4]{\beta} + {\cal O}(10^{-13})\sqrt{\beta} + {\cal O}(10^{-10})\beta^{\frac{3}{4}} + {\cal O}(1)\beta$. Hence, by
choosing $\beta\sim{\cal O}(10^{-19})$, one sees that the most dominant terms are the $\beta^{\frac{1}{4}}$ and the $\beta$ terms which are both of the same order. This is hence the reason why we will work with corrections in the co-frames up to ${\cal O}(\beta^{\frac{1}{4}})$ and as explained above, this will capture the essence of the exact calculation up to
${\cal O}(\beta)$. Further one notes that in (\ref{x01234}), the very large numerical factors in one of the two ${\cal O}(\beta^0)$ terms in the each of the five $x_i$'s are compensated by very large $N$-suppression factors in the denominators.
\end{itemize}

Denoting the matrix with entries $u_{ij}$ ($i=1,...,5$ indexing the eigenvector and $j=1,...,5$ indexing the column vector element of the $i$th eigenvector $u_i$) and embedding the same in a $6\times6$ matrix ${\cal U}$ with ${\cal U}_{ab}, a/b=r,i$  then:
\begin{eqnarray}
\label{ea}
& & \hskip -0.3in \sum_{a=1}^6 \left( e^a\right)^2 = \left(\begin{array}{cccccc}
dr & d\theta_1 & d\theta_2 & dx & dy & dz
\end{array}
\right) {\cal U}\left(\begin{array}{cccccc}
G_{rr} & 0 & 0 & 0 & 0 & 0 \\
0 & x_0 & 0 & 0 & 0 & 0 \\
0 & 0 & x_1 & 0 & 0 & 0 \\
0 & 0 & 0 & x_2 & 0 & 0 \\
0 & 0 & 0 & 0 & x_3 & 0 \\
0 & 0 & 0 & 0 & 0 & x_4
\end{array}
\right) {\cal U}^T \left(\begin{array}{c}
dr \\
d\theta_1 \\
d\theta_2 \\
dx \\
dy \\
dz
\end{array}
\right). \nonumber\\
& &
\end{eqnarray}
The co-frames are hence given by:
{\scriptsize
\begin{eqnarray}
\label{e12}
& & \hskip -0.87in e^1 = \sqrt{G_{rr}}dr,\nonumber\\
& & \nonumber\\
& &\hskip -0.87in e^2 = \sqrt{ \kappa^2_{1; \beta^0}\csc ^2\left(\theta _2\right) + \kappa^2_{2, \beta^0}\frac{ r^6 \sin ^{61}\left(\theta _2\right) \sin ^{64}\left(\theta _1\right)}{{g_s}^6 \log N ^3 M^3
   {N_f}^3 \log ^3(r)}
+ \kappa^2_{1; \beta}\frac{ \sqrt[4]{\beta } \sqrt{C_{zz}^{(1)}} {\cal C}_q }{\sin ^{\frac{7}{2}}\left(\theta
   _2\right)}} \nonumber\\
& &\hskip -0.87in \times\Biggl[\frac{d\theta_1 \left(\kappa^2_{\theta_1, 1; \beta^0}\frac{  {g_s}^{7/4} \log N  M {N_f} \left(0.25 a^2-0.06 r^2\right) \sin \left(\theta _1\right) \csc
   \left(\theta _2\right) \log (r)}{r^2} + \kappa^2_{\theta_1; 2; \beta}\frac{\sqrt[4]{\beta }
   \sqrt{C_{zz}^{(1)}} {\cal C}_q  {g_s}^{7/4} \log N  M {N_f} \left({1.8}
   a^2+{2.5} r^2\right) \csc \left(\theta _2\right) \log (r)}{r^2 \sqrt{\sin \left(\theta
   _1\right)}}\right)}{\sqrt[4]{N}}\nonumber\\
& &\hskip -0.87in +d\theta_2 \left(\kappa^2_{\theta_2, 1; \beta^0}\frac{ {g_s}^{7/4} M {N_f} \sin ^2\left(\theta _1\right) \csc ^2\left(\theta _2\right) \log (r) \left(3 a^2 \log (r)+0.08
   r\right)}{\sqrt[4]{N} r} + \kappa^2_{\theta_2; 2; \beta}\frac{\sqrt[4]{\beta } \sqrt{C_{zz}^{(1)}}
   {\cal C}_q   {g_s}^{21/4} M^3 {N_f}^3 \sin ^{\frac{9}{2}}\left(\theta _1\right) \csc ^6\left(\theta _2\right) \log
   ^3(r) \left({a^2 r^2 \log (r)}+{r^3}\right)}{N^{3/4}
   r^3}\right)\nonumber\\
& &\hskip -0.87in +{dx} \left(\kappa^2_{x, 1; \beta^0} \sin ^2\left(\theta _1\right)
   \csc \left(\theta _2\right) + \kappa^2_{x; 2; \beta}\frac{ \sqrt[4]{\beta } \sqrt{C_{zz}^{(1)}} {\cal C}_q  {g_s}^{7/2} M^2
   {N_f}^2 \log ^2(r) \sin ^{\frac{9}{2}}\left({\theta _1}\right) \csc ^5\left({\theta _2}\right) \left({3.6} a^2 r \log (r)+{4.8} r^2\right)}{\sqrt{N} r^2}\right)\nonumber\\
& &\hskip -0.87in +{dy} \left(1-\kappa^2_{y, 1; \beta^0}\frac{ {g_s}^{7/2} M^2 {N_f}^2 \sin ^4\left(\theta _1\right) \csc
   ^4\left(\theta _2\right) \log ^2(r) \left(3 a^2 \log (r)+0.08 r\right)^2}{\sqrt{N} r^2}-\kappa^2_{y; 1; \beta}\frac{
   \sqrt[4]{\beta } \sqrt{C_{zz}^{(1)}} {\cal C}_q }{\sin ^{\frac{3}{2}}\left(\theta _1\right)}\right)\nonumber\\
& &\hskip -0.87in +{dz} \left( \kappa^2_{z, 1; \beta^0}\sin
   \left(\theta _2\right)-\kappa^2_{z; 1; \beta}\frac{ \sqrt[4]{\beta } \sqrt{C_{zz}^{(1)}} {\cal C}_q  \sin
   \left(\theta _2\right)}{\sin ^{\frac{3}{2}}\left(\theta _1\right)}\right)\Biggr]\nonumber\\
& & \nonumber\\
& & \nonumber\\
& & \nonumber\\
& &\hskip -0.87in e^3 = \sqrt{\kappa^3_{1; \beta^0} \csc ^2\left(\theta _2\right) + \kappa^3_{2; \beta^0}\frac{ r^6 \sin ^{61}\left(\theta _2\right) \sin ^{64}\left(\theta _1\right)}{{g_s}^6
   \log N ^3 M^3 N^{311/10} {N_f}^3 \log ^3(r)}-\kappa^3_{1; \beta}{\sqrt[4]{\beta } \sqrt{C_{zz}^{(1)}} {\cal C}_q  \sqrt{\sin \left(\theta _1\right)} \csc ^2\left(\theta
   _2\right)}}\nonumber\\
& &\hskip -0.87in \times \Biggl[\frac{d\theta_1 \left(\kappa^3_{\theta_1; 1; \beta^0}\frac{4 {g_s}^{7/4} \log N  M {N_f} \left(0.2 a^2-0.1 r^2\right) \sin \left(\theta
   _1\right) \csc \left(\theta _2\right) \log (r)}{N^{3/20} r^2} + \kappa^3_{\theta_1; 1; \beta}\frac{16.
   \sqrt[4]{\beta } \sqrt{C_{zz}^{(1)}} {\cal C}_q  {g_s}^{7/4} \log N  M {N_f} \left({1.56} a^2-{0.522} r^2\right) \csc \left(\theta _2\right) \log (r) \sin
   ^{\frac{3}{2}}({\theta_1})}{r^2}\right)}{\sqrt[4]{N}}\nonumber\\
& &\hskip -0.87in +\frac{d\theta_2 \left(\kappa^3_{\theta_2; 1; \beta^0}\frac{{g_s}^{7/4} M {N_f} \sin ^2\left(\theta _1\right) \csc ^2\left(\theta _2\right) \log (r) \left(0.0005
   a^2 \log (r)+0.000014 r\right)}{\sqrt[20]{N} r} + \kappa^3_{\theta_2; 1; \beta}\frac{16\sqrt[4]{\beta }
   \sqrt{C_{zz}^{(1)}} {\cal C}_q   {g_s}^{7/4} M {N_f} \sin ^{\frac{5}{2}}\left(\theta _1\right) \csc ^2\left(\theta
   _2\right) \log (r) \left({2.7} a^2 \log (r)+{0.081}
   r\right)}{r}\right)}{\sqrt[4]{N}}\nonumber\\
& &\hskip -0.87in +{dy} \left(1-\kappa^3_{y; 1; \beta^0}\frac{\frac{{g_s}^{7/2} M^2 {N_f}^2 \sin
   ^4\left(\theta _1\right) \csc ^4\left(\theta _2\right) \log ^2(r) \left(4.32 a^2 \log (r)+0.064 r\right)}{r}-\kappa^3_{y; 1; \beta}\frac{16. \sqrt[4]{\beta }
   \sqrt{C_{zz}^{(1)}} {\cal C}_q  {g_s}^{7/2} M^2 {N_f}^2 \sin ^{\frac{9}{2}}\left(\theta _1\right) \csc ^4\left(\theta
   _2\right) \log ^2(r) \left(-{7 a^2 \log (r)}-{0.1r}\right)}{\sqrt[5]{N}
   r}}{\sqrt{N}}\right)\nonumber\\
& &\hskip -0.87in +{dx} \left(\kappa^3_{x; 1; \beta^0} \sin ^2\left(\theta _1\right) \csc \left(\theta _2\right)-\kappa^3_{x; 1; \beta}
   \sqrt[4]{\beta } \sqrt{C_{zz}^{(1)}} {\cal C}_q  \sin ^{\frac{5}{2}}\left(\theta _1\right) \csc \left(\theta
   _2\right)\right) +{dz} \left(\kappa^3_{z; 1; \beta^0}\sin\theta_2 + {\kappa^3_{z; 1; \beta} \sqrt[4]{\beta } \sqrt{C_{zz}^{(1)}} {\cal C}_q
   \sqrt{\sin{\theta _1}}\sin{\theta _2}}
\right)\Biggr]\nonumber\\
\end{eqnarray}
\begin{eqnarray}
\label{e^4}
& &\hskip -0.87in e^4 = \sqrt{\kappa^4_{1;\beta^0} \csc ^2\left(\theta _2\right) -\kappa^4_{2;\beta^0}\frac{ r^6 \sin ^{61}\left(\theta _2\right) \sin ^{64}\left(\theta _1\right)}{{g_s}^6 \log N ^3 M^3
   {N_f}^3 \log ^3(r)} -\kappa^4_{1;\beta}\frac{ \sqrt[4]{\beta } \sqrt{C_{zz}^{(1)}} {\cal C}_q }{\sin ^{\frac{7}{2}}\left(\theta
   _2\right)}}\nonumber\\
& &\hskip -0.87in \times \Biggl[d\theta_1 \sqrt[4]{N} \left(\kappa^4_{\theta_1; 1; \beta^0}\frac{\left(0.006 a^2 r^2-0.002
   r^4\right) \csc \left(\theta _1\right)}{{g_s}^{7/4} \log N  M {N_f} \left(0.02 a^4-0.01 a^2 r^2+0.002 r^4\right) \log
   (r)}+\kappa^4_{\theta_1; 1; \beta}\frac{\sqrt[4]{\beta } \sqrt{C_{zz}^{(1)}} {\cal C}_q   \csc ^3\left(\theta
   _2\right)}{{g_s}^{15/4} \log N ^3 M^2 {N_f}^2 \sqrt{\sin \left(\theta _1\right)} \log ^2(r)}\right)\nonumber\\
& &\hskip -0.87in +d\theta_2
  \left(\kappa^4_{\theta_2; 1; \beta^0}\frac{ \sqrt[4]{\beta } \sqrt{C_{zz}^{(1)}} {\cal C}_q  \sqrt[4]{N} \csc ^2\left(\theta
  _2\right)}{{g_s}^{15/4} \log N ^2 M^2 {N_f}^2 \sin ^{\frac{3}{2}}\left(\theta _1\right) \log ^2(r)}-\kappa^4_{\theta_2; 2; \beta^0}\frac{
 r^{10} \sin ^{\frac{127}{2}}\left(\theta _1\right) \sin ^{59}\left(\theta _2\right)}{\sqrt{C_{zz}^{(1)}} {\cal C}_q  {g_s}^{25/4}
 \log N ^3 M^3 \sqrt[4]{N} {N_f}^3 \left(0.022 a^4-0.015 a^2 r^2+0.002 r^4\right) \log ^3(r)}\right)\nonumber\\
& &\hskip -0.87in +{dx}
 \left(\kappa^4_{x; 1; \beta^0}\frac{\left(0.033 a^4-0.02 a^2 r^2+0.004 r^4\right) \sin ^2\left(\theta _1\right)}{0.022 a^4-0.015 a^2 r^2+0.002
 r^4}-\kappa^4_{x; 1; \beta}\frac{ \sqrt[4]{\beta } \sqrt{C_{zz}^{(1)}} {\cal C}_q  \sqrt{N}}{{g_s}^{7/2}
 \log N ^2 M^2 {N_f}^2 \sin ^{\frac{3}{2}}\left(\theta _1\right) \log ^2(r)}\right)\nonumber\\
& &\hskip -0.87in +{dy} \left(\kappa^4_{y; 1; \beta^0}\frac{\pi ^{5/2}  r^9
   \sin ^{57}\left(\theta _2\right) \sin ^{66}\left(\theta _1\right) \left(9.54 a^2 \log (r)+0.265
   r\right)}{{g_s}^8 \log N ^5 M^4 {N_f}^4 \left(0.022 a^4-0.015 a^2 r^2+0.002 r^4\right) \log ^4(r)}+\kappa^4_{y; 1; \beta}\frac{ \sqrt[4]{\beta } \sqrt{C_{zz}^{(1)}} {\cal C}_q   \sqrt{\sin \left(\theta _1\right)} \csc
   ^4\left(\theta _2\right)}{{g_s}^2 \log N ^2 M {N_f} \log (r)}\right)\nonumber\\
& &\hskip -0.87in +{dz} \left(1-\sqrt{N} \left(\kappa^4_{z; 1; \beta^0}\frac{ r^{20} \sin ^{118}\left(\theta _2\right) \sin ^{128}\left(\theta _1\right)}{{g_s}^{39/2} \log N ^{10} M^{10}
   {N_f}^{10} \left(0.022 a^4-0.015 a^2 r^2+0.002 r^4\right)^2 \log ^{10}(r)}+\kappa^4_{z; 1; \beta}\frac{
   \sqrt[4]{\beta } \sqrt{C_{zz}^{(1)}} {\cal C}_q  \sqrt{\sin \left(\theta _1\right)}}{{g_s}^{7/2} \log N ^2 M^2 {N_f}^2
   \log ^2(r)}\right)\right)\Biggr]\nonumber\\
& & \nonumber\\
& & \nonumber\\
& & \nonumber\\
& &\hskip -0.87in e^5 = \sqrt{\kappa^5_{1;\beta^0}\csc ^2\left(\theta _2\right) +\kappa^5_{2;\beta^0}\frac{ r^6 \sin ^{61}\left(\theta _2\right) \sin ^{64}\left(\theta _1\right)}{{g_s}^6 \log N ^3 M^3
   {N_f}^3 \log ^3(r)}+ \kappa^5_{1;\beta} \frac{ \sqrt[4]{\beta } \sqrt{C_{zz}^{(1)}} {\cal C}_q }{\sin ^{\frac{7}{2}}\left(\theta
   _2\right)}}\nonumber\\
& & \hskip -0.87in\times \Biggl[d\theta_1 \biggl( -\kappa^5_{\theta_1;1;\beta^0}\frac{  {g_s}^{7/4} \log N  M {N_f}
   \left(0.00002 r^2-0.00004 a^2\right) \sin \left(\theta _1\right) \csc \left(\theta _2\right) \log (r)}{\sqrt[4]{N}
   r^2}\nonumber\\
& & \hskip -0.87in + \kappa^5_{\theta_1;1;\beta}\frac{ \sqrt[4]{\beta }
   \sqrt{C_{zz}^{(1)}} {\cal C}_q   {g_s}^{21/4} \log N  M^3 {N_f}^3 \left({0.592}
   r^2-{2.37} a^2\right) \sin ^{\frac{7}{2}}\left(\theta _1\right) \csc ^5\left(\theta _2\right) \log ^3(r)
   \left(0.00049 a^2 \log (r)+0.000014 r\right)^2}{N^{3/4} r^4}\biggr)\nonumber\\
   & & \hskip -0.87in \nonumber\\
& &\hskip -0.87in +d\theta_2 \biggl(-\kappa^5_{\theta_2;1;\beta^0}
\frac{ {g_s}^{7/4} M {N_f} \sin ^2\left(\theta _1\right) \csc ^2\left(\theta _2\right) \log (r) \left(-0.00049 a^2
   \log (r)-0.000014 r\right)}{\sqrt[4]{N} r}\nonumber\\
& & \hskip -0.87in +\kappa^5_{\theta_2;1;\beta}\frac{ \sqrt[4]{\beta } \sqrt{C_{zz}^{(1)}} {\cal C}_q  {g_s}^{21/4} M^3
   {N_f}^3 \sin ^{\frac{9}{2}}\left(\theta _1\right) \csc ^6\left(\theta _2\right) \log ^3(r) \left(-{1.89} a^2 \log (r)-{0.0197} r\right) \left(0.00049 a^2 \log (r)+0.000014 r\right)^2}{N^{3/4}
   r^3}\biggr)\nonumber\\
& & \hskip -0.87in+{dx} \left(\kappa^5_{x;1;\beta}\frac{ \sqrt[4]{\beta }
   \sqrt{C_{zz}^{(1)}} {\cal C}_q  {g_s}^{7/2} M^2 {N_f}^2 \sin ^{\frac{9}{2}}\left(\theta _1\right) \csc ^5\left(\theta
   _2\right) \log ^2(r) \left(0.00049 a^2 \log (r)+0.000014 r\right)^2}{N^{2/5} r^2}+\kappa^5_{x;1;\beta^0} \sin ^2\left(\theta _1\right) \csc \left(\theta_2\right)\right)\nonumber\\
& &\hskip -0.87in +{dy} \Biggl(1-\frac{1}{\sqrt{N}}\Biggl\{\kappa^5_{y;1;\beta^0}\frac{ {g_s}^{7/2} M^2 {N_f}^2 \sin ^4\left(\theta
   _1\right) \csc ^4\left(\theta _2\right) \log ^2(r) \left(0.00049 a^2 \log (r)+0.000014 r\right)^2}{r^2} \nonumber\\
  & & \hskip -0.87in+ \kappa^5_{y;1;\beta}\frac{ \sqrt[4]{\beta } \sqrt{C_{zz}^{(1)}}
   {\cal C}_q  {g_s}^{7/2} M^2 \sqrt[5]{N} {N_f}^2 \sin ^{\frac{5}{2}}\left(\theta _1\right) \csc ^4\left(\theta _2\right) \log
   ^2(r) \left(0.00049 a^2 \log (r)+0.000014 r\right)^2}{r^2}\Biggr\}\Biggr) + {dz}
   \left(\kappa^5_{z;1;\beta^0} \sin \left(\theta _2\right) + \kappa^5_{z;1;\beta}\frac{ \sqrt[4]{\beta } \sqrt{C_{zz}^{(1)}} {\cal C}_q }{\sqrt{\sin \left(\theta
   _2\right)}}\right)\Biggr]\nonumber\\
\end{eqnarray}
\begin{eqnarray}
\label{e^6}
& &\hskip -0.87in e^6 = \sqrt{\kappa^6_{1;\beta^0}\csc ^2\left(\theta _2\right) + \kappa^6_{2;\beta^0}\frac{ r^6 \sin ^{61}\left(\theta _2\right) \sin ^{64}\left(\theta _1\right)}{{g_s}^6 \log N ^3 M^3
   {N_f}^3 \log ^3(r)}-\kappa^6_{1;\beta}\frac{ \sqrt[4]{\beta } \sqrt{C_{zz}^{(1)}} {\cal C}_q }{\sin ^{\frac{7}{2}}\left(\theta
   _2\right)}}\nonumber\\
& &\hskip -0.87in \times \Biggl[d\theta_1 \left(\kappa^6_{\theta_1;1;\beta^0}\frac{
   {g_s}^{7/4} \log N  M {N_f} \left(34.9 a^2-11.6 r^2\right) \sin \left(\theta _1\right) \csc \left(\theta _2\right) \log
   (r)}{\sqrt[4]{N} r^2} + \kappa^6_{\theta_1;1;\beta}\frac{ \sqrt[4]{\beta } \sqrt{C_{zz}^{(1)}}
   {\cal C}_q  {g_s}^{7/4} \log N  M {N_f} \left({0.064} r^2-{5.8} a^2\right) \csc \left(\theta _2\right) \log (r)}{\sqrt[4]{N} r^2 \sqrt{\sin \left(\theta _1\right)}}\right)\nonumber\\
& &\hskip -0.87in +\frac{d\theta_2 \left(\kappa^6_{\theta_2;1;\beta^0}\frac{ {g_s}^{7/4} M {N_f} \sin ^2\left(\theta _1\right) \csc ^2\left(\theta
   _2\right) \log (r) \left(0.5 a^2 \log (r)+0.014 r\right)}{r} +\kappa^6_{\theta_2;1;\beta} \frac{ \sqrt[4]{\beta } \sqrt{C_{zz}^{(1)}} {\cal C}_q  {g_s}^{7/4} M
   {N_f} \sqrt{\sin \left(\theta _1\right)} \csc ^2\left(\theta _2\right) \log (r) \left({2.9} a^2 \log
   (r)+{0.1r}\right)}{r}\right)}{\sqrt[4]{N}}\nonumber\\
& &\hskip -0.87in +{dy} \biggl(1 - \kappa^6_{y;1;\beta^0}\frac{ {g_s}^{7/2} M^2 {N_f}^2 \sin ^4\left(\theta
   _1\right) \csc ^4\left(\theta _2\right) \log ^2(r) \left(15.625 a^4 \log ^2(r)+0.875 a^2 r \log (r)+0.01225 r^2\right)}{\sqrt{N}
   r^2}\nonumber\\
& & \hskip -0.87in + \kappa^6_{y;1;\beta}\frac{ \sqrt[4]{\beta }
   \sqrt{C_{zz}^{(1)}} {\cal C}_q  {g_s}^{7/2} M^2 {N_f}^2 \sin ^{\frac{5}{2}}\left(\theta _1\right) \csc ^4\left(\theta
   _2\right) \log ^2(r) \left(- a^4 \log ^2(r)-{0.1} a^2 r \log
   (r)\right)}{\sqrt{N} r^2} \biggr)\nonumber\\
& &\hskip -0.87in +{dx} \left(\kappa^6_{x;1;\beta^0} \sin ^2\left(\theta _1\right) \csc \left(\theta _2\right)-\kappa^6_{x;1;\beta}
   \sqrt[4]{\beta } \sqrt{C_{zz}^{(1)}} {\cal C}_q  \sqrt{\sin \left(\theta _1\right)} \csc \left(\theta _2\right)\right)+{dz}
   \left(\kappa^6_{z;1;\beta^0} \sin \left(\theta _2\right) + \kappa^6_{z;1;\beta}\frac{ \sqrt[4]{\beta } \sqrt{C_{zz}^{(1)}} {\cal C}_q  \sin \left(\theta
   _2\right)}{\sin ^{\frac{3}{2}}\left(\theta _1\right)}\right)\Biggr].
\end{eqnarray}
}
In (\ref{e12}) - (\ref{e^6}), $\kappa^{a=1,...,6}_{\theta_{1,2}/x/y/z;1;\beta}\ll1$. Except for $e^4$, however, all the rest have an IR-enhancement factor involving some power of $\log r$ appearing in the contributions picked up from the ${\cal O}(R^4)$ terms. Further, these contributions also receive near-Ouyang-embedding enhancements around small $\theta_{1,2}$ - which also provide the most dominant contributions to all the terms of the action.  Also, $\kappa^a_{1;\beta^0}\gg1$ but are accompanied by IR-suppression factors involving exponents of $r$ along with near-Ouyang-embedding enhancements around small $\theta_{1,2}$.

Now, (\ref{e12}) - (\ref{e^6}) can be inverted - in Section {\bf 3}, for simplicity, one restricts to the Ouyang embedding $\left(r^6 + 9 a^2 r^4\right)^{\frac{1}{4}}e^{\frac{i}{2}\left(\psi - \phi_1 - \phi_2\right)}\sin \frac{\theta_1}{2}\sin\frac{\theta_2}{2} = \mu,\ \mu$ being the Ouyang embedding parameter assuming $|\mu|\ll r^{\frac{3}{2}}$, effected, e.g., by working near the $\theta_1=\frac{\alpha_{\theta_1}}{N^{\frac{1}{5}}}, \theta_2 = \frac{\alpha_{\theta_2}}{N^{\frac{3}{10}}}$-coordinate patch (wherein an explicit $SU(3)$-structure for the type IIB dual of \cite{metrics} and its delocalized SYZ  type IIA mirror \cite{MQGP}, and an explicit $G_2$-structure for its ${\cal M}$-Theory uplift \cite{MQGP} was worked out in \cite{NPB}).

\section{$\Omega_{ab}$s appearing in (\ref{Omega_ab})}
\setcounter{equation}{0} \seceqdd

The components of the ``structure constants" of  (\ref{Omega_ab}) $\Omega_{ab}$s after a small-$\beta$ large-$N$ small-$a$ expansion are given as under:
{\scriptsize
\begin{eqnarray}
\label{Omega_2a}
& & \Omega_{22} = \frac{\left(\frac{1}{|\log r|}\right)^{4/3} \sqrt[4]{\frac{1}{N}}
   \left(\frac{32.9 a^2}{r^2}+9.4\right)}{g_s^{\frac{1}{4}} N_f^{\frac{1}{3}}}\sqrt{1-\frac{r_h^4}{r^4}}+\frac{\omega_{22} \sqrt[3]{\frac{1}{|\log r|}}
   \sqrt{{\cal C}_{zz}^{(1)}} {\cal C}_q  {g_s}^{3/2} M \sqrt[4]{N}
   {N_f} \sqrt[4]{\beta } \alpha _{\theta _1}^{5/2}}{r \alpha
   _{\theta _2}^5 {g_s^{\frac{1}{4}} N_f^{\frac{1}{3}}}}\sqrt{1-\frac{r_h^4}{r^4}},\nonumber\\
& & \Omega_{23} = \frac{\left(\frac{1}{|\log r|}\right)^{4/3} \sqrt[4]{\frac{1}{N}}
   \left(\frac{23.6 a^2}{r^2}+15.7\right)}{g_s^{\frac{1}{4}} N_f^{\frac{1}{3}}}\sqrt{1-\frac{r_h^4}{r^4}}+\frac{\omega_{23} \sqrt[3]{\frac{1}{|\log r|}}
   \sqrt{{\cal C}_{zz}^{(1)}} {\cal C}_q  {g_s}^{3/2} M \sqrt[4]{N}
   {N_f} \sqrt[4]{\beta } \alpha _{\theta _1}^{5/2}}{\alpha _{\theta
   _2}^5 {g_s^{\frac{1}{4}} N_f^{\frac{1}{3}}}}\sqrt{1-\frac{r_h^4}{r^4}},\nonumber\\
   & &  \Omega_{24} = \frac{|\log r|^{2/3} {g_s}^{7/2} \log N ^2 M^2
   \left(\frac{1}{N}\right)^{17/20} {N_f}^2 \alpha _{\theta _1}^2
   \left(3.97-\frac{4 a^2}{r^2}\right)}{\alpha _{\theta _2}
   {g_s^{\frac{1}{4}} N_f^{\frac{1}{3}}}}\sqrt{1-\frac{r_h^4}{r^4}}\nonumber\\
   & & +\frac{\omega_{24}
   |\log r|^{2/3} \sqrt{{\cal C}_{zz}^{(1)}} {\cal C}_q
   {g_s}^{7/2} \log N ^2 M^2 \left(\frac{1}{N}\right)^{11/20}
   {N_f}^2 \sqrt[4]{\beta } \sqrt{\alpha _{\theta _1}}}{\alpha
   _{\theta _2} {g_s^{\frac{1}{4}} N_f^{\frac{1}{3}}}}\sqrt{1-\frac{r_h^4}{r^4}},\nonumber\\
   & & \Omega_{25} = -\frac{\left(\frac{1}{|\log r|}\right)^{4
   /3} \sqrt[4]{\frac{1}{N}} \left(\frac{36.6 a^2}{r^2}+24.4\right)}{
   {g_s^{\frac{1}{4}} N_f^{\frac{1}{3}}}}\sqrt{1-\frac{r_h^4}{r^4}} + \frac{\omega_{25}
   \sqrt[3]{\frac{1}{|\log r|}} \sqrt{{\cal C}_{zz}^{(1)}}
   {\cal C}_q  {g_s}^{3/2} M \sqrt[4]{N} {N_f} \sqrt[4]{\beta }
   \alpha _{\theta _1}^{5/2}}{\alpha _{\theta _2}^5
   {g_s^{\frac{1}{4}} N_f^{\frac{1}{3}}}}\sqrt{1-\frac{r_h^4}{r^4}},\nonumber\\
   & & \Omega_{26} = -\frac{|\log r|^{2/3}
   {g_s}^{7/2} M^2 \left(\frac{1}{N}\right)^{7/20} {N_f}^2
   \left(\frac{0.2 a^2}{r^2}+0.1\right) \alpha _{\theta _1}^4}{ \alpha
   _{\theta _2}^4 {g_s^{\frac{1}{4}} N_f^{\frac{1}{3}}}}\sqrt{1-\frac{r_h^4}{r^4}} \nonumber\\
   & & + \frac{\omega_{26} |\log r|^{2/3}
   \sqrt{{\cal C}_{zz}^{(1)}} {\cal C}_q {g_s}^{7/2} M^2
   \sqrt[20]{\frac{1}{N}} {N_f}^2 \sqrt[4]{\beta } \alpha _{\theta
   _1}^{5/2}}{g_s^{\frac{1}{4}} N_f^{\frac{1}{3}}}\sqrt{1-\frac{r_h^4}{r^4}};\nonumber\\
& & \nonumber\\
& & \nonumber\\
& & \Omega_{32} = \frac{\left(\frac{1}{|\log r|}\right)^{4/3} \sqrt[4]{\frac{1}{N}}
   \left(\frac{43.3 a^2}{r^2}+12.4\right)}{g_s^{\frac{1}{4}} N_f^{\frac{1}{3}}}\sqrt{1-\frac{r_h^4}{r^4}}+\frac{\omega_{32} \sqrt[3]{\frac{1}{|\log r|}}
   \sqrt{{\cal C}_{zz}^{(1)}} {\cal C}_q  {g_s}^{3/2} M \sqrt[4]{N}
   {N_f} \sqrt[4]{\beta } \alpha _{\theta _1}^{5/2}}{\alpha _{\theta
   _2}^5 {g_s^{\frac{1}{4}} N_f^{\frac{1}{3}}}}\sqrt{1-\frac{r_h^4}{r^4}},\nonumber\\
   & & \Omega_{33} = \frac{\left(\frac{1}{|\log r|}\right)^{4/3} \sqrt[4]{\frac{1}{N}}
   \left(\frac{31 a^2}{r^2}+20.7\right)}{g_s^{\frac{1}{4}} N_f^{\frac{1}{3}}}\sqrt{1-\frac{r_h^4}{r^4}}+\frac{\omega_{33} \sqrt[3]{\frac{1}{|\log r|}}
   \sqrt{{\cal C}_{zz}^{(1)}} {\cal C}_q  {g_s}^{3/2} M \sqrt[4]{N}
   {N_f} \sqrt[4]{\beta } \alpha _{\theta _1}^{5/2}}{\alpha _{\theta
   _2}^5 {g_s^{\frac{1}{4}} N_f^{\frac{1}{3}}}}\sqrt{1-\frac{r_h^4}{r^4}},\nonumber\\
   & & \Omega_{34} = \frac{|\log r|^{2/3} {g_s}^{7/2} \log N ^2 M^2
   \left(\frac{1}{N}\right)^{17/20} {N_f}^2 \left(\frac{28.1
   a^2}{r^2}+5.2\right) \alpha _{\theta _1}^2 }{ {g_s^{\frac{1}{4}} N_f^{\frac{1}{3}}}\alpha
   _{\theta _2}}\sqrt{1-\frac{r_h^4}{r^4}}\nonumber\\
   & & -\frac{\omega_{34}
   \sqrt[3]{\frac{1}{|\log r|}} \sqrt{{\cal C}_{zz}^{(1)}}
   {\cal C}_q  {g_s}^{3/2} M \sqrt[20]{\frac{1}{N}} {N_f}
   \sqrt[4]{\beta } \alpha _{\theta _1}^{5/2}}{ \alpha _{\theta _2}^4
   {g_s^{\frac{1}{4}} N_f^{\frac{1}{3}}}}\sqrt{1-\frac{r_h^4}{r^4}},\nonumber\\
   & & \Omega_{35} = -\frac{\left(\frac{1}{|\log r|}\right)^{4/3} \sqrt[4]{\frac{1}{N}}
   \left(\frac{48.1 a^2}{r^2}+32.1\right)}{g_s^{\frac{1}{4}} N_f^{\frac{1}{3}}}\sqrt{1-\frac{r_h^4}{r^4}}-\frac{\omega_{35} \sqrt[3]{\frac{1}{|\log r|}}
   \sqrt{{\cal C}_{zz}^{(1)}} {\cal C}_q  {g_s}^{3/2} M \sqrt[4]{N}
   {N_f} \sqrt[4]{\beta } \alpha _{\theta _1}^{5/2}}{\alpha _{\theta
   _2}^5 {g_s^{\frac{1}{4}} N_f^{\frac{1}{3}}}}\sqrt{1-\frac{r_h^4}{r^4}};
\end{eqnarray}
\begin{eqnarray}
\label{Omega_4a}
& & \Omega_{42} = -\frac{0.02 a^2 {g_s}^{7/4} \log N  M \sqrt[10]{\frac{1}{N}}
   {N_f} \alpha _{\theta _1}}{\sqrt[3]{|\log r|} r^2 \alpha
   _{\theta _2}^2 {g_s^{\frac{1}{4}} N_f^{\frac{1}{3}}}}\sqrt{1-\frac{r_h^4}{r^4}}-\frac{\omega_{42}\sqrt{{\cal C}_{zz}^{(1)}} {\cal C}_q
   N^{17/20} \sqrt[4]{\beta } \sqrt{\alpha _{\theta _1}}}{
   |\log r|^2 {g_s}^2 \log N ^2 M {N_f} \alpha _{\theta
   _2}^4 g_s^{\frac{1}{4}} N_f^{\frac{1}{3}}}\sqrt{1-\frac{r_h^4}{r^4}}\nonumber\\
   & & \Omega_{43} = -\frac{\left(\frac{1}{|\log r|}\right)^{4/3} \sqrt[20]{N}
   \left(\frac{0.49 a^2}{r^2}+0.49\right)}{\alpha _{\theta _2}
   {g_s^{\frac{1}{4}} N_f^{\frac{1}{3}}}}\sqrt{1-\frac{r_h^4}{r^4}}\nonumber\\
   & & \Omega_{45} = -\frac{0.2 a^2
   \sqrt[3]{\frac{1}{|\log r|}} {g_s}^{7/4} \log N  M
   \left(\frac{1}{N}\right)^{7/10} {N_f} \alpha _{\theta _1}}{r^2
   {g_s^{\frac{1}{4}} N_f^{\frac{1}{3}}}}\sqrt{1-\frac{r_h^4}{r^4}} + \frac{\omega_{45}
   \left(\frac{1}{|\log r|}\right)^{7/3} \sqrt{{\cal C}_{zz}^{(1)}}
   {\cal C}_q  N^{11/20} \sqrt[4]{\beta } \sqrt{\alpha _{\theta
   _1}}}{{g_s}^2 \log N ^2 M {N_f} r \alpha _{\theta _2}^3
   {g_s^{\frac{1}{4}} N_f^{\frac{1}{3}}}}\sqrt{1-\frac{r_h^4}{r^4}}\nonumber\\
   & & \Omega_{46} = \frac{\left(\frac{1}{|\log r|}\right)^{4/3} \sqrt[20]{N}
   \left(0.29-\frac{0.44 a^2}{r^2}\right)}{\alpha _{\theta _2}
   {g_s^{\frac{1}{4}} N_f^{\frac{1}{3}}}}\sqrt{1-\frac{r_h^4}{r^4}}+\frac{\omega_{46}
   \sqrt{{\cal C}_{zz}^{(1)}} {\cal C}_q  N^{17/20} \sqrt[4]{\beta }
   \sqrt{\alpha _{\theta _1}}}{|\log r|^{7/3} {g_s}^{7/3} \log N ^2
   M {N_f} \alpha _{\theta _2}^4 }\sqrt{1-\frac{r_h^4}{r^4}};\nonumber\\
& & \nonumber\\
& & \nonumber\\
& & \Omega_{52} = \frac{\left(\frac{1}{|\log r|}\right)^{4/3} \sqrt[4]{\frac{1}{N}}
   \left(11.4-\frac{40.1 a^2}{r^2}\right)}{\sqrt{\frac{\sqrt{{g_s}}
   {N_f}^{2/3}
   r^4}{r^4-{r_h}^4}}}+\frac{\sqrt[3]{\frac{1}{|\log r|}}
   \sqrt{{\cal C}_{zz}^{(1)}} {\cal C}_q  {g_s}^{3/2} M \sqrt[4]{N}
   {N_f} \sqrt[4]{\beta } \alpha _{\theta _1}^{5/2}}{10^{13}
   \alpha _{\theta _2}^5 {g_s^{\frac{1}{4}} N_f^{\frac{1}{3}}}}\sqrt{1-\frac{r_h^4}{r^4}},\nonumber\\
& & \Omega_{53} = \frac{\left(\frac{1}{|\log r|}\right)^{4/3} \sqrt[4]{\frac{1}{N}}
   \left(\frac{28.7 a^2}{r^2}+19.1\right)}{g_s^{\frac{1}{4}} N_f^{\frac{1}{3}}}\sqrt{1-\frac{r_h^4}{r^4}}+{\omega_{53} \sqrt[3]{\frac{1}{|\log r|}}
   \sqrt{{\cal C}_{zz}^{(1)}} {\cal C}_q  {g_s}^{19/12} M \sqrt[4]{N}
   {N_f}^{2/3} \sqrt[4]{\beta } \alpha _{\theta _1}^{5/2}}\sqrt{1 - \frac{r_h^4}{r^4}}\nonumber\\
   & & \Omega_{54} = \frac{|\log r|^{2/3} {g_s}^{7/2} \log N ^2 M^2
   \left(\frac{1}{N}\right)^{17/20} {N_f}^2 \left(\frac{4.8
   a^2}{r^2}+4.8\right) \alpha _{\theta _1}^2 \sqrt{\frac{1}{\alpha
   _{\theta _2}^2}}}{r {g_s^{\frac{1}{4}} N_f^{\frac{1}{3}}}}\sqrt{1-\frac{r_h^4}{r^4}}\nonumber\\
   & & -\frac{\omega_{54}
   \sqrt[3]{\frac{1}{|\log r|}} \sqrt{{\cal C}_{zz}^{(1)}}
   {\cal C}_q  {g_s}^{3/2} M \sqrt[20]{\frac{1}{N}} {N_f}
   \sqrt[4]{\beta } \alpha _{\theta _1}^{5/2}}{r \alpha _{\theta _2}^4
   {g_s^{1/4}N_f^{1/3}}}\sqrt{1-\frac{r_h^4}{r^4}}\nonumber\\
   & & \Omega_{55} = -\frac{\left(\frac{1}{|\log r|}\right)^{4/3} \sqrt[4]{\frac{1}{N}}
   \left(\frac{44.5 a^2}{r^2}+29.7\right)}{g_s^{1/4}N_f^{1/3}}\sqrt{1-\frac{r_h^4}{r^4}}-\frac{\omega_{55}\sqrt[3]{\frac{1}{|\log r|}}
   \sqrt{{\cal C}_{zz}^{(1)}} {\cal C}_q  {g_s}^{3/2} M \sqrt[4]{N}
   {N_f} \sqrt[4]{\beta } \alpha _{\theta _1}^{5/2}}{r
   \alpha _{\theta _2}^5 {g_s^{\frac{1}{4}} N_f^{\frac{1}{3}}}}\sqrt{1-\frac{r_h^4}{r^4}}\nonumber\\
& & \Omega_{56} = -\frac{|\log r|^{2/3} {g_s}^{7/2} M^2
   \left(\frac{1}{N}\right)^{7/20} {N_f}^2 \left(\frac{0.27
   a^2}{r^2}+0.18\right) \alpha _{\theta _1}^4}{\alpha _{\theta _2}^4
   {g_s^{\frac{1}{4}} N_f^{\frac{1}{3}}}}\sqrt{1-\frac{r_h^4}{r^4}} + \frac{\omega_{56}
   \sqrt[3]{\frac{1}{|\log r|}} \sqrt{{\cal C}_{zz}^{(1)}}
   {\cal C}_q  {g_s}^{3/2} M \sqrt[4]{N} {N_f} \sqrt[4]{\beta }
   \alpha _{\theta _1}^{5/2}}{\alpha _{\theta _2}^5
   {g_s^{\frac{1}{4}} N_f^{\frac{1}{3}}}}\sqrt{1-\frac{r_h^4}{r^4}};\nonumber\\
   & &
\end{eqnarray}
\begin{eqnarray}
\label{Omega_6a}
& & \Omega_{62} = \frac{\left(\frac{1}{|\log r|}\right)^{4/3} \sqrt[4]{\frac{1}{N}}
   \left(12.5-\frac{43.6 a^2}{r^2}\right)}{g_s^{1/4}N_f^{1/3}}\sqrt{1-\frac{r_h^4}{r^4}}+\frac{\omega_{62} \sqrt[3]{\frac{1}{|\log r|}}
   \sqrt{{\cal C}_{zz}^{(1)}} {\cal C}_q  {g_s}^{3/2} M \sqrt[4]{N}
   {N_f} \sqrt[4]{\beta } \alpha _{\theta _1}^{5/2}}{r \alpha
   _{\theta _2}^5 {g_s^{\frac{1}{4}} N_f^{\frac{1}{3}}}}\sqrt{1-\frac{r_h^4}{r^4}}\nonumber\\
& & \Omega_{63} = \frac{\left(\frac{1}{|\log r|}\right)^{4/3} \sqrt[4]{\frac{1}{N}}
   \left(\frac{31.3 a^2}{r^2}+20.8\right)}{g_s^{1/4}N_f^{1/3}}\sqrt{1-\frac{r_h^4}{r^4}}+\frac{\omega_{63} \sqrt[3]{\frac{1}{|\log r|}}
   \sqrt{{\cal C}_{zz}^{(1)}} {\cal C}_q  {g_s}^{3/2} M \sqrt[4]{N}
   {N_f} \sqrt[4]{\beta } \alpha _{\theta _1}^{5/2}}{r \alpha
   _{\theta _2}^5 {g_s^{\frac{1}{4}} N_f^{\frac{1}{3}}}}\sqrt{1-\frac{r_h^4}{r^4}}\nonumber\\
   & & \Omega_{64} = \frac{|\log r|^{2/3} {g_s}^{7/2} \log N ^2 M^2
   \left(\frac{1}{N}\right)^{17/20} {N_f}^2 \alpha _{\theta _1}^2
   \sqrt{\frac{1}{\alpha _{\theta _2}^2}} \left(5.3-\frac{5.2
   a^2}{r^2}\right)}{ {g_s^{\frac{1}{4}} N_f^{\frac{1}{3}}}}\sqrt{1-\frac{r_h^4}{r^4}}\nonumber\\
   & & -\frac{\omega_{64}
   \sqrt[3]{\frac{1}{|\log r|}} \sqrt{{\cal C}_{zz}^{(1)}}
   {\cal C}_q  {g_s}^{3/2} M \sqrt[20]{\frac{1}{N}} {N_f}
   \sqrt[4]{\beta } \alpha _{\theta _1}^{5/2}}{\alpha _{\theta _2}^4
   {g_s^{1/4}N_f^{1/3}}}\sqrt{1-\frac{r_h^4}{r^4}}\nonumber\\
& & \Omega_{65} = -\frac{\left(\frac{1}{|\log r|}\right)^{4/3} \sqrt[4]{\frac{1}{N}}
   \left(\frac{48.5 a^2}{r^2}+32.3\right)}{g_s^{1/4}N_f^{1/3}}\sqrt{1-\frac{r_h^4}{r^4}}-\frac{\omega_{65} \sqrt[3]{\frac{1}{|\log r|}}
   \sqrt{{\cal C}_{zz}^{(1)}} {\cal C}_q  {g_s}^{3/2} M \sqrt[4]{N}
   {N_f} \sqrt[4]{\beta } \alpha _{\theta _1}^{5/2}}{r \alpha
   _{\theta _2}^5 {g_s^{\frac{1}{4}} N_f^{\frac{1}{3}}}}\sqrt{1-\frac{r_h^4}{r^4}}\nonumber\\
   & & \Omega_{66} = -\frac{|\log r|^{2/3} {g_s}^{7/2} M^2
   \left(\frac{1}{N}\right)^{7/20} {N_f}^2 \left(\frac{0.3
   a^2}{r^2}+0.2\right) \alpha _{\theta _1}^4}{ \alpha _{\theta _2}^4 {g_s^{1/4}N_f^{1/3}}}\sqrt{1-\frac{r_h^4}{r^4}} + \frac{\omega_{66}
   \sqrt[3]{\frac{1}{|\log r|}} \sqrt{{\cal C}_{zz}^{(1)}}
   {\cal C}_q  {g_s}^{3/2} M \sqrt[4]{N} {N_f} \sqrt[4]{\beta }
   \alpha _{\theta _1}^{5/2}}{r \alpha _{\theta _2}^5
   {g_s^{\frac{1}{4}} N_f^{\frac{1}{3}}}}\sqrt{1-\frac{r_h^4}{r^4}}.\nonumber\\
   & &
\end{eqnarray}
}
wherein the numerical constants $\omega_{ab}\ll1$; it turns out that in $\Omega_{4a}, a=2, 3, 4, 5, 6$, these very small numerical pre-factors arising in the higher derivative contribution to the ``structure constants", are maximally compensated by the highest positive powers of $N$ (amongst the other $\Omega_{ab}'s, a\neq4$).

\section{ Equation (\ref{C1=0-III}) Relevant to Proving Result 3 }
\setcounter{equation}{0} \seceqee

In this appendix, we will show that the equation (\ref{C1=0-III}): \\${\scriptsize \sum_{N,P\in\left\{t,x^{1,2,3},r,\theta_{1,2},\phi_{1,2},\psi,x^{10}\right\}}\beta\partial_r\left(\sqrt{-g^{(0)}}g^{(0)NP}g_{NP}^{(0)}f_{NP}g_{(0)}^{rr}\partial_rC^{M_1M_2 x^{10}}_{(0)}\right) \delta^{M_3}_{x^{10}}=0}$, \\ is satisfied by setting the constant of integration $C_{\theta_1x}^{(1)}$, to zero up to LO in $\beta$ and $N$; up to ${\cal O}(\beta)$ and NLO in $N$,
one would additionally require $C_{\theta_1z}^{(1)} - 2  C_{zz}^{(1)} = C_{\theta_2z}^{(1)}= 0$.

\newpage
\begin{itemize}
\item $(M_1, M_2) = (\theta_1,\theta_2)$:

One can show that near $r = \chi r_h, \chi={\cal O}(1)$, (\ref{C1=0-III}) reduces to:
\begin{eqnarray}
\label{67-2}
& & \beta\frac{1}{4 \pi ^{19/4} {g_s}^{5/4} \chi^5 \alpha _{\theta _1}^8 \alpha
   _{\theta _2}^4}\Biggl\{27 \log N  \left(\frac{1}{N}\right)^{3/20} {N_f}^2 {r_h}^3 \Biggl(\left(\frac{1}{N}\right)^{2/5} \left(-162
   b^4+9 b^2 \left(\chi^6+\chi^2\right)+4 \chi^8\right)\nonumber\\
& & \times \left(0.01 \alpha _{\theta _2}^7 ({C_{zz}}^{(1)}-2 C_{\theta_1z}^{(1)})+0.09
   C_{\theta_2z}^{(1)} {g_s}^{3/2} \log r_h  M {N_f} \alpha _{\theta _1}^8\right)-\sqrt{3} \pi ^{3/2} b^4 C_{\theta_1x}^{(1)}
   \alpha _{\theta _1}^4 \alpha _{\theta _2}^3\Biggr)\Biggr\}\nonumber\\
& & \sim \beta\frac{\log N r_h^3}{N^{\frac{3}{20}}} \frac{N_f^2}{g_s^{\frac{5}{4}}}C_{\theta_1x}^{(1)},
\end{eqnarray}
where the ``$\sim$" in (\ref{67-2}) and henceforth implies equality up to NLO-in-$N$ terms. Therefore by setting:
\begin{equation}
\label{C68=0}
C_{\theta_1x}^{(1)} = 0,
\end{equation}
(\ref{C1=0-III}) is satisfied in the IR \footnote{If one wishes to also consider the NLO-in-$N$ term in (\ref{67-2}), one sees that one needs to impose the additional constraints: $0.01 \alpha _{\theta _2}^7 ({C_{zz}}^{(1)}-2 C_{\theta_1z}^{(1)})+0.09
   C_{\theta_2z}^{(1)} {g_s}^{3/2} \log r_h  M {N_f} \alpha _{\theta _1}^8 = 0$ - see (\ref{constraints-C1_MN-NLO-N}).}.

\item $(M_1, M_2) = (\theta_1,x)$:

Working in the IR, i.e., $r = \chi r_h, \chi={\cal O}(1)$, one an show that (\ref{C1=0-III}) yields:
{\footnotesize
\begin{eqnarray}
\label{68-constraint-CMNs}
& & \beta \frac{N_f^2 r_h^3}{g_s}\left({\cal F}_{\theta_1x}^{\theta_1x}(b, \chi, \alpha_{\theta_{1,2}}, r_h)C_{\theta_1x}^{(1)}
\frac{\log N }{ N^{\frac{7}{10}}}+ \frac{1}{N^{\frac{11}{10}}}\sum_{(M,N)=(z,z),(\theta_1,z),(\theta_2,z)}{\cal F}_{MN}^{\theta_1x}(b, \chi, \alpha_{\theta_{1,2}})C_{MN}^{(1)}\right)\nonumber\\
& & \sim  \beta\frac{N_f^2 r_h^3}{g_s}{\cal F}_{\theta_1x}^{\theta_1x}(b, \chi, \alpha_{\theta_{1,2}}, r_h)C_{\theta_1x}^{(1)}\frac{\log N }{ N^{\frac{7}{10}}},
\end{eqnarray}
}
[wherein the notation used is ${\cal F}^{M_1M_2}_{MN}, (M,N)\in\left\{(\theta_1,x), (\theta_1,z), (\theta_2,z), (z,z)\right\}$], which again would vanish at (\ref{C68=0}).

We will, for the remaining eight equations, given only the equivalents of (\ref{68-constraint-CMNs}) below (assuming one is working in the IR, i.e., $r = \chi r_h, \chi={\cal O}(1)$)

\item $(M_1, M_2) = (\theta_1,y)$:

One obtains:
{\footnotesize
\begin{eqnarray}
\label{69-constraint-CMNs}
& & \beta\frac{N_f^2r_h^3 \log N}{g_s}\left(N^{\frac{2}{5}}{\cal F}_{\theta_1x}^{\theta_1y}(b, \chi, \alpha_{\theta_{1,2}}, r_h)C_{\theta_1x}^{(1)}
+\sum_{(M,N)=(z,z),(\theta_1,z),(\theta_2,z)}{\cal F}_{MN}^{\theta_1y}(b, \chi, \alpha_{\theta_{1,2}},r_h)C_{MN}^{(1)}\right)\nonumber\\
& & \sim  \beta\frac{N_f^2 r_h^3 N^{\frac{2}{5}}}{g_s}{\cal F}_{\theta_1x}^{\theta_1y}(b, \chi, \alpha_{\theta_{1,2}}, r_h)C_{\theta_1x}^{(1)},
\end{eqnarray}
}
which again would vanish at (\ref{C68=0}).

\item $(M_1, M_2) = (\theta_1,z)$

One obtains:
{\footnotesize
\begin{eqnarray}
\label{610-constraint-CMNs}
& & \beta\frac{N_f^2r_h^3 \log N}{g_s}\left(N^{\frac{1}{10}}{\cal F}_{\theta_1x}^{\theta_1z}(b, \chi, \alpha_{\theta_{1,2}}, r_h)C_{\theta_1x}^{(1)}
+\frac{1}{N^{\frac{3}{10}}}\sum_{(M,N)=(z,z),(\theta_1,z),(\theta_2,z)}{\cal F}_{MN}^{\theta_1z}(b, \chi, \alpha_{\theta_{1,2}},r_h)C_{MN}^{(1)}\right)\nonumber\\
& & \sim  \beta\frac{N_f^2 r_h^3 N^{\frac{1}{10}}}{g_s}{\cal F}_{\theta_1x}^{\theta_1z}(b, \chi, \alpha_{\theta_{1,2}}, r_h)C_{\theta_1x}^{(1)},
\end{eqnarray}
}
which again would vanish at (\ref{C68=0}).

\item $(M_1, M_2) = (\theta_2,x)$

One obtains:
{\footnotesize
\begin{eqnarray}
\label{78-constraint-CMNs}
& & \beta M N_f^3 \sqrt{g_s} \left(\log N\right)^2\left(\frac{1}{N^{\frac{7}{5}}}{\cal F}_{\theta_1x}^{\theta_2x}(b, \chi, \alpha_{\theta_{1,2}}, r_h)C_{\theta_1x}^{(1)}
+\frac{1}{N^{\frac{9}{5}}}\sum_{(M,N)=(z,z),(\theta_1,z),(\theta_2,z)}{\cal F}_{MN}^{\theta_2x}(b, \chi, \alpha_{\theta_{1,2}},r_h)C_{MN}^{(1)}\right)\nonumber\\
& & \sim   \beta M N_f^3 \sqrt{g_s} \left(\log N\right)^2\frac{1}{N^{\frac{7}{5}}}{\cal F}_{\theta_1x}^{\theta_2x}(b, \chi, \alpha_{\theta_{1,2}}, r_h)C_{\theta_1x}^{(1)},
\end{eqnarray}
}
which again would vanish at (\ref{C68=0}).

\item $(M_1, M_2) = (\theta_2,y)$

One obtains:
{\footnotesize
\begin{eqnarray}
\label{79-constraint-CMNs}
& & \beta M N_f^3 \sqrt{g_s} \left(\log N\right)^2\left(\frac{1}{N^{\frac{3}{10}}}{\cal F}_{\theta_1x}^{\theta_2y}(b, \chi, \alpha_{\theta_{1,2}}, r_h)C_{\theta_1x}^{(1)}
+\frac{1}{N^{\frac{7}{10}}}\sum_{(M,N)=(z,z),(\theta_1,z),(\theta_2,z)}{\cal F}_{MN}^{\theta_2x}(b, \chi, \alpha_{\theta_{1,2}},r_h)C_{MN}^{(1)}\right)\nonumber\\
& & \sim   \beta M N_f^3 \sqrt{g_s} \left(\log N\right)^2\frac{1}{N^{\frac{3}{10}}}{\cal F}_{\theta_1x}^{\theta_2y}(b, \chi, \alpha_{\theta_{1,2}}, r_h)C_{\theta_1x}^{(1)},
\end{eqnarray}
}
which again would vanish at (\ref{C68=0}).

\item $(M_1, M_2) = (\theta_2,z)$

One obtains:
{\footnotesize
\begin{eqnarray}
\label{710-constraint-CMNs}
& & \beta M N_f^3 \sqrt{g_s}r_h^3 \left(\log N\right)^2\left(\frac{1}{N^{\frac{3}{5}}}{\cal F}_{\theta_1x}^{\theta_2z}(b, \chi, \alpha_{\theta_{1,2}}, r_h)C_{\theta_1x}^{(1)}
+\frac{1}{N}\sum_{(M,N)=(z,z),(\theta_1,z),(\theta_2,z)}{\cal F}_{MN}^{\theta_2z}(b, \chi, \alpha_{\theta_{1,2}},r_h)C_{MN}^{(1)}\right)\nonumber\\
& & \sim  \beta M N_f^3 \sqrt{g_s}r_h^3 \left(\log N\right)^2\frac{1}{N^{\frac{3}{5}}}{\cal F}_{\theta_1x}^{\theta_2z}(b, \chi, \alpha_{\theta_{1,2}}, r_h)C_{\theta_1x}^{(1)},
\end{eqnarray}
}
which again would vanish at (\ref{C68=0}).

\item $(M_1, M_2) = (x,y)$

One obtains:
{\footnotesize
\begin{eqnarray}
\label{89-constraint-CMNs}
& & \beta M N_f^3 {g_s}^{\frac{9}{4}} \left(\log N\right)^2\left(\frac{1}{N^{\frac{5}{4}}}{\cal F}_{\theta_1x}^{xy}(b, \chi, \alpha_{\theta_{1,2}}, r_h)C_{\theta_1x}^{(1)}
+\frac{1}{N^{\frac{33}{20}}}\sum_{(M,N)=(z,z),(\theta_1,z),(\theta_2,z)}{\cal F}_{MN}^{xy}(b, \chi, \alpha_{\theta_{1,2}},r_h)C_{MN}^{(1)}\right)\nonumber\\
& & \sim   \beta M N_f^3 {g_s}^{\frac{9}{4}} \left(\log N\right)^2\frac{1}{N^{\frac{5}{4}}}{\cal F}_{\theta_1x}^{xy}(b, \chi, \alpha_{\theta_{1,2}}, r_h)C_{\theta_1x}^{(1)},
\end{eqnarray}
}
which again would vanish at (\ref{C68=0}).

\item $(M_1, M_2) = (x,z)$

One obtains:
{\footnotesize
\begin{eqnarray}
\label{810-constraint-CMNs}
& & \beta M N_f^3 \left(\log N\right)^2\left(\frac{1}{N^{\frac{23}{20}}}{\cal F}_{\theta_1x}^{xz}(b, \chi, \alpha_{\theta_{1,2}}, r_h)C_{\theta_1x}^{(1)}
+\frac{1}{N^{\frac{31}{20}}}\sum_{(M,N)=(z,z),(\theta_1,z),(\theta_2,z)}{\cal F}_{MN}^{xy}(b, \chi, \alpha_{\theta_{1,2}},r_h)C_{MN}^{(1)}\right)\nonumber\\
& & \sim   \beta M N_f^3 \left(\log N\right)^2\frac{1}{N^{\frac{23}{20}}}{\cal F}_{\theta_1x}^{xz}(b, \chi, \alpha_{\theta_{1,2}}, r_h)C_{\theta_1x}^{(1)},
\end{eqnarray}
}
which again would vanish at (\ref{C68=0}).

\item $(M_1, M_2) = (y,z)$

One obtains:
{\footnotesize
\begin{eqnarray}
\label{910-constraint-CMNs}
& & \beta M N_f^3 g_s^{\frac{3}{4}} \left(\log N\right)^2\left(\frac{1}{N^{\frac{1}{20}}}{\cal F}_{\theta_1x}^{yz}(b, \chi, \alpha_{\theta_{1,2}}, r_h)C_{\theta_1x}^{(1)}
+\frac{1}{N^{\frac{9}{20}}}\sum_{(M,N)=(z,z),(\theta_1,z),(\theta_2,z)}{\cal F}_{MN}^{yz}(b, \chi, \alpha_{\theta_{1,2}},r_h)C_{MN}^{(1)}\right)\nonumber\\
& & \sim   \beta M N_f^3 g_s^{\frac{3}{4}} \left(\log N\right)^2\frac{1}{N^{\frac{1}{20}}}{\cal F}_{\theta_1x}^{yz}(b, \chi, \alpha_{\theta_{1,2}}, r_h)C_{\theta_1x}^{(1)},
\end{eqnarray}
}
which again would vanish at (\ref{C68=0}).

\end{itemize}

 \section{Details of $G_2$ Structure of \\$\left.M_7(r, \theta_{1,2}, \phi_{1,2}, \psi, x^{10})\right|_{{\rm Ouyang-embedding[parent\ type\ IIB]}\cap|\mu_{\rm Ouyang}|\ll1}$}
\label{tau01M7r}
\setcounter{equation}{0} \seceqee

The details of the evaluation of $\tau_{0,1}$ in the neighborhood of the Ouyang embedding of the flavor $D7$-branes in the parent type IIB theory assuming a infinitesimal modulus of the Ouyang embedding parameter $\mu_{\rm Ouyang}$,  are provided below. 

The four intrinsic $G_2$-structure torsion classes are then given by \cite{J. G. J. Held's thesis [2012]}
\begin{eqnarray}
\label{tau-i}
& & W_1 = \frac{1}{7}d\Phi\lrcorner*_7\Phi,\nonumber\\
& & W_7 = -\frac{1}{12}d\Phi\lrcorner\Phi = \frac{1}{12}\Phi\wedge*_7d\Phi,\nonumber\\
& & W_{14} = \frac{1}{2}\left(d*_7\Phi\lrcorner\Phi - *_7d*_7\Phi\right) - 2 W_7\lrcorner\Phi
 = - *_7d*_7\Phi + 4 W_7\lrcorner\Phi,
\nonumber\\
& & W_{27} = *_7d\Phi - W_1\Phi + 3W_7\lrcorner*_7\Phi.
\end{eqnarray}
 
\subsection{$W_7(M_7(r, \theta_{1,2}, \phi_{1,2}, \psi, x^{10})$}

Utilizing,
$W_7 = *_7\left(\Phi\wedge*_7d\Phi\right)$, let us first evaluate $\Phi\wedge*_7d\Phi$. One sees that,
\begin{eqnarray}
\label{ta_1-i}
& & \Phi\wedge*_7d\Phi = e^{-2\Phi^{\rm IIA}}\left(\Omega_{25}e^{135247}
- \Omega_{23}e^{135267} -(\Omega_{43}+\Omega_{65})e^{135467} -\Omega_{24}e^{135257}-\Omega_{26}e^{146237}\right.\nonumber\\
& & \left. +(\Omega_{34}+\Omega_{56})e^{146357} -( \Omega_{66}+\Omega_{22}+\Omega_{33})e^{236457} - (\Omega_{22}+\Omega_{44}+\Omega_{55})e^{245367}\right)\nonumber\\
& &  -\frac{e^{-\frac{5}{3}\Phi^{\rm IIA}}}{\sqrt{G^{\cal M}_{x^{10}x^{10}}}}\left(-(\Omega_{45}+\Omega_{63})e^{135246} - (\Omega_{36}+\Omega_{54})e^{146235}\right) -\frac{4}{3}\frac{e^{-\frac{5}{3}\Phi^{\rm IIA}}}{\sqrt{\left(G^{\cal M}_{x^{10}x^{10}}\right)^2G^{\cal M}_{rr}}} e^{234567}\nonumber\\
& & + \frac{e^{-\frac{4}{3}\Phi^{\rm IIA}}}{G^{\cal M}_{x^{10}x^{10}}}\biggl(\Omega_{33}+\Omega_{44})e^{347256}-(\Omega_{55}+\Omega_{66})e^{567234} + \Omega_{32}e^{127356} + \Omega_{42}e^{127456}- \Omega_{52}e^{127345}\nonumber\\
& &  + \Omega_{62}e^{127346}\biggr)
\end{eqnarray}

From appendix {\bf C}, the most dominant $\Omega_{ij}$s are given as under:
\begin{eqnarray}
\label{Omega-OR4}
& & \Omega_{34} = \frac{|\log r|^{2/3} {g_s}^{7/2} \log N ^2 M^2
   \left(\frac{1}{N}\right)^{17/20} {N_f}^2 \left(\frac{28.1
   a^2}{r^2}+5.2\right) \alpha _{\theta _1}^2 }{ {g_s^{\frac{1}{4}} N_f^{\frac{1}{3}}}\alpha
   _{\theta _2}}\sqrt{1-\frac{r_h^4}{r^4}}\nonumber\\
   & & -\frac{\omega_{34}
   \sqrt[3]{\frac{1}{|\log r|}} \sqrt{{\cal C}_{zz}^{(1)}}
   {\cal C}_q  {g_s}^{3/2} M \sqrt[20]{\frac{1}{N}} {N_f}
   \sqrt[4]{\beta } \alpha _{\theta _1}^{5/2}}{ \alpha _{\theta _2}^4
   {g_s^{\frac{1}{4}} N_f^{\frac{1}{3}}}}\sqrt{1-\frac{r_h^4}{r^4}},\nonumber\\
& & \Omega_{36} = -\frac{|\log r|^{2/3}
   {gs}^{13/4} M^2 \left(\frac{1}{N}\right)^{7/20} {Nf}^{5/3} \alpha _{\theta _1}^4
   \sqrt{\frac{1}{\alpha _{\theta _2}^2}} \sqrt{1-\frac{{rh}^4}{r^4}} \left(0.2\, -\frac{0.3
   a^2}{r^2}\right)}{\alpha _{\theta _2}^3}\nonumber\\
   & & + \frac{\omega_{36} \sqrt[3]{\frac{1}{|\log r|}} \sqrt[4]{\beta }
   \sqrt{{\cal C}_{zz}^{(1)}} {\cal C}_q{gs}^{5/4} M \sqrt[4]{N} {Nf}^{2/3} \alpha _{\theta
   _1}^{5/2} \sqrt{1-\frac{{rh}^4}{r^4}}}{\alpha _{\theta _2}^5};\nonumber\\
  & & \Omega_{54} = \frac{|\log r|^{2/3} {g_s}^{7/2} \log N ^2 M^2
   \left(\frac{1}{N}\right)^{17/20} {N_f}^2 \left(\frac{4.8
   a^2}{r^2}+4.8\right) \alpha _{\theta _1}^2 \sqrt{\frac{1}{\alpha
   _{\theta _2}^2}}}{r {g_s^{\frac{1}{4}} N_f^{\frac{1}{3}}}}\sqrt{1-\frac{r_h^4}{r^4}}\nonumber\\
   & & -\frac{\omega_{54}
   \sqrt[3]{\frac{1}{|\log r|}} \sqrt{{\cal C}_{zz}^{(1)}}
   {\cal C}_q  {g_s}^{3/2} M \sqrt[20]{\frac{1}{N}} {N_f}
   \sqrt[4]{\beta } \alpha _{\theta _1}^{5/2}}{r \alpha _{\theta _2}^4
   {g_s^{1/4}N_f^{1/3}}}\sqrt{1-\frac{r_h^4}{r^4}}\nonumber\\
   & & \Omega_{56} = -\frac{|\log r|^{2/3} {g_s}^{7/2} M^2
   \left(\frac{1}{N}\right)^{7/20} {N_f}^2 \left(\frac{0.27
   a^2}{r^2}+0.18\right) \alpha _{\theta _1}^4}{\alpha _{\theta _2}^4
   {g_s^{\frac{1}{4}} N_f^{\frac{1}{3}}}}\sqrt{1-\frac{r_h^4}{r^4}} +\nonumber\\
   & &  \frac{\omega_{56}
   \sqrt[3]{\frac{1}{|\log r|}} \sqrt{{\cal C}_{zz}^{(1)}}
   {\cal C}_q  {g_s}^{3/2} M \sqrt[4]{N} {N_f} \sqrt[4]{\beta }
   \alpha _{\theta _1}^{5/2}}{\alpha _{\theta _2}^5
   {g_s^{\frac{1}{4}} N_f^{\frac{1}{3}}}}\sqrt{1-\frac{r_h^4}{r^4}},   
\end{eqnarray}
where $\omega_{ij}\ll1$. The ${\cal M}$-theory metric components $G^{\cal M}_{x^{10}x^{10}},\ G^{\cal M}_{rr}$ are given by:
\begin{eqnarray}
& & G^{\cal M}_{rr}= \left\{\frac{\sqrt{g_s} \left(6 a^2+r^2\right)}{2 \sqrt[3]{3} \sqrt[6]{\pi } \sqrt[3]{\cal A} r^2 \left(9 a^2+r^2\right) \left(1-\frac{r_h^4}{r^4}\right)}\right\} + \Biggl[3 {\cal A} \sqrt{N}-\frac{96 \pi  a^2 g_s M^2 \sqrt{\frac{1}{N}} N_f (c_1+c_2 \log (r_h))}{9 a^2+r^2} \nonumber\\
& &  +\frac{9}{32} {\cal A} g_s M^2 \sqrt{\frac{1}{N}} \left(-\frac{64 a^2 r^2 (c_1+c_2 \log (r_h))}{\left(6 a^2+r^2\right) \left(9 a^2+r^2\right)}-\frac{\log (r) (A g_s+12 g_s
   N_f+8 \pi )}{\pi ^2}\right))\Biggr] \nonumber\\
   & & 
    + \beta \left\{\frac{3^{2/3} {\cal A}^{2/3} \sqrt{g_s} \sqrt{N} r^2 \left(6 a^2+r^2\right) ({\cal C}_{zz}^{bh}-2 {\cal C}_{\theta_1 z}^{bh}+2 {\cal C}_{\theta_1 x}^{bh})}{2 \sqrt[6]{\pi } \left(9 a^2+r^2\right)
   \left(r^4-r_h^4\right)}\right\} 
\end{eqnarray}
\begin{eqnarray}
& & G^{\cal M}_{x_{10}x_{10}}=\left\{\frac{16 \pi ^{4/3} \left({64 \pi  a^2 g_s M^2 N_f (c_1+c_2 \log (r_h))} + {\cal A} N \left(9 a^2+r^2\right)\right)}{3 \sqrt[3]{3} {\cal A}^{7/3} {N \left(9 a^2+r^2\right)}}\right\}\nonumber\\
& &  -\beta \Biggl[ \left\{\frac{-19683 \sqrt{6} \alpha _{\theta _1}^6-6642 \alpha _{\theta _2}^2 \alpha _{\theta _1}^3+40 \sqrt{6} \alpha _{\theta _2}^4}{{\cal A}^{4/3} \left(3 b^2-1\right)^5 N_f r_h^4 \alpha _{\theta
   _2}^3 \left(9 a^2+r_h^2\right) }\right\} \nonumber\\
    &&   \left( \frac{48\ 3^{2/3} \sqrt[3]{\pi } b^{10} \left(9 b^2+1\right)^4 M \left(\frac{1}{N}\right)^{5/4} r \left(6 a^2+r_h^2\right) (r-2 r_h) \log ^3(r_h)}{\left(6 b^2 \log N+\log N\right)^4} \right) \Biggr]\nonumber 
\end{eqnarray}
where using \cite{Bulk-Viscosity-McGill-IIT-Roorkee},
\begin{equation}
{\cal A} \equiv 4\left[- N_f \log\left(\frac{4\sqrt{N}}{ \left(9 a^2
   r^4+r^6\right)^{\frac{1}{4}}\alpha_{\theta_1}\alpha_{\theta_2}}\right)+\frac{2 \pi }{g_s}\right]\stackrel{r\in{\rm IR}}{\longrightarrow} N_f\log r\stackrel{r\in{\rm IR}}{\longrightarrow}N_fN^{\frac{1}{3}}.
\end{equation}
Replacing $\alpha_{\theta_1}\rightarrow N^{\frac{1}{5}}\sin\theta_1$ and $\alpha_{\theta_2}\rightarrow N^{\frac{3}{10}}\sin\theta_2$:
\begin{eqnarray}
\label{Omega-ij-global}
& & \Omega_{22}^{\beta^0}\sim\Omega_{23}^{\beta^0}\sim\Omega_{25^{\beta^0}}\sim\Omega_{32}^{\beta^0}\sim\Omega_{33}^{\beta^0}\sim\Omega_{35}^{\beta^0}\sim-\Omega_{42}^{\beta^0}\sim\Omega_{43}^{\beta^0}\sim\Omega_{45}^{\beta^0}\sim\Omega_{46}^{\beta^0}\sim\Omega_{52}^{\beta^0}\sim\Omega_{53}^{\beta^0}\sim-\Omega_{55}\nonumber\\
& & \sim\Omega_{62}^{\beta^0}\sim\Omega_{63}^{\beta^0}\sim-\Omega_{65}^{\beta^0}\sim\frac{1}{N^{\frac{1}{4}}|\ln r|^{\frac{4}{3}}}\stackrel{r\in{\rm IR}}{\longrightarrow}{\frac{1}{N_f^{\frac{4}{3}}N^{\frac{25}{36}}}};\nonumber\\
& & \Omega_{24}^{\beta^0}\sim-\Omega_{26}^{\beta^0}\sim-\Omega_{36}^{\beta^0}\sim\Omega_{56}^{\beta^0}\sim\Omega_{66}^{\beta^0}\sim \frac{|\log r|^{\frac{2}{3}}}{N^{\frac{3}{4}}}\stackrel{r\in{\rm IR}}{\longrightarrow}\frac{N_f^{\frac{2}{3}}}{N^{\frac{19}{36}}};\nonumber\\
& & \Omega_{34}\sim\Omega_{54}^{\beta^0}\sim \frac{\left(\log N\right)^2|\log r|^{\frac{2}{3}}}{N^{\frac{3}{4}}}\stackrel{r\in{\rm IR}}{\longrightarrow}\frac{\left(\log N\right)^2N_f^{\frac{2}{3}}}{N^{\frac{19}{36}}}.
\end{eqnarray}
We thus see:
\begin{eqnarray}
\label{*tau1}
& & \Phi\wedge*_7d\Phi (r\in{\rm IR})) \sim e^{-2\Phi^{\rm IIA}}\left(\Omega_{34} + \Omega_{56}\right)e^{146357} - \frac{e^{-\frac{5}{3}\Phi^{\rm IIA}}}{\sqrt{G^{\cal M}_{x^{10}x^{10}}}}\left(\Omega_{36}-\Omega_{54}\right)e^{146235}
\stackrel{r\in{\rm IR}}{\longrightarrow} \frac{\left(\log N\right)^2}{N^{\frac{3}{4}}|\log r|^{\frac{4}{3}}}\nonumber\\
& & \stackrel{r\in{\rm IR}}{\longrightarrow}\frac{1}{N^{\alpha_{W_7}}}e^{146357} + \frac{\alpha_{\Phi^{\rm IIA}G_{x^{10}x^{10}}}}{|\log r|}\left(\Omega_{36}-\Omega_{54}\right)e^{146235},
\end{eqnarray}
where $\alpha_{W_7}>1$. Now, from (\ref{Omega-OR4}), one can show that:
\begin{eqnarray}
\label{Omega36-Omega54}
& & \left(\Omega_{36} - \Omega_{54}\right)^{\beta^0} = 0, 
\end{eqnarray}
for $\forall r\sim\sqrt{\frac{3}{2}}a\in$IR - note that $r=\sqrt{3}a$ is the interface of the UV and IR-UV interpolating regions. Therefore,
\begin{eqnarray}
\label{tau1=0}
& & \hskip -0.3in \left.W_7 \left[= *_7\left(\Phi\wedge*_7d\Phi\right)\right]\right|_{{\rm Ouyang-embedding[parent\ type\ IIB]}\cap|\mu_{\rm Ouyang}|\ll1}\nonumber\\
& & \hskip -0.3in  = - \left.d\Phi\lrcorner\Phi\right|_{{\rm Ouyang-embedding[parent\ type\ IIB]}\cap|\mu_{\rm Ouyang}|\ll1} = {\cal O}\left(\frac{1}{N^{\alpha>1}}\right)\sim0\left({\rm as\ work\ only\ up\ to}\ {\cal O}\left(\frac{1}{N}\right)\right),\nonumber\\
& & 
\end{eqnarray}
as stated in (\ref{G2-Ws}).

\subsection{$W_1(M_7(r, \theta_{1,2}, \phi_{1,2}, \psi, x^{10}))$}

One can show that:
\begin{eqnarray}
\label{tau0}
& & W_1 = \frac{1}{7}d\Phi\lrcorner*_7\Phi = \frac{2}{7} \frac{e^{-\frac{5}{3}\Phi^{\rm IIA}}}{\sqrt{G^{\cal M}_{x^{10}x^{10}}}}\left(\Omega_{53} - \Omega_{64}\right)\stackrel{r\in{\rm IR}}{\longrightarrow}\sim\frac{1}{|\log r|}\left(\Omega_{53} - \Omega_{64}\right).
\end{eqnarray}
Using results of appendix {\bf C}, 
\begin{eqnarray}
\label{Omega53-Omega64}
& & \left(\Omega_{53} - \Omega_{64}\right)^{\beta^0} \nonumber\\
& & =
\frac{\sqrt{1-\frac{{r_h}^4}{r^4}} \left(\left(\frac{1}{{|\log r|}}\right)^{4/3}
   \sqrt[4]{\frac{1}{N}} \left(\frac{28.7 a^2}{r^2}+19.1\right)-\frac{{|\log r|}^{2/3}
   {g_s}^{7/2} {\log N}^2 M^2 \left(\frac{1}{N}\right)^{17/20} {N_f}^2 \alpha _{\theta _1}^2
   \left(5.3\, -\frac{5.2 a^2}{r^2}\right)}{\alpha _{\theta _2}}\right)}{\sqrt[4]{{g_s}}
   \sqrt[3]{{N_f}}}\nonumber\\
 \end{eqnarray}
and replacing $\alpha_{\theta_2}\rightarrow N^{\frac{3}{10}}\sin\theta_2$ to get the conjectured result $\forall\theta_{1,2}, \phi_{1,2}, \psi$ but in the neighborhood of the Ouyuang embedding in the parent type IIB theoryone can show
\begin{eqnarray}  
 & & \hskip -0.5in  \left(\Omega_{53} - \Omega_{64}\right)^{\beta^0}\nonumber\\
   & & \hskip -0.5in = \frac{\left(5.2 {|\log r|}^{5/3} {g_s}^{\frac{7}{2}} {\log N}^2 M^2 \left(\frac{1}{N}\right)^{3/5} {N_f}^2
   \alpha _{\theta _1}^2 \left( a^2-1.02 r^2\right)+28.7 \sqrt[3]{\frac{1}{|\log r|}} \alpha
   _{\theta _2} \left(a^2+0.67 r^2\right)\right)}{N^{\frac{1}{4}+\frac{3}{10}}(\sin\theta_2)\log r}\sqrt{1-\frac{{r_h}^4}{r^4}} \nonumber\\
\end{eqnarray}
One can further show that (guided by \cite{Bulk-Viscosity-McGill-IIT-Roorkee}) assuming $|\log r|\sim\alpha_{|\log r|_{\rm IR}} N^{\frac{1}{3}}$, for
\begin{eqnarray}
\label{r-IR-tau0=0}
& & r =\Biggl[ 0.99 +\frac{0.095  |\kappa_r| }{{\alpha_{|\log r|}}^{5/3} {g_s}^{7/2} {\log N}^2 M^2
   {N_f}^2 \alpha _{\theta _1}^2} + {\cal O}\left(\left(\frac{1}{{\log N}}\right)^4\right)\Biggr]a\in{\rm IR},
\end{eqnarray}
\begin{eqnarray}
\label{tau=0-ii}
& & W_1 \sim \frac{\sqrt{1-\frac{{r_h}^4}{r^4}}}{|\log r|^2 N^{\frac{11}{20}}\sin\theta_2}\left(\frac{|\kappa_r|}{N^{\frac{2}{25}}}\right)a^2\stackrel{r\in{\rm IR}}{\longrightarrow}{\cal O}\left(\frac{1}{N^{\alpha_{W_1}>1}}\right)a^2\sim 0\left({\rm as\ work\ only\ up\ to}\ {\cal O}\left(\frac{1}{N}\right)\right),\nonumber\\
& & 
\end{eqnarray}
for $\kappa_r<0$.


\begin{thebibliography}{99}
\bibitem{MQGP}M.~Dhuria and A.~Misra, {\it Towards MQGP}, JHEP 1311 (2013) 001, [arXiv:hep-th/1306.4339].
\bibitem{metrics}  M.~Mia, K.~Dasgupta, C.~Gale and S.~Jeon, {\it Five Easy Pieces: The Dynamics of Quarks in Strongly Coupled Plasmas}, Nucl.\ Phys.\ B {\bf 839}, 187 (2010) [arXiv:hep-th/0902.1540].
\bibitem{previous-higher-ders} S.~Grozdanov and A.~O.~Starinets, {\it Second-order transport, quasinormal modes and zero-viscosity limit in the Gauss-Bonnet holographic fluid},
JHEP \textbf{03}, 166 (2017) [arXiv:1611.07053 [hep-th]]; J.~Casalderrey-Solana, S.~Grozdanov and A.~O.~Starinets, {\it Transport Peak in the Thermal Spectral Function of $\mathcal N=4$ Supersymmetric Yang-Mills Plasma at Intermediate Coupling}, Phys. Rev. Lett. \textbf{121}, no.19, 191603 (2018) [arXiv:1806.10997 [hep-th]].
\bibitem{Vikas+Gopal+Aalok} V.~Yadav, G.~Yadav and A.~Misra,
{\it (Phenomenology/Lattice-Compatible) $SU(3)$ M$\chi$PT HD up to ${\cal O}(p^4)$ and the ${\cal O}\left(R^4\right)$-Large-$N$ Connection}, JHEP 08 (2021) 151 [arXiv:2011.04660 [hep-th]].
\bibitem{Gopal+Vikas+Aalok} G.~Yadav, V.~Yadav and A.~Misra, {\it McTEQ (M chiral perturbation theory-compatible deconfinement Temperature and Entanglement entropy up to terms Quartic in curvature) and FM (Flavor Memory)},   JHEP 10 (2021) 220, arXiv:2108.05372 [hep-th].
\bibitem{Tc-EE} I.~R.~Klebanov, D.~Kutasov and A.~Murugan, {\it Entanglement as a probe of confinement},
Nucl.\ Phys.\ B {\bf 796}, 274 (2008) [arXiv:0709.2140 [hep-th]].
\bibitem{hybrid-fermions} C.~A.~Ballon Bayona, H.~Boschi-Filho and N.~R.~F.~Braga,
{\it Deep inelastic scattering from gauge string duality in the soft wall model}, JHEP \textbf{03}, 064 (2008)
[arXiv:0711.0221 [hep-th]].

 \bibitem{O(R^4)}  M. J. Duff, J. T. Liu and R. Minasian, {\it Eleven-dimensional origin of string-string duality:
A One loop test}, Nucl. Phys. B 452, 261 (1995) [hep-th/9506126];  M. B. Green and P. Vanhove, {\it D instantons, strings and ${\cal M}$ theory}, Phys. Lett. B 408,
122 (1997) [hep-th/9704145];  M. B. Green, M. Gutperle and P. Vanhove, {\it One loop in eleven-dimensions}, Phys. Lett. B 409, 177 (1997) [hep-th/9706175];
E. Kiritsis and B. Pioline, {\it On R**4 threshold corrections in IIb string theory and (p, q)
string instantons}, Nucl. Phys. B 508, 509 (1997) [hep-th/9707018]; J. G. Russo and A. A. Tseytlin, {\it One loop four graviton amplitude in eleven-dimensional
supergravity}, Nucl. Phys. B 508, 245 (1997) [hep-th/9707134]; I. Antoniadis, S. Ferrara, R. Minasian and K. S. Narain, {\it R**4 couplings in M and type
II theories on Calabi-Yau spaces}, Nucl. Phys. B 507, 571 (1997) [hep-th/9707013]; A. A. Tseytlin, {\it R**4 terms in 11 dimensions and conformal anomaly of (2,0) theory}, Nucl. Phys. B 584, 233 (2000) [hep-th/0005072].
\bibitem{O(R^3G^2)} J. T. Liu and R. Minasian, {\it Higher-derivative couplings in string theory: dualities and the
B-field}, arXiv:1304.3137 [hep-th].
\bibitem{Berlin-torsion-classes} G.~Lopes Cardoso, G.~Curio, G.~Dall'Agata, D.~Lust, P.~Manousselis and G.~Zoupanos,
{\it Non-Kahler string backgrounds and their five torsion classes}, Nucl. Phys. B {\bf 652}, 5-34 (2003) [arXiv:hep-th/0211118 [hep-th]].
\bibitem{SYZ-free-delocalization}M.~Becker, K.~Dasgupta, A.~Knauf and R.~Tatar, {\it Geometric transitions, flops and nonKahler manifolds. I.},  Nucl.\ Phys.\ B {\bf 702}, 207 (2004) [hep-th/0403288].
\bibitem{Dasgupta+Tatar-et-al} M.~Becker, K.~Dasgupta, S.~H.~Katz, A.~Knauf and R.~Tatar,
{\it Geometric transitions, flops and non-Kahler manifolds. II.},
Nucl. Phys. B {\bf 738}, 124-183 (2006) [arXiv:hep-th/0511099 [hep-th]]; F.~Chen, K.~Dasgupta, P.~Franche, S.~Katz and R.~Tatar,
{\it Supersymmetric Configurations, Geometric Transitions and New Non-Kahler Manifolds}, Nucl. Phys. B {\bf 852}, 553-591 (2011) [arXiv:1007.5316 [hep-th]].
\bibitem{theta0-theta} K.~Dasgupta, M.~Grisaru, R.~Gwyn, S.~H.~Katz, A.~Knauf and R.~Tatar, {\it Gauge-Gravity Dualities, Dipoles and New Non-Kahler Manifolds}, Nucl. Phys. B {\bf 755}, 21-78 (2006) [arXiv:hep-th/0605201 [hep-th]].
 \bibitem{Butti et al [2004]}A.~Butti, M.~Grana, R.~Minasian, M.~Petrini and A.~zaffaroni, {\it The baryonic branch of Klebanov-Strassler solution: A supersymmetric family of SU(3) structure backgrounds}, JHEP 0503, 069
(2005) [arXiv:hep-th/0412187].
\bibitem{EPJC-2}K.~Sil and A.~Misra,
  {\it New Insights into Properties of Large-N Holographic Thermal QCD at Finite Gauge Coupling at (the Non-Conformal/Next-to) Leading Order in N}, Eur.\ Phys.\ J.\ C {\bf 76}, no. 11, 618 (2016)
  [arXiv:1606.04949 [hep-th]]; M.~Dhuria and A.~Misra, {\it Transport Coefficients of Black MQGP M3-Branes}, Eur.\ Phys.\ J.\ C {\bf 75}, no. 1, 16 (2015)  [arXiv:1406.6076 [hep-th]].
\bibitem{NPB}K.~Sil and A.~Misra, {\it On Aspects of Holographic Thermal QCD at Finite Coupling},
  Nucl.\ Phys.\ B {\bf 910}, 754 (2016) [arXiv:1507.02692 [hep-th]].
\bibitem{KW} Igor R. Klebanov and Edward Witten, {\it Superconformal Field Theory on Threebranes at a Calabi-Yau Singularity},  Nucl. Phys. B 536, 199 (1998)[arXiv:hep-th/9807080].
\bibitem{KT} I.R. Klebanov and A. Tseytlin, {\it Gravity Duals of Supersymmetric $SU(N)\times SU(M+N)$ Gauge Theories}, [hep-th/0002159].
\bibitem{KS} I.~R.~Klebanov and  M.~J.~Strassler, {\it Supergravity and a Confining Gauge Theory: Duality Cascades and $X$SB-Resolution of Naked Singularities}, JHEP 0008:052,2000 [arXiv:hep-th/0007191].
\bibitem{ouyang} P.~Ouyang, {\it Holomorphic D7-Branes and Flavored N=1 Gauge Theories}, Nucl.Phys.B 699:207-225 (2004), [arXiv:hep-th/0311084].
\bibitem{Buchel} A.~Buchel, {\it Finite temperature resolution of the Klebanov-Tseytlin singularity}, Nucl.\ Phys.\ B {\bf 600}, 219 (2001) [hep-th/0011146].
\bibitem{Gubser-et-al-finitetemp}  S.~S.~Gubser, C.~P.~Herzog, I.~R.~Klebanov and A.~A.~Tseytlin, {\it Restoration of chiral symmetry: A Supergravity perspective},  JHEP {\bf 0105}, 028 (2001) [hep-th/0102172]; A.~Buchel, C.~P.~Herzog, I.~R.~Klebanov, L.~A.~Pando Zayas and A.~A.~Tseytlin, {\it Nonextremal gravity duals for fractional D-3 branes on the conifold},   JHEP {\bf 0104}, 033 (2001)[hep-th/0102105].
\bibitem{Leo-i} B.~A.~Burrington, J.~T.~Liu, L.~A.~Pando Zayas and D.~Vaman, {\it Holographic duals of flavored N=1 super Yang-mills: Beyond the probe approximation}, JHEP {\bf 0502}, 022 (2005)[hep-th/0406207].
\bibitem{Leo-ii} M.~Mahato, L.~A.~Pando Zayas and C.~A.~Terrero-Escalante, {\it Black Holes in Cascading Theories: Confinement/Deconfinement Transition and other Thermal Properties}, JHEP {\bf 0709}, 083 (2007)  [arXiv:0707.2737 [hep-th]].
 \bibitem{M(r)N_f(r)-Dasgupta_et_al} M.~Mia, K.~Dasgupta, C.~Gale and S.~Jeon,
  {\it Toward Large N Thermal QCD from Dual Gravity: The Heavy Quarkonium Potential},''
  Phys.\ Rev.\ D {\bf 82}, 026004 (2010)  [arXiv:1004.0387 [hep-th]].
\bibitem{syz} A.~Strominger, S.~T.~Yau and E.~Zaslow, {\it Mirror symmetry is T duality},  Nucl.\ Phys.\ B {\bf 479}, 243 (1996)  [hep-th/9606040].
\bibitem{M.Ionel and M.Min-OO (2008)}M.~Ionel and M.~Min-OO, {\it Cohomogeneity One Special Lagrangian 3-Folds in the Deformed and the Resolved Conifolds},  Illinois Journal of Mathematics, Vol. 52, Number 3 (2008).
\bibitem{Zayas-Tseytlin}L.~A.~Pando Zayas and A.~A.~Tseytlin, {\it 3-branes on resolved conifold},
  JHEP {\bf 0011}, 028 (2000)  [hep-th/0010088].
\bibitem{Franche-thesis}P.~Franche, {\it Towards New Classes of Flux Compactifications},  arXiv:1303.6726 [hep-th].
\bibitem{Misra+Gale_Conformal_Anomaly} A.~Misra and C.~Gale, {\it The QCD Trace Anomaly at Strong Coupling from ${\cal M}$-Theory}, Eur. Phys. J. C {\bf 80}, no.7, 620 (2020) [arXiv:1909.04062 [hep-th]].
  \bibitem{Sil+Yadav+Misra-glueball} K.~Sil, V.~Yadav and A.~Misra,
 {\it Top-down holographic G-structure glueball spectroscopy at (N)LO in $N$ and finite coupling},
  Eur.\ Phys.\ J.\ C {\bf 77}, no. 6, 381 (2017)
  [arXiv:1703.01306 [hep-th]].
\bibitem{mesons_0E++-to-mesons-decays} V.~Yadav and A.~Misra, {\it ${\cal M}$-Theory Exotic Scalar Glueball Decays to Mesons at Finite Coupling},  JHEP {\bf 1809}, 133 (2018)  [arXiv:1808.01182 [hep-th]].
\bibitem{Green and Gutperle} M.~B.~Green and M.~Gutperle, {\it Effects of D instantons},
Nucl. Phys. B bf{498}, 195-227 (1997) [arXiv:hep-th/9701093 [hep-th]].
\bibitem{Green and Vanhove} M.~B.~Green and P.~Vanhove, {\it D instantons, strings and ${\cal M}$ theory},
Phys. Lett. B bf{408}, 122-134 (1997) [arXiv:hep-th/9704145 [hep-th]].
\bibitem{Horava and Witten} P.~Horava and E.~Witten,
{\it Eleven-dimensional supergravity on a manifold with boundary},
Nucl. Phys. B {\bf 475}, 94-114 (1996)
[arXiv:hep-th/9603142 [hep-th]].
\bibitem{Vafa and Witten} C.~Vafa and E.~Witten, {\it A One loop test of string duality}, Nucl. Phys. B bf{447}, 261-270 (1995) [arXiv:hep-th/9505053 [hep-th]].
\bibitem{Tseytlin} A.~A.~Tseytlin, {\it Heterotic type I superstring duality and low-energy effective actions},
Nucl. Phys. B bf{467}, 383-398 (1996) [arXiv:hep-th/9512081 [hep-th]].
\bibitem{Becker-sisters-O(R^4)}K.~Becker and M.~Becker, {\it Supersymmetry breaking, ${\cal M}$ theory and fluxes}, JHEP bf{07}, 038 (2001)
doi:10.1088/1126-6708/2001/07/038 [arXiv:hep-th/0107044 [hep-th]].
\bibitem{Tseytlin-epsilonD^2R^4-kroneckerdeltaR^4}
A.~A.~Tseytlin, {\it R**4 terms in 11 dimensions and conformal anomaly of (2,0) theory},
Nucl.\ Phys.\ B bf{584}, 233-250 (2000)
doi:10.1016/S0550-3213(00)00380-1
[arXiv:hep-th/0005072 [hep-th]].
\bibitem{Knauf-thesis} A.~Knauf, {\it Geometric Transitions on non-Kaehler Manifolds}, Fortsch. Phys. {\bf 55}, 5-107 (2007)
[arXiv:hep-th/0605283 [hep-th]].
\bibitem{Bulk-Viscosity}A.~Czajka, K.~Dasgupta, C.~Gale, S.~Jeon, A.~Misra, M.~Richard and K.~Sil, {\it Bulk Viscosity at Extreme Limits: From Kinetic Theory to Strings},  JHEP {\bf 1907}, 145 (2019)
  [arXiv:1807.04713 [hep-th]].
\bibitem{VA-Glueball-decay}V.~Yadav and A.~Misra, {\it M-Theory Exotic Scalar Glueball Decays to Mesons at Finite Coupling}, JHEP {\bf 09}, 133 (2018) [arXiv:1808.01182 [hep-th]].
\bibitem{Kruczenski et al-2003} M.~Kruczenski, D.~Mateos, R.~C.~Myers and D.~J.~Winters,
{\it Towards a holographic dual of large N(c) QCD}, JHEP {\bf 05}, 041 (2004) [arXiv:hep-th/0311270].
\bibitem{DM-transport-2014}M.~Dhuria and A.~Misra, {\it Transport Coefficients of Black MQGP M3-Branes},
Eur. Phys. J. C {\bf 75}, 16 (2015) [arXiv:1406.6076 [hep-th]].
\bibitem{Armoni et al-2020} R.~Argurio, A.~Armoni, M.~Bertolini, F.~Mignosa and P.~Niro,
{\it Vacuum structure of large $N$ $QCD_{3}$ from holography}, JHEP {\bf 07}, 134 (2020) [arXiv:hep-th/2006.01755].
\bibitem{Herzog_T_c}C.~P.~Herzog, {\it A Holographic Prediction of the Deconfinement Temperature}, Phys. Rev. Lett. \textbf{98}, 091601 (2007) [arXiv:hep-th/0608151 [hep-th]].
\bibitem{GLF} J.~Gasser and H.~Leutwyler, {\it Chiral perturbation theory to one loop}, Annals of Physics  158, 142 (1984).
\bibitem{Yadav+Misra+Sil-Mesons}V.~Yadav, A.~Misra and K.~Sil, {\it Delocalized SYZ Mirrors and Confronting Top-Down $SU(3)$-Structure Holographic Meson Masses at Finite $g$ and $N_c$ with P(article) D(ata) G(roup) Values}, Eur. Phys. J. C bf{77}, no.10, 656 (2017) [arXiv:1707.02818 [hep-th]].
\bibitem{HARADA}  M.~Harada, S.~Matsuzaki and K.~Yamawaki, {\it Holographic QCD Integrated back to Hidden Local Symmetry}, Phys.Rev.D82:076010,2010 [arXiv:hep-th/1007.4715].
\bibitem{Witten-Hawking-Page-Tc} E.~Witten, {\it Anti-de Sitter space, thermal phase transition, and confinement in gauge theories}, Adv. Theor. Math. Phys. {\bf 2}, 505 (1998) [arXiv:hep-th/9803131].
\bibitem{Glueball-Roorkee} K.~Sil, V.~Yadav and A.~Misra, {\it Top-down holographic G-structure glueball spectroscopy at (N)LO in $N$ and finite coupling}, Eur. Phys. J. C {\bf 77}, no.6, 381 (2017) [arXiv:hep-th/1703.01306].
\bibitem{RT} S.~Ryu, T.~Takayanagi, {\it Holographic Derivation of Entanglement Entropy from AdS/CFT}, Phys. Rev. Lett. {\bf 96}, 181602 (2006) [arXiv:hep-th/0603001].
\bibitem{Magdalena-2013}M.~Larfors, {\it Revisiting toric SU(3) structures}, Fortsch. Phys. \textbf{61}, 1031-1055 (2013) [arXiv:1309.2953 [hep-th]].
\bibitem{GKP} S.~B.~Giddings, S.~Kachru and J.~Polchinski, {\it Hierarchies from fluxes in string compactifications}, Phys. Rev. D \textbf{66}, 106006 (2002) [arXiv:hep-th/0105097 [hep-th]].
\bibitem{Camara+Grana} P.~G.~Camara and M.~Grana, No-scale supersymmetry breaking vacua and soft terms with torsion, JHEP 0802 (2008) 017, [arXiv:0710.4577].
\bibitem{Luest-et-al_2008}D.~Lust, F.~Marchesano, L.~Martucci, and D.~Tsimpis, {\it Generalized non-supersymmetric flux vacua}, JHEP 0811 (2008) 021, [arXiv:0807.4540]
\bibitem{Luest-et-al_2010} J.~Held, D.~Lust, F.~Marchesano, and L.~Martucci,  {\it DWSB in heterotic flux compactifications}, JHEP 1006 (2010) 090, [arXiv:1004.0867].
\bibitem{thesis-Held} J.~G.~J.~Held, {\it Non-supersymmetric flux compactifications of heterotic string- and M-theory}, Ph.D. thesis 2012, Max-Planck-Institut f\"{u}r Physik and the Ludwig–
Maximilians–Universit, Munich.
\bibitem{Bedulli+Vezzoni} L.~Bedulli and L.~Vezzoni, {\it The Ricci tensor of SU(3)-manifolds}, Journal of Geometry and Physics 57 (Mar., 2007) 1125–1146, [math/0606786].
\bibitem{Bryant-G2} R. L. Bryant, {\it Some remarks on G(2)-structures}, arXiv:math/0305124 [math-dg].
\bibitem{Spin7-Ivanov} S.~Ivanov, {\it Connection with torsion, parallel spinors and geometry of spin(7) manifolds}, [arXiv:math/0111216 [math.DG]].
\bibitem{Chiossi-Salamon} S.~Chiossi and S.~Salamon, {\it The Intrinsic torsion of SU(3) and G(2) structures}, Contribution to: International Conference on Differential Geometry held in honor of the 60th Birthday of A.M. Naveira [arXiv:math/0202282 [math.DG]].
 \bibitem{Umemura}H.~Umemura, {\it Resolution of algebraic equations by theta constants}, in D. Mumford, Tata Lectures
on Theta II, Progress in Math. Vol. 43, Birkh\"{a}user, 1984, 261-272.
\bibitem{Zhivkov}A.~Zhivkov, {\it Resolution of degree$\leq$6 algebraic
equations by genus two theta constants}, Journal of Geometry and Symmetry in Physics, {\bf 11}, 77 (2008).
\bibitem{Bruce-King-Beyond-Quartic} {\it Beyond the Quartic Equation}, R. B. King, Birkh\"{a}user (1996).
\bibitem{Keshav_et_al_SUSY_type-II_torsion} F.~Chen, K.~Dasgupta, P.~Franche and R.~Tatar, {\it Toward the Gravity Dual of Heterotic Small Instantons},
Phys.\ Rev.\ D bf{83}, 046006 (2011) [arXiv:1010.5509 [hep-th]].
\bibitem{Phi-G_2}P.~Kaste, R.~Minasian, M.~Petrini and A.~Tomasiello,
{\it Kaluza-Klein bundles and manifolds of exceptional holonomy},
  JHEP {\bf 0209}, 033 (2002)
  [hep-th/0206213].
\bibitem{Gauntlett_et_al} J.~P.~Gauntlett, D.~Martelli and D.~Waldram, {\it Superstrings with intrinsic torsion},
  Phys.\ Rev.\ D {\bf 69}, 086002 (2004) [hep-th/0302158]; D.~Prins and D.~Tsimpis, {\it Type IIA supergravity and M -theory on manifolds with SU(4) structure},
  Phys.\ Rev.\ D {\bf 89}, 064030 (2014)   [arXiv:1312.1692 [hep-th]].
\bibitem{Puhle-Spin(7)}C.~Puhle, {\it Spin(7) Manifolds with Parallel Torsion Form}, arXiv:0807.4875[math.DG] .
\bibitem{Moore+Saremi} G.~D.~Moore and O.~Saremi, {\it Bulk viscosity and spectral functions in QCD}, JHEP \textbf{09}, 015 (2008) [arXiv:0805.4201 [hep-ph]].
\bibitem{ACMS-1} M.~F.~Arikan, H.~Cho and S.~Salur, {\it Contact Structures on G2-Manifolds and Spin 7-Manifolds}, arXiv:1207.2046; {\it Existence of Compatible Contact Structures on G2-manifolds}, Asian J. Math. 17 (2013) 321 [1112.2951]; A.~J.~Todd, {\it An Almost Contact Structure on G2-Manifolds}, 1501.06966.
\bibitem{ACMS-SU2} T.~Friedrich, I.~Kath, A.~Moroianu and U.~Semmelmann, {\it On nearly parallel G2-structures}, Journal of Geometry and Physics 23 (1997) 259.
\bibitem{Cvetic_et_al_G2_SU3_SU2} K.~Behrndt, M.~Cvetic and T.~Liu, {\it Classification of supersymmetric flux vacua in M theory}, Nucl. Phys. B \textbf{749}, 25-68 (2006) [arXiv:hep-th/0512032 [hep-th]].
\bibitem{ACMS-2} X. ~de la Ossa, M.~Larfors, M.~Magill {\it Almost contact structures on manifolds with a G2 structure}, arXiv: 2101.12605.
\bibitem{AC3MS+SU3} A.~Misra, G.~Yadav, To appear.
\bibitem{J. G. J. Held's thesis [2012]} J.~Held, Ph.D. thesis, {\it Non-Supersymmetric Flux Compactications of Heterotic String- and M-theory} , 2012.
\bibitem{Bulk-Viscosity-McGill-IIT-Roorkee}A.~Czajka, K.~Dasgupta, C.~Gale, S.~Jeon, A.~Misra, M.~Richard and K.~Sil,
{\it Bulk Viscosity at Extreme Limits: From Kinetic Theory to Strings}, JHEP bf{07}, 145 (2019)[arXiv:1807.04713 [hep-th]].
\end{thebibliography}
\end{document}